\documentclass[aps, pra, superscriptaddress, 10pt, twocolumn, longbibliography, floatfix, notitlepage, nofootinbib]{revtex4-1}
\usepackage{silence}
\usepackage[nointlimits]{amsmath}
\usepackage{amssymb, cancel}
\usepackage{exscale}
\usepackage{graphicx, nicefrac}
\usepackage{color}
\usepackage{stackengine}
\usepackage{txfonts}
\usepackage{algorithm}
\usepackage{mathtools}
\usepackage{hyperref}
\hypersetup{colorlinks=true, citecolor=magenta, linkcolor=magenta, urlcolor=magenta}
\usepackage{footnote}
\usepackage{subfigure}
\usepackage{color}
\usepackage{blkarray}
\usepackage{dsfont}
\usepackage{mathrsfs}
\usepackage{bbold}
\usepackage{bbm}
\usepackage{overpic}

\makeatletter
\DeclareFontFamily{OMX}{MnSymbolE}{}
\DeclareSymbolFont{MnLargeSymbols}{OMX}{MnSymbolE}{m}{n}
\SetSymbolFont{MnLargeSymbols}{bold}{OMX}{MnSymbolE}{b}{n}
\DeclareFontShape{OMX}{MnSymbolE}{m}{n}{
    <-6>  MnSymbolE5
   <6-7>  MnSymbolE6
   <7-8>  MnSymbolE7
   <8-9>  MnSymbolE8
   <9-10> MnSymbolE9
  <10-12> MnSymbolE10
  <12->   MnSymbolE12
}{}
\DeclareFontShape{OMX}{MnSymbolE}{b}{n}{
    <-6>  MnSymbolE-Bold5
   <6-7>  MnSymbolE-Bold6
   <7-8>  MnSymbolE-Bold7
   <8-9>  MnSymbolE-Bold8
   <9-10> MnSymbolE-Bold9
  <10-12> MnSymbolE-Bold10
  <12->   MnSymbolE-Bold12
}{}

\let\llangle\@undefined
\let\rrangle\@undefined
\DeclareMathDelimiter{\llangle}{\mathopen}%
                     {MnLargeSymbols}{'164}{MnLargeSymbols}{'164}
\DeclareMathDelimiter{\rrangle}{\mathclose}%
                     {MnLargeSymbols}{'171}{MnLargeSymbols}{'171}
\makeatother

\graphicspath{{./figures/}}

\def\sbra#1{{\left\llangle #1 \right|}}
\def\sket#1{{\left| #1 \right\rrangle}}

\usepackage{amsthm}

\newcommand{\tocless}[1]{%
\let\oldaddcontentsline\addcontentsline
\renewcommand{\addcontentsline}[3]{}
#1%
\let\addcontentsline\oldaddcontentsline
}

\newcommand{\tr}{\operatorname{tr}}
\newcommand{\Tr}{\operatorname{Tr}}

\usepackage{microtype} 
\usepackage[T1]{fontenc} 
\allowdisplaybreaks

\begin{document}

\affiliation{Quantum Performance Laboratory, Sandia National Laboratories, Albuquerque, NM 87185}
\affiliation{Quantum Performance Laboratory, Sandia National Laboratories, Livermore, CA 94550}
\affiliation{HRL Laboratories, LLC, Malibu, CA 90265}

\title{Randomized Benchmarking with Synthetic Quantum Circuits}
\author{Yale Fan}
\email{yalefan@gmail.com}
\affiliation{Quantum Performance Laboratory, Sandia National Laboratories, Albuquerque, NM 87185}
\affiliation{Department of Physics, University of Idaho, Moscow, ID 83844}
\author{Riley Murray}
\affiliation{Quantum Performance Laboratory, Sandia National Laboratories, Livermore, CA 94550}
\author{Thaddeus D. Ladd}
\affiliation{HRL Laboratories, LLC, Malibu, CA 90265}
\author{Kevin Young}
\affiliation{Quantum Performance Laboratory, Sandia National Laboratories, Livermore, CA 94550}
\author{Robin Blume-Kohout}
\affiliation{Quantum Performance Laboratory, Sandia National Laboratories, Albuquerque, NM 87185}

\begin{abstract}
Noise characterization methods such as randomized benchmarking (RB) are critical for the development of scalable quantum computers. Modern RB protocols for multiqubit systems extract physically relevant error rates by exploiting the structure of the group representation generated by the set of benchmarked operations. However, existing techniques become prohibitively inefficient for representations that are highly reducible yet decompose into irreducible subspaces of high dimension. These situations prevail when benchmarking high-dimensional systems such as qudits or bosonic modes, where experimental control is limited to implementing a small subset of all possible unitary operations. We introduce a broad framework for enhancing the sample efficiency of RB that is sufficiently powerful to extend the practical reach of RB beyond the multiqubit setting. Our strategy, which applies to any benchmarking group, uses ``synthetic'' quantum circuits with classical post-processing of both input and output data to leverage the full structure of reducible superoperator representations. To demonstrate the efficacy of our approach, we develop a detailed theory of RB for systems with rotational symmetry. Such systems carry a natural action of the group $\text{SU}(2)$, and they form the basis for several novel quantum error-correcting codes. We show that, for experimentally accessible high-spin systems, synthetic RB protocols can reduce the complexity of measuring rotationally invariant error rates by two orders of magnitude relative to standard approaches such as character RB.
\end{abstract}

\maketitle

\tocless{\section{Introduction}}

The bedrock of fault-tolerant quantum computation is the threshold theorem, which states that sufficiently low physical error rates can be traded for arbitrarily low logical error rates in computations on encoded quantum information.  The importance of measuring physical error rates---both for comparison to fault tolerance thresholds and for optimizing hardware performance---has catalyzed the development of methods to characterize noise in quantum gates.  Among such methods, randomized benchmarking (RB) \cite{RB-Emerson, RB-Dankert} enjoys widespread use for its simplicity and scalability.

In its most common formulation, RB employs gates sampled uniformly at random from the Clifford group \cite{CRB-Emerson, CRB-Knill, CRB-Magesan-1, CRB-Magesan-2} to extract measures of quality for multiqubit quantum processors in a way that is both scalable and robust to errors in state preparation and measurement (SPAM).  The key principle of standard RB is that the group of gates to be benchmarked should form a unitary 2-design via its representation on the Hilbert space of the system.  This property ensures that the representation of the gate group on the space of operators (the ``superoperator representation'') is maximally irreducible, facilitating the extraction of a single average error rate.

On the other hand, benchmarking groups beyond 2-designs can yield more fine-grained information than a single figure of merit while also allowing for more accurate assessment of a processor's native gates.  Clifford RB can be generalized to other finite and compact benchmarking groups via character RB \cite{Helsen, Claes} and the framework of filtered RB \cite{FilteredRB-1, Kong}, as well as to gate sets without a group structure \cite{Chen, FilteredRB-2}.  The reliability and interpretability of an RB protocol depend sensitively on the representation theory of the group generated by the chosen gate set.  When this group acts reducibly on the state space, RB experiments may reveal multiple error rates corresponding to distinct classes of errors, which can be measured individually using filtered RB methods.

At the same time, the road to fault-tolerant quantum computation extends well beyond the qubit paradigm, traversing qudits \cite{Sanders}, fermions \cite{Bravyi-Kitaev}, continuous variables \cite{Braunstein}, and subsystem encodings \cite{Andrews}.  The proliferation of quantum computing platforms calls for benchmarking methods tailored to the native operations of each \cite{Jafarzadeh, BosonicRB-2019, BosonicRB-2020, BosonicRB-2024-1, BosonicRB-2024-2, BosonicRB-2024-3}.

In this broader setting, traditional RB approaches face substantial challenges.  The efficiency of RB is limited by the size of the dimensions of the irreducible representations (irreps) that occur in the superoperator representation of the gate group.  For sufficiently ``large'' benchmarking groups such as $\text{SU}(d)$ for a $d$-level system or the $n$-qubit Clifford group, non-filtered RB protocols work well because the superoperator representation contains a single nontrivial irrep, eliminating the need to distinguish signals from different irreps.  Conversely, for sufficiently ``small'' benchmarking groups such as the $n$-qubit Pauli group (whose superoperator representation is abelian), filtered RB protocols work well because the superoperator representation splits into many irreps of small dimension, making the corresponding signals easy to distinguish.  However, in the ``intermediate'' regime where the superoperator representation decomposes into many irreps of large dimension, existing filtered RB methods such as character RB become infeasible.  This problem is especially stark for benchmarking groups that are both infinite and nonabelian, which have infinitely many candidate irreps of potentially unbounded dimension.  Unfortunately, these ``problematic'' groups are precisely the most physically interesting ones for benchmarking operations on high-dimensional quantum systems.

To solve this problem, we introduce a family of \emph{synthetic randomized benchmarking} schemes that isolate irrep-specific RB signals via highly targeted \emph{synthetic quantum circuits}.  Each synthetic circuit incorporates some combination of \emph{synthetic gates} and \emph{synthetic SPAM}, where each synthetic operation is a linear combination of physical operations that is realized in classical post-processing.  The core of our approach consists of two new classes of powerful and extensible RB primitives, synthetic-SPAM RB and rank-1 RB, that generalize aspects of both filtered RB and character RB.  The full power of these primitives emerges when they are combined: in conjunction, they preserve the characteristic SPAM-robustness of RB while drastically increasing its efficiency.  Our new synthetic RB protocols apply to general groups, including finite benchmarking groups beyond the Pauli and Clifford groups.  Furthermore, our analysis offers new insights into the conditions under which character RB is feasible even for qubits, although the principles apply across platforms.

Our framework has broad applications to benchmarking continuous gate groups for quantum information processors with continuous symmetry.  We illustrate our framework by applying it to the benchmarking group $\text{SU}(2)$, the group of natural operations on a rotationally invariant system.  This rotational symmetry can be either naturally occurring or engineered, as manifested in platforms ranging from spin qudits \cite{Yu} to superconducting circuits \cite{Roy, Champion} to atomic ensembles \cite{Anderson}.  We develop a suite of protocols to measure rotationally invariant error rates and investigate their performance in detail.  In this setting, synthetic RB is provably and dramatically superior in sample complexity to character RB in the zero-noise limit.  We further discuss the physical interpretation of these rotationally invariant error rates as they pertain to the implementation of novel single-spin quantum error-correcting codes \cite{Gross, Omanakuttan-1, Omanakuttan-2} and fault-tolerant logical operations therein.  A final motivation for our focus on $\text{SU}(2)$ is fundamental in nature: $\text{SU}(2)$ not only underlies the theory of angular momentum, but is additionally the prototypical nonabelian compact Lie group whose representation theory underlies that of all semisimple Lie groups.  As such, it provides a starting point for the analysis of arbitrary continuous gate groups.

After briefly reviewing the theory of traditional RB (Section \ref{sec:background}), we explain in generality how synthetic RB both extends and improves upon existing methods (Section \ref{sec:synthetic}).  We then specialize to the rotationally invariant context (Section \ref{sec:RIRB}), where we examine the performance of synthetic RB both analytically (Section \ref{sec:samplecomplexity}) and numerically (Section \ref{sec:numerics}).  We aim for a high-level overview throughout, leaving details and derivations to the appendices.

\tocless{\section{Background}} \label{sec:background}

To establish key concepts and notation, we first review existing RB protocols (see Appendices \ref{app:preliminaries} and \ref{app:review} for more details).  In what follows, given a finite-dimensional Hilbert space $\mathscr{H}$, let $\sket{\rho}$ denote the vector representation of the linear operator $\rho$ in a chosen basis for the Hilbert-Schmidt space $\mathcal{B}(\mathscr{H})$.  A superoperator is a linear operator on $\mathcal{B}(\mathscr{H})$.

We restrict our attention to the case where the native gates to be benchmarked comprise a group $G$, the ``benchmarking group.''  We regard $G$ as a subgroup of the unitary group $\text{U}(\mathscr{H})$ and therefore do not distinguish between group elements $g\in G$ and their unitary representations on $\mathscr{H}$.  For simplicity of presentation, we assume a gate-independent and Markovian error model throughout, although both assumptions can be lifted \cite{Proctor, Wallman, Merkel}.  In particular, we model gates as quantum channels and perfect gates as unitary channels.  For a given $g\in G$, we denote its superoperator (or process matrix) representation by $\mathcal{G} : \rho\mapsto g\rho g^\dag$ and its noisy implementation by $\widetilde{\mathcal{G}} = \mathcal{E}\mathcal{G}$ for some gate-independent noise channel $\mathcal{E}$.

We write $\mathbb{E}_{g\in G}$ for a group average, which should be interpreted as a sum $|G|^{-1} \linebreak[1] \sum_{g\in G}$ for finite $G$ and as a Haar integral $|G|^{-1} \linebreak[1] \int_G \linebreak[1] dg$ for compact $G$.

\tocless{\subsection{Randomized Benchmarking}}

In its simplest form, RB aims to estimate a single average gate error rate.  The idea is to run many circuits of $m$ uniformly randomly sampled gates $\vec{g}\equiv (g_1, \ldots, g_m)\in G^m$ followed by a global inversion gate $g_\text{inv} = (g_m\cdots g_1)^\dag$ to estimate the average survival probability
\begin{equation}
p_m = \mathbb{E}_{\vec{g}\in G^m}\big[\sbra{E}\widetilde{\mathcal{G}}_\text{inv}\widetilde{\mathcal{G}}_m\cdots \widetilde{\mathcal{G}}_1\sket{\rho}\big],
\end{equation}
where $\rho$ is a fixed initial state and $E$ is a fixed final measurement effect.  With perfect gates, $p_m$ would be the constant $\llangle E|\rho\rrangle$.  In practice, $p_m$ decays exponentially with $m$.  The decay rate is interpreted as an average gate error rate.

More precisely, let us assume for simplicity that the superoperator representation $\mathcal{G}$ of $G$ is multiplicity-free.  Then it decomposes as
\begin{equation}
\mathcal{G} = \bigoplus_\lambda \phi_\lambda,
\label{decomposition}
\end{equation}
where the $\phi_\lambda$ denote inequivalent irreps indexed by $\lambda$.  Under our assumptions, the corresponding average survival probability can be shown to take the form
\begin{equation}
p_m = \sum_\lambda A_\lambda f_\lambda^m.
\label{pmsum}
\end{equation}
The coefficients $A_\lambda$ measure the overlap of the SPAM operations $|\rho\rrangle$ and $\llangle E|$ within irrep $\lambda$.  The quality parameters $f_\lambda$ measure the overlap between the gate error channel $\mathcal{E}$ and the projector onto the support of irrep $\lambda$.  The trivial representation, corresponding to $f_\lambda = 1$, always occurs in \eqref{decomposition} and shows up in \eqref{pmsum} because $\mathcal{E}$ is trace-preserving.  Thus we see that RB is not SPAM-invariant for general $G$: it will produce a linear combination of exponential decays with different rates that is difficult to fit numerically, and the precise linear combination will depend strongly on the chosen SPAM operations.  The exceptions (those $G$ with only one nontrivial $\phi_\lambda)$ are unitary 2-designs such as the Clifford group on $n$ qubits, which comprise the gate set in the simplest RB protocols.

The basic principle at play is that RB is only sensitive to the $G$-twirled error channel
\begin{equation}
\mathbb{E}_{g\in G}\big[\mathcal{G}^\dag\mathcal{E}\mathcal{G}\big] = \sum_\lambda f_\lambda\Pi_\lambda,
\label{Gtwirl}
\end{equation}
where $\Pi_\lambda$ is the projector onto the support of the irrep $\phi_\lambda$.  An extreme case of the expression \eqref{Gtwirl} is that twirling a channel over a unitary 2-design (e.g., the entire unitary group) converts it to a depolarizing channel $\rho\mapsto (1 - \gamma)\rho + \frac{\gamma}{d}I$.

We draw the following lesson: the more reducible the su\-per\-op\-er\-a\-tor representation of the benchmarking group, the more information that the RB procedure outputs, but the harder it is to extract and analyze the individual decay rates.

\tocless{\subsection{Character Randomized Benchmarking}}

The goal of character RB is to isolate the individual quality parameters associated with each irrep \cite{Helsen, Claes}.  Character RB is designed to implement the standard formula
\begin{equation}
(\dim\phi_{\lambda'})\mathbb{E}_{h\in H}\big[\chi_{\lambda'}(h)^\ast\mathcal{H}\big] = \Pi_{\lambda'},
\label{standardformula}
\end{equation}
where $\chi_{\lambda'}$ is the character of the irrep $\phi_{\lambda'}$.  The idea is to choose a ``character group''\footnote{Not to be confused with mathematicians' use of this term.} $H\subset G$ whose superoperator representation $\mathcal{H}$ contains an irrep $\phi_{\lambda'}$ with support inside the desired irrep $\phi_\lambda$ of $G$.  For instance, one can always choose $H = G$ and $\lambda' = \lambda$.  One then estimates the average \emph{weighted} survival probability
\begin{equation}
p_m^{\lambda'} = (\dim\phi_{\lambda'})\mathbb{E}_{\vec{g}\in G^m, h\in H}\big[\chi_{\lambda'}(h)^\ast\sbra{E}\widetilde{\mathcal{G}}_\text{inv}\widetilde{\mathcal{G}}_m\cdots \widetilde{\mathcal{G}}_2\widetilde{\mathcal{G}_1\mathcal{H}}\sket{\rho}\big],
\label{chiRBweighted}
\end{equation}
which is proportional to $f_\lambda^m$.  Consequently, we can extract $f_\lambda$ by fitting $p_m^{\lambda'}$ to a single exponential decay.  Note that $h$ and $g_1$ are compiled into a single gate in \eqref{chiRBweighted}.

The above procedure can alternatively be viewed as sampling uniformly at random from the set of all gate sequences of length $m + 1$ that compile to an element of the character group $H$.  Sampling over this distribution of sequences in $G^{m+1}$ eliminates the need for an inversion gate.  This alternative strategy is particularly easy to implement when using the benchmarking group as its own character group (in which case we sample each gate uniformly at random from $G$ with no restrictions).

\tocless{\subsection{Filtered Randomized Benchmarking}}

The latter point of view on character RB brings it into the framework of filtered RB \cite{FilteredRB-1, FilteredRB-2}.  Filtered RB does not employ an end-of-circuit inversion gate, but instead uses classical post-processing of output data streams to isolate the signal from a fixed irrep.  Let the ``ending gate'' $g_\text{end}$ denote the net operation performed in a given RB sequence (this would be the identity in standard RB but chosen at random from the character group in character RB).  Filtered RB subsumes all RB protocols that employ the following post-processing technique: estimate the outcome probabilities $p(i, g_\text{end}, m)$ for all measurement effects $i$ and different choices of $g_\text{end}\in G$, and then correlate the resulting vector of signals with a dual vector of values $f_\lambda(i, g_\text{end})$, where $f_\lambda$ is a scalar function (the \emph{filter function}) that depends on the target irrep $\lambda$ of $G$.  In other words, filtered RB estimates
\begin{equation}
p_m^\lambda = \sum_i \mathbb{E}_{g_\text{end} | g_{m+1}\cdots g_1 = g_\text{end}}\big[f_\lambda(i, g_\text{end})\llangle E_i|\widetilde{\mathcal{G}}_{m+1}\cdots \widetilde{\mathcal{G}}_1|\rho\rrangle\big]
\label{filteredRB}
\end{equation}
for a fixed set of measurement effects indexed by $i$ and a fixed set of ending gates $g_\text{end}$ by running many circuits composed of $m + 1$ random gates $g_1, \ldots, g_{m+1}$ that satisfy $g_{m+1} \linebreak[1] \cdots \linebreak[1] g_1 = g_\text{end}$.  The filter function is chosen so that $p_m^\lambda\propto f_\lambda^m$.

Character RB can be described in the language of filtered RB as follows.  For a fixed sequence length $m$, we prepare the state $\rho$, apply gates $g_1, \ldots, g_{m+1}$ drawn uniformly at random from the set of all sequences in $G^{m+1}$ that satisfy $h\equiv g_{m+1}\cdots g_1\in H$, and then measure the POVM $\{E, \mathds{1} - E\}$.  A single run produces the output $(i, g_1, \ldots, g_{m+1})$ where $i\in \{0, 1\}$.  We define the filter function
\begin{equation}
f_{\lambda'}(i, g_1, \ldots, g_{m+1}) = (\dim\phi_{\lambda'})\chi_{\lambda'}(h)^\ast,
\end{equation}
which depends on a chosen irrep $\lambda'$ of $H$ and is independent of $i$.  We then average the values of this filter function over shots with weight $i$.

For completeness, we note that an RB protocol is commonly called \emph{uniform} or \emph{non-uniform} depending on whether it draws gates uniformly or non-uniformly at random from a compact benchmarking group.  Non-uniform RB includes the case where gates are drawn at random from a gate set that does not form a group.  The existing theory of uniform RB with finite \cite{FilteredRB-1} and compact \cite{Kong} groups provides guarantees on the multi-ex\-po\-nen\-tial form of the RB signal in the presence of sufficiently small gate noise.\footnote{If the irreps in the superoperator (or reference) representation of the benchmarking group appear with nontrivial multiplicities, then the RB signal includes matrix exponential decays with potentially \emph{complex} exponentials arising from complex eigenvalue pairs.}  A corresponding theory has been developed for non-uniform RB using random circuits that form approximate unitary 2-designs \cite{Chen} and, more generally, using the language of filtered RB \cite{FilteredRB-2}. (Common methods for benchmarking native gate sets directly \cite{DRB-Proctor, DRB-Polloreno}, bypassing possible compilation into Clifford gates, still require that the native gates generate a group that forms a unitary 2-design.)

\tocless{\section{Synthetic Randomized Benchmarking}} \label{sec:synthetic}

We propose a family of ``synthetic RB'' schemes that all accomplish the goal of disambiguating the irrep-specific decay rates in the RB signal.  Our starting point is the simple observation that a circuit encompasses not only a sequence of gates, but also state preparation and measurement.  We use the adjective ``synthetic'' to refer to an operation (gate or SPAM) that is a linear combination of physical operations.  This is a mathematical abstraction.  To construct synthetic operations, we use the linearity of quantum mechanics to interpret linear combinations of (estimated) probabilities as individual probabilities corresponding to unphysical linear combinations of states, measurement effects, or gates (quantum processes).  Concretely, given collections of physical states $\{\rho_i\}$, measurement effects $\{E_j\}$, and superoperators $\{\mathcal{S}_k\}$, we write
\begin{equation}
\llangle E'|\mathcal{S}'|\rho'\rrangle = \sum_{i, j, k} c_i^\rho c_j^E c_k^{\mathcal{S}}\llangle E_j|\mathcal{S}_k|\rho_i\rrangle,
\label{synthesis}
\end{equation}
where we have defined the synthetic operations $\rho' = \sum_i c_i^\rho\rho_i$, $E' = \sum_j c_j^E E_j$, and $\mathcal{S}' = \sum_k c_k^{\mathcal{S}}\mathcal{S}_k$.  We have written discrete sums for illustration, but these could be continuous integrals.  Importantly, there are no \emph{a priori} restrictions on the signs or magnitudes of the coefficients.  Therefore, the synthetic operations $\rho', E', \mathcal{S}'$ need not satisfy the normalization and positivity conditions of physical states, measurements, or channels; only the individual probabilities $\llangle E_j|\mathcal{S}_k|\rho_i\rrangle$ on the right side of \eqref{synthesis} are associated with experimental outcomes.  All of our protocols use \emph{synthetic circuits} that incorporate some combination of \emph{synthetic gates} and \emph{synthetic SPAM}.\footnote{A note on terminology: as in most of the RB literature, we use the term ``gate'' to refer to a circuit layer (a unitary operation on the entire Hilbert space) rather than an elementary quantum gate on a quantum register.  However, we \emph{also} use the term ``gate'' more flexibly to refer to the entire sequence of gates that comprises the part of a quantum circuit other than state preparation and measurement---i.e., what most other literature would call a ``circuit.''  We hope the meaning is clear from context.}

\begin{figure}[!htb]
\centering
\includegraphics[scale=0.33]{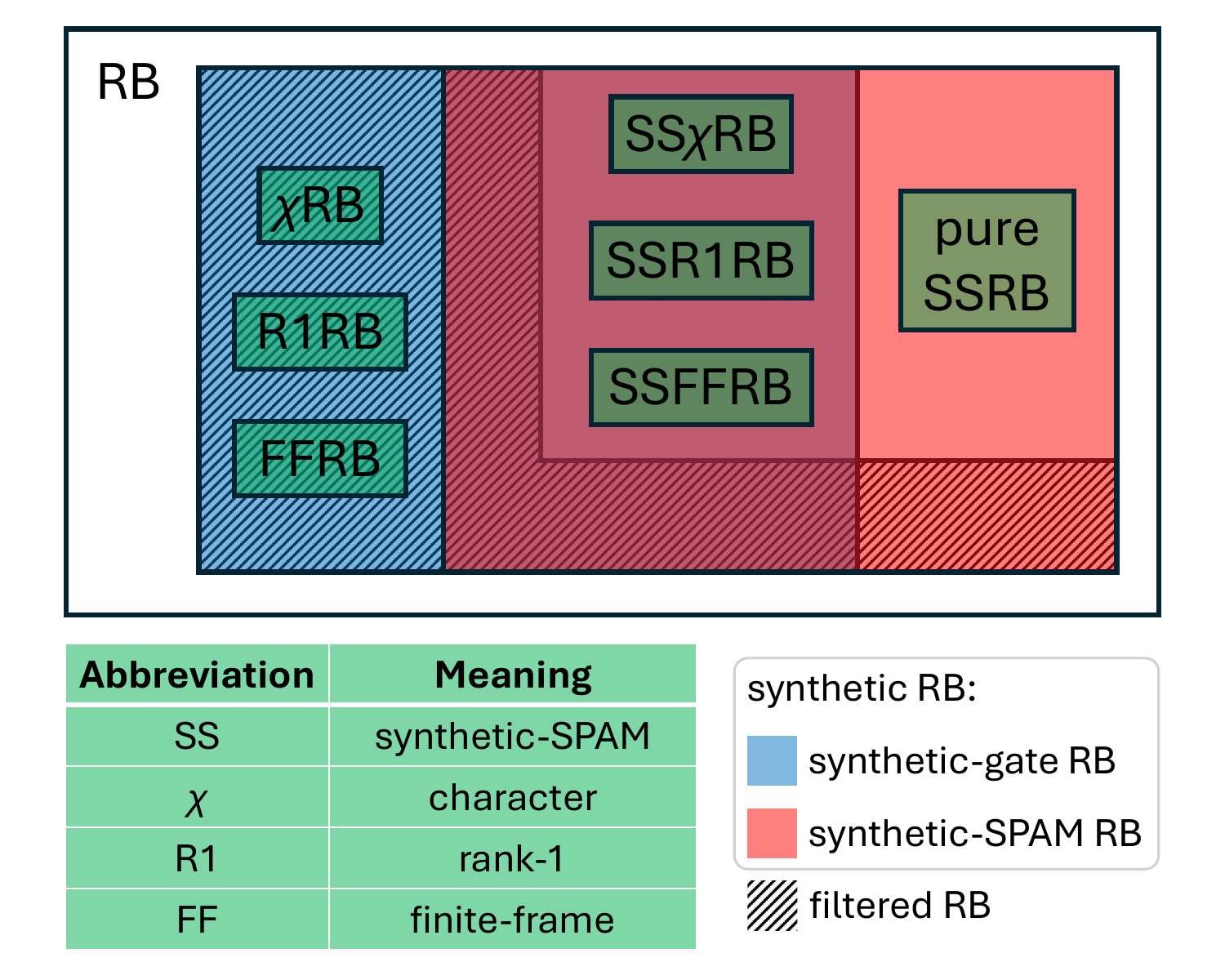}
\caption{A taxonomy of several RB protocols discussed here and elsewhere (abbreviations are defined in the main text).  By synthetic-SPAM RB or synthetic-gate RB, we mean any protocol that incorporates synthetic SPAM or gates rather than a protocol that uses them exclusively.  Filtered RB uses both synthetic gates and measurements but only physical states.  By finite-frame RB, we mean any protocol that uses a non-group-based finite frame.}
\label{taxonomy}
\end{figure}

Our extremely simple framework unifies and extends existing constructions in various ways.  Character RB is a special case of synthetic RB in which the synthetic circuits involve only synthetic gates and those synthetic gates are constructed according to the formula \eqref{standardformula}.  More broadly, synthetic RB is a generalization of filtered RB (which itself encompasses many existing RB protocols) in which we process data streams over inputs as well as outputs to isolate a target signal.  That is, we allow the filter function to depend on \emph{both} the initial state and the final measurement effect, thereby estimating the quantity
\begin{equation}
p_m^\lambda = \sum_{i, j} \mathbb{E}_{g_\text{end} | g_{m+1}\cdots g_1 = g_\text{end}}\big[f_\lambda(i, j, g_\text{end})\llangle E_j|\widetilde{\mathcal{G}}_{m+1}\cdots \widetilde{\mathcal{G}}_1|\rho_i\rrangle\big]
\end{equation}
---which, like \eqref{filteredRB}, is constructed to be proportional to $f_\lambda^m$.  For sufficiently good SPAM, this simple but significant augmentation to \eqref{filteredRB} can be used to great effect to enhance the RB signal.  See Figure \ref{taxonomy} for a contextualization of synthetic RB protocols in the form of a Venn diagram.

We begin by describing the conceptual ingredients of synthetic RB.  We then discuss its quantitative advantages over existing protocols.

\tocless{\subsection{Synthetic SPAM}}

Our end goal is to measure gate error rates as effectively as possible.  Existing RB protocols use synthetic gates both to isolate RB signals and to ensure SPAM-robustness.  However, regardless of the efficiency of any ``irrep projection'' scheme realized by synthetic gates, it needs to be bracketed by SPAM operations that have significant overlap with the desired irrep to ensure a good exponentially decaying signal.

Restricting ourselves to physically realizable states and measurement effects limits our ability to optimize this overlap.  For example, any physical state has constant overlap with the trivial irrep due to the unit trace condition and therefore cannot be made to lie entirely inside any nontrivial superoperator irrep.

One can circumvent the limitations of physical states and effects by using \emph{synthetic} states and effects (collectively, synthetic SPAM).  The optimal initial states and measurement effects are supported exclusively in a target irrep.  However, these states are unphysical, as their density matrices have negative eigenvalues.  We can nonetheless simulate the preparation of these unphysical states by preparing different physical states and then taking linear combinations of the results of the corresponding RB experiments.  This is accomplished by classical post-processing.  The same holds for measurement.  If the resulting synthetic SPAM has sufficiently good overlap with the target irrep, then it may even obviate the need to use \emph{any} irrep projection scheme derived from synthetic gates.  This is the idea of pure synthetic-SPAM RB (SSRB).\footnote{SSRB can be interpreted as minimal process tomography on the $G$-twirl of the error process $\mathcal{E}$, which has large degenerate eigenspaces: if there are $\Lambda\equiv |\{\lambda\}|$ irreps, then we only need to construct a $\Lambda\times \Lambda$ superoperator (regardless of the system's dimension).  See Appendix \ref{app:improving}.}

The goal of SSRB is to construct a single operator $\Xi$ that lies inside the target irrep $\lambda$ of the superoperator representation $\mathcal{G}$.  We then use this operator as both our synthetic state and our synthetic measurement effect, taking $\rho' = E' = \Xi$ in \eqref{synthesis}.  Thus we effectively prepare $\Xi$ and measure the ``POVM'' $\{\Xi, \mathds{1} - \Xi\}$.

A precursor of the idea of synthetic states can be found in the modification to standard Clifford RB introduced in \cite{RB-samples}, which uses a traceless input ``state,'' obtained as a difference of two density matrices, that has support only in the nontrivial irrep of the superoperator representation of the Clifford group.\footnote{A possible reason that this approach has not been pursued more systematically in the literature is the simple irrep structure of the superoperator representation of the Clifford group.}  Methods for Pauli noise learning \cite{Pauli-1, Pauli-2} also implicitly use synthetic states in the form of non-identity Pauli operators.

Of course, obtaining reliable results from SSRB requires reliable physical SPAM operations.  However, even SSRB without synthetic gates is more robust to SPAM error than one might expect.  Indeed, we still isolate individual exponential decays if either the state preparations or the measurement operations are supported entirely on the target irrep---for example, if arbitrary errors afflict either the states or the measurements alone.  This is because such SPAM errors would reduce the overlap between synthetic states and effects within the same irrep (thus decreasing the signal strength) without mixing different irreps.  For this reason, SSRB is fairly robust to SPAM error: it fails only when \emph{both} state preparation and measurement are noisy.  Furthermore, it fails only when they are both noisy \emph{in the same way}: if the noise affecting (synthetic) states and the corresponding (synthetic) effects causes them to leak into distinct irreps, then it does not corrupt the RB signal.

\tocless{\subsection{Synthetic Gates}}

The purpose of synthetic gates is to build a projection superoperator onto a target irrep as a linear combination of superoperator representations of elements of the benchmarking group $G$.  For this purpose, we use a frame (spanning set) of superoperators for the subspace of superoperators that we wish to synthesize.  This frame can be infinite or finite.  The first option requires random sampling and is the approach taken by character RB, which constructs projectors as weighted integrals over the character group, if the groups in question are infinite.  For finite frames, such as those coming from finite groups, random sampling may still be useful if the order of the group is large or the circuit depth is large.

The following fact provides a foundation for RB protocols that use group-based frames to construct synthetic gates, including but not limited to character RB: a unitary representation of a compact group furnishes a complete set of operators within each irrep subspace.  In other words, the superoperator representation matrices of a compact group $G$ span the space of all superoperators with the same block structure.  In particular, any superoperator $\mathcal{A}$ that acts within a particular irrep $\lambda$ occurring in the superoperator representation $\mathcal{G}$ of $G$ (meaning $\Pi_\lambda\mathcal{A}\Pi_\lambda = \mathcal{A}$) can be expressed as a linear combination of superoperator representation matrices:
\begin{equation}
(\dim\phi_\lambda)\mathbb{E}_{g\in G}\big[(\mathcal{G}|\mathcal{A})\mathcal{G}\big] = \mathcal{A},
\label{completeness}
\end{equation}
where we define the Hilbert-Schmidt inner product on superoperators by $(\mathcal{A}|\mathcal{B}) = \Tr(\mathcal{A}^\dag\mathcal{B})$.  If $\mathcal{A}$ itself is a projector, $\mathcal{A} = \Pi_\lambda$, then this reduces to the standard formula
\begin{equation}
(\dim\phi_\lambda)\mathbb{E}_{g\in G}\big[\chi_\lambda(g)^\ast\mathcal{G}\big] = \Pi_\lambda
\end{equation}
by virtue of $\chi_\lambda(g) = \Tr(\Pi_\lambda\mathcal{G})$.  See Appendix \ref{app:completeness} for details.

We consider three examples of synthetic-gate RB protocols.

\tocless{\subsubsection{Character RB}}

In character RB ($\chi$RB), we fix a benchmarking group $G$ and a character group $H\subset G$.  For simplicity, suppose that $H = G$ and $\lambda' = \lambda$.  Let $\vec{g}\equiv (g_1, \ldots, g_m)$.  We estimate the average weighted survival probability
\begin{equation}
p_m^\lambda\equiv \mathbb{E}_{(\vec{g}, g)\in G^{m+1}}\big[(\dim\phi_\lambda)\chi_\lambda(g)^\ast\sbra{E}\widetilde{\mathcal{G}}_\text{inv}\widetilde{\mathcal{G}}_m\cdots \widetilde{\mathcal{G}}_2\widetilde{\mathcal{G}_1\mathcal{G}}\sket{\rho}\big]
\end{equation}
to a desired level of precision by running many circuits with $\vec{g}, g$ sampled randomly before every shot\footnote{This strategy of using a single shot per circuit makes sense if settings are cheap and shots are not.} and averaging the rescaled measurement outcomes.  We thus sample from a random variable with mean $p_m^\lambda$.  One can show that the variance is bounded above as $\sigma^2\leq (\dim\phi_\lambda)^2$ (see Appendix \ref{app:syntheticgatebounds}).  It follows from the central limit theorem that for $N$ random samples, the variance of the sample mean is bounded above by $(\dim\phi_\lambda)^2/N$ as $N\to\infty$.  The fact that this bound increases with the dimension of the target irrep already suggests that it may be more efficient to build a rank-deficient projector than a full projector onto the irrep.  The most extreme option is to build a projector onto a 1D subspace of the target irrep.

\tocless{\subsubsection{Rank-1 RB}}

The idea of rank-1 RB (R1RB) is to use the infinite frame of all $G$ superoperators to construct projectors of the smallest possible dimension---one---rather than full-rank irrep projectors, as in character RB.  In this case, the weighting factor in \eqref{completeness} is a single matrix element of a given $G$ superoperator determined by the desired projector rather than $\chi_\lambda(g)^\ast = \Tr(\mathcal{G}^\dag\Pi_\lambda)$ for a given $g\in G$.  We argue that this procedure has significantly better sample complexity than character RB due to the reduced variance in this matrix element compared to that of the character function, as the former is bounded in absolute value by a constant rather than by the dimension of the irrep.

An important question remains: what is the best choice of projector?  Any circuit outcome probability can be written as $\llangle E|\mathcal{S}|\rho\rrangle = \Tr(|\rho\rrangle\llangle E|\cdot \mathcal{S})$, so to avoid degrading the RB signal, we would like to maximize the overlap of this projector with the rank-1 SPAM superoperator $|E\rrangle\llangle\rho|$.  We thus encounter a tension between the desire to maximize this overlap and the desire to minimize the rank of the projector.  In the ideal case that $\Pi_\lambda \linebreak[1] |E\rrangle\llangle\rho| \linebreak[1] \Pi_\lambda \linebreak[1] = \linebreak[1] |E\rrangle\llangle\rho|$, we would synthesize the SPAM superoperator itself (which is not necessarily a projector) by using the weighting factor $\Tr(\mathcal{G}^\dag|E\rrangle\llangle\rho|)$ in \eqref{completeness}.  While this is too much to expect from physical SPAM, precisely this situation can be arranged by using \emph{synthetic} SPAM.  By constructing synthetic SPAM operations $\rho' = E' = \Xi$ inside the target irrep, we single out the optimal rank-1 projector $|\Xi\rrangle\llangle\Xi|$.  This is the key observation behind our most effective protocols.

\tocless{\subsubsection{Finite-Frame RB}}

Because Hilbert-Schmidt space is finite-dimensional, a finite superoperator frame suffices to build any desired superoperator.  Furthermore, unlike in character RB, we need not require a group structure on the subset of group elements whose superoperator representations are used to synthesize a projector.  This approach has the advantage that it requires no random sampling from an infinite set, which eliminates finite-sample approximation error.

The idea of finite-frame RB (FFRB) is to use the superoperator representations $\mathcal{G}_i'$ of a finite set of $N_\lambda$ group elements $g_i'$ to synthesize a superoperator $\mathcal{P}_\lambda$ whose image lies inside the irrep $\phi_\lambda$:
\begin{equation}
\sum_{i=1}^{N_\lambda} c_i^\lambda\mathcal{G}_i' = \mathcal{P}_\lambda,
\end{equation}
where the $c_i^\lambda$ are real coefficients.  For example, $\mathcal{P}_\lambda$ could be the full projector $\Pi_\lambda$ or a projector of smaller rank.  Correspondingly, we estimate
\begin{equation}
p_m^\lambda\equiv \sum_{i=1}^{N_\lambda} c_i^\lambda\mathbb{E}_{\vec{g}\in G^m}\big[\sbra{E}\widetilde{\mathcal{G}}_\text{inv}\widetilde{\mathcal{G}}_m\cdots \widetilde{\mathcal{G}}_2\widetilde{\mathcal{G}_1\mathcal{G}_i'}\sket{\rho}\big]
\end{equation}
by sampling from a linear combination of $N_\lambda$ independent random variables.  By sampling optimally from the components, one can bound the variance of the sample mean as
\begin{equation}
\overline{\sigma}^2\leq \frac{|\vec{c}^\lambda|_1^2}{4N},
\end{equation}
where $N$ is the total number of samples across all of the component random variables and we write $|\vec{c}^\lambda|_1\equiv \sum_{i=1}^{N_\lambda} |c_i^\lambda|$ for the 1-norm of the coefficient vector. (See Appendix \ref{app:syntheticgatebounds}.)

FFRB can offer drastic sample complexity advantages over character RB and is particularly useful for finite benchmarking groups.  For example, consider character RB with the $n$-qubit Clifford group as the benchmarking group (``Clifford character RB'').  Character RB is not strictly necessary in this case because the 2-design property of the Clifford group implies that the standard RB signal takes the form of a single exponential function with a constant offset ($m\mapsto Af^m + B$), but the principles of our discussion apply more generally.

The sample complexity of Clifford character RB with Clifford character group, which synthesizes full projectors onto the nontrivial irrep in the Clifford superoperator representation to extract the signal $Af^m$, is exponential in the number of qubits $n$.  On the other hand, Clifford character RB with Pauli character group constructs rank-1 projectors onto the nontrivial Clifford irrep with constant sample complexity, but at the cost of a reduced signal.  FFRB with optimal sampling combines the strengths of both of these character-based approaches: by using the Pauli superoperators (the purely diagonal Clifford superoperators) as a finite frame for the space of all diagonal superoperators, we can construct the \emph{full} exponentially large projector onto the nontrivial Clifford irrep with \emph{constant} sample complexity. (See Appendix \ref{app:comparisontoPC}.)

What we call finite-frame RB can be seen as a special case of the ``subset RB'' version of uniform RB.  How to choose a non-group-based finite frame that is optimal with respect to sample complexity is an interesting open problem.  We focus on group-based (hence potentially infinite) frames because they allow for better mathematical control.

\tocless{\subsection{Synthetic Circuits}}

Of the protocols that we have described so far, those that incorporate synthetic gates are SPAM-robust.  We focus on the synthetic-gate protocols $\chi$RB and R1RB, which use a group-based frame to construct full-rank and rank-1 projectors, respectively, onto any desired irrep.  These protocols can easily be adapted to synthesize projectors of any rank.  We also consider synthetic-SPAM protocols, the simplest of which (SSRB) is highly efficient but not SPAM-robust under the most general circumstances.  Synthetic RB protocols use any combination of synthetic gates and synthetic SPAM.  We refer to them by abbreviations such as SS$\chi$RB and SSR1RB.

In general, we wish to optimize an RB protocol according to three criteria:
\begin{enumerate}
\item Maximize precision, as quantified by the variance of the estimator of an RB quality parameter.
\item Maximize accuracy, as quantified by the closeness of the mean of the estimator to the true answer.
\item Maximize efficiency (minimize sample complexity).
\end{enumerate}
To analytically quantify the performance of synthetic RB protocols, we make the assumption of perfect SPAM.  Under this assumption, any such protocol perfectly isolates the signal from a single irrep: for a fixed irrep $\lambda$ and a fixed sequence length $m$, it samples from a random variable with mean $A_\lambda f_\lambda^m$, where $A_\lambda$ is a SPAM-dependent coefficient that measures the strength of the signal and $f_\lambda$ is a quality parameter.  Any synthetic-SPAM protocol has $A_\lambda = 1$ in the limit of vanishing gate noise, while any synthetic-gate protocol that uses physical SPAM generally has $A_\lambda < 1$.  Normalizing by $A_\lambda$, the performance of the protocol is determined by the variance of a random variable with mean $f_\lambda^m$.  Therefore, in the absence of SPAM error, we address all three criteria by minimizing the variance of this normalized random variable, which determines the number of samples required to estimate the mean to a given precision.

With noisy SPAM, a non-SPAM-robust protocol such as pure synthetic-SPAM RB cannot isolate signals from specific irreps, leading to systematic error in the mean of the estimator.  The goal of any synthetic RB protocol that uses both synthetic gates and SPAM is to achieve greater accuracy than pure synthetic-SPAM RB, at moderate cost in the sample complexity needed to achieve a desired level of precision.

We estimate the sample complexity of the various synthetic RB protocols by deriving exact expressions for the variance of the associated random variables in the limit of vanishing gate noise, where $f_\lambda = 1$.  We summarize our results as follows.  For sufficiently small gate noise, the SSRB variance will always be less than that of the SPAM-robust protocols.  In order of zero-noise variance (greatest to least), the SPAM-robust protocols are as follows:
\begin{itemize}
\item $\chi$RB has SPAM-dependent signal strength and high sample complexity from constructing full-rank projectors.
\item R1RB reduces the sample complexity of constructing projectors relative to $\chi$RB.
\item SS$\chi$RB improves the signal strength of $\chi$RB by ensuring that the support of the SPAM operations is as large as possible within irreps.
\item SSR1RB both greatly improves the signal strength and reduces the sample complexity of constructing projectors relative to $\chi$RB.
\end{itemize}
Using synthetic SPAM counterbalances the potential loss of RB signal from using rank-1 projectors, leading to the superior performance of SSR1RB.  Since the optimization of the finite frame is left as an open problem, we do not make quantitative comparisons between (SS)FFRB and the other protocols.

In the remainder of the paper, we examine in detail the case $G = \text{SU}(2)$, where we substantiate these claims.

\tocless{\section{Rotationally Invariant Randomized Benchmarking}} \label{sec:RIRB}

To illustrate the application of our protocols, we establish theoretical foundations for efficiently benchmarking rotationally invariant quantum systems.

Any rotationally invariant system is built from basic units that carry irreducible actions of $\text{SU}(2)$.  When used as computational primitives, we refer to such units as ``spin qudits.''  Qudits (local degrees of freedom with Hilbert spaces of dimension $d \linebreak[1] \geq \linebreak[1] 2$) have different physical realizations and modes of operation than collections of qubits, and are therefore subject to different kinds of error.  In particular, spin qudits (realizable in diverse platforms such as nuclear spins) allow for precise characterization and experimental control, and they form the basis for several recently proposed quantum error-correcting codes.  Realizing the potential of spin qudits for fault-tolerant quantum computation requires methods for benchmarking operations on quantum information encoded therein.

The defining characteristic of a spin is that its Hilbert space carries an irrep of $\text{SU}(2)$, which is also the group of single-qubit unitaries.  This observation suggests that it may be natural to encode a logical qubit in the Hilbert space of a high-spin qudit in such a way that global $\text{SU}(2)$ rotations on the physical qudit implement the corresponding logical $\text{SU}(2)$ operations on the encoded qubit (thus providing a notion of ``transversal'' gate in this context).  This is the idea underlying recently constructed \emph{spin codes} \cite{Gross}.

The effectiveness of spin codes relies on the assumption that $\text{SU}(2)$ rotations on the physical spin are both physically natural operations \emph{and} errors.  We make the simplifying assumption that the native experimental operations are global $\text{SU}(2)$ rotations inside the unitary group $\text{SU}(2j + 1)$ on a spin-$j$ Hilbert space and develop protocols for measuring the error rates of their physical implementations.

This is not a straightforward task because the spin-$j$ representation of the group of global $\text{SU}(2)$ rotations does not form a unitary 2-design for $\mathbb{C}^{2j + 1}$ when $j > 1/2$.  Consequently, the superoperator representation of $\text{SU}(2)$ on the state space of a $j > 1/2$ system is highly reducible.  This implies that standard $\text{SU}(2)$ RB on a $j > 1/2$ system will not be SPAM-invariant: it will produce a linear combination of up to $2j + 1$ exponential decays with different rates, and the precise linear combination that occurs will depend strongly on the initial state.  Our task is therefore to measure a separate rotationally invariant error rate for each class of $\text{SU}(2)$-twirled error.

\tocless{\subsection{\texorpdfstring{$\text{SU}(2)$}{SU(2)} Operations}}

To set the stage, we review the ``kinematics'' of our problem.  The generators of $\text{SU}(2)$ satisfy $[J_a, J_b] = i\epsilon_{abc}J_c$ for $a, b, c\in \{x, y, z\}$.  We use $\ell$ to denote $J_z$ eigenvalues, reserving $m$ for sequence length, and we often leave the spin $j$ implicit when writing $J_z$ eigenstates: $|\ell\rangle\equiv |j, \ell\rangle$.  We also use the shorthand notation $\sket{\ell}\equiv \sket{|\ell\rangle\langle\ell|}$.

We wish to benchmark global $\text{SU}(2)$ rotations on a spin-$j$ Hilbert space.\footnote{While $\text{SU}(2)$ is an infinite group, it is compact and therefore behaves in many ways like the finite groups used in standard RB protocols.  For example, the Peter-Weyl theorem (which generalizes Maschke's theorem for finite groups) guarantees that the unitary representations are completely reducible.}  In the spin-$j$ representation of $\text{SU}(2)$, we have (with $J_\pm\equiv J_x\pm iJ_y$)
\begin{equation}
J_\pm|\ell\rangle = \sqrt{j(j + 1) - \ell(\ell\pm 1)}|\ell\pm 1\rangle, \quad J_z|\ell\rangle = \ell|\ell\rangle.
\end{equation}
A global $\text{SU}(2)$ rotation takes the form $e^{-i\theta\vec{n}\cdot\vec{J}}$, where $\vec{n}$ is a unit vector.

While global $\text{SU}(2)$ rotations act irreducibly on the $(2j + 1)$-dimensional Hilbert space, the superoperator representation of $\text{SU}(2)$ on the $(2j + 1)^2$-dimensional space of linear operators on a spin-$j$ system is highly reducible: $j\otimes j^\ast\cong 0\oplus 1\oplus \cdots\oplus 2j$, where we label each irrep by its spin.  Note that all irreps of $\text{SU}(2)$ are real or pseudoreal, so $j^\ast\cong j$.

\begin{figure}[!htb]
\centering
\includegraphics[scale=0.22]{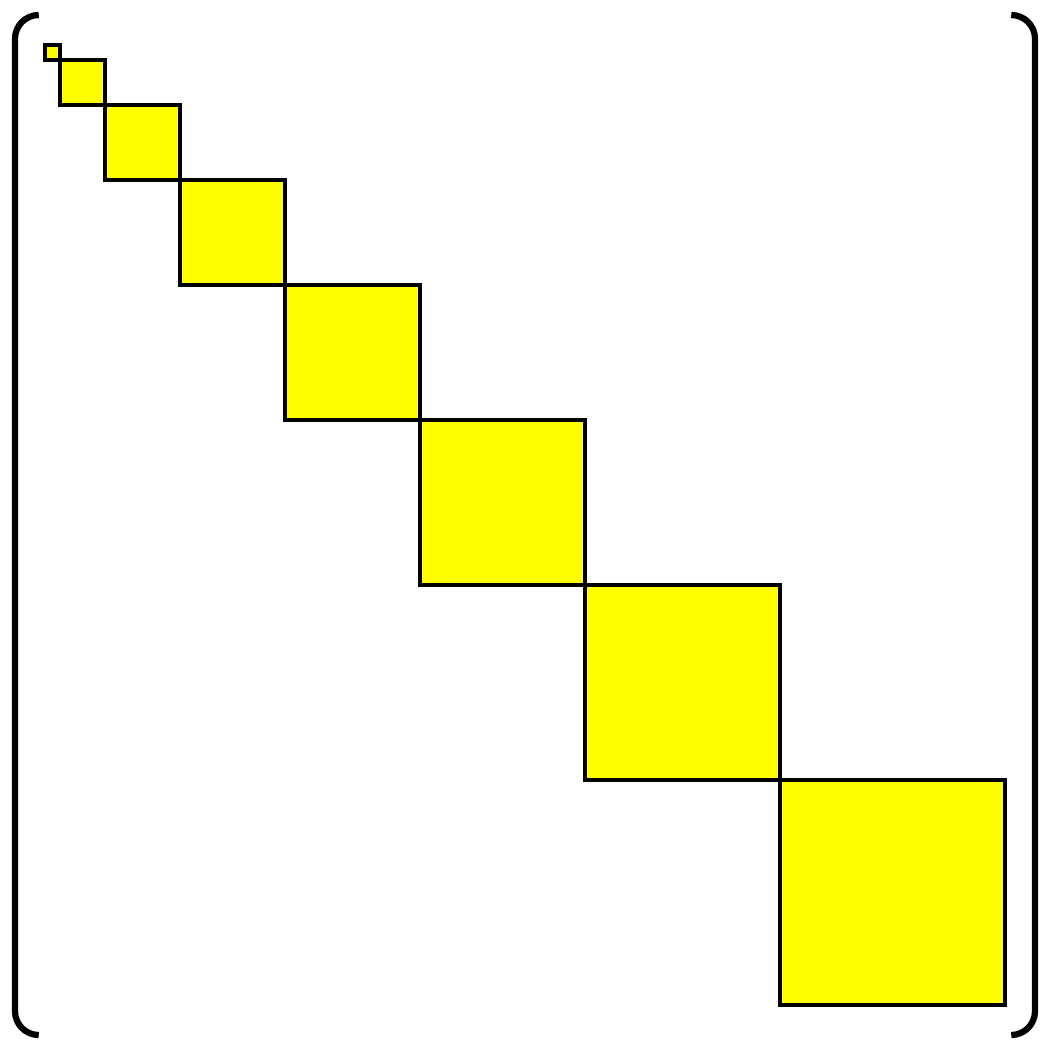}
\caption{An $\text{SU}(2)$ superoperator has a characteristic ``kite'' structure, shown here for $j = 7/2$.}
\label{kite}
\end{figure}

To make this irrep decomposition manifest, note that any spin-$j$ operator can be expanded in irreducible spherical tensors of rank $k = 0, 1, \ldots, 2j$ (see Appendices \ref{app:preliminaries} and \ref{app:derivations} for details):
\begin{equation}
T^{(k)}_q = \sqrt{\frac{2k + 1}{2j + 1}}\sum_{\ell, \ell' = -j}^j C^{j, \ell}_{j, \ell'; k, q}|\ell\rangle\langle\ell'|.
\end{equation}
A rank-$k$ spherical tensor operator is a homogeneous degree-$k$ polynomial in the angular momentum operators $J_x$, $J_y$, $J_z$ and has $2k + 1$ components indexed by $q = -k, \ldots, k$.  These operators are orthonormal with respect to the Hilbert-Schmidt inner product:
\begin{equation}
\tr((T^{(k)}_q)^\dag T^{(k')}_{q'}) = \delta_{kk'}\delta_{qq'}.
\end{equation}
In this basis, the superoperator representation of $\text{SU}(2)$ is block diagonal with the blocks being Wigner $D$-matrices $D^k(g)$:
\begin{equation}
g\mapsto \mathcal{G}(g) = \operatorname{diag}(D^0(g), D^1(g), \ldots, D^{2j}(g)).
\end{equation}
The dimensions of the blocks satisfy $1 + 3 + \cdots + (4j + 1) = (2j + 1)^2$.  See Figure \ref{kite} for an illustration.

Any spin-$j$ superoperator can be expanded in Choi units:
\begin{equation}
\rho\mapsto T^{(k)}_q\rho(T^{(k')}_{q'})^\dag.
\end{equation}
We generally expect errors of low rank to be more prevalent than those of high rank, particularly for large $j$.  We use the term ``weight'' for the rank of an error channel, by analogy to Pauli errors on multiqubit systems.  Spin codes protect against errors of weight $k = 1$ and $k = 2$ \cite{Gross, Omanakuttan-1, Omanakuttan-2}.

\tocless{\subsection{\texorpdfstring{$\text{SU}(2)$}{SU(2)} Error Rates}}

With these ingredients in place, we come to a precise understanding of the outputs of any RB protocol that uses $\text{SU}(2)$ as the benchmarking group.

When twirled over $\text{SU}(2)$, spherical tensor Choi units become uniformly random weight-$k$ errors:
\begin{equation}
\rho\mapsto T^{(k)}_q\rho(T^{(k')}_{q'})^\dag \xrightarrow{\text{twirl}} \begin{cases} \mathcal{G}^{(k)} & \text{if $k = k'$, $q = q'$}, \\ 0 & \text{otherwise}, \end{cases}
\end{equation}
where we have defined the linear maps
\begin{equation}
\mathcal{G}^{(k)} : \rho\mapsto \frac{1}{2k + 1}\sum_{q=-k}^k T_q^{(k)}\rho(T_q^{(k)})^\dag.
\end{equation}
By Schur's lemma, any $\text{SU}(2)$-twirled error channel can be expressed either as a linear combination of $\text{SU}(2)$-twirled Choi units or as a linear combination of projectors:
\begin{equation}
\mathbb{E}_{g\in \text{SU}(2)}\big[\mathcal{G}^\dag\mathcal{E}\mathcal{G}\big] = \sum_{k=0}^{2j} p_k\mathcal{G}^{(k)} = \sum_{k=0}^{2j} f_k\Pi_k.
\end{equation}
The maps $\mathcal{G}^{(k)}$ are not properly normalized channels.  Consequently, the coefficients $p_k$ are \emph{unnormalized} probabilities of weight-$k$ $\text{SU}(2)$ errors, whereas the $f_k\in [0, 1]$ are RB quality parameters describing the decay rates of the components of the initial state within each irrep.  There exists a linear relationship between these two sets of coefficients because each $\mathcal{G}^{(k)}$ must itself be expressible as a linear combination of projectors:
\begin{equation}
\mathcal{G}^{(k)} = \sum_{k'=0}^{2j} g^{(k)}_{k'}\Pi_{k'}.
\end{equation}
Therefore, we have
\begin{equation}
f_{k'} = \sum_{k=0}^{2j} g^{(k)}_{k'} p_k, \qquad \mathcal{G}^{(k)} : T_{q'}^{(k')}\mapsto g^{(k)}_{k'}T_{q'}^{(k')}
\end{equation}
(no sum on $k'$).

The transformation coefficients $g^{(k)}_{k'}$ take the form
\begin{equation}
g^{(k)}_{k'} = (-1)^{2j + k + k'}\left\{\begin{array}{ccc} k & j & j \\ k' & j & j \end{array}\right\},
\end{equation}
as we work out explicitly in Appendix \ref{app:derivations}.  For convenience, we package them into a discrete ``Fourier'' matrix
\begin{equation}
F = [F_{kk'}]\equiv g^{(k)}_{k'},
\label{Fmaintext}
\end{equation}
which is the $\text{SU}(2)$ analogue of the familiar Walsh-Hadamard transform between Pauli error rates and Pauli eigenvalues.  To obtain the $p_k$, we multiply the vector of $f_k$ by $F^{-1}$.  Since the maps $(2j + 1)\mathcal{G}^{(k)}$ are properly normalized channels, to obtain normalized error rates with $\sum_{k=0}^{2j} p_k = 1$, we omit the overall prefactor of $1/(2j + 1)$ in the matrix $F$.

Using an appropriate RB protocol to obtain the quality parameters $f_k$ for any $\text{SU}(2)$-twirled error channel, we can infer the corresponding $\text{SU}(2)$ error rates $p_k$ and thereby distinguish rates of correctable (low-$k$) errors from those of uncorrectable (high-$k$) errors.  While standard RB yields only the sum of all $2j$ nontrivial error rates, we wish to extract all nontrivial rates individually.

This would seem to be a job for $\text{SU}(2)$ character RB.  However, ordinary character RB is prohibitively inefficient in this case.  In contrast to the case of character RB with Pauli benchmarking group, which works well because the superoperator representation of the Pauli group has 1D irreps, the large irrep dimensions in the superoperator representation of $\text{SU}(2)$ incur large sampling variance when using $\text{SU}(2)$ as its own character group.  Moreover, the only viable character group for $\text{SU}(2)$ character RB is $\text{SU}(2)$ itself (except for very small $j$).  Indeed, no proper subgroup of $\text{SU}(2)$ furnishes a character group that is suitable for implementing character RB with multiplicities \cite{Claes} because the isotypic components of the superoperator representations of both discrete and continuous proper subgroups of $\text{SU}(2)$ do not have support inside individual $\text{SU}(2)$ irreps (see Appendix \ref{app:su2chiRB}).

We now describe how to overcome these obstacles using synthetic RB protocols with synthetic SPAM.

\tocless{\subsection{\texorpdfstring{$\text{SU}(2)$}{SU(2)} Randomized Benchmarking}}

For the following discussion, we fix a spin $j$.  Superoperator irreps are labeled by $k\in \{0, \ldots, 2j\}$, while states are labeled by $\ell \linebreak[1] \in \linebreak[1] \{j, \linebreak[1] \ldots, \linebreak[1] -j\}$.  We restrict our attention to preparing $J_z$ eigenstates and to measuring the operator $J_z$.  For a given RB experiment, we fix a sample size $N$ (a large positive integer) and a set of sequence lengths (positive integers) $\mathcal{M}$.  For notational convenience, let $G = \text{SU}(2)$, $\vec{g}\equiv (g_1, \ldots, g_m)\in G^m$, and $g_\text{inv} = (g_m\cdots g_1)^\dag$ for any given $m\in \mathcal{M}$.

Physical SPAM is of limited utility for isolating RB signals.  For example, a natural choice for $\rho$ is a $J_z$ eigenstate.  The only component of a pure $J_z$ eigenstate $\rho = |\ell\rangle\langle\ell|$ within the $k^\text{th}$ irrep is that in the $T^{(k)}_0$ direction, with amplitude
\begin{equation}
M_{k, \ell}\equiv \langle\ell|T^{(k)}_0|\ell\rangle = \sqrt{\frac{2k + 1}{2j + 1}}C^{j, \ell}_{j, \ell; k, 0}.
\label{Mmaintext}
\end{equation}
Another natural choice for $\rho$ is a spin coherent state.  Spin coherent states include the highest- and lowest-weight $J_z$ eigenstates with eigenvalues $\ell = \pm j$.  Since any two states in the same $\text{SU}(2)$ orbit have the same support within irreps, all spin coherent states have the same support within irreps as the states $|{\pm j}\rangle$.  This support exhibits a single peak and a long tail as a function of the irrep label, becoming progressively smaller in high-spin irreps as $j\to\infty$.

By contrast, it is possible to construct synthetic states and effects that not only lie entirely in a given irrep, but are also diagonal in the basis of $J_z$ eigenstate matrix units.  These are precisely the diagonal spherical tensor operators $T^{(k)}_0$ for each $k = 0, \ldots, 2j$, which are related to the $|\ell\rangle\langle\ell|$ via the coefficients $M_{k, \ell}$ in \eqref{Mmaintext}.  Thus, by using the $2j + 1$ matrix units $|\ell\rangle\langle\ell|$ as initial states and measurement effects and taking linear combinations of the corresponding survival probabilities, we can simulate both the preparation and the measurement of $T^{(k)}_0$.

To extract the signal from irrep $k$, we implement SSRB using $T^{(k)}_0$ as both the synthetic state and the synthetic measurement effect.  The corresponding rank-1 SPAM superoperator is $\mathcal{Q}^{(k)} = |T^{(k)}_0\rrangle\llangle T^{(k)}_0|$, which satisfies
\begin{equation}
\Tr(\mathcal{G}^\dag\mathcal{Q}^{(k)}) = D^k_{00}(g) = d^k_{00}(g)
\end{equation}
where $D^k_{00}$ and $d^k_{00}$ are Wigner $D$-matrix and $d$-matrix elements.  Via \eqref{completeness}, we therefore have
\begin{equation}
(2k + 1)\mathbb{E}_{g\in \text{SU}(2)}\big[d^k_{00}(g)\mathcal{G}\big] = \mathcal{Q}^{(k)}.
\end{equation}
The weighting factor $d^k_{00}(g)$, a single diagonal entry of $\mathcal{G}$ in the spherical tensor basis, is real and bounded in absolute value by 1.  By contrast, character RB uses the weighting factor
\begin{equation}
\Tr(\mathcal{G}^\dag\Pi_k) = \chi_k(g) = \sum_{q=-k}^k D_{qq}^k(g),
\end{equation}
a sum of diagonal entries of $\mathcal{G}$ in the spherical tensor basis, to synthesize the full irrep projector $\Pi_k = \sum_{q=-k}^k |T^{(k)}_q\rrangle\llangle T^{(k)}_q|$.

We now describe the implementation details of our $\text{SU}(2)$ RB protocols.  In all cases, the orthogonal matrix $M = [M_{k, \ell}]$ in \eqref{Mmaintext} relates physical states ($J_z$ eigenstates) labeled by $\ell$ to synthetic states labeled by $k$.  The ``Fourier'' matrix $F$ in \eqref{Fmaintext} relates $\text{SU}(2)$ error rates to RB quality parameters: $\vec{f} = F\vec{p}$.

\tocless{\subsubsection{\texorpdfstring{$\text{SU}(2)$}{SU(2)} Synthetic-SPAM RB}}

SSRB uses both synthetic states and synthetic measurement effects that lie within individual irreps.  In the data collection phase, we run $N$ random circuits to obtain estimates $p_{\ell, \ell', m}$ for the averaged survival probabilities
\begin{equation}
\mu_{\ell, \ell', m}\equiv \mathbb{E}_{\vec{g}\in G^m}\big[\llangle\ell'|\widetilde{\mathcal{G}}_\text{inv}\widetilde{\mathcal{G}}_m\cdots \widetilde{\mathcal{G}}_1\sket{\ell}\big].
\end{equation}
We assemble these estimates into a list of $|\mathcal{M}|$ matrices $P_m = [p_{\ell, \ell', m}]$.  In the post-processing phase, we do the following:
\begin{itemize}
\item For each $m\in \mathcal{M}$, let $\vec{d}_m $ be the diagonal of $MP_m M^T$.  We denote the components of $\vec{d}_m$ by $d_{k, m}$ for $k = 0, \ldots, 2j$.
\item For each $k$, fit the $|\mathcal{M}|$ numbers $d_{k, m}$ to a single decaying exponential function of the form $m\mapsto A_k f_k^m$, where $A_k$ and $f_k$ are constants. (For small gate noise, we should have $A_k\approx 1$ for all $k$.) Assemble the $f_k$ into a vector $\vec{f}$.
\item Compute $\vec{p}\equiv F^{-1}\vec{f}$.  The components $p_k$ ($k = 0, \ldots, 2j$) are the rates of random weight-$k$ $\text{SU}(2)$ errors in the implemented $\text{SU}(2)$ rotations.
\end{itemize}
Note that, under the assumptions of gate-independent noise and perfect SPAM, each matrix $MP_m M^T$ is diagonal in the limit of perfect twirling ($N\to\infty$).  Therefore, after subtracting finite-sample fluctuations, the off-diagonal entries of $MP_m M^T$ can be used to estimate the degree to which SPAM noise contaminates the results of SSRB experiments.

\tocless{\subsubsection{\texorpdfstring{$\text{SU}(2)$}{SU(2)} Synthetic-SPAM Character RB}}

SS$\chi$RB uses synthetic SPAM in combination with synthetic gates to ensure SPAM-robustness.  In the data collection phase, we run $N$ random circuits to obtain estimates $p_{\ell, \ell', m}^k$ for the averaged weighted survival probabilities
\begin{equation}
\mu_{\ell, \ell', m}^k\equiv (2k + 1)\mathbb{E}_{(\vec{g}, g)\in G^{m+1}}\big[\chi_k(g)\llangle\ell'|\widetilde{\mathcal{G}}_\text{inv}\widetilde{\mathcal{G}}_m\cdots \widetilde{\mathcal{G}}_2\widetilde{\mathcal{G}_1\mathcal{G}}\sket{\ell}\big].
\end{equation}
We assemble these estimates into a list of $(2j + 1)|\mathcal{M}|$ matrices $P_m^k = [p_{\ell, \ell', m}^k]$.  In the post-processing phase, for each $m\in \mathcal{M}$ and $k\in \{0, \ldots, 2j\}$, we let $d_{k, m}$ be the {$(k, k)$ component of $MP_m^k M^T$.  The remaining steps are the same as for SSRB.

\tocless{\subsubsection{\texorpdfstring{$\text{SU}(2)$}{SU(2)} Synthetic-SPAM Rank-1 RB}}

The procedure for SSR1RB is identical to that for SS$\chi$RB, except that we obtain estimates $p_{\ell, \ell', m}^k$ for the averaged weight\-ed survival probabilities
\begin{equation}
\tilde{\mu}_{\ell, \ell', m}^k\equiv (2k + 1)\mathbb{E}_{(\vec{g}, g)\in G^{m+1}}\big[d_{00}^k(g)\llangle\ell'|\widetilde{\mathcal{G}}_\text{inv}\widetilde{\mathcal{G}}_m\cdots \widetilde{\mathcal{G}}_2\widetilde{\mathcal{G}_1\mathcal{G}}\sket{\ell}\big]
\end{equation}
in the data collection phase.

\tocless{\subsubsection{Finite Frames}}

For FFRB, the space of superoperators with the same block structure as $\mathcal{G}$ has dimension
\begin{equation}
N_j\equiv \sum_{\ell=0}^{2j} (2\ell + 1)^2 = \frac{1}{3}(2j + 1)(4j + 1)(4j + 3).
\end{equation}
If one chooses $N_j$ random $\text{SU}(2)$ elements $g_i$ for $i = 1, \ldots, N_j$, then their superoperator representations will be linearly in\-de\-pen\-dent with probability 1.  This allows one to construct ar\-bi\-trary superoperators $\mathcal{A}$ with the same block structure, including irrep projectors, as linear combinations of the form
\begin{equation}
\mathcal{A} = \sum_{i=1}^{N_j} a_i\mathcal{G}_i.
\end{equation}
Unlike the character-based approach, this procedure does not require randomly sampling from $\text{SU}(2)$ and is exact.  Determining the coefficients $a_i$ involves inverting a large matrix, and the specific choice of $N_j$ group elements (superoperator frame) determines how well-conditioned that matrix is.  Rather than choosing the $N_j$ elements randomly, one could fix them at the outset by optimizing that choice with respect to the sample complexity of the protocol.

\tocless{\subsubsection{Random Variables and Estimators}}

To quantify the resource costs of our protocols, we take the useful point of view that RB experiments sample from prob\-ability distributions, allowing us to estimate their means up to fluctuations due to finite sample size.

A regular RB protocol with physical ($J_z$ eigenstate) SPAM is associated with a random variable $X_{\ell, \ell', m}$ with mean
\begin{equation}
\langle X_{\ell, \ell', m}\rangle = \mu_{\ell, \ell', m}.
\end{equation}
To sample from $X_{\ell, \ell', m}$, we do the following: (1) sample $\vec{g}\in G^m$ uniformly at random; (2) apply the noisy gates corresponding to $g_1, \ldots, g_m, g_\text{inv}$ (in that order) to the initial state $|\ell\rangle\langle\ell|$; (3) measure $J_z$.  Then $X_{\ell, \ell', m}$ takes the value 1 if the outcome of the $J_z$ measurement is $|\ell'\rangle\langle\ell'|$ and 0 otherwise.  An RB estimator is the average of $X_{\ell, \ell', m}$ over $N$ shots.

A synthetic-gate RB protocol with physical ($J_z$ eigenstate) SPAM such as $\chi$RB or R1RB, which we denote by $x$RB for $x\in \{\chi, \text{R1}\}$, is associated with a random variable $X_{\ell, \ell', m}^k$ with mean
\begin{equation}
\langle X_{\ell, \ell', m}^k\rangle = \begin{cases} \mu_{\ell, \ell', m}^k & \text{if $x = \chi$}, \\[2 pt] \tilde{\mu}_{\ell, \ell', m}^k & \text{if $x = \text{R1}$}. \end{cases}
\end{equation}
To sample from $X_{\ell, \ell', m}^k$, we do the following: (1) sample $\vec{g}\in G^m$ and $g\in G$ uniformly at random; (2) apply the noisy gates corresponding to $g_1 g, g_2, \ldots, g_m, g_\text{inv}$ (in that order) to the initial state $|\ell\rangle\langle\ell|$; (3) measure $J_z$.  Then $X_{\ell, \ell', m}^k$ takes the value
\begin{equation}
(2k + 1)\times \begin{cases} \chi_k(g) & \text{if $x = \chi$} \\[2 pt] d^k_{00}(g) & \text{if $x = \text{R1}$} \end{cases}
\end{equation}
if the outcome of the $J_z$ measurement is $|\ell'\rangle\langle\ell'|$ and 0 otherwise.  An $x$RB estimator is the average of $X_{\ell, \ell', m}^k$ over $N$ shots.

Finally, in all $\text{SU}(2)$ RB protocols with synthetic SPAM, we use the diagonal spherical tensor operators as both synthetic states and measurement effects.  An SS$x$RB estimator is a linear combination of $(2j + 1)^2$ $x$RB estimators, according to
\begin{equation}
X_{k, m}\equiv \sum_{\ell, \ell' = -j}^j M_{k, \ell}M_{k, \ell'}X_{\ell, \ell', m}^k.
\end{equation}
(If $x$ is the empty string, corresponding to pure SSRB, then we replace $X_{\ell, \ell', m}^k$ with $X_{\ell, \ell', m}$ in the above.) A single SS$x$RB shot corresponds to $2j + 1$ $x$RB shots, distributed over the $2j + 1$ distinct initial states.  We aim to estimate the mean of the random variable $X_{k, m}$.  The variance of the sample mean is determined by $\operatorname{Var}(X_{k, m})$: this is a measure of the sample complexity of the protocol.  To compute or estimate $\operatorname{Var}(X_{k, m})$, we must account for the fact that the component random variables $X_{\ell, \ell', m}^k$ have a nontrivial covariance matrix because they are not all independent.

\tocless{\section{Sample Complexity}} \label{sec:samplecomplexity}

In Appendix \ref{app:SCbounds}, we derive bounds on the sample complexity of synthetic RB in the presence of gate noise.  However, it is more illuminating to consider exact results for zero gate noise (as we do in Appendix \ref{app:SCexact}).  This analysis also holds general lessons for the feasibility of character RB.\footnote{To justify our analysis, we point out a subtle but critical difference between synthetic or filtered RB protocols and standard RB.  Synthetic and filtered RB estimators construct weighted averages over measurement outcomes with both positive and negative coefficients to get a mean value that, in the zero-noise case, equals 1.  Standard RB also constructs such averages to get a mean value of 1 in the zero-noise case, but with strictly positive coefficients.  This has a significant consequence: the Fisher information of the binomial distribution implies that in the zero-noise limit, the variance of the ensemble over which standard RB averages goes to zero.  In other words, the variance of standard RB is extremely dependent on the amount of noise, and the variance vanishes (uniquely) in the zero-noise limit.  This is not true of synthetic or filtered RB, for which the variance is dominated by the variance of the distribution of weights assigned to circuits rather than by the Fisher information of individual circuit probabilities.}

Before doing so, we remark that existing work on the sample complexity of standard Clifford RB \cite{RB-confidence, RB-samples} takes as its foundation the representation theory of the Clifford group (in particular, the irrep decomposition of the tensor square of its superoperator representation \cite{RB-representations}), leading to rigorous bounds that are tighter than those derivable via our na\"ive methods.  The generalization of this fine-grained analysis to other families of benchmarking groups, which can be adapted to either the standard RB or the character RB context, is posed as an open problem in \cite{Helsen}.  We do not solve this problem here.

A lesson from \cite{RB-samples} is that for a slight modification of Clifford RB, sample complexity bounds can be derived that are independent of the number of qubits.  Similarly, under certain conditions on the measurement POVM, the sample complexity of a filtered RB protocol (which includes data collection and post-processing) is asymptotically independent of the dimension of the underlying Hilbert space, regardless of the chosen benchmarking group \cite{FilteredRB-1}.  In our exact results for the variance of synthetic RB in the zero-noise limit, we find only a weak dependence on the Hilbert space dimension (or $j$).

To obtain sample complexity estimates for our various protocols, we work in the zero-noise limit, which allows us to derive the leading contributions to estimator variance exactly.  These exact results provide a starting point for perturbative treatments of noise.  We assume perfect SPAM.

In the limit of zero noise, these protocols are insensitive to the sequence length $m$.  Consequently, we replace $X_{\ell, \ell', m}$ by a random variable $X_{\ell, \ell'}$ with mean
\begin{equation}
\langle X_{\ell, \ell'}\rangle = \delta_{\ell\ell'},
\end{equation}
$X_{\ell, \ell', m}^k$ by a random variable $X_{\ell, \ell'}^k$ with mean
\begin{equation}
\langle X_{\ell, \ell'}^k\rangle = M_{k, \ell}M_{k, \ell'},
\end{equation}
and $X_{k, m}$ by a random variable $X_k$ with mean $\langle X_k\rangle = 1$.

\tocless{\subsection{Synthetic-Gate Protocols}}

Consider $x$RB with $x\in \{\chi, \text{R1}\}$ and physical SPAM $\rho = E = |\ell\rangle\langle\ell|$, for which we sample from $X_{\ell, \ell}^k$.  The mean is
\begin{equation}
\langle X_{\ell, \ell}^k\rangle = M_{k, \ell}^2.
\end{equation}
Since the mean is not 1 (as would be the case with synthetic SPAM), we must normalize the variance by the squared mean to obtain the variance of the random variable corresponding to the estimator for $f_k = 1$, which is
\begin{equation}
\frac{\operatorname{Var}(X_{\ell, \ell}^k)}{\langle X_{\ell, \ell}^k\rangle^2} = \frac{(2k + 1)^2}{M_{k, \ell}^4}\sum_{k'=0}^{2k} \frac{C(k, k')}{2k' + 1}M_{k', \ell}^2 - 1,
\label{normalizedvariance}
\end{equation}
where we have defined
\begin{equation}
C(k, k')\equiv \begin{cases} 1 & \text{if $x = \chi$}, \\ (C^{k', 0}_{k, 0; k, 0})^2 & \text{if $x = \text{R1}$}. \end{cases}
\end{equation}
The normalized variance \eqref{normalizedvariance} is invariant under taking $\ell\to -\ell$.

\tocless{\subsection{Synthetic-SPAM Protocols}}

For SSRB, we have $\operatorname{Var}(X_k) = 0$, which makes SSRB the most efficient option if one is willing to sacrifice guaranteed SPAM-robustness.  For SS$x$RB with $x\in \{\chi, \text{R1}\}$, we have
\begin{equation}
\operatorname{Var}(X_k) = (2k + 1)^2\sum_{k'=0}^{2k} \frac{C(k, k')}{2k' + 1}\left(\sum_{\ell=-j}^j M_{k, \ell}^2 M_{k', \ell}\right)^2 - \sum_{\ell=-j}^j M_{k, \ell}^4.
\label{SSvariance}
\end{equation}
The (SS)R1RB variance is manifestly less than the (SS)$\chi$RB variance because $(C^{k', 0}_{k, 0; k, 0})^2\leq \sum_{q=-k}^k (C^{k', 0}_{k, q; k, -q})^2 = 1$.

\tocless{\subsection{Examples}}

To gain some intuition for \eqref{normalizedvariance} and \eqref{SSvariance}, consider the physically germane example of a spin qudit with $j = 7/2$, which can be realized as the nuclear spin of an antimony atom implanted in a silicon device \cite{Asaad, Yu} (among other ways \cite{Champion}).  The irreps for this eight-level system are labeled by $k = 0, 1, \ldots, 7$.\footnote{For FFRB, we would have $N_{7/2} = \sum_{\ell=0}^7 (2\ell + 1)^2 = 680$.}

In Table \ref{variancewithk}, we display the zero-noise variance for various protocols as a function of irrep $k$.  For the $x$RB protocols with non-synthetic SPAM, we use the optimal $J_z$ eigenstate SPAM for each irrep, this being $\ell = \pm 7/2, \linebreak[1] \pm 7/2, \linebreak[1] \pm 3/2, \linebreak[1] \pm 5/2, \linebreak[1] \pm 5/2, \linebreak[1] \pm 3/2, \linebreak[1] \pm 1/2$ for $k = 1, \ldots, 7$, respectively.

\begin{table}[!htb]
\centering
\begin{tabular}{|c|c|c|c|c|} \hline
& $\chi$RB & R1RB & SS$\chi$RB & SSR1RB \\ \hline
$k = 0$ & 7 & 7 & 0 & 0 \\ \hline
$k = 1$ & 28.6816 & 7.52245 & 1.07619 & 0.269048 \\ \hline
$k = 2$ & 91.8386 & 12.5807 & 3.23842 & 0.540816 \\ \hline
$k = 3$ & 308.139 & 42.3744 & 6.15572 & 0.773292 \\ \hline
$k = 4$ & 268.103 & 21.0241 & 10.4498 & 1.02387 \\ \hline
$k = 5$ & 514.734 & 32.779 & 15.668 & 1.28994 \\ \hline
$k = 6$ & 404.56 & 23.2173 & 23.0531 & 1.62223 \\ \hline
$k = 7$ & 381.656 & 21.6442 & 34.0697 & 2.11888 \\ \hline
\end{tabular}
\caption{$\text{SU}(2)$ synthetic RB variance as a function of $k$ with $j$ fixed to $7/2$ ($k = \linebreak[1] 0, \linebreak[1] 1, \linebreak[1] \ldots, \linebreak[1] 7$).}
\label{variancewithk}
\end{table}

Now consider the highest irrep $k = 2j$.  For various protocols, the zero-noise variance as a function of spin $j$ (up to $j = 7/2$) is given in Table \ref{variancewithj}.  For the $x$RB protocols with non-synthetic SPAM, we use the optimal $J_z$ eigenstate SPAM for each spin, this being $\ell = 0$ for integer $j$ and $\ell = \pm 1/2$ for half-integer $j$.

\begin{table}[!htb]
\centering
\begin{tabular}{|c|c|c|c|c|} \hline
& $\chi$RB & R1RB & SS$\chi$RB & SSR1RB \\ \hline
$j = 0$ & 0 & 0 & 0 & 0 \\ \hline
$j = 1/2$ & 23 & 5 & 4 & 1 \\ \hline
$j = 1$ & 25.25 & 4.89286 & 8.66667 & 1.40476 \\ \hline
$j = 3/2$ & 91.1811 & 9.9465 & 13.408 & 1.63867 \\ \hline
$j = 2$ & 95.25 & 11.163 & 18.4047 & 1.80578 \\ \hline
$j = 5/2$ & 209.672 & 15.5894 & 23.5132 & 1.9322 \\ \hline
$j = 3$ & 215.636 & 18.0822 & 28.7441 & 2.03407 \\ \hline
$j = 7/2$ & 381.656 & 21.6442 & 34.0697 & 2.11888 \\ \hline
\end{tabular}
\caption{$\text{SU}(2)$ synthetic RB variance as a function of $j$ with $k$ fixed to $2j$ ($j = \linebreak[1] 0, \linebreak[1] 1/2, \linebreak[1] \ldots, \linebreak[1] 7/2$).}
\label{variancewithj}
\end{table}

In each case, the SSR1RB variance improves upon the $\chi$RB variance by about two orders of magnitude.  If $\chi$RB had not been optimized with respect to SPAM, this improvement could be as high as eight orders of magnitude.  Note that, by sampling from $2j + 1$ $x$RB random variables in each experimental run, SS$x$RB takes advantage of measurement data that would otherwise be discarded in a single run of $x$RB.  However, SS$x$RB also requires $2j + 1$ input states (physical circuit executions) for each shot, while $x$RB requires only one.  In Figure \ref{shots}, we translate the above variance comparison into a comparison of physical shot counts, finding that SSR1RB consistently outperforms $\chi$RB by one to two orders of magnitude even with the experimental overhead of synthetic SPAM.

\begin{figure}[!htb]
\centering
\includegraphics[scale=0.55]{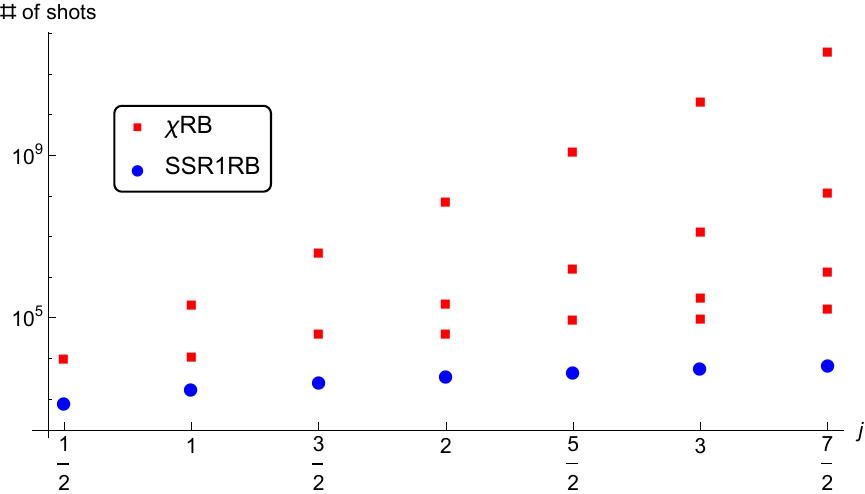}
\caption{Number of physical shots required to estimate the quality parameter $f_{2j}$ to an uncertainty of 0.05 for various spins $j$.  For $\chi$RB, results are shown for all possible choices of $J_z$ eigenstate SPAM.}
\label{shots}
\end{figure}

\tocless{\section{Numerical Results}} \label{sec:numerics}

To obtain a more realistic picture of how synthetic RB protocols compare, we numerically examine their performance in the presence of noisy SPAM.  Due to their advantages over physical-SPAM protocols, we focus on the synthetic-SPAM protocols SSRB, SS$\chi$RB, and SSR1RB.

\begin{figure}[!htb]
\centering
\begin{overpic}[width=\linewidth]{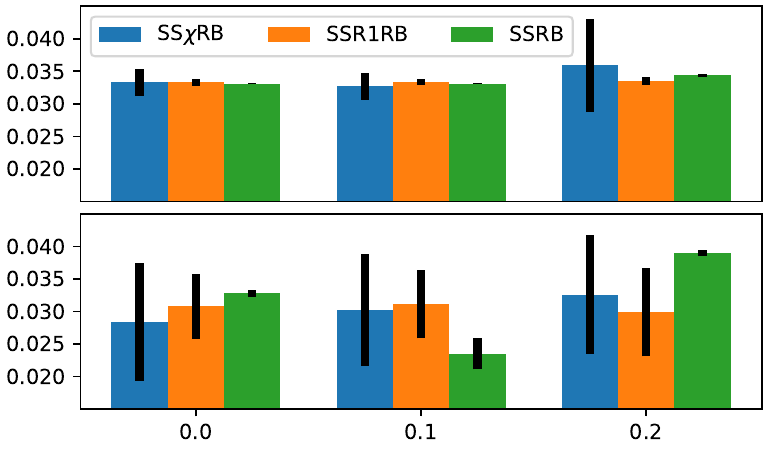}
\end{overpic}
\caption{Estimated weight-2 $\text{SU}(2)$ error rates $p_2$ obtained by simulation under various models of SPAM error and a common per-gate error of $\rho\mapsto U\rho U^\dag$ for $U = \exp(-i 0.04 J_z^2)$.  For each $\ell \in \{0,\ldots,7\}$, state preparation error is modeled by a random $\text{SU}(2)$ rotation through an angle $\phi \in \{0, 0.1, 0.2\}$ (left to right).  In the top row, measurement error is modeled by a single coherent rotation of all POVM effects through the angle $\phi$.  In the bottom row, measurement error is modeled by randomly permuting the POVM effects.}
\label{fig:spam_robustness_main_text}
\end{figure}

Fixing $j = 7/2$, we simulate the execution of these synthetic RB protocols with $10^4$ randomly chosen circuits and infinitely many shots per circuit.  We consider coherent gate noise of the form $\rho\mapsto U\rho U^\dag$ with $U = \exp(-i\gamma J_z^2)$.  For small $\gamma$, such noise produces nontrivial $\text{SU}(2)$ errors localized to the irrep $k = 2$.  We set $\gamma = 0.04$, for which the theoretically expected $\text{SU}(2)$ error rates in the limit of infinite sample size are
\begin{equation}
(p_0, p_2, p_4, p_6)\approx (0.967, 0.0330, 1.43\times 10^{-4}, 1.11\times 10^{-7})
\end{equation}
and $p_{1, 3, 5, 7} = 0$.  Any SPAM-robust protocol should produce error rates that converge to these.

Our models for SPAM error are chosen to test the performance of our protocols.  For each $\ell\in \{0, \ldots, 7\}$, state preparation error is modeled by taking $|\ell\rangle\langle\ell|\mapsto V_\ell|\ell\rangle\langle\ell|V_\ell^\dag$, where $V_\ell = \exp(-i\phi\vec{n}_{\ell}\cdot \vec{J})$ for a fixed angle $\phi$ and a random unit vector $\vec{n}_{\ell}\in \mathbb{R}^3$.  We model measurement error by either
\begin{enumerate}
\item a coherent rotation of all measurement effects about a single random axis by the same angle $\phi$ as before or
\item a random (but fixed) permutation of the measurement effects.
\end{enumerate}
Note that measurement errors preserve the condition that the POVM effects sum to the identity.

\begin{figure}[!htb]
\centering
\begin{overpic}[width=0.925\linewidth, trim={0.65cm 3.2cm 1.1cm 3cm}, clip]{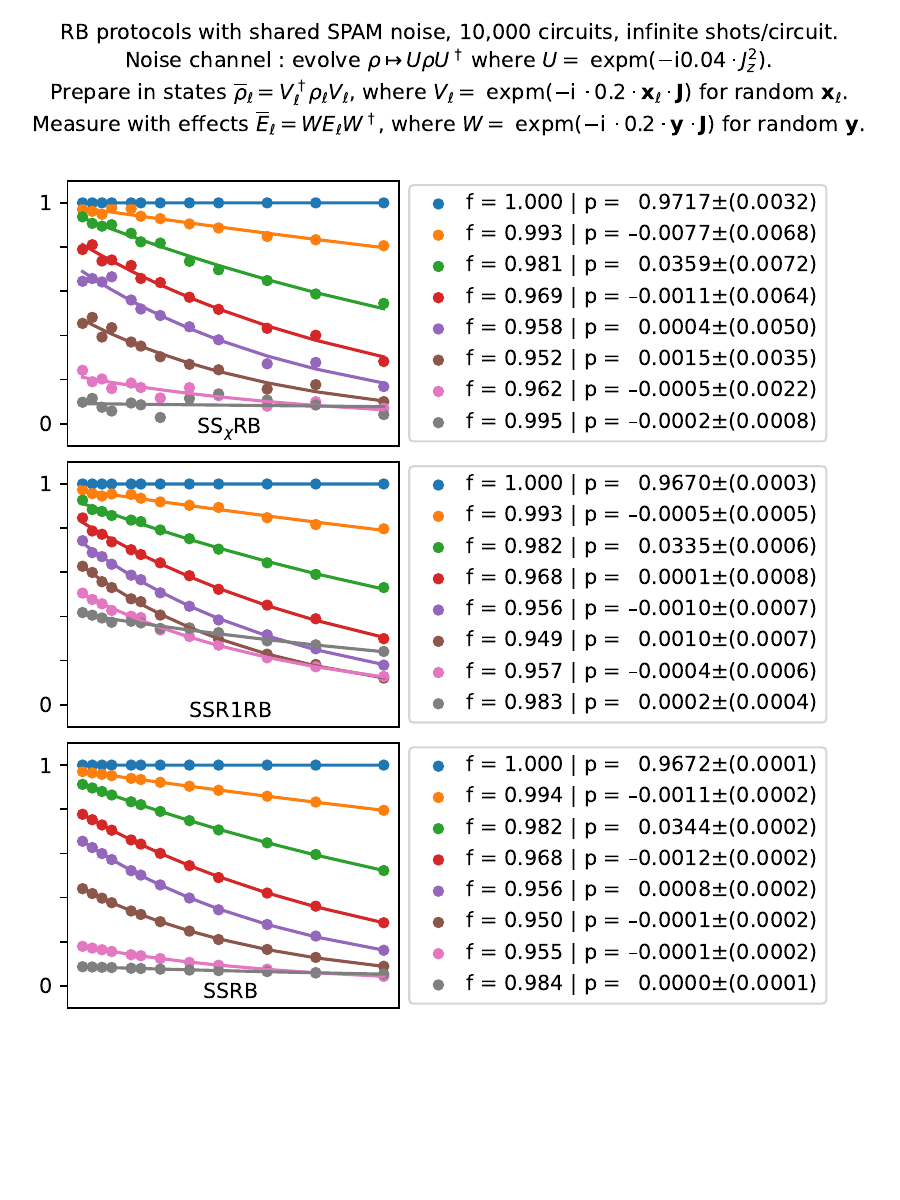}
\end{overpic}
\caption{Synthetic survival probabilities (and exponential fits thereof) for three synthetic RB protocols using the same gate and SPAM errors as the top panel of Figure \ref{fig:spam_robustness_main_text}, with $\phi = 0.2$.}
\label{fig:syn_prob_decays_rotational_meas_err_main_text}
\end{figure}

In Figure \ref{fig:spam_robustness_main_text}, we plot estimates for the nontrivial error rate $p_2$ across three protocols.  The corresponding uncertainties satisfy $\sigma_\text{SS$\chi$RB} > \sigma_\text{SSR1RB} > \sigma_\text{SSRB}$, as expected from analytical results in the noiseless case.  SSRB produces both accurate and precise results in the absence of SPAM error (top left of Figure \ref{fig:spam_robustness_main_text}), but becomes less reliable than the SPAM-robust protocols as $\phi$ increases.

For measurement error corresponding to a \emph{coherent rotation} of POVM effects (top row of Figure \ref{fig:spam_robustness_main_text}), $\phi$ quantifies the amount of error in both state preparation and measurement.  For $\phi = 0.0$, all protocols estimate the correct $p_2$ to within error bars, but for $\phi = 0.1$ and $\phi = 0.2$, SSRB does not.\footnote{While $\text{SU}(2)$ rotations that act identically on all measurement effects leave the synthetic effect inside the target irrep, there is still some corruption of the signal due to gate noise.}  Hence the high precision of SSRB can be misleading.  The uncertainty of SSR1RB is smaller than that of SS$\chi$RB by a factor of 4 or more, and that of SSRB is smaller by a factor of 20 or more.  Hence SS$\chi$RB needs roughly an order of magnitude more shots than SSR1RB to achieve similar precision.

For measurement error corresponding to a \emph{random permutation} of POVM effects (bottom row of Figure \ref{fig:spam_robustness_main_text}), $\phi$ quantifies the amount of error in state preparation only.  As $\phi$ increases, the results of SSRB become substantially less accurate.  This loss in accuracy is mitigated by SS$\chi$RB and to an even greater extent by SSR1RB, at a cost in precision.  SSRB produces both underestimates ($\phi = 0.1$) and overestimates ($\phi = 0.2$) for $p_2$.  The discrepancy between the uncertainties is smaller in this case: relative to that of SS$\chi$RB, that of SSR1RB is smaller by an $O(1)$ factor and that of SSRB is smaller by an $O(10)$ factor.

\begin{figure}[!htb]
\centering
\begin{overpic}[width=0.925\linewidth, trim={0.65cm 3.2cm 1.1cm 3cm}, clip]{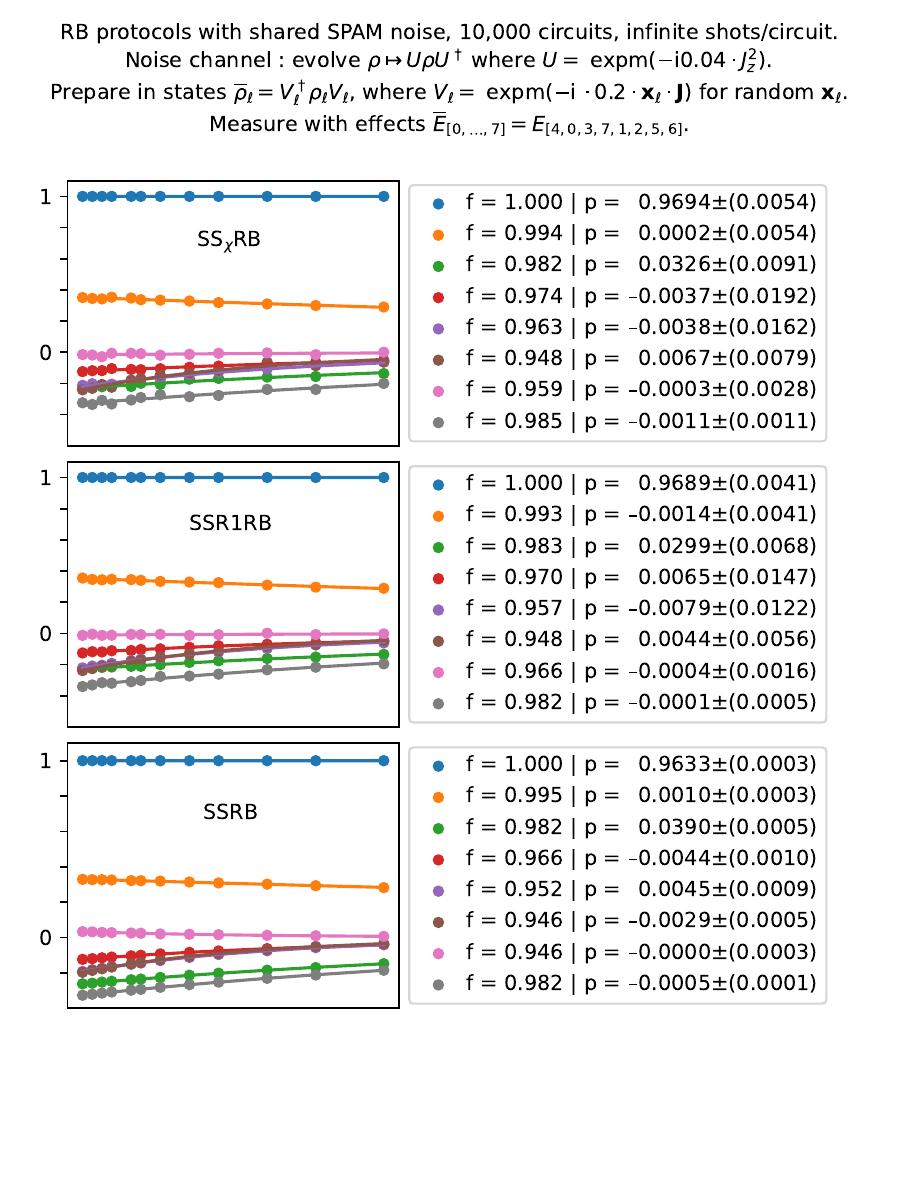}
\end{overpic}
\caption{Synthetic survival probabilities (and exponential fits thereof) for three synthetic RB protocols using the same gate and SPAM errors as the bottom panel of Figure \ref{fig:spam_robustness_main_text}, with $\phi = 0.2$.}
\label{fig:syn_prob_decays_permutation_meas_err_main_text}
\end{figure}

In Figures \ref{fig:syn_prob_decays_rotational_meas_err_main_text} and \ref{fig:syn_prob_decays_permutation_meas_err_main_text}, we show examples of exponential curve fits used to compute the $f_k$ and their corresponding uncertainties, which are then propagated to the $p_k$.  Colors in the legend (top to bottom) indicate irreps 0 through 7.  Negative error rates are spurious and are implicitly cut off at 0.

We provide additional examples in Appendix \ref{app:numerics}.

\tocless{\section{Discussion}}

In this paper, we have developed a comprehensive framework for randomized benchmarking---synthetic RB---that both subsumes many existing protocols and provides mechanisms for increasing their efficiency.  Synthetic RB extends the regime of feasibility of filtered RB methods to high-dimensional quantum systems and expands the range of gate groups that can be reliably benchmarked, eliminating the need to find appropriate character groups on a case-by-case basis.  When applied to rotationally invariant systems, this framework yields sample complexity improvements of several orders of magnitude over character RB.  Synthetic RB thus provides the first practical RB tools for high-dimensional spin qudits.

We have used $G = \text{SU}(2)$ as our primary case study.  This choice was motivated both by its fundamental role in quantum physics and by its direct relevance to spin-based quantum error-correcting codes, where global $\text{SU}(2)$ rotations realize logical operations.  Benchmarking with respect to $\text{SU}(2)$ allows us to extract precisely the rotationally invariant error rates that spin codes are designed to mitigate, suggesting an important role for synthetic RB protocols in the broader effort toward fault-tolerant quantum computation with spin qudits.  At the same time, realistic logical operations on spin codes are drawn not from the full group $\text{SU}(2)$, but rather from finite subgroups such as the binary icosahedral group $2I$.  Performing logical randomized benchmarking \cite{Combes} of spin codes therefore requires sampling from such discrete subgroups, which has the downside that the physical meaning of the resulting error rates becomes less transparent.  Balancing interpretability against operational relevance in the logical setting remains an open problem.  Another possible extension of $\text{SU}(2)$ RB stems from the fact that our working assumption about the native operations on spin qudits being $\text{SU}(2)$ is not strictly true: for example, transitions between neighboring levels are individually addressable in nuclear spins \cite{Yu}.  Application-specific benchmarking using the actual native operations of a given platform (see Appendix \ref{app:native}) may be fruitful.  Such an approach may comprise a ``direct'' version of $\text{SU}(2)$ rather than Clifford RB \cite{DRB-Proctor, DRB-Polloreno}.  Finally, as we sketch in Appendix \ref{app:contraction}, it would be fascinating to draw connections between $\text{SU}(2)$ RB and bosonic RB \cite{BosonicRB-2019, BosonicRB-2020, BosonicRB-2024-1, BosonicRB-2024-2, BosonicRB-2024-3} by considering the large-spin limit of a spin qudit to obtain a noncompact group of phase space symmetries via Wigner-\.In\"on\"u contraction.  Such a correspondence could help to unify RB techniques across discrete and continuous systems.

Beyond application-specific issues, it would be interesting to extend our techniques to handle multiplicities in the superoperator representation of the benchmarking group.  Other auxiliary technical problems include:
\begin{itemize}
\item Finding a way to optimize the frame (either analytically or numerically) to minimize the variance of the sample mean in finite-frame RB.  We comment briefly on some possible approaches in Appendix \ref{app:syntheticgates}.
\item Finding a way to optimize non-synthetic (i.e., physical) SPAM to maximize overlap with a target irrep. (This strategy is rendered unnecessary by synthetic SPAM.)
\end{itemize}
We leave the resolution of these problems for future work.

\tocless{\section*{Acknowledgements}}

We thank Andrew Baczewski, Jonathan Gross, Jordan Hines, Corey Ostrove, and Timothy Proctor for interesting discussions and pointers to the literature.  We thank the members of the Fundamental Quantum Technologies Laboratory at the University of New South Wales, including Andrea Morello, Rocky Su, and Xi Yu, for discussions about the experimental implementation of our protocols.  We especially thank Winton Brown for suggesting that rank-1 projectors might yield lower variance and therefore better sample complexity.

This work was performed in part at the Center for Integrated Nanotechnologies, an Office of Science User Facility operated for the U.S. Department of Energy (DOE) Office of Science, and partially funded by the U.S. Army Research Office (contract no.\ W911NF-23-1-0113).


This article has been coauthored by employees of National Technology \& Engineering Solutions of Sandia, LLC under Contract No.\ DE-NA0003525 with the U.S. Department of Energy (DOE). The employees own all right, title, and interest in and to the article and are solely responsible for its contents. The U.S. Government retains, and the publisher, by accepting the article for publication, acknowledges that the U.S. Government retains, a non-exclusive, paid-up, irrevocable, worldwide license to publish or reproduce the published form of this article or allow others to do so, for U.S. Government purposes. The DOE will provide public access to these results of federally sponsored research in accordance with the DOE Public Access Plan.


This paper describes objective technical results and analysis. Any subjective views or opinions that might be expressed in the paper do not necessarily represent the views of the U.S. Department of Energy or the U.S. Government.

\tocless{\bibliography{\jobname}}

\newpage
\appendix
\clearpage
\onecolumngrid

\tableofcontents

\section{Preliminaries} \label{app:preliminaries}

We set $\hbar = 1$ unless otherwise noted.  We use $\tr$ for the operator trace and $\Tr$ for the superoperator trace.

\subsection{Hilbert-Schmidt Space} \label{app:HS}

Consider a $d$-level qudit with Hilbert space $\mathbb{C}^d$.  We refer to the space of linear operators $\mathcal{B}(\mathbb{C}^d)$, of which the set of density matrices forms a convex subset, as Hilbert-Schmidt space.  An element of $\mathcal{B}(\mathbb{C}^d)$ can be represented as a $d^2$-component vector (superket), while a linear operator on $\mathcal{B}(\mathbb{C}^d)$ (superoperator) can be represented as a $d^2\times d^2$ transfer matrix.  The (unnormalized) Hilbert-Schmidt inner product on operators/superkets is defined as
\begin{equation}
\llangle A|B\rrangle\equiv \tr(A^\dag B),
\end{equation}
while the (unnormalized) Hilbert-Schmidt inner product on superoperators $X, Y$ can be written as
\begin{equation}
\Tr(X^\dag Y) = \sum_O \llangle O|X^\dag Y|O\rrangle = \sum_O \llangle X[O]|Y[O]\rrangle = \sum_O \tr(X[O]^\dag Y[O]),
\end{equation}
where $O$ runs over any orthonormal operator basis.  The normalized Pauli basis $\{P\}$ and the basis of matrix units $|i\rangle\langle j|$ are both examples of orthonormal bases for Hilbert-Schmidt space in which operators (including density matrices) and superoperators can be written; the latter corresponds to vectorization.  The Frobenius norm is given by $||A||\equiv \llangle A|A\rrangle^{1/2}$.

The Choi representation of a superoperator results from writing the corresponding Kraus operators in an orthogonal basis of choice.  A superoperator is completely positive (CP) if and only if the matrix of Choi coefficients is positive-semidefinite.

One often encounters three desiderata for an (operator) error basis $\{O_i\}$ on $\mathbb{C}^d$, where $i = 0, \ldots, d^2 - 1$:
\begin{enumerate}
\item The (normalized) identity matrix should be the only traceful element: $O_0 = \mathds{1}/\sqrt{d}$, while $\tr(O_i) = 0$ for $i > 0$.
\item Orthonormality: $\tr(O_i O_j) = \delta_{ij}$.
\item Hermiticity: $O_i^\dag = O_i$.
\end{enumerate}
(Note that the second criterion, along with $O_0 = \mathds{1}/\sqrt{d}$, implies the first.) These three criteria imply that $\{O_{i>0}\}$ comprises an orthonormal basis for the Lie algebra of $\text{SU}(d)$.

The first criterion allows for a clean description of trace-preserving (TP) superoperators.  In particular, an error basis that respects the decomposition of the tensor product of the fundamental and antifundamental representations of $\text{SU}(d)$ into the trivial and adjoint representations makes clear which subspaces are invariant under unitary transformations of basis elements.  The second criterion is natural for the following reason.  Let us define the Choi units
\begin{equation}
X_{A, B}[\rho] = A\rho B^\dag,
\end{equation}
where $A, B$ are basis elements.  Using an intermediate basis of matrix units, we find that
\begin{equation}
\Tr(X_{A, B}^\dag X_{A', B'}) = \sum_{i, j = 0}^{d-1} \tr(B|j\rangle\langle i|A^\dag A'|i\rangle\langle j|B'^\dag) = \sum_{i, j = 0}^{d-1} \langle i|A^\dag A'|i\rangle\langle j|B'^\dag B|j\rangle = \tr(A^\dag A')\tr(B'^\dag B).
\end{equation}
Hence the Choi units are mutually orthogonal with respect to the Hilbert-Schmidt inner product on superoperators if and only if the basis elements are mutually orthogonal with respect to the Hilbert-Schmidt inner product on operators.  Finally, the third criterion is natural if these operators are thought of as parametrizing small errors that exponentiate to finite unitary errors.

Examples of operator bases satisfying these criteria include the $n$-qubit Pauli basis with $d = 2^n$ (which is also unitary) and the Gell-Mann basis.  Non-examples include the basis of matrix units (which violates the first criterion), the generalized Pauli basis $\{X^a Z^b/\sqrt{d} \,|\, a, b = 0, \ldots, d - 1\}$ generated by $d$-dimensional clock ($Z$) and shift ($X$) matrices (which is a unitary rather than a Hermitian operator basis for $d > 2$), and the spherical tensor basis (which is real but not Hermitian).

In any preferred basis of the above form, Hermitian operators are represented by real superkets, so CP superoperators (which preserve Hermiticity) are represented by real matrices with $d^4$ mostly free parameters.  Moreover, any TP superoperator has top row $(1, 0, \ldots, 0)$.  Therefore, any CPTP superoperator has $d^2(d^2 - 1)$ mostly free parameters.

\subsection{Spherical Tensor Basis}

We work primarily in the spherical tensor basis, which finds use in applications from spin state tomography to error correction \cite{Perlin, Omanakuttan-1, Omanakuttan-2}.  The Cayley-Hamilton theorem implies that in the spin-$j$ representation, any function of the angular momentum operators $J_x$, $J_y$, $J_z$ is a polynomial of degree at most $2j$.  Thus any operator, and in particular any density matrix, can be expressed as a sum of irreducible tensors of rank $0, 1, \ldots, 2j$ constructed from the $J_i$.

To couple two angular momenta, we use the Clebsch-Gordan coefficients
\begin{equation}
C^{j, m}_{j_1, m_1; j_2, m_2}\equiv \langle j_1, m_1; j_2, m_2|j, m\rangle\equiv \langle j_1, m_1; j_2, m_2|j, m, j_1, j_2\rangle,
\end{equation}
where $\vec{J} = \vec{J}_1 + \vec{J}_2$.  These coefficients can all be chosen to be real.  For fixed $j$, we define the spherical tensor operators
\begin{equation}
T^{(k)}_q = \sqrt{\frac{2k + 1}{2j + 1}}\sum_{m, m' = -j}^j C^{j, m}_{j, m'; k, q}|j, m\rangle\langle j, m'| = \sqrt{\frac{2k + 1}{2j + 1}}\sum_{m' = -j}^j C^{j, m' + q}_{j, m'; k, q}|j, m' + q\rangle\langle j, m'|
\end{equation}
of rank $k = 0, 1, \ldots, 2j$ (where $q = -k, \ldots, k$), which are homogeneous polynomials of degree $k$ in the $J_i$.  These operators are orthonormal with respect to the Hilbert-Schmidt inner product:
\begin{equation}
\tr((T^{(k)}_q)^\dag T^{(k')}_{q'}) = \delta_{kk'}\delta_{qq'}.
\label{Tnormalization}
\end{equation}
Given the $\text{SU}(2)$ transformation properties of spherical tensor operators, the Wigner-Eckart theorem fixes their form up to a $k$-dependent constant; we have chosen this constant to respect the normalization condition \eqref{Tnormalization}.

Let $D(g)$ denote the spin-$j$ representation of an $\text{SU}(2)$ group element $g$, with $\smash{D_{mm'}^j}(g) = \langle j, m|D(g)|j, m'\rangle$.  Spherical tensor operators transform irreducibly under $\text{SU}(2)$:
\begin{equation}
D(g)^\dag T_q^{(k)}D(g) = \sum_{q'} D_{qq'}^k(g)^\ast T_{q'}^{(k)}.
\label{sphericaltensorrotation}
\end{equation}
In infinitesimal form, this rule gives the commutation relations
\begin{align}
[J_\pm, T_q^{(k)}] &= \sqrt{k(k + 1) - q(q\pm 1)}T_{q\pm 1}^{(k)}, \label{Tcommutation1} \\
[J_z, T_q^{(k)}] &= qT_q^{(k)}. \label{Tcommutation2}
\end{align}
More succinctly, $[J_\mu, T_q^{(k)}] = (-1)^{(\mu + |\mu|)/2}\sqrt{1 + |\mu|}\sqrt{k(k + 1)}C_{k, q; 1, \mu}^{k, q + \mu}T_{q + \mu}^{(k)}$ for $\mu\in \{0, \pm 1\}$.  In spherical notation, the scalar product of two irreducible tensor operators of the same rank takes the form
\begin{equation}
U^{(k)}\cdot V^{(k)} = \sum_q (-1)^q U_q^{(k)}\otimes V_{-q}^{(k)}.
\label{scalarproduct}
\end{equation}
This construction is a special case of the fact that the tensor product of two irreducible tensors $U^{(k_1)}$ and $V^{(k_2)}$ of rank $k_1$ and $k_2$ decomposes into irreducible tensors $\{U^{(k_1)}\otimes V^{(k_2)}\}^{(k)}$ of rank $k = |k_1 - k_2|, \ldots, k_1 + k_2$ with components
\begin{equation}
\{U^{(k_1)}\otimes V^{(k_2)}\}_q^{(k)} = \sum_{q_1=-k_1}^{k_1}\sum_{q_2=-k_2}^{k_2} C_{k_1, q_1; k_2, q_2}^{k, q}U_{q_1}^{(k_1)}\otimes V_{q_2}^{(k_2)} \Longleftrightarrow U_{q_1}^{(k_1)}\otimes V_{q_2}^{(k_2)} = \sum_{k = |k_1 - k_2|}^{k_1 + k_2}\sum_{q=-k}^k C_{k_1, q_1; k_2, q_2}^{k, q}\{U^{(k_1)}\otimes V^{(k_2)}\}_q^{(k)}.
\end{equation}
Indeed, when $k_1 = k_2 = k$, the rank-0 irreducible tensor product reduces to
\begin{equation}
\{U^{(k)}\otimes V^{(k)}\}_0^{(0)} = \sum_{q, q' = -k}^k C_{k, q; k, q'}^{0, 0}U_q^{(k)}\otimes V_{q'}^{(k)} = \frac{(-1)^k(U^{(k)}\cdot V^{(k)})}{\sqrt{2k + 1}},
\end{equation}
where we have used $C_{k, q; k, q'}^{0, 0} = \frac{(-1)^{k - q}}{\sqrt{2k + 1}}\delta_{q, -q'}$.  Other useful properties of spherical tensor operators include $(T^{(k)}_q)^\ast = T^{(k)}_q$ and
\begin{equation}
(T^{(k)}_q)^\dag = (T^{(k)}_q)^T = (-1)^q T^{(k)}_{-q}.
\label{daggersign}
\end{equation}
In terms of Wigner 6-$j$ symbols, the multiplication rule for spherical tensor operators is
\begin{equation}
T_{q_1}^{(k_1)}T_{q_2}^{(k_2)} = \sum_{k = |k_1 - k_2|}^{k_1 + k_2} (-1)^{2j + k}\sqrt{(2k_1 + 1)(2k_2 + 1)}\left\{\begin{array}{ccc} k_1 & k_2 & k \\ j & j & j \end{array}\right\}C^{k, q_1 + q_2}_{k_1, q_1; k_2, q_2}T_{q_1 + q_2}^{(k)}.
\label{multiplication}
\end{equation}
The trace of a product of spherical tensor operators takes the following form:
\begin{equation}
\tr(T^{(k_1)}_{q_1}\cdots T^{(k_n)}_{q_n}) = \left[\frac{\prod_{i=1}^n (2k_i + 1)}{(2j + 1)^n}\right]^{1/2}\delta_{\sum_{i=1}^n q_i, 0}\sum_{m=-j}^j \prod_{i=1}^n C^{j, m + \mu_i}_{k_i, q_i; j, m + \mu_{i-1}},
\label{traceproduct}
\end{equation}
where $\mu_i = \sum_{j=1}^i q_j$.  We recommend \cite{Varshalovich} as a resource for formulas such as those collected here.

\subsection{Parametrizations of \texorpdfstring{$\text{SU}(2)$}{SU(2)}}

In the Euler angle parametrization
\begin{equation}
D(g) = D_z(\alpha)D_y(\beta)D_z(\gamma) = e^{-i\alpha J_z}e^{-i\beta J_y}e^{-i\gamma J_z},
\end{equation}
the Wigner $D$-matrices and $d$-matrices are defined by
\begin{equation}
D_{mm'}^j(g) = \langle j, m|D(g)|j, m'\rangle = e^{-im\alpha}d_{mm'}^j(\beta)e^{-im'\gamma}, \qquad d_{mm'}^j(\beta) = \langle j, m|e^{-i\beta J_y}|j, m'\rangle,
\end{equation}
with $d_{mm'}^j$ real.  Any $\text{SO}(3)$ rotation can be written in terms of Euler angles
\begin{equation}
\alpha\in [0, 2\pi), \qquad \beta\in [0, \pi], \qquad \gamma\in [0, 2\pi).
\label{EAdomain}
\end{equation}
To parametrize the $\text{SU}(2)$ group manifold, we may extend the range of $\alpha$ or $\gamma$ to $[0, 4\pi)$.  On the other hand, in the axis-angle parametrization $D(g) = e^{-i\theta\vec{n}\cdot \smash{\vec{J}}}$, any $\text{SO}(3)$ rotation is uniquely specified by a rotation angle $\theta$ and a unit vector $\vec{n} = (\sin\Theta\cos\Phi, \linebreak[1] \sin\Theta\sin\Phi, \linebreak[1] \cos\Theta)$, where
\begin{equation}
\theta\in [0, \pi], \qquad \Theta\in [0, \pi], \qquad \Phi\in [0, 2\pi).
\label{AAdomain}
\end{equation}
To cover $\text{SU}(2)$, we may extend the range of $\theta$ to $[0, 2\pi)$.  The volume of the group corresponding to the domain \eqref{EAdomain} or \eqref{AAdomain} is $8\pi^2$, while the ``double'' domain has volume $16\pi^2$.  In practice, we restrict our attention to the domain \eqref{EAdomain} or \eqref{AAdomain} because our primary interest lies in representations of $\text{SU}(2)$ that are also representations of $\text{SO}(3)$.  Writing the orthogonality relations (and other integral relations) for Wigner $D$-matrices in arbitrary $\text{SU}(2)$ representations requires integrating over the double domain \cite{Varshalovich}; for our applications, however, the upper index $k$ of the Wigner $D$-matrix $D^k$ is always an integer, so the single domain suffices.

The Euler angles, rotation angles, etc., of an $\text{SU}(2)$ group element $g = g(\alpha, \beta, \gamma) = g(\theta, \Theta, \Phi)$ determine its image under any desired representation.  Working in the fundamental representation of $\text{SU}(2)$ allows for easy interconversion between different parametrizations.  For instance, we have
\begin{equation}
D^{1/2}(\alpha, \beta, \gamma) = e^{-i\alpha\sigma_3/2}e^{-i\beta\sigma_2/2}e^{-i\gamma\sigma_3/2} = \left[\begin{array}{cc} e^{-i(\alpha + \gamma)/2}\cos\frac{\beta}{2} & -e^{-i(\alpha - \gamma)/2}\sin\frac{\beta}{2} \\[2 pt] e^{i(\alpha - \gamma)/2}\sin\frac{\beta}{2} & e^{i(\alpha + \gamma)/2}\cos\frac{\beta}{2} \end{array}\right].
\end{equation}
On the other hand,
\begin{equation}
D^{1/2}(\theta, \Theta, \Phi) = e^{-i\theta(\sin\Theta\cos\Phi\sigma_1 + \sin\Theta\sin\Phi\sigma_2 + \cos\Theta\sigma_3)/2} = \left[\begin{array}{cc} \cos\frac{\theta}{2} - i\cos\Theta\sin\frac{\theta}{2} & -ie^{-i\Phi}\sin\Theta\sin\frac{\theta}{2} \\[2 pt] -ie^{i\Phi}\sin\Theta\sin\frac{\theta}{2} & \cos\frac{\theta}{2} + i\cos\Theta\sin\frac{\theta}{2} \end{array}\right].
\end{equation}
Inspecting the above expressions yields the following relations between rotation angles $(\theta, \Theta, \Phi)$ and Euler angles $(\alpha, \beta, \gamma)$:
\begin{equation}
\cos\frac{\theta}{2} = \cos\frac{\beta}{2}\cos\frac{\alpha + \gamma}{2}, \qquad \tan\Theta = \frac{\tan\frac{\beta}{2}}{\sin\frac{\alpha + \gamma}{2}}, \qquad \Phi = \frac{\pi}{2} + \frac{\alpha - \gamma}{2}.
\label{EAtoAA}
\end{equation}
The inverse relations are
\begin{equation}
\sin\frac{\beta}{2} = \sin\Theta\sin\frac{\theta}{2}, \qquad \tan\frac{\alpha + \gamma}{2} = \cos\Theta\tan\frac{\theta}{2}, \qquad \frac{\alpha - \gamma}{2} = \Phi - \frac{\pi}{2}.
\end{equation}
In terms of Euler angles, a normalized Haar integral over $\text{SU}(2)$ takes the form
\begin{equation}
\langle f(g)\rangle_{\text{SU}(2)}\equiv \frac{1}{8\pi^2}\int_0^{2\pi} d\alpha\int_0^\pi d\beta\sin\beta\int_0^{2\pi} d\gamma\, f(\alpha, \beta, \gamma) = \frac{1}{8\pi^2}\int_0^{2\pi} d\alpha\int_{-1}^1 d(\cos\beta)\int_0^{2\pi} d\gamma\, f(\alpha, \beta, \gamma).
\label{normalizedHaar}
\end{equation}
With respect to the Haar measure, the Euler angles $\alpha$ and $\gamma$ are distributed uniformly from 0 to $2\pi$ while $\cos\beta$ is distributed uniformly from $-1$ to 1.  Therefore, to sample elements uniformly at random from $\text{SU}(2)$, we may set $\alpha = 2\pi a$, $\beta = \arccos(1 - 2b)$, $\gamma = 2\pi c$ where $a, b, c$ are drawn uniformly at random from $[0, 1]$.  Note that we have chosen the argument of the $\arccos$ so that $b = 0$ corresponds to $\beta = 0$.  As long as $\beta\neq 0, \pi$, the Euler angles $\alpha, \beta, \gamma$ uniquely determine the rotation.  Under \eqref{EAtoAA}, the Haar measure transforms as follows:
\begin{equation}
\sin\beta\, |d\alpha\wedge d\beta\wedge d\gamma| = 4\sin^2\frac{\theta}{2}\sin\Theta\, |d\theta\wedge d\Theta\wedge d\Phi|.
\end{equation}
Correspondingly, we have
\begin{equation}
\int_0^{2\pi} d\alpha\int_0^\pi d\beta\sin\beta\int_0^{2\pi} d\gamma = 4\int_0^\pi d\theta\sin^2\frac{\theta}{2}\int_0^\pi d\Theta\sin\Theta\int_0^{2\pi} d\Phi = 8\pi^2.
\end{equation}
In other words, any $g\in \text{SU}(2)$ can be parametrized by its rotation axis $\vec{n}\in S^2$ and its rotation angle $\theta\in [0, \pi]$; to sample $g$ uniformly at random, we sample $\vec{n}$ uniformly at random from $S^2$ and $\theta$ according to the measure $\frac{2}{\pi}\sin^2\frac{\theta}{2}\, d\theta$.

Wigner $D$-matrix elements are orthogonal with respect to the Haar inner product on $\text{SU}(2)$:
\begin{equation}
\left\langle D_{qr}^k(g)^\ast D_{q'r'}^{k'}(g)\right\rangle_{\text{SU}(2)} = \frac{1}{2k + 1}\delta_{kk'}\delta_{qq'}\delta_{rr'}.
\label{Dmatrixorthogonality}
\end{equation}
In terms of Wigner $d$-matrices, we have
\begin{equation}
\frac{1}{2}\int_0^\pi d\beta\sin\beta\, d_{qr}^k(\beta)d_{qr}^{k'}(\beta) = \frac{1}{2k + 1}\delta_{kk'}.
\end{equation}
Since $\text{SU}(2)$ characters are traces of Wigner $D$-matrices,
\begin{equation}
\chi_k(\alpha, \beta, \gamma) = \sum_{q=-k}^k D_{qq}^k(\alpha, \beta, \gamma) = \sum_{q=-k}^k e^{-iq(\alpha + \gamma)}d_{qq}^k(\beta),
\end{equation}
we have the following orthogonality relation for characters: $\langle\chi_k(g)^\ast\chi_{k'}(g)\rangle_{\text{SU}(2)} = \delta_{kk'}$.  This relation can be simplified by noting that the character of a group element depends only on its angle of rotation and not on its axis:
\begin{equation}
\chi_k(\alpha, \beta, \gamma) = \chi_k(\theta) = \frac{\sin((2k + 1)\theta/2)}{\sin(\theta/2)},
\end{equation}
where we identify parameters via \eqref{EAtoAA}. (Indeed, any group element can be conjugated to a convenient maximal torus, where $g = g(0, \beta, 0)$ and $\chi_k(0, \beta, 0) = \sin((2k + 1)\beta/2)/\sin(\beta/2)$.) We then obtain\footnote{Note that \eqref{charorthog}, as written, holds for arbitrary (half-)integers $k$ and $k'$.  Had we started directly from $\langle\chi_k(g)^\ast\chi_{k'}(g)\rangle_{\text{SU}(2)} = \delta_{kk'}$ and our definition \eqref{normalizedHaar}, we would have obtained a relation with a truncated range of integration, $\frac{2}{\pi} \linebreak[1] \int_0^{\pi} \linebreak[1] d\theta \linebreak[1] \sin^2\frac{\theta}{2} \linebreak[1] \chi_k(\theta)^\ast \linebreak[1] \chi_{k'}(\theta) = \delta_{kk'}$, which holds when $k$ and $k'$ are both integers or half-integers.}
\begin{equation}
\frac{1}{\pi}\int_0^{2\pi} d\theta\sin^2\frac{\theta}{2}\chi_k(\theta)^\ast\chi_{k'}(\theta) = \delta_{kk'}.
\label{charorthog}
\end{equation}
The characters are real, so no complex conjugation is necessary.

\section{Derivations} \label{app:derivations}

\subsection{\texorpdfstring{$\text{SU}(2)$}{SU(2)} Operations}

A global $\text{SU}(2)$ rotation has a $(2j + 1)^2\times (2j + 1)^2$ block diagonal transfer matrix with blocks of size $1, 3, \ldots, 4j + 1$.  In terms of the adjoint action on operators, these irreps correspond to the spans of the independent polynomials of degree $0, 1, \ldots, 2j$ in the angular momentum operators. (One can write the projectors $\Pi_k$ onto the irrep subspaces in the superoperator representation using Haar integration.) To demonstrate these statements, it is convenient to work in the spherical tensor basis.  Using the properties $D(\alpha, \beta, \gamma)^\dag = D(-\gamma, -\beta, -\alpha)$ and $D^k_{qq'}(\alpha, \beta, \gamma)^\ast = D^k_{q'q}(-\gamma, -\beta, -\alpha)$, \eqref{sphericaltensorrotation} is equivalent to
\begin{equation}
D(g)T_q^{(k)}D(g)^\dag = \sum_{q'} D_{q'q}^k(g)T_{q'}^{(k)}.
\end{equation}
Given any $g\in \text{SU}(2)$, we thus compute the matrix elements of its superoperator representation $\mathcal{G}$:
\begin{equation}
\llangle T_q^{(k)}|\mathcal{G}|T_{q'}^{(k')}\rrangle = \llangle T_q^{(k)}|D(g)T_{q'}^{(k')}D(g)^\dag\rrangle = \sum_{q''} D_{q''q'}^{k'}(g)\llangle T_q^{(k)}|T_{q''}^{(k')}\rrangle = \delta_{kk'}D_{qq'}^k(g).
\end{equation}
Therefore, if $\rho\mapsto g\rho g^\dag$, then the superoperator representation matrix is simply $\mathcal{G}(g) = \operatorname{diag}(D^0(g), D^1(g), \ldots, D^{2j}(g))$.  Explicitly, we may expand an arbitrary operator $\rho$ as
\begin{equation}
\rho = \sum_{k=0}^{2j}\sum_{q=-k}^k a_{k, q}T_q^{(k)}, \qquad a_{k, q}\equiv \llangle T_q^{(k)}|\rho\rrangle,
\end{equation}
where the coefficients are the components of the superket $|\rho\rrangle$.  Thus, for instance, $\Pi_k|\rho\rrangle = \sum_{q=-k}^k a_{k, q}|T_q^{(k)}\rrangle$.  Then we have
\begin{equation}
D(g)\rho D(g)^\dag = \sum_{k=0}^{2j}\sum_{m=-k}^k \left(\sum_{q=-k}^k D_{mq}^k(g)a_{k, q}\right)T_m^{(k)}.
\end{equation}
In other words, if rotations act contravariantly on the basis vectors, then they act covariantly on the coefficients.

The spherical tensor Choi units
\begin{equation}
\rho\mapsto T^{(k)}_q\rho(T^{(k')}_{q'})^\dag
\label{Choiunit}
\end{equation}
form an orthonormal basis for the space of spin-$j$ superoperators.  When twirled over $\text{SU}(2)$, \eqref{Choiunit} becomes
\begin{align}
\rho &\mapsto \left\langle[D(g)^\dag T^{(k)}_q D(g)]\rho[D(g)^\dag T^{(k')}_{q'}D(g)]^\dag\right\rangle_{\text{SU}(2)} \\
&= \sum_{r=-k}^k\sum_{r'=-k'}^{k'} \left\langle D_{qr}^k(g)^\ast D_{q'r'}^{k'}(g)\right\rangle_{\text{SU}(2)}T_r^{(k)}\rho(T_{r'}^{(k')})^\dag \\
&= \frac{1}{2k + 1}\smash[t]{\delta_{kk'}\delta_{qq'}\sum_{r=-k}^k T_r^{(k)}\rho(T_r^{(k)})^\dag},
\end{align}
where we have used \eqref{sphericaltensorrotation} and \eqref{Dmatrixorthogonality}.  Thus \eqref{Choiunit} twirls either to zero or to a uniformly random weight-$k$ error $\mathcal{G}^{(k)}$.

The $\mathcal{G}^{(k)}$ are completely positive maps because they have diagonal Kraus representations, but they are generally not trace-preserving.  Indeed, the Kraus operators of $\mathcal{G}^{(k)}$ are not properly normalized because
\begin{equation}
\frac{1}{2k + 1}\sum_{q=-k}^k (T_q^{(k)})^\dag T_q^{(k)} = \frac{1}{2k + 1}\sum_{q=-k}^k (-1)^q T_{-q}^{(k)}T_q^{(k)} = \frac{\mathds{1}}{2j + 1},
\end{equation}
which implies that $\tr(\mathcal{G}^{(k)}[\rho]) = \tr(\rho)/(2j + 1)$.  On the other hand, $(2j + 1)\mathcal{G}^{(k)}$ is a properly normalized channel.  For example, the $\text{SU}(2)$-twirled weight-1 error map is proportional to the Landau-Streater channel, or the spin analogue of the depolarizing channel:
\begin{equation}
\mathcal{G}^{(1)}[\rho] = \frac{1}{3}\sum_{q=-1}^1 T_q^{(1)}\rho(T_q^{(1)})^\dag = \frac{1}{2j + 1}\mathcal{E}_\text{LS}[\rho], \qquad \mathcal{E}_\text{LS}[\rho]\equiv \frac{1}{j(j + 1)}(J_x\rho J_x + J_y\rho J_y + J_z\rho J_z).
\end{equation}
To see this, note that
\begin{align}
C^{j, m}_{j, m'; 1, \pm 1} = \mp\sqrt{\frac{(j\mp m + 1)(j\pm m)}{2j(j + 1)}}\delta_{m, m'\pm 1} &\implies T^{(1)}_{\pm 1} = \mp\sqrt{\frac{3}{2j(j + 1)(2j + 1)}}J_\pm, \\
C^{j, m}_{j, m'; 1, 0} = \frac{m}{\sqrt{j(j + 1)}}\delta_{mm'} &\implies T^{(1)}_0 = \sqrt{\frac{3}{j(j + 1)(2j + 1)}}J_z,
\end{align}
for which the commutation relations \eqref{Tcommutation1} and \eqref{Tcommutation2} reduce to the angular momentum algebra $[J_z, J_\pm] = \pm J_\pm$ and $[J_\pm, J_\mp] = \pm 2J_z$.  Another special case of the $T^{(k)}_q$ is $T^{(0)}_0 = \mathds{1}/\sqrt{2j + 1}$, which follows from $C^{j, m}_{j, m'; 0, 0} = \delta_{mm'}$.

\subsection{\texorpdfstring{$\text{SU}(2)$}{SU(2)} ``Fourier'' Transform}

We now derive for $\text{SU}(2)$ the Fourier-like linear transformation $F$ between RB quality parameters and $\text{SU}(2)$ error rates, which in the case of Pauli channels is the Walsh-Hadamard transform. (This is not to be confused with the generalized Fourier transform between the group element basis and the representation basis of $L^2(G)$ for a compact group $G$.) The simplest way to view $F$ is as a matrix of angular momentum recoupling coefficients:
\begin{equation}
F_{kk'} = (-1)^{2j + k + k'}\left\{\begin{array}{ccc} k & j & j \\ k' & j & j \end{array}\right\},
\label{recoupling}
\end{equation}
for $k, k'$ ranging from 0 to $2j$. (Here, $F_{kk'}$ is what we formerly called $g^{(k)}_{k'}$.) By the symmetries of the 6-$j$ symbol, this matrix is manifestly symmetric: $F_{kk'} = F_{k'k}$.

A heuristic derivation is the following.  By vectorization in the basis of angular momentum ($J_z$ eigenstate) matrix units, the spin-$j$ operator $\rho$ can be viewed as a state in the tensor product Hilbert space of two spins.  By the identity $\operatorname{vec}(ABC) = (C^T\otimes A)\operatorname{vec}(B)$, the action of $\mathcal{G}^{(k)}$ on $\rho$ can be related to the action of a scalar product of two spin-$j$ tensor operators on the corresponding bipartite state, as in \eqref{scalarproduct}.  We now use the following characterization of the 6-$j$ symbol via the Wigner-Eckart theorem.  In the absence of non-rotational quantum numbers, if the tensor operators $U^{(k)}$ and $V^{(k)}$ act on spin-$j_1$ and spin-$j_2$ subsystems, then the matrix elements of their scalar product in the coupled basis $\vec{J} = \vec{j}_1 + \vec{j}_2$ are given by
\begin{equation}
\langle J', M', j_1', j_2'|U^{(k)}\cdot V^{(k)}|J, M, j_1, j_2\rangle = \delta_{JJ'}\delta_{MM'}(-1)^{J + j_1 + j_2'}\left\{\begin{array}{ccc} J & j_2' & j_1' \\ k & j_1 & j_2 \end{array}\right\}\langle j_1'||U^{(k)}||j_1\rangle\langle j_2'||V^{(k)}||j_2\rangle.
\end{equation}
In our case, $j_1 = j_2 = j_1' = j_2' = j$, and the coupled basis for the tensor product Hilbert space is given by vectorized spherical tensor operators.  Setting $J = k'$ and $M = q'$ and using the fact that our normalization corresponds to the reduced matrix elements $\langle j||T^{(k)}||j\rangle = \sqrt{2k + 1}$, which cancel the normalization of $\mathcal{G}^{(k)}$, yields \eqref{recoupling}.

Here is a more careful derivation.  The entries of the matrix $F$ can be written as
\begin{equation}
F_{kk'} = \frac{(-1)^{q'}}{2k + 1}\sum_{q=-k}^k (-1)^q\tr(T_q^{(k)}T_{q'}^{(k')}T_{-q}^{(k)}T_{-q'}^{(k')}),
\end{equation}
where $q'$ is arbitrary.  It will be convenient to average over $q'$:
\begin{equation}
F_{kk'} = \frac{1}{(2k + 1)(2k' + 1)}\sum_{q=-k}^k\sum_{q'=-k'}^{k'} (-1)^{q + q'}\tr(T_q^{(k)}T_{q'}^{(k')}T_{-q}^{(k)}T_{-q'}^{(k')}).
\label{Faveraged}
\end{equation}
In any orthonormal basis for the spin-$j$ Hilbert space, we have
\begin{align}
\tr(T_q^{(k)}T_{q'}^{(k')}T_{-q}^{(k)}T_{-q'}^{(k')}) &= \sum_{a, b, c, d} \langle a|T_q^{(k)}|b\rangle\langle c|T_{-q}^{(k)}|d\rangle\langle b|T_{q'}^{(k')}|c\rangle\langle d|T_{-q'}^{(k')}|a\rangle \\
&= \sum_{a, b, c, d} (\langle a|\otimes \langle c|)(T_q^{(k)}\otimes T_{-q}^{(k)})(|b\rangle\otimes |d\rangle)(\langle b|\otimes \langle d|)(T_{q'}^{(k')}\otimes T_{-q'}^{(k')})(|c\rangle\otimes |a\rangle) \\
&= \sum_{a, c} (\langle a|\otimes \langle c|)(T_q^{(k)}\otimes T_{-q}^{(k)})(T_{q'}^{(k')}\otimes T_{-q'}^{(k')})(|c\rangle\otimes |a\rangle).
\end{align}
Working in the $J_z$ eigenbasis, we have
\begin{equation}
\tr(T_q^{(k)}T_{q'}^{(k')}T_{-q}^{(k)}T_{-q'}^{(k')}) = \sum_{m, n = -j}^j (\langle j, m|\otimes \langle j, n|)(T_q^{(k)}\otimes T_{-q}^{(k)})(T_{q'}^{(k')}\otimes T_{-q'}^{(k')})(|j, n\rangle\otimes |j, m\rangle) = \tr_{j\otimes j}[S(T_q^{(k)}\otimes T_{-q}^{(k)})(T_{q'}^{(k')}\otimes T_{-q'}^{(k')})],
\end{equation}
where we write $\tr_{j\otimes j}$ for the trace in the tensor product Hilbert space (in contrast to $\tr\equiv \tr_j$, the trace in a single copy) and $S$ is the swap operator that acts in the uncoupled basis by exchanging the two tensor factors:
\begin{equation}
S(|j, m\rangle\otimes |j, n\rangle) = |j, n\rangle\otimes |j, m\rangle.
\end{equation}
Writing $|j, m; j, n\rangle = \sum_{\ell=0}^{2j} C^{\ell, m + n}_{j, m; j, n}|\ell, m + n, j, j\rangle$ and using the symmetry property $C^{\ell, m + n}_{j, m; j, n} = (-1)^{2j - \ell}C^{\ell, m + n}_{j, n; j, m}$, we see that $S$ acts in the coupled basis as follows:
\begin{equation}
S|\ell, m, j, j\rangle = (-1)^{2j - \ell}|\ell, m, j, j\rangle.
\end{equation}
Therefore, by tracing over the coupled basis instead, we may write
\begin{equation}
\tr(T_q^{(k)}T_{q'}^{(k')}T_{-q}^{(k)}T_{-q'}^{(k')}) = \sum_{\ell=0}^{2j}\sum_{m=-\ell}^\ell (-1)^{2j - \ell}\langle\ell, m, j, j|(T_q^{(k)}\otimes T_{-q}^{(k)})(T_{q'}^{(k')}\otimes T_{-q'}^{(k')})|\ell, m, j, j\rangle.
\label{tracecoupled}
\end{equation}
Substituting \eqref{tracecoupled} into \eqref{Faveraged} gives
\begin{equation}
F_{kk'} = \frac{1}{(2k + 1)(2k' + 1)}\sum_{\ell, \ell' = 0}^{2j}\sum_{m=-\ell}^\ell\sum_{m'=-\ell'}^{\ell'} (-1)^{2j - \ell}\langle\ell, m, j, j|T^{(k)}\cdot T^{(k)}|\ell', m', j, j\rangle\langle\ell', m', j, j|T^{(k')}\cdot T^{(k')}|\ell, m, j, j\rangle,
\end{equation}
where we have inserted another complete set of coupled states.  We now use the Wigner-Eckart theorem in the form
\begin{equation}
\langle\ell, m, j, j|T^{(k)}\cdot T^{(k)}|\ell', m', j, j\rangle = \delta_{\ell\ell'}\delta_{mm'}(-1)^{2j + \ell}\left\{\begin{array}{ccc} \ell & j & j \\ k & j & j \end{array}\right\}\langle j||T^{(k)}||j\rangle^2,
\end{equation}
with $\langle j||T^{(k)}||j\rangle = \sqrt{2k + 1}$, to write
\begin{align}
F_{kk'} &= \sum_{\ell, \ell' = 0}^{2j}\sum_{m=-\ell}^\ell\sum_{m'=-\ell'}^{\ell'} (-1)^{2j - \ell}\delta_{\ell\ell'}\delta_{mm'}\left\{\begin{array}{ccc} \ell & j & j \\ k & j & j \end{array}\right\}\left\{\begin{array}{ccc} \ell' & j & j \\ k' & j & j \end{array}\right\} \\
&= \sum_{\ell=0}^{2j} (-1)^{2j - \ell}(2\ell + 1)\left\{\begin{array}{ccc} \ell & j & j \\ k & j & j \end{array}\right\}\left\{\begin{array}{ccc} \ell & j & j \\ k' & j & j \end{array}\right\} \\
&= (-1)^{2j + k + k'}\left\{\begin{array}{ccc} k & j & j \\ k' & j & j \end{array}\right\},
\end{align}
where we used a standard orthogonality relation for 6-$j$ symbols \cite{Varshalovich} in the last step.

There are many equivalent (but less elegant) expressions for the $F_{kk'}$, as all recoupling coefficients can be expressed as sums of products of Clebsch-Gordan coefficients.  For example, we may use \eqref{daggersign} to write
\begin{equation}
\mathcal{G}^{(k)} : \rho\mapsto \frac{1}{2k + 1}\sum_{q=-k}^k (-1)^q T_q^{(k)}\rho T_{-q}^{(k)}.
\end{equation}
Using \eqref{multiplication} to combine factors successively, we compute that
\begin{align}
F_{kk'}T_{q'}^{(k')} &= \frac{1}{2k + 1}\sum_{q=-k}^k (-1)^q T_q^{(k)}T_{q'}^{(k')}T_{-q}^{(k)} \label{startingpoint} \\
&= \frac{1}{2k + 1}\sum_{q=-k}^k\sum_{\ell = |k - k'|}^{k + k'} (-1)^{2j + q + \ell}\sqrt{(2k + 1)(2k' + 1)}\left\{\begin{array}{ccc} k & k' & \ell \\ j & j & j \end{array}\right\}C^{\ell, q + q'}_{k, q; k', q'}T_{q + q'}^{(\ell)}T_{-q}^{(k)} \\
&= \sum_{q=-k}^k\sum_{\ell = |k - k'|}^{k + k'}\sum_{\ell' = |k - \ell|}^{k + \ell} (-1)^{q + \ell + \ell'}\sqrt{(2k' + 1)(2\ell + 1)}\left\{\begin{array}{ccc} k & k' & \ell \\ j & j & j \end{array}\right\}\left\{\begin{array}{ccc} \ell & k & \ell' \\ j & j & j \end{array}\right\}C^{\ell, q + q'}_{k, q; k', q'}C^{\ell', q'}_{\ell, q + q'; k, -q}T_{q'}^{(\ell')}.
\end{align}
On general grounds, only the terms with $\ell' = k'$ should survive (which requires $|k - \ell|\leq k'\leq k + \ell$), so we find that
\begin{equation}
F_{kk'} = \sum_{q=-k}^k\sum_{\ell = |k - k'|}^{k + k'} (-1)^{q + \ell + k'}\sqrt{(2k' + 1)(2\ell + 1)}\left\{\begin{array}{ccc} k & k' & \ell \\ j & j & j \end{array}\right\}^2 C^{\ell, q + q'}_{k, q; k', q'}C^{k', q'}_{\ell, q + q'; k, -q},
\end{equation}
where we have used the invariance of the 6-$j$ symbol under permutations of its columns.  Since this coefficient is independent of $q'$, we can set $q'$ to a fixed ($k'$-dependent) value, with the stipulation that $-k'\leq q'\leq k'$ (e.g., $q' = 0$ or $q' = k'$).  Alternatively, we can average over all $q'$ to get
\begin{equation}
F_{kk'} = \sum_{q=-k}^k\sum_{q'=-k'}^{k'}\sum_{\ell = |k - k'|}^{k + k'} (-1)^{q + \ell + k'}\sqrt{\frac{2\ell + 1}{2k' + 1}}\left\{\begin{array}{ccc} k & k' & \ell \\ j & j & j \end{array}\right\}^2 C^{\ell, q + q'}_{k, q; k', q'}C^{k', q'}_{\ell, q + q'; k, -q}.
\end{equation}
This expression has the downside of being asymmetric between $k$ and $k'$.

Alternatively, starting from \eqref{startingpoint}, we can multiply both sides by $T_{-q'}^{(k')}$ and take the trace using $\tr(T_{q'}^{(k')}T_{-q'}^{(k')}) = (-1)^{q'}$ and
\begin{equation}
\tr(T_q^{(k)}T_{q'}^{(k')}T_{-q}^{(k)}T_{-q'}^{(k')}) = \frac{(2k + 1)(2k' + 1)}{(2j + 1)^2}\sum_{m=-j}^j C^{j, m + q}_{k, q; j, m}C^{j, m + q + q'}_{k', q'; j, m + q}C^{j, m + q'}_{k, -q; j, m + q + q'}C^{j, m}_{k', -q'; j, m + q'},
\end{equation}
which follows from \eqref{traceproduct}.  This gives
\begin{align}
F_{kk'} &= \frac{(-1)^{q'}}{2k + 1}\sum_{q=-k}^k (-1)^q\tr(T_q^{(k)}T_{q'}^{(k')}T_{-q}^{(k)}T_{-q'}^{(k')}) \\
&= \frac{2k' + 1}{(2j + 1)^2}\sum_{q=-k}^k (-1)^{q + q'}\sum_{m=-j}^j C^{j, m + q}_{k, q; j, m}C^{j, m + q + q'}_{k', q'; j, m + q}C^{j, m + q'}_{k, -q; j, m + q + q'}C^{j, m}_{k', -q'; j, m + q'}.
\end{align}
Again, this expression is independent of $q'$, which may be fixed to a definite value.  A more symmetric option is to average over all $q'$, yielding
\begin{equation}
F_{kk'} = \frac{1}{(2j + 1)^2}\sum_{q=-k}^k\sum_{q'=-k'}^{k'} (-1)^{q + q'}\sum_{m=-j}^j C^{j, m + q}_{k, q; j, m}C^{j, m + q + q'}_{k', q'; j, m + q}C^{j, m + q'}_{k, -q; j, m + q + q'}C^{j, m}_{k', -q'; j, m + q'}.
\end{equation}
Clearly, this matrix is real.  It is also symmetric, but not manifestly so: $F_{kk'} = F_{k'k}$.

\subsection{Completeness Relations} \label{app:completeness}

Fix a finite-dimensional Hilbert space.  Let $\{|i\rangle\}$ be an orthonormal basis of states, so that $\{e_{ij}\equiv |i\rangle\langle j|\}$ is an orthonormal basis of operators.  Let $\{O\}$ be a complete set of operators (not necessarily an orthonormal basis).  Then we have the completeness relation
\begin{equation}
\sum_O |O\rrangle\llangle O| = \hat{\mathds{1}},
\label{completeness1}
\end{equation}
where $\hat{\mathds{1}}$ is the identity superoperator.  Multiplying on the left and right by $\llangle e_{ij}|$ and $|e_{i'j'}\rrangle$ shows that \eqref{completeness1} is equivalent to
\begin{equation}
\sum_O \langle i|O|j\rangle\langle j'|O^\dag|i'\rangle = \delta_{ii'}\delta_{jj'}.
\label{completeness2}
\end{equation}
Dropping the $\langle i|$ and $|i'\rangle$ shows that \eqref{completeness2} is in turn equivalent to
\begin{equation}
\sum_O Oe_{ij}O^\dag = \delta_{ij}\mathds{1},
\label{completeness3}
\end{equation}
where $\mathds{1}$ is the identity operator.  Finally, expanding an arbitrary operator as $X = \sum_{i, j} \langle i|X|j\rangle e_{ij}$ shows that \eqref{completeness3} is equivalent to
\begin{equation}
\sum_O OXO^\dag = \tr(X)\mathds{1}
\label{completeness4}
\end{equation}
for all operators $X$.  So we have an equivalence between the superoperator identity \eqref{completeness1} and the operator identity \eqref{completeness4}.

Now suppose that $\{U_g\}$ is a unitary representation of a finite group: $g\mapsto U_g$ for $g\in G$.  By Schur's lemma, this representation is irreducible if and only if
\begin{equation}
\frac{1}{|G|}\sum_{g\in G} U_g XU_g^\dag = \frac{\tr(X)}{d}\mathds{1}
\end{equation}
for all operators $X$, where $d$ is the dimension of the Hilbert space.  In view of condition \eqref{completeness4}, such an irrep furnishes the following (properly normalized) complete set of operators:
\begin{equation}
O_g = \sqrt{\frac{d}{|G|}}U_g, \qquad \sum_{g\in G} |O_g\rrangle\llangle O_g| = \hat{\mathds{1}}.
\end{equation}
In this case, $\{U_g\}$ can be said to form a unitary 1-design.  We thus arrive at a corollary of Burnside's theorem \cite{Burnside}: the span of the representation matrices of a unitary irrep of a finite group is the entire space of linear operators on the (finite-dimensional) carrier space of that representation.  This statement generalizes readily to compact groups.

Suppose instead that $\{\mathcal{O}\}$ is a complete set of superoperators.  We refer to a linear map on the space of superoperators as a superduperoperator.  Introducing the superduperket notation $|\mathcal{O})$, in terms of which the Hilbert-Schmidt inner product on superoperators is given by $(\mathcal{A}|\mathcal{B}) = \Tr(\mathcal{A}^\dag\mathcal{B})$, we have the superduperoperator identity
\begin{equation}
\sum_{\mathcal{O}} |\mathcal{O})(\mathcal{O}| = \widehat{\mathds{1}},
\end{equation}
where $\widehat{\mathds{1}}$ is the identity superduperoperator (note the bigger hat).  This is equivalent to the superoperator identity
\begin{equation}
\sum_{\mathcal{O}} \mathcal{O}\mathcal{X}\mathcal{O}^\dag = \Tr(\mathcal{X})\hat{\mathds{1}}
\end{equation}
for all superoperators $\mathcal{X}$.

Now let $g\mapsto \mathcal{G}$ be a unitary superoperator representation of a finite group $G$.  Since a unitary (hence CPTP) superoperator representation is always reducible, we must generally write
\begin{equation}
\frac{1}{|G|}\sum_{g\in G} \mathcal{G}\mathcal{X}\mathcal{G}^\dag = \sum_\lambda \frac{\Tr(\mathcal{X}\Pi_\lambda)}{\Tr(\Pi_\lambda)}\Pi_\lambda
\label{initial}
\end{equation}
for some irreps $\{\lambda\}$, assuming no multiplicity for simplicity.  Note that the order of $\mathcal{G}$ and $\mathcal{G}^\dag$ in \eqref{initial} is immaterial; one could choose to sum over $g^{-1}$ rather than $g$.  To determine the modified ``completeness'' relation corresponding to \eqref{initial}, we reverse the steps in our previous reasoning.  Let $\{I\}$ be an orthonormal basis of operators, so that $\{\mathcal{E}_{IJ}\equiv |I\rrangle\llangle J|\}$ is an orthonormal basis of superoperators.  Since \eqref{initial} must hold for $\mathcal{X} = \mathcal{E}_{IJ}$, it is equivalent to
\begin{equation}
\frac{1}{|G|}\sum_{g\in G} \llangle I|\mathcal{G}|J\rrangle\llangle J'|\mathcal{G}^\dag|I'\rrangle = \sum_\lambda \frac{\llangle I|\Pi_\lambda|I'\rrangle\llangle J'|\Pi_\lambda|J\rrangle}{\Tr(\Pi_\lambda)}
\label{next}
\end{equation}
for all $I, I', J, J'$.  We can rewrite \eqref{next} as
\begin{equation}
\frac{1}{|G|}\sum_{g\in G} (\mathcal{E}_{IJ}|\mathcal{G})(\mathcal{G}|\mathcal{E}_{I'J'}) = \sum_\lambda \frac{(\mathcal{E}_{IJ}|\Pi_\lambda\mathcal{E}_{I'J'}\Pi_\lambda)}{\Tr(\Pi_\lambda)},
\end{equation}
which, by linearity, is equivalent to
\begin{equation}
\frac{1}{|G|}\sum_{g\in G} |\mathcal{G})(\mathcal{G}|\mathcal{X}) = \sum_\lambda \frac{|\Pi_\lambda\mathcal{X}\Pi_\lambda)}{\Tr(\Pi_\lambda)}
\label{newcompleteness}
\end{equation}
for all superoperators $\mathcal{X}$.  So we see that if a superoperator $\mathcal{A}$ acts within a single irrep, meaning $\Pi_\lambda\mathcal{A}\Pi_\lambda = \mathcal{A}$ for some fixed $\lambda$, then we can effectively resolve the identity superduperoperator within the subspace of such $\mathcal{A}$ as follows:
\begin{equation}
\sum_{g\in G} |\mathcal{G}_\lambda)(\mathcal{G}_\lambda|\mathcal{A}) = |\mathcal{A}), \qquad \mathcal{G}_\lambda\equiv \sqrt{\frac{\dim\phi_\lambda}{|G|}}\mathcal{G}.
\end{equation}
In other words, any such $\mathcal{A}$ can be constructed as a linear combination of group representation matrices.  As a special case, we can take $\mathcal{A}$ itself to be a projector, $\mathcal{A} = \Pi_\lambda$, for which \eqref{newcompleteness} gives
\begin{equation}
\frac{\Tr(\Pi_\lambda)}{|G|}\sum_{g\in G} \Tr(\mathcal{G}^\dag\Pi_\lambda)\mathcal{G} = \Pi_\lambda.
\end{equation}
Since $\chi_\lambda(g) = \Tr(\Pi_\lambda\mathcal{G})$ and $\Pi_\lambda^\dag = \Pi_\lambda^2 = \Pi_\lambda$, this is equivalent to the standard formula
\begin{equation}
\frac{\dim\phi_\lambda}{|G|}\sum_{g\in G} \chi_\lambda(g)^\ast\mathcal{G} = \Pi_\lambda.
\end{equation}
More generally, let $\{I_\lambda\}$ be an orthonormal basis of operators for the superoperator irrep $\phi_\lambda$.  Then any superoperator matrix unit $|I_\lambda\rrangle\llangle J_\lambda|$ can be constructed as follows:
\begin{equation}
\frac{\dim\phi_\lambda}{|G|}\sum_{g\in G} \mathcal{G}_{I_\lambda J_\lambda}^\ast\mathcal{G} = |I_\lambda\rrangle\llangle J_\lambda|,
\end{equation}
where $\mathcal{G}_{I_\lambda J_\lambda}^\ast = \llangle I_\lambda|\mathcal{G}|J_\lambda\rrangle^\ast = \Tr(\mathcal{G}^\dag|I_\lambda\rrangle\llangle J_\lambda|)$.  Even more generally, suppose we demand only that the superoperator $\mathcal{A}$ respect the irrep structure of $\mathcal{G}$, meaning that it takes the following block diagonal form: $\sum_\lambda \Pi_\lambda\mathcal{A}\Pi_\lambda = \mathcal{A}$.  By \eqref{newcompleteness}, any such superoperator can be expressed as a linear combination of group representation matrices, with the coefficients determined by a superoperator with appropriately rescaled blocks:
\begin{equation}
\frac{1}{|G|}\sum_{g\in G} (\mathcal{G}|\mathcal{A}')\mathcal{G} = \mathcal{A}, \qquad \mathcal{A}'\equiv \sum_\lambda \Tr(\Pi_\lambda)\Pi_\lambda\mathcal{A}\Pi_\lambda.
\label{genburnside}
\end{equation}
This is a generalization of (the aforementioned corollary of) Burnside's theorem to reducible representations.

The above considerations, suitably extended to compact groups, imply that the set of $\text{SU}(2)$ superoperator representation matrices is complete for the subspace of all superoperators with the same block structure.  Let us prove this statement directly.  The subspace of such superoperators is spanned by
\begin{equation}
|k, q_1, q_2)\equiv |T^{(k)}_{q_1}\rrangle\llangle T^{(k)}_{q_2}|,
\end{equation}
where $k = 0, \ldots, 2j$ and $q_1, q_2 = -k, \ldots, k$.  In terms of the superoperator representation matrices $\mathcal{G}$, we compute that
\begin{align}
\int_{\text{SU}(2)} dg\, (k, q_1, q_2|\mathcal{G})(\mathcal{G}|k', q_1', q_2') &= \int_{\text{SU}(2)} dg\, \llangle T^{(k)}_{q_1}|\mathcal{G}|T^{(k)}_{q_2}\rrangle\llangle T^{(k')}_{q_2'}|\mathcal{G}^\dag|T^{(k')}_{q_1'}\rrangle \\
&= \int_{\text{SU}(2)} dg\, D^k_{q_1 q_2}(g)D^{k'}_{q_1' q_2'}(g)^\ast \\
&= \frac{8\pi^2}{2k + 1}\delta_{kk'}\delta_{q_1 q_1'}\delta_{q_2 q_2'}, \label{superoperatorcompleteness}
\end{align}
where $dg$ is the Haar measure on $\text{SU}(2)$.  So the properly normalized complete set of superoperators is given by rescaling the blocks of $\mathcal{G}$ by factors of $\sqrt{(2k + 1)/8\pi^2}$, where $8\pi^2 = \operatorname{vol}(\text{SU}(2))$.  A nearly identical argument shows that the set of $\text{SU}(2)$ \emph{operator} representation matrices is complete for the space of operators on a spin-$j$ Hilbert space:
\begin{equation}
\int_{\text{SU}(2)} dg\, \langle j, m_1|D(g)|j, m_2\rangle\langle j, m_2'|D(g)^\dag|j, m_1'\rangle = \int_{\text{SU}(2)} dg\, D^j_{m_1 m_2}(g)D^j_{m_1' m_2'}(g)^\ast = \frac{8\pi^2}{2j + 1}\delta_{m_1 m_1'}\delta_{m_2 m_2'}.
\label{operatorcompleteness}
\end{equation}
Again, one obtains the corresponding normalized set of operators by rescaling by the same representation-dependent factor.  Both statements \eqref{superoperatorcompleteness} and \eqref{operatorcompleteness}, which follow from Burnside's theorem, come down to the orthogonality property of Wigner $D$-matrix elements.

\section{Review of Randomized Benchmarking} \label{app:review}

Here, we review some common randomized benchmarking protocols both to set notation and to lay the groundwork for our analysis.  For useful background on the representation theory behind traditional RB, we recommend that the reader consult the supplementary information of \cite{Gambetta} or a standard reference such as \cite{Fulton-Harris}.

\subsection{Standard Randomized Benchmarking}

The goal of randomized benchmarking (RB) is to estimate an average gate error rate.  The simplest RB protocol uses a gate set that forms a unitary 2-design and runs many circuits of varying length $m$ that (1) begin with the same initial state, (2) end with a measurement that verifies preservation of that initial state, and (3) consist of a sequence of uniformly random gates followed by a short inversion circuit of constant length.  Any such circuit would implement the identity operation if the gates were perfect.  In practice, the average final measurement success probability is fit to an exponentially decaying function of $m$.  The decay rate is the estimated gate error rate.

We assume that the gate set to be benchmarked forms a group $G$ (the ``benchmarking group'').  For the purpose of modeling noise, we represent gates as quantum channels and perfect gates as unitary channels.  Therefore, we are concerned not with the unitary representation $U$ of the gate group on the Hilbert space of the system, but rather with its superoperator representation $U\otimes U^\ast$ on the space of density matrices, given by $\rho\mapsto U\rho U^\dag$.  The superoperator representation always contains a copy of the trivial irrep (the span of the identity matrix) because the action of unitary gates on density matrices is trace-preserving.  For simplicity, we assume gate-independent noise and multiplicity-freeness of the superoperator representation, although both assumptions can be relaxed \cite{Helsen, Claes}.

$G$ is usually assumed to be a finite subgroup of $\text{SU}(d)$ (for $n$ qubits, $d = 2^n$).  Given an element (gate) $g\in G$, we denote its superoperator representation by $\mathcal{G} : \rho\mapsto g\rho g^\dag$ and its noisy implementation by $\widetilde{\mathcal{G}} = \mathcal{E}\mathcal{G}$ for some gate-independent noise channel (CPTP map) $\mathcal{E}$.  We write the superoperator representation of $G$ as
\begin{equation}
\mathcal{G} = \bigoplus_\lambda \phi_\lambda(g),
\end{equation}
where the irreps $\phi_\lambda$ of $G$ are mutually inequivalent.  We denote the input state by $\rho$ and the measurement POVM by $\{E, \mathds{1} - E\}$, where $\Tr(\rho E)$ is as large as possible. ($E$ stands for ``effect,'' while $\mathcal{E}$ stands for ``error.'')

For a given sequence length $m$, we sample gates $g_1, \ldots, g_m$ uniformly at random from $G$ and set $g_\text{inv} = (g_m\cdots g_1)^\dag$.  We apply these gates in the stated order to $\rho$ and then measure the POVM.  The corresponding survival probability, averaged over all random sequences, is given by
\begin{align}
p_m &= \frac{1}{|G|^m}\sum_{g_1, \ldots, g_m} \sbra{E}\widetilde{\mathcal{G}}_\text{inv}\widetilde{\mathcal{G}}_m\cdots \widetilde{\mathcal{G}}_1\sket{\rho} \\
&= \frac{1}{|G|^m}\sum_{g_1, \ldots, g_m} \sbra{E}(\mathcal{E}\mathcal{G}_1^\dag\cdots \mathcal{G}_m^\dag)(\mathcal{E}\mathcal{G}_m)\cdots (\mathcal{E}\mathcal{G}_1)\sket{\rho} \\
&= \frac{1}{|G|^m}\sum_{g_1, \ldots, g_m} \sbra{E}\mathcal{E}(\mathcal{G}_m^\dag\mathcal{E}\mathcal{G}_m)\cdots (\mathcal{G}_1^\dag\mathcal{E}\mathcal{G}_1)\sket{\rho} \\
&= \sbra{\smash{\mathcal{E}^\dag}(E)}\left(\frac{1}{|G|}\sum_{g\in G} \mathcal{G}^\dag\mathcal{E}\mathcal{G}\right)^m\sket{\rho}, \label{pmsimplified}
\end{align}
where we have used that the twirled error channel $|G|^{-1}\sum_{g'\in G} \mathcal{G}'^\dag\mathcal{E}\mathcal{G}'$ commutes with $\mathcal{G}$ for all $g\in G$.  By Schur's lemma, the twirl simplifies to give
\begin{equation}
p_m = \sbra{\smash{\mathcal{E}^\dag}(E)}\left(\sum_\lambda f_\lambda\Pi_\lambda\right)^m\sket{\rho} = \sum_\lambda A_\lambda f_\lambda^m.
\label{sumexponentials}
\end{equation}
Here, $\Pi_\lambda$ denotes the orthogonal projector onto the support of $\phi_\lambda$ and
\begin{equation}
A_\lambda\equiv \sbra{\smash{\mathcal{E}^\dag}(E)}\Pi_\lambda\sket{\rho}, \qquad f_\lambda\equiv \frac{\Tr(\Pi_\lambda\mathcal{E})}{\Tr(\Pi_\lambda)},
\end{equation}
where the trace is taken over superoperators.  While the index set $\{\lambda\}$ depends on $G$, the trivial representation spanned by $\sket{\mathds{1}}$ always appears; the corresponding $f_\lambda = 1$.  The $f_\lambda$ are quality parameters that depend only on the noisy gates being implemented, while the coefficients $A_\lambda$ depend only on state preparation and measurement (SPAM).  Thus for generic finite $G$, standard RB produces a weighted sum of decaying exponentials \eqref{sumexponentials}.

While the interpretation of the individual $f_\lambda$ depends on $G$, the $f_\lambda$ collectively yield the average fidelity of the gates in $G$ via
\begin{equation}
F_\text{ave} = \frac{d^{-1}\sum_\lambda \Tr(\Pi_\lambda)f_\lambda + 1}{d + 1}.
\label{averagefidelity}
\end{equation}
To demonstrate this, recall that twirling a channel over the unitary group converts it to a uniform depolarizing channel \cite{Nielsen}: for any $\mathcal{E}$, there exists a $p\in [0, 1 + (d^2 - 1)^{-1}]$ such that
\begin{equation}
\int dU\, \mathcal{U}^\dag\mathcal{E}\mathcal{U}\sket{X} = (1 - p)\sket{X} + \frac{p}{d}\tr(X)\sket{\mathds{1}}
\label{depolarizing}
\end{equation}
for all $d\times d$ matrices $X$, where $U\in \text{SU}(d)$.  Now, given a fiducial pure state $\rho_0 = |\psi_0\rangle\langle\psi_0|$, the average fidelity of $\mathcal{E}$ with respect to the identity channel can be written as
\begin{equation}
F_\text{ave}(\mathcal{E}) = \int dU \tr(\mathcal{U}(\rho_0)\mathcal{E}(\mathcal{U}(\rho_0))) = \int dU \sbra{\rho_0}\mathcal{U}^\dag\mathcal{E}\mathcal{U}\sket{\rho_0}.
\end{equation}
Using \eqref{depolarizing}, as well as $\tr(\rho_0^2) = \llangle\rho_0|\rho_0\rrangle = 1$ and $\tr(\rho_0) = \llangle\rho_0|\mathds{1}\rrangle = 1$, this becomes
\begin{equation}
F_\text{ave}(\mathcal{E}) = (1 - p)\llangle\rho_0|\rho_0\rrangle + \frac{p}{d}\tr(\rho_0)\llangle\rho_0|\mathds{1}\rrangle = 1 - p + \frac{p}{d}.
\label{faveofE}
\end{equation}
On the other hand, working in an operator basis $\{O_i\} = \{O_0\}\cup \{O_{i>0}\}$ of the form in Appendix \ref{app:HS} and using \eqref{depolarizing} gives
\begin{align}
\Tr(\mathcal{E}) &= \Tr\left(\int dU\, \mathcal{U}^\dag\mathcal{E}\mathcal{U}\right) \\
&= \sum_i \left((1 - p)\llangle O_i|O_i\rrangle + \frac{p}{d}\tr(O_i)\llangle O_i|\mathds{1}\rrangle\right) \\
&= d^2(1 - p) + \frac{p}{d}\tr(O_0)\llangle O_0|\mathds{1}\rrangle = d^2 - (d^2 - 1)p, \label{trofE}
\end{align}
where we have used $\tr(O_0) = \llangle O_0|\mathds{1}\rrangle = d^{1/2}$.  Combining \eqref{faveofE} and \eqref{trofE} gives
\begin{equation}
F_\text{ave}(\mathcal{E}) = \frac{d^{-1}\Tr(\mathcal{E}) + 1}{d + 1}.
\end{equation}
Substituting $\Tr(\mathcal{E}) = \Tr(\langle\mathcal{G}^\dag\mathcal{E}\mathcal{G}\rangle_G) = \sum_\lambda \Tr(\Pi_\lambda)f_\lambda$ then yields \eqref{averagefidelity}.

We note that, for a general multiplicity-free benchmarking group $G$, the relation between normalized $G$-invariant error rates $p_\lambda$ ($\sum_\lambda p_\lambda = 1$) and RB quality parameters $f_\lambda$ is as follows.  The $G$-twirl of an error channel $\mathcal{E}$ takes the form
\begin{equation}
\mathbb{E}_{g\in G}[\mathcal{G}^\dag\mathcal{E}\mathcal{G}] = \sum_\lambda p_\lambda\mathcal{G}^{(\lambda)} = \sum_\lambda f_\lambda\Pi_\lambda,
\label{Gtwirled}
\end{equation}
where the $\mathcal{G}^{(\lambda)}$ are mutually orthogonal channels and the $\Pi_\lambda$ are irrep projectors.  The channel $\mathcal{G}^{(0)}$ corresponding to the trivial irrep is the $d^2\times d^2$ identity superoperator, so taking Hilbert-Schmidt inner products with the identity channel in \eqref{Gtwirled} gives $p_0 = d^{-2}\sum_\lambda \Tr(\Pi_\lambda)f_\lambda$.  This is the ``zero-error rate,'' which coincides with the entanglement (or process) fidelity $F_e$ \cite{Hashim}.  On the other hand, the average gate fidelity of $G$ is given by $F_\text{ave} = \frac{dp_0 + 1}{d + 1}$, as shown above.  Since $p_0\leq 1$, it follows that $p_0\leq F_\text{ave}\leq 1$.

For standard RB to be SPAM-invariant (and thus to yield a single nontrivial $f_\lambda$ regardless of the initial state), the superoperator representation of $G$ should be irreducible on the orthogonal complement of the trivial irrep (which is spanned by traceless matrices).  This is certainly the case if $G$ is all of $\text{SU}(d)$, for which $\Box\otimes \overline{\Box}\cong \operatorname{trivial}\oplus \operatorname{adjoint}$.  More generally, this holds if $G$ is a unitary 2-design.  For example, Clifford randomized benchmarking (CRB) samples randomly from the $n$-qubit Clifford group, which is a unitary 3-design for $\text{SU}(2^n)$.  CRB is difficult to scale up because it requires compiling $n$-qubit Clifford gates into large circuits over native gates.  On the other hand, direct randomized benchmarking (DRB) uses native gates directly and requires that they generate a unitary 2-design \cite{DRB-Proctor, DRB-Polloreno}.  The tetrahedral, octahedral, and icosahedral groups form unitary 2-, 3-, and 5-designs for $\text{SU}(2)$, respectively, and are therefore viable benchmarking groups for standard single-qubit RB \cite{Barends}.

\subsection{Character Randomized Benchmarking}

Character RB \cite{Helsen} relaxes the requirement that the gate set (group) form a unitary 2-design and therefore allows its superoperator representation to be less than maximally irreducible.  Character RB isolates the individual quality parameters $f_\lambda$ by exploiting the standard formula
\begin{equation}
\frac{\dim\phi_\lambda}{|G|}\sum_{g\in G} \chi_\lambda(g)^\ast\mathcal{G} = \Pi_\lambda,
\label{projector}
\end{equation}
where $\chi_\lambda$ is the character of the irrep $\phi_\lambda$. [Strictly speaking, $\Pi_\lambda$ in \eqref{projector} should be replaced by the projector onto the support of all irreps in $\mathcal{G}$ isomorphic to $\phi_\lambda$ (the isotypic component of $\phi_\lambda$), but we have used the assumption of multiplicity-freeness.]

Given a benchmarking group $G$, we fix an irrep label $\lambda$.  We then choose a ``character group'' $H\subset G$ (in the sense of \cite{Helsen}, not in the mathematical sense) whose superoperator representation $\mathcal{H}$ has an irreducible subrepresentation $\phi_{\lambda'}$ with support inside the irrep $\phi_\lambda$ of $G$: $\Pi_{\lambda'}\Pi_\lambda = \Pi_{\lambda'}$ (note that $\lambda'$ does not necessarily belong to $\{\lambda\}$).  For example, one can always use the benchmarking group itself as the character group by choosing $H = G$ and $\phi_{\lambda'} = \phi_\lambda$.  Finally, we choose an initial state and a measurement POVM such that $\Tr(\Pi_{\lambda'}(\rho)E)$ is large.

For a given sequence length $m$, we sample gates $g_1, \ldots, g_m$ uniformly at random from $G$ and let $g_\text{inv} = (g_m\cdots g_1)^\dag$; we also sample the gate $h$ uniformly at random from the character group $H$.  We apply the gates $g_1 h, g_2, \ldots, g_m, g_\text{inv}$ sequentially to $\rho$.  Note that $h$ and $g_1$ are compiled into a single gate, as with $g_\text{inv}$, and thus incur only one application of the noise channel.  We then estimate the average \emph{weighted} survival probability
\begin{equation}
p_m = \frac{\dim\phi_{\lambda'}}{|H||G|^m}\sum_h\sum_{g_1, \ldots, g_m} \chi_{\lambda'}(h)^\ast\sbra{E}\widetilde{\mathcal{G}}_\text{inv}\widetilde{\mathcal{G}}_m\cdots \widetilde{\mathcal{G}}_2\widetilde{\mathcal{G}_1\mathcal{H}}\sket{\rho}.
\end{equation}
By manipulations similar to those leading to \eqref{pmsimplified}, we have
\begin{equation}
p_m = \sbra{\smash{\mathcal{E}^\dag}(E)}\left(\frac{1}{|G|}\sum_{g\in G} \mathcal{G}^\dag\mathcal{E}\mathcal{G}\right)^m\left(\frac{\dim\phi_{\lambda'}}{|H|}\sum_{h\in H} \chi_{\lambda'}(h)^\ast\mathcal{H}\right)\sket{\rho}.
\end{equation}
By \eqref{projector}, Schur's lemma, and $\Pi_\Lambda\Pi_{\lambda'} = \delta_{\Lambda\lambda}\Pi_{\lambda'}$, we can then write
\begin{equation}
p_m = \llangle\mathcal{E}^\dag(E)|\left(\sum_\Lambda f_\Lambda\Pi_\Lambda\right)^m|\Pi_{\lambda'}(\rho)\rrangle = \llangle\mathcal{E}^\dag(E)|\Pi_{\lambda'}(\rho)\rrangle f_\lambda^m.
\end{equation}
In this way, we can extract $f_\lambda$ by fitting to a single exponential decay.

\subsection{Filtered Randomized Benchmarking}

Character RB fits into the filtered RB framework \cite{FilteredRB-1, FilteredRB-2} as follows.  For a fixed sequence length $m$, we prepare the state $\rho$, apply gates $g_1, \ldots, g_{m+1}$ drawn uniformly at random from the set of all sequences in $G^{m+1}$ that satisfy $g_{m+1}\cdots g_1\in H$, and then measure the POVM $\{E, \mathds{1} - E\}$.  A single run produces the output $(i, g_1, \ldots, g_{m+1})$ where $i\in \{0, 1\}$.  We define the filter function
\begin{equation}
f_{\lambda'}(i, g_1, \ldots, g_{m+1}) = (\dim\phi_{\lambda'})\chi_{\lambda'}(h)^\ast, \qquad h\equiv g_{m+1}\cdots g_1,
\end{equation}
which depends on a chosen irrep $\lambda'$ of $H$ and is independent of $i$.  If the gates are noiseless, then averaging the values of this filter function over shots is tantamount to averaging $f_{\lambda'}(h)$ over $H$ with weight $\llangle E|\mathcal{H}|\rho\rrangle$ (by the Born rule), yielding
\begin{equation}
\frac{\dim\phi_{\lambda'}}{|H|}\int_H dh\, \chi_{\lambda'}(h)^\ast\llangle E|\mathcal{H}|\rho\rrangle = \llangle E|\Pi_{\lambda'}|\rho\rrangle,
\end{equation}
as desired.

The non-uniform filtered RB protocol of \cite{FilteredRB-2} employs a particular family of filter functions and works as follows.  For a fixed sequence length $m$, we define the following procedural primitive: prepare the state $\rho$, apply gates $g_1, \ldots, g_m$ drawn independently from a specified probability measure $\nu$ on $G$, and measure in the basis $\{|i\rangle\}$ where $i\in \{1, \ldots, d\}$.  A single run of this primitive produces the output $(i, g_1, \ldots, g_m)$ for some measurement outcome $i$.  For post-processing, let $\sket{i}\equiv \sket{|i\rangle\langle i|}$ and
\begin{equation}
S\equiv \frac{1}{|G|}\int_G dg\, \mathcal{G}^\dag D\mathcal{G}, \qquad D\equiv \sum_{i=1}^d |i\rrangle\llangle i|.
\end{equation}
Note that we use the unnormalized Haar measure, hence the factor of $1/|G|$.  For a fixed irrep $\lambda$ of $G$, we define the filter function $f_\lambda : \{1, \ldots, d\}\times G^m\to \mathbb{R}$ by
\begin{equation}
f_\lambda(i, g_1, \ldots, g_m) = \llangle\rho|\Pi_\lambda S^{-1}\mathcal{G}_\text{inv}|i\rrangle,
\end{equation}
where $g_\text{inv} = (g_m\cdots g_1)^\dag$.  We have assumed that the operators $\{\mathcal{G}|i\rrangle\}$ span Hilbert-Schmidt space, which ensures that $S$ is invertible (otherwise, $S^{-1}$ may be replaced with $S^+$, the Moore-Penrose pseudoinverse).  We then compute the average of this filter function over $N$ samples, $\frac{1}{N}\sum_{n=1}^N f_\lambda(i^{(n)}, g_1^{(n)}, \ldots, g_m^{(n)})$, to estimate the filtered RB signal.  The motivation for this particular filter function is as follows.  Suppose the gates are noiseless and that $\nu$ is the uniform Haar measure.  Then the probability of producing a given output $(i, g_1, \ldots, g_m)$ in a single run is
\begin{equation}
\llangle i|\mathcal{G}_m\cdots \mathcal{G}_1|\rho\rrangle = \llangle i|\mathcal{G}_\text{inv}^\dag|\rho\rrangle,
\end{equation}
so the average value of the filter function is
\begin{equation}
\frac{1}{|G|}\sum_{i=1}^d \int_G dg_\text{inv}\, \llangle i|\mathcal{G}_\text{inv}^\dag|\rho\rrangle\llangle\rho|\Pi_\lambda S^{-1}\mathcal{G}_\text{inv}|i\rrangle = \Tr\left[|\rho\rrangle\llangle\rho|\Pi_\lambda S^{-1}\left(\frac{1}{|G|}\int_G dg_\text{inv}\, \mathcal{G}_\text{inv}D\mathcal{G}_\text{inv}^\dag\right)\right] = \llangle\rho|\Pi_\lambda|\rho\rrangle.
\end{equation}
The authors of \cite{FilteredRB-2} give conditions under which the filtered RB signal takes the form of a single exponential decay in the presence of noise and provide bounds on the sample complexity of estimating the signal.

\section{\texorpdfstring{$\text{SU}(2)$}{SU(2)} Character Randomized Benchmarking} \label{app:su2chiRB}

\subsection{Character Groups}

Suppose we use $\text{SU}(2)$ as the benchmarking group.  We argue that for generic $j$, the only viable character group for $\text{SU}(2)$ character RB is $\text{SU}(2)$ itself.

To determine the irreps that appear in the superoperator representation of discrete subgroups of $\text{SU}(2)$, we use the decomposition of the superoperator representation of $\text{SU}(2)$ into $\text{SU}(2)$ irreps in combination with the branching rules tabulated in \cite{Gross}.  Labeling the spin-$j$ irrep of $\text{SU}(2)$ by its dimension as $(2j + 1)$, we have $(2j + 1)\otimes (2j + 1) = (1)\oplus (3)\oplus \cdots\oplus (4j + 1)$.  Only integer spins occur in this decomposition, so we focus on the odd-dimensional irreps of $\text{SU}(2)$.  We ask: for any given $\text{SU}(2)$ irrep within $(2j + 1)\otimes (2j + 1)$, can we identify an appropriate character group, i.e., one whose superoperator representation has an irrep with isotypic component supported inside the given $\text{SU}(2)$ irrep?

The answer is no.  One can immediately rule out the polyhedral subgroups of $\text{SU}(2)$ as viable character groups except for very small spins because for sufficiently large $j$, any irrep of such a subgroup occurs within multiple $\text{SU}(2)$ irreps in $(2j + 1)\otimes (2j + 1)$.  Specifically, the subgroups $2T, 2O, 2I$ can only be used as character groups for spins of size at most $j = 1/2, 1, 3/2$, respectively:
\begin{itemize}
\item For $2T$, $(2)\otimes (2)$ is a direct sum of $(1)\cong \rho_1$ and $(3)\cong \rho_7$.
\item For $2O$, $(3)\otimes (3)$ is a direct sum of $(1)\cong \rho_1$, $(3)\cong \rho_6$, and $(5)\cong \rho_3\oplus \rho_7$.
\item For $2I$, $(4)\otimes (4)$ is a direct sum of $(1)\cong \rho_1$, $(3)\cong \rho_4$, $(5)\cong \rho_8$, and $(7)\cong \rho_5\oplus \rho_6$.
\end{itemize}
(We use the notation of \cite{Gross}, in which irreps of $2T$, $2O$, $2I$ are labeled by $\rho_{1, \ldots, 7}$, $\rho_{1, \ldots, 8}$, $\rho_{1, \ldots, 9}$, respectively.) For any $j$ higher than $1/2, 1, 3/2$, some irrep of $2T, 2O, 2I$, respectively, will occur with multiplicity in $(2j + 1)\otimes (2j + 1)$ such that different copies are distributed between different $\text{SU}(2)$ irreps.

One could also consider using the cyclic and binary dihedral (or dicyclic) subgroups of $\text{SU}(2)$, which come in infinite families, as character groups.  Up to conjugation, the cyclic subgroups of order $2n$ ($2n + 1$) are generated by $e^{-2\pi iJ_z/n}$ ($e^{-4\pi iJ_z/(2n + 1)}$), while the binary dihedral subgroups of order $4n$ are generated by $e^{-i\pi J_x}$ and $e^{-2\pi iJ_z/n}$.  The character group itself could even be infinite.  For example, consider the $\text{U}(1)$ subgroup generated by $J_z$.  The 1D irreps are multiplicity-free on Hilbert space and spanned by $|m\rangle$ for each $m = -j, \ldots, j$.  However, the irreps have multiplicity on Hilbert-Schmidt space, being spanned by $|k\rangle\langle k + m|$ for all $k$ and fixed $m$.  The isotypic components do not have support inside individual $\text{SU}(2)$ irreps.  Similar reasoning rules out the cyclic subgroups of $\text{U}(1)$ and their dicyclic counterparts.

\subsection{Discrete Benchmarking Groups}

One could simplify the protocol by using one of the finite subgroups of $\text{SU}(2)$ as the entire benchmarking group rather than merely the character group (so that one would only have to sample gates from one of these subgroups)---for instance, ``icosahedral benchmarking.''  These subgroups have a finite number of irreps, so the multiplicities of these irreps inside the superoperator representation will grow indefinitely as the spin $j$ grows (unlike for $\text{SU}(2)$, where there are no multiplicities).

For general character RB with multiplicities (i.e., when the superoperator representation of $G$ is not multiplicity-free), we require the isotypic component of some irrep of the character group to be contained within the isotypic component of a given irrep of the benchmarking group \cite{Claes}.  A given irrep of the benchmarking group can have multiple decay rates, with the number of rates equal to its multiplicity.

\section{Synthetic Randomized Benchmarking}

Assuming that native operations are global $\text{SU}(2)$ rotations and that logical operations form a discrete subgroup thereof, our ultimate goal is to compare the following quantities:
\begin{enumerate}
\item Error rates of native operations on a spin-$j$ system (physical error rates).
\item Error rates of logical operations on a spin code, i.e., a protected encoding of a qubit into the spin-$j$ system (logical error rates).
\end{enumerate}
Theoretically, the goal is to predict the logical error rates for a given spin code and a given set of physical error rates.  Experimentally, the goal is to show that measured logical error rates are lower than measured physical error rates.\footnote{Note that there is no experimental reason to expect that discrete rotations can be implemented with higher fidelity than generic rotations, so the reduction in error rates must be a property of the encoded \emph{states}.}  In this work, we focus on a single subgoal: that of developing randomized benchmarking protocols (along with theoretical estimates of their efficiency) to measure physical error rates of global $\text{SU}(2)$ rotations inside the group of unitaries $\text{SU}(2j + 1)$ on a spin-$j$ qudit.

The group of global $\text{SU}(2)$ rotations does not form a 2-design for $\text{SU}(2j + 1)$ when $j > 1/2$.\footnote{More precisely, a unitary design is defined with respect to a Hermitian vector space.  One can regard $\mathbb{C}^{2j+1}$ as the carrier space of either the fundamental representation of $\text{SU}(2j + 1)$ or the spin-$j$ representation of $\text{SU}(2)$.  The spin-$j$ representation of a subgroup $2T, 2O, 2I\subset \text{SU}(2)$ is a unitary 2-design for the spin-$j$ representation of $\text{SU}(2)$ but is not a unitary 2-design for the fundamental representation of $\text{SU}(2j + 1)$ when $j > 1/2$ (in the former case, we average over $\text{SU}(2)$ elements represented as $(2j + 1)\times (2j + 1)$ matrices; in the latter case, we average over $\text{SU}(2j + 1)$ elements represented as $(2j + 1)\times (2j + 1)$ matrices).  In the latter sense, the spin-$j$ representation of a subgroup $2T, 2O, 2I\subset \text{SU}(2)$ is not a unitary 2-design for $\mathbb{C}^{2j+1}$ when $j > 1/2$.  The same is true of $\text{SU}(2)$ itself.}  Consequently, the superoperator representation of $\text{SU}(2)$ on the state space of a $j > 1/2$ system is highly reducible.  This implies that standard $\text{SU}(2)$ RB on a $j > 1/2$ system will not be SPAM-invariant: it will produce a linear combination of up to $2j + 1$ exponential decays with different rates, and the precise linear combination that occurs will depend strongly on the initial state.

We propose a family of ``synthetic RB'' schemes that all accomplish the goal of disambiguating the irrep-specific decay rates in the $\text{SU}(2)$ RB signal.  All of our protocols use \textbf{synthetic circuits} that incorporate some combination of \textbf{synthetic gates} and \textbf{synthetic SPAM}.  Each operation (gate or SPAM) is a linear combination of physical operations.  Synthetic RB encompasses ``character RB'' as a special case, where the synthetic circuits involve only synthetic gates and those synthetic gates are constructed in a particular way.

\subsection{Synthetic Gates} \label{app:syntheticgates}

The purpose of synthetic gates is to build a projection superoperator onto or into a target irrep as a linear combination of superoperator representations of elements of the benchmarking group (in this case, $\text{SU}(2)$).  We have two basic options:
\begin{enumerate}
\item Write a projector as an infinite linear combination of superoperators.
\item Write a projector as a finite linear combination of superoperators.
\end{enumerate}
The first option requires random sampling because an infinite linear combination is impossible to construct in practice.  The second option may or may not involve random sampling.  More generally, one may consider using a spanning set of superoperators that is overcomplete, complete, or possibly even undercomplete for a subspace of superoperators that contains the superoperators that we wish to synthesize.

The first option is the approach taken by character RB with $\text{SU}(2)$ as the character group, which constructs projectors as weighted integrals over $\text{SU}(2)$ group elements and requires randomly sampling from an infinite character group.  One may worry that this approach is potentially wasteful because Hilbert-Schmidt space is finite-dimensional.

For the second option, one can either (1) use a Monte Carlo method to choose a finite frame (i.e., a possibly overcomplete spanning set) of superoperators, corresponding to a finite set of group elements, or (2) fix a frame analytically.  Either way, the superoperator frame should be chosen to optimize the efficiency of constructing the projector.  We generally focus on the case where the frame is a basis.

Importance sampling with respect to the absolute value of the coefficients in the projector sum can potentially be used to reduce the sample complexity of any of these approaches.

\subsubsection{Infinite Frame}

$\text{SU}(2)$ character RB, for which the frame is indexed by the entire group $\text{SU}(2)$, employs the following random sampling procedure.  Fix an irrep $k$ between 0 and $2j$.  Choose $N$ random group elements $g_i$ with superoperator representations $\mathcal{G}_i$, where $i = 1, \ldots, N$.  We approximate the projection superoperator by
\begin{equation}
\widetilde{\Pi}_k = \frac{2k + 1}{N}\sum_{i=1}^N \chi_k(g_i)\mathcal{G}_i, \qquad \chi_k(g_i) = \sum_{q=-k}^k D_{qq}^k(g_i).
\end{equation}
This na\"ive random sampling leads to slow convergence of $\widetilde{\Pi}_k$ to the exact projector $\Pi_k$ as a function of sample size $N$ (with respect to, e.g., the Frobenius norm).  In contrast to the character-based approach, which only gives the exact answer in the limit of infinite sample size, fixing a finite frame can give the exact projector (up to numerical error) with exactly $N_j$ samples.

In practice, we absorb experimental uncertainty in the implementation of a specified $\text{SU}(2)$ rotation into gate error.  While random sampling does not aim to implement a predetermined $\text{SU}(2)$ element in a given shot (unlike the finite frame), it is still important for the target $\text{SU}(2)$ element (as opposed to the noisy $\text{SU}(2)$ element that ends up being implemented) to be known precisely so that the correct value of the character function can be computed.

Instead of building the full projector onto the target irrep $k$, one could build a projector of smaller rank whose image lies within the target irrep.  The most parsimonious option, which leverages the fact that any circuit outcome probability can be written as $\llangle E|S|\rho\rrangle = \Tr(|\rho\rrangle\llangle E|\cdot S)$, is not to build a projector at all, but rather to build the rank-1 SPAM superoperator $|E\rrangle\llangle\rho|$.  Correspondingly, instead of the weight function $\chi_k(g) = \Tr(\mathcal{G}^\dag\Pi_k)$ for a given $g\in \text{SU}(2)$, one would use $\Tr(\mathcal{G}^\dag|E\rrangle\llangle\rho|)$ (under the assumption that $\Pi_k|E\rrangle\llangle\rho|\Pi_k = |E\rrangle\llangle\rho|$).  This weight function takes smaller values than the character function and should therefore have smaller variance.  This observation will be important for us.

\subsubsection{Finite Frame}

One only needs a spanning set of superoperator representation matrices to construct a projector.  This observation motivates a generalization of the character RB protocol to use subsets rather than subgroups of the benchmarking group: it does not require a group structure on the set of group elements whose superoperator representations are used to synthesize a projector.

In the spherical tensor basis, the superoperator representation of an $\text{SU}(2)$ group element $g(\alpha, \beta, \gamma)$ is block diagonal:
\begin{equation}
\mathcal{G}(\alpha, \beta, \gamma) = \operatorname{diag}(D^0(\alpha, \beta, \gamma), D^1(\alpha, \beta, \gamma), \ldots, D^{2j}(\alpha, \beta, \gamma)),
\end{equation}
where $\sum_{\ell=0}^{2j} (2\ell + 1) = (2j + 1)^2$.  The space of superoperators with the same block structure as $\mathcal{G}(\alpha, \beta, \gamma)$ has dimension
\begin{equation}
N_j\equiv \sum_{\ell=0}^{2j} (2\ell + 1)^2 = \frac{1}{3}(2j + 1)(4j + 1)(4j + 3).
\end{equation}
Suppose we choose $N_j$ group elements $g_i\in \text{SU}(2)$ ($i = 1, \ldots, N_j$) with linearly independent superoperator representations $\mathcal{G}_i$.  Then the $\mathcal{G}_i$ form a basis for the space of superoperators with the same block structure.  Any superoperator $\mathcal{A}$ with this block structure, and in particular any irrep projector, can be written as a finite linear combination of the $\mathcal{G}_i$:
\begin{equation}
\mathcal{A} = \operatorname{diag}(A^0, A^1, \ldots, A^{2j}) = \sum_{i=1}^{N_j} a_i\mathcal{G}_i.
\end{equation}
To determine the coefficients $a_i$, it is convenient to vectorize all of the superoperators block by block.  We order the spherical tensor basis as
\begin{equation}
T^{(0)}_0; T^{(1)}_1, T^{(1)}_0, T^{(1)}_{-1}; \ldots; T^{(2j)}_{2j}, T^{(2j)}_{2j - 1}, \ldots, T^{(2j)}_{-(2j - 1)}, T^{(2j)}_{-2j},
\end{equation}
so that the vectorization of a single block takes the form
\begin{equation}
\operatorname{vec}(A^k) = [A_{k, k}^k, \ldots, A_{-k, k}^k; A_{k, k - 1}^k, \ldots, A_{-k, k - 1}^k; \ldots; A_{k, -k}^k, \ldots, A_{-k, -k}^k]^T.
\end{equation}
We then solve
\begin{equation}
\left[\begin{array}{ccc} \operatorname{vec}(\mathcal{G}_1) & \cdots & \operatorname{vec}(\mathcal{G}_{N_j}) \end{array}\right]\left[\begin{array}{c} a_1 \\ \vdots \\ a_{N_j} \end{array}\right] = \operatorname{vec}(\mathcal{A}), \qquad \operatorname{vec}(\mathcal{A})\equiv \left[\begin{array}{c} \operatorname{vec}(A^0) \\ \operatorname{vec}(A^1) \\ \vdots \\ \operatorname{vec}(A^{2j}) \end{array}\right].
\end{equation}
The frame matrix (the matrix of vectorized $\mathcal{G}_i$) is invertible by assumption.  The choice of group elements $g_i$ determines how well-conditioned this matrix is.

To build intuition, consider the following two examples.

\paragraph{Example 1.} Since our goal is to build projectors, it suffices to find a basis for the space of diagonal superoperators.  For the sake of analogy, if we could realize a direct sum of $\text{U}(1)$ representations of the form $\theta\mapsto \operatorname{diag}(e^{i\theta}, \ldots, e^{ik\theta})$ for $\theta\in [0, 2\pi)$ and some positive integer $k$, then taking linear combinations of the images would suffice to construct $\operatorname{diag}(c_1, \ldots, c_k)$ for any $c_j\in \mathbb{C}$.  Indeed, fix distinct $\theta_j$ for $j = 1, \ldots, k$ and set $u_j = e^{i\theta_j}$.  Then the $k$ vectors $(u_j, \ldots, u_j^k)$ are linearly independent because the determinant of the corresponding Vandermonde matrix is nonzero.  In our case, this strategy doesn't work directly.  A diagonal Wigner $D$-matrix corresponds to a pure $z$-rotation via $D_{qq'}^k(\alpha, 0, 0) = e^{-iq\alpha}\delta_{qq'}$, which has the superoperator representation
\begin{equation}
\operatorname{diag}(1; e^{-i\alpha}, 1, e^{i\alpha}; \ldots; e^{-i(2j)\alpha}, e^{-i(2j - 1)\alpha}, \ldots, e^{i(2j - 1)\alpha}, e^{i(2j)\alpha}).
\end{equation}
The repeated entries mean that these operators do not span the space of diagonal superoperators.  However, the phases $e^{\pm ik\alpha}$ appear only in irreps $k, k + 1, \ldots, 2j$.  Therefore, in principle, one could realize a superoperator with support only in irrep $2j$ and use it to extract the signal from that irrep; one could then realize a superoperator with support only in irreps $2j$ and $2j - 1$ and use it to extract the signal from irrep $2j - 1$ by subtracting the previous signal; and so on.

\paragraph{Example 2.} The foundation of $\text{SU}(2)$ character RB is the projector formula
\begin{equation}
\Pi_k = (2k + 1)\mathbb{E}_{g\in \text{SU}(2)}[\chi_k(g)\mathcal{G}].
\end{equation}
Using the axis-angle parametrization and the fact that the character function depends only on the rotation angle, we can split the group average into an average over $\theta\in [0, \pi]$ and an average over $\vec{n} = (\sin\Theta\cos\Phi, \sin\Theta\sin\Phi, \cos\Theta)\in S^2$:
\begin{equation}
\mathbb{E}_{g\in \text{SU}(2)}[\chi_k(\theta)\mathcal{G}(\theta, \Theta, \Phi)] = \frac{2}{\pi}\int_0^\pi d\theta\sin^2\frac{\theta}{2}\chi_k(\theta)\underbrace{\left[\frac{1}{4\pi}\int_0^\pi d\Theta\sin\Theta\int_0^{2\pi} d\Phi\, \mathcal{G}(\theta, \Theta, \Phi)\right]}_{\mathbb{E}_{(\Theta, \Phi)\in S^2}[\mathcal{G}(\theta, \Theta, \Phi)]}.
\label{splitaverage}
\end{equation}
In turn, we can write the average over all $\mathcal{G}$ with a fixed rotation angle (i.e., the average over axes) as an average over all images of a fiducial $\mathcal{G}$ under conjugation by $\text{SU}(2)$ superoperators (i.e., an $\text{SU}(2)$ twirl), which reduces to a weighted sum of projectors by Schur's lemma:
\begin{equation}
\Pi(\theta)\equiv \mathbb{E}_{(\Theta, \Phi)\in S^2}[\mathcal{G}(\theta, \Theta, \Phi)] = \mathbb{E}_{u\in\text{SU}(2)}[\mathcal{U}\mathcal{G}\mathcal{U}^\dag] = \sum_{\ell=0}^{2j} \frac{\Tr(\Pi_\ell\mathcal{G})}{2\ell + 1}\Pi_\ell = \sum_{\ell=0}^{2j} \frac{\chi_\ell(\theta)}{2\ell + 1}\Pi_\ell.
\end{equation}
This is because conjugating a given rotation by another rotation can be used to effect a rotation through the same angle but about any desired axis.  In standard character RB, we would construct a projector as a weighted average over \emph{all} rotation angles:
\begin{equation}
\mathbb{E}_{g\in \text{SU}(2)}[\chi_k(\theta)\mathcal{G}(\theta, \Theta, \Phi)] = \frac{2}{\pi}\int_0^\pi d\theta\sin^2\frac{\theta}{2}\chi_k(\theta)\Pi(\theta) = \sum_{\ell=0}^{2j} \frac{\delta_{k\ell}}{2\ell + 1}\Pi_\ell = \frac{\Pi_k}{2k + 1},
\end{equation}
where we have used character orthogonality.  However, we can improve upon this character weighting (which yields a pseudoinverse) by choosing a discrete subset of angles $\theta$ that is optimized for a given irrep.  To do so, we fix $k$ and choose $2j + 1$ angles $\theta_0, \ldots, \theta_{2j}$.  Generically, the superoperators $\Pi(\theta_0), \ldots, \Pi(\theta_{2j})$ are linearly independent and hence form a basis for the $(2j + 1)$-dimensional subspace of superoperators spanned by $\Pi_0, \ldots, \Pi_{2j}$.  We then minimize the absolute values of the coefficients $c_i$ in the equation
\begin{equation}
\left[\begin{array}{ccc}
\frac{\chi_0(\theta_0)}{2\cdot 0 + 1} & \cdots & \frac{\chi_0(\theta_{2j})}{2\cdot 0 + 1} \\
\vdots & \ddots & \vdots \\
\frac{\chi_{2j}(\theta_0)}{2\cdot 2j + 1} & \cdots & \frac{\chi_{2j}(\theta_{2j})}{2\cdot 2j + 1}
\end{array}\right]
\left[\begin{array}{c}
c_0 \\ \vdots \\ c_{2j}
\end{array}\right]
=
\left[\begin{array}{c}
\vdots \\ 1 \\ \vdots
\end{array}\right],
\end{equation}
where the vector on the right has 1 in the $k^\text{th}$ position and 0 elsewhere.  Equivalently, we minimize the entries in the $k^\text{th}$ column of the inverse of the matrix on the left.  Empirically, one can do better than character weighting, subject to the limitation that the columns of the matrix on the left are biased toward positive values, regardless of the values of $\theta_i$ (this is a property of the character function).  This is an obstacle to choosing a mutually orthogonal set of columns, and therefore to making the matrix well-conditioned.  We emphasize that this axis-angle approach to constructing a finite frame, which simplifies the problem to inverting $(2j + 1)\times (2j + 1)$ matrices, is only a mild generalization of character RB because it assumes that the weighting factors are (axis-independent) class functions and that \emph{every} $\text{SU}(2)$ group element is included in the projector sum.  Its motivation is that the ``inner'' average over rotation axes in \eqref{splitaverage} is not a weighted average and is therefore uninteresting from the perspective of importance sampling.  If we consider only the one-dimensional average over rotation angles and assume that for each $\theta$, the average over $\vec{n}$ has already been done, then we are left with taking linear combinations of the $\Pi(\theta)$, which restricts us to constructing superoperators of the form $\sum_{\ell=0}^{2j} a_\ell\Pi_\ell$, each of which is specified by its $2j + 1$ eigenvalues $a_0, \ldots, a_{2j}$.  Consequently, this procedure is only capable of constructing full-rank irrep projectors $\Pi_k$ for which $a_\ell = \delta_{k\ell}$.  With a less restrictive frame, we can construct rank-1 projectors. (We have already observed that infinite frames can be used to construct rank-1 projectors.  But this can be done with finite frames as well.)

How should one choose a finite frame that is optimal with respect to some criteria (e.g., the variation in the coefficients necessary to construct a desired projector)?  One can take either a random or a deterministic approach.

\paragraph{Random approach.} If we choose $N_j$ group elements $g_i$ uniformly at random, then their superoperator representations $\mathcal{G}_i$ will be linearly independent with probability 1.  If the frame matrix is ill-conditioned, then the coefficients in the linear combinations needed to construct projectors may have high variance.  One could mitigate this possibility by iteratively refining the set of $N_j$ group elements so as to maximize the condition number.  Alternatively, one could choose more than $N_j$ group elements such that their overcomplete frame matrix is well-conditioned and use it to build the desired projector.  The coefficients would not be unique, but could be optimized according to different criteria.  To minimize their variance, one might want to minimize their $L^2$ norm by taking the Moore-Penrose pseudoinverse.

\paragraph{Deterministic approach.} One could make the frame superoperators maximally ``spread out'' by minimizing their mutual inner products.  Inner products between Wigner $D$-matrices and $\text{SU}(2)$ superoperators are given by
\begin{equation}
\tr(D^k(g_1)^\dag D^k(g_2)) = \tr(D^k(g_1^{-1}g_2)) = \chi_k(g_1^{-1}g_2) \implies \Tr(\mathcal{G}(g_1)^\dag\mathcal{G}(g_2)) = \sum_{k=0}^{2j} \chi_k(g_1^{-1}g_2),
\end{equation}
which can be further simplified by composing the corresponding axis-angle rotations. (Note that the absolute value of the character function $\chi_k(\theta)$ is maximized at $\theta = 0$.) One heuristic way to maximize this spread is to choose a spherical code of size $N_j$ on $S^3$, the group manifold of $\text{SU}(2)$.  Another is to choose a regularly spaced grid of at least $N_j$ points in $[0, 1]^3$ and then map them to $\text{SU}(2)$ according to the Haar measure in either the Euler angle or the axis-angle parametrization.  There are at least two caveats to doing so:
\begin{itemize}
\item The resulting points are not evenly distributed on the sphere $S^3$, as would be the case for a spherical code.
\item One must carefully choose the original grid to avoid linear dependence among the corresponding $\text{SU}(2)$ superoperators.  For example, given a set of rotations about a single axis, at most $4j + 1$ of them can be linearly independent.  This is because, in a basis where the rotation axis is the quantization axis, such rotations yield purely diagonal superoperators, and of the $(2j + 1)^2$ entries on the diagonal of such a superoperator, only $4j + 1$ of them are \emph{distinct} for generic rotation angles.
\end{itemize}

In this work, we do not pursue finite frames in detail.  When choosing a finite frame at random, there is no guarantee that the resulting variance will be smaller than that of character RB (in fact, we find empirically that it varies over many orders of magnitude and is generally larger).

\subsection{(Synthetic) SPAM}

Our end goal is to measure gate error rates as effectively as possible.  Regardless of the efficiency of the ``irrep projection'' scheme realized by synthetic gates, it needs SPAM operations that have significant overlap with the desired irrep to ensure a good exponentially decaying signal.

Restricting ourselves to physically realizable states and measurement effects limits our ability to optimize this overlap.  In particular, any physical state has constant overlap with the trivial irrep due to the unit trace condition and therefore cannot be made to lie entirely inside any nontrivial irrep.  More specifically, we would like to maximize the support of the initial (physical) state $\rho$ within any chosen irrep subspace.  Let $a_{k, q}\equiv \llangle T_q^{(k)}|\rho\rrangle$ denote the $(2j + 1)^2$ complex coefficients of $\rho$ in the spherical tensor basis.  Since $\tr T_q^{(k)} = \delta_{k0}\delta_{q0}\sqrt{2j + 1}$, normalization and hermiticity require that
\begin{equation}
a_{0, 0} = \frac{1}{\sqrt{2j + 1}}, \qquad a_{k, q} = (-1)^q a_{k, -q}^\ast.
\label{akqconditions}
\end{equation}
The component of $\rho$ within the $k^\text{th}$ irrep has squared norm
\begin{equation}
\llangle\rho|\Pi_k|\rho\rrangle = \sum_{q=-k}^k |a_{k, q}|^2.
\end{equation}
We thus wish to maximize $\llangle\rho|\Pi_k|\rho\rrangle$ for a given $k$ while minimizing $\llangle\rho|\Pi_{k'}|\rho\rrangle$ for $k'\neq k$, subject to \eqref{akqconditions} and the positive-semidefiniteness of $\rho$.  In general, this is a constrained optimization problem over $(2j + 1)^2 - 1$ real parameters.  This problem simplifies somewhat if $\rho$ is pure ($\rho = |\psi\rangle\langle\psi|$), in which case $4j$ real parameters suffice.

We do not attempt to solve this problem in generality, but let us comment on some simple cases.

A natural choice for $\rho$ is a $J_z$ eigenstate. (Here and below, we often leave $j$ implicit when writing such states: $|\ell\rangle\equiv |j, \ell\rangle$.) The only component of a pure $J_z$ eigenstate $\rho = |\ell\rangle\langle\ell|$ within the $k^\text{th}$ irrep is that in the $T^{(k)}_0$ direction, with amplitude
\begin{equation}
M_{k, \ell}\equiv \langle\ell|T^{(k)}_0|\ell\rangle = \sqrt{\frac{2k + 1}{2j + 1}}C^{j, \ell}_{j, \ell; k, 0}.
\label{Mdefinition}
\end{equation}
Note that $M_{k, -\ell} = (-1)^k M_{k, \ell}$.  As a function of $k$, the squared norm $M_{k, \ell}^2$ has $j - |\ell| + 1$ peaks.  The peaks move toward higher $k$ as $|\ell|$ decreases.

Another natural choice for $\rho$ is a spin coherent state.  Spin coherent states are informationally complete for process tomography: as state vectors, they span the $(2j + 1)$-dimensional Hilbert space, and as rank-one density matrices, they span the $(2j + 1)^2$-dimensional Hilbert-Schmidt space.  Global $\text{SU}(2)$ rotations act transitively on the set of spin coherent states, which includes the ``stretched'' $J_z$ eigenstates with eigenvalues $\ell = \pm j$.  Since any two states in the same $\text{SU}(2)$ orbit have the same support within irreps, all spin coherent states have the same support within irreps as the states $|{\pm j}\rangle$.  This support exhibits a single peak and a long tail as a function of the irrep label, becoming progressively smaller in high-spin irreps as $j\to\infty$.

One can circumvent the limitations of physical states and effects by using \emph{synthetic} states and effects (collectively, synthetic SPAM).  The optimal initial states and measurement effects are supported exclusively in a target irrep.  However, these states are unphysical, as their density matrices have negative eigenvalues.  We can nonetheless simulate the preparation of these unphysical states by preparing different physical states (e.g., $J_z$ eigenstates) and then taking linear combinations of the results of the corresponding RB experiments.  This is accomplished by classical post-processing.  The same goes for measurement.  If the resulting synthetic SPAM has sufficiently good overlap with the target irrep, then it may even obviate the need to use \emph{any} irrep projection scheme derived from synthetic gates.

It is possible to construct synthetic states and effects that not only lie entirely in a given irrep, but are also diagonal in the basis of $J_z$ eigenstate matrix units.  These are precisely the diagonal spherical tensor operators $T^{(k)}_0$ for each $k = 0, \ldots, 2j$, which are related to the $|\ell\rangle\langle\ell|$ via the coefficients \eqref{Mdefinition}:
\begin{equation}
T^{(k)}_0 = \sum_{\ell = -j}^j M_{k, \ell}|\ell\rangle\langle\ell| \Longleftrightarrow |\ell\rangle\langle\ell| = \sum_{k=0}^{2j} M_{k, \ell}T_0^{(k)},
\label{byorthogonality}
\end{equation}
where we have used orthogonality of the matrix $M = [M_{k, \ell}]$:
\begin{equation}
\sum_{k=0}^{2j} M_{k, \ell}M_{k, \ell'} = \delta_{\ell\ell'}, \qquad \sum_{\ell=-j}^j M_{k, \ell}M_{k', \ell} = \delta_{kk'}.
\end{equation}
Thus, by using the $2j + 1$ matrix units $|\ell\rangle\langle\ell|$ as initial states and measurement effects and taking linear combinations of the corresponding survival probabilities, we can simulate both the preparation and the measurement of $T^{(k)}_0$.

\section{Sample Complexity: Bounds} \label{app:SCbounds}

We first derive rigorous bounds on the sample complexity of our various protocols in the presence of arbitrary gate noise, taking the opportunity to discuss generalities of the sample complexity analysis.  Later, we improve these estimates by deriving exact results for the sample complexity in the limit of zero noise.

The basic idea is that experiments sample from probability distributions, allowing us to estimate their means up to fluctuations due to finite sample size.  The time cost of an experiment can be quantified by both ``shots'' and ``settings.''  A shot is a single execution of a quantum circuit that yields a specific outcome.  Shots are cheap if the system is fast.  Settings are cheap if the control system can be reconfigured quickly.  The optimal approach will depend on the relative cost of shots and settings.  For simplicity, we focus on the cost of shots and assume that settings are cheap.

We first consider separately the sample complexity of RB protocols that use synthetic gates and synthetic SPAM.  We then combine these analyses to discuss RB protocols that use both.

\subsection{Synthetic-Gate RB} \label{app:syntheticgatebounds}

\subsubsection{Infinite Frame}

Consider character RB (which we sometimes abbreviate as $\chi$RB to avoid confusion with Clifford RB, or CRB), the prototypical example of a protocol that uses an infinite frame to construct projection superoperators.  For a given benchmarking group $G$ and character group $H$ (both possibly continuous), we assume accurate SPAM and noisy gates, with the noise channel $\mathcal{E}$ unknown but fixed.  Given a fixed number of shots, we would like to bound the variance of the estimated (weighted) survival probabilities corresponding to each irrep.  For a given irrep, the weighted success probability is an average of some function over the character group.  The primary case of interest is $\text{SU}(2)$ character RB with $\text{SU}(2)$ character group: $G = H = \text{SU}(2)$ with representation label $\lambda' = k$ and $\dim\phi_k = 2k + 1$.  We assume that the characters of $G$ are real-valued, i.e., that the representations are self-conjugate.  This is true of $\text{SU}(2)$ because the representations of $\text{SU}(2)$ are (pseudo)real (concretely, one can conjugate group elements to a maximal torus where they are manifestly real).  For compact character and benchmarking groups, we write
\begin{equation}
\mathbb{E}_{h\in H}[f(h)] = \frac{1}{|H|}\int dh\, f(h), \qquad \mathbb{E}_{\vec{g}\in G^m}[f(\vec{g})] = \frac{1}{|G|^m}\int d\vec{g}\, f(\vec{g}),
\end{equation}
where $dh$ and $d\vec{g}\equiv dg_1\cdots dg_m$ (with $\vec{g}\equiv (g_1, \ldots, g_m)$) are the relevant (unnormalized) Haar measures.

We may consider three contributions to estimator variance in character RB \cite{Helsen}: (1) variance from sampling many different circuits (random sequences of $\text{SU}(2)$ operations), (2) variance from repeating each circuit finitely many times, (3) variance from sampling many different end-of-circuit $\text{SU}(2)$ character group elements.  Fixing a finite frame would eliminate the third source of variance; similarly, random sampling would be unnecessary for a finite character group.

The ``na\"ive'' procedure of \cite{Helsen} is as follows.  First sample $\vec{g}\in G^m$ and $h\in H$ uniformly at random.  Then estimate the weighted survival probability
\begin{equation}
p_m^{\lambda'}(\vec{g}, h)\equiv (\dim\phi_{\lambda'})\chi_{\lambda'}(h)\sbra{E}\widetilde{\mathcal{G}}_\text{inv}\widetilde{\mathcal{G}}_m\cdots \widetilde{\mathcal{G}}_2\widetilde{\mathcal{G}_1\mathcal{H}}\sket{\rho}
\end{equation}
to within the desired precision by running the corresponding circuit and measuring the binary outcome $\in \{0, 1\}$ sufficiently many times.  Next, repeat for sufficiently many $h$ to estimate the $H$-average
\begin{equation}
p_m^{\lambda'}(\vec{g})\equiv \mathbb{E}_{h\in H}[p_m^{\lambda'}(\vec{g}, h)].
\end{equation}
Finally, repeat for sufficiently many $\vec{g}$ to estimate the $G$-average
\begin{equation}
p_m^{\lambda'}\equiv \mathbb{E}_{\vec{g}\in G^m}[p_m^{\lambda'}(\vec{g})].
\end{equation}
If $H$ is a small finite group, then one can simply iterate over all $h\in H$ without random sampling; this is inefficient if $H$ is large and impossible if $H$ is infinite.  The same is true of $G$.

The authors of \cite{Helsen} propose a ``shortcut'' to this procedure: directly estimate the $H$-averaged weighted survival probability $p_m^{\lambda'}(\vec{g})$ for a given sequence $\vec{g}$ without first estimating $p_m^{\lambda'}(\vec{g}, h)$ for fixed $h$ by
\begin{enumerate}
\item randomly sampling $h$,
\item running and measuring the corresponding circuit to obtain a result $\in \{0, 1\}$,
\item multiplying the result by $(\dim\phi_{\lambda'})\chi_{\lambda'}(h)$,
\item and averaging these rescaled results over many trials.
\end{enumerate}
This shortcut entails viewing $p_m^{\lambda'}(\vec{g})$ as the mean of a random variable.  The absolute value of the character function of a finite-dimensional unitary representation of a group is bounded above by the dimension of the corresponding representation.  Therefore, the random variable in question takes values in the interval $[-(\dim\phi_{\lambda'})^2, (\dim\phi_{\lambda'})^2]$.  For a bounded, real-valued probability distribution, Hoeffding's inequality gives a lower bound on the number of samples required to achieve a given confidence interval about the mean. (Given a probability distribution with mean $\mu$ and an estimate $\mu_N$ for the mean based on $N$ samples, we define a confidence interval as a pair of real numbers $(\epsilon, \delta)$ such that $\operatorname{Pr}(|\mu_N - \mu|\geq \epsilon)\leq 1 - \delta$.) However, it is not determined in \cite{Helsen} when this shortcut is actually a shortcut, nor is the estimation of $p_m^{\lambda'}$ given the estimate(s) for $p_m^{\lambda'}(\vec{g})$ discussed.

We note that there is an even shorter shortcut to this procedure: we need not view the estimation of $p_m^{\lambda'}$ as a two-step process at all.  One can simply randomly sample $\vec{g}, h$ anew \emph{before every shot} and average the rescaled measurement outcomes.  This strategy, which we take, makes sense if settings are cheap and shots are not.

To proceed, we need a few elementary facts.  In the limit of large sample size $N$, the central limit theorem states that the sample mean of a random variable with mean $\mu$ and variance $\sigma^2$ is a Gaussian random variable with mean $\mu$ and variance $\sigma^2/N$.  Also, a mixture of distributions differs from a linear combination of random variables as follows:
\begin{itemize}
\item To sample from a mixture distribution, an element of a set of random variables is selected according to some probability distribution, and then a value of that random variable is realized.  The probability density function of the mixture is a convex combination of the probability density functions of the components.
\item To sample from a linear combination of random variables, each constituent random variable is sampled separately and then the corresponding linear combination of the results is taken.  The probability density function of the linear combination is a convolution of the probability density functions of the summands.
\end{itemize}
In particular, the total mean and variance of a mixture of one-dimensional distributions with weights $w_i$, means $\mu_i$, and variances $\sigma_i^2$ are
\begin{equation}
\mu = \sum_i w_i\mu_i, \qquad \sigma^2 = \sum_i w_i(\sigma_i^2 + \mu_i^2) - \mu^2.
\end{equation}
If all of the components are Bernoulli distributions, then we have $\sigma_i^2 = \mu_i(1 - \mu_i)$ and therefore
\begin{equation}
\sigma^2 = \mu(1 - \mu).
\end{equation}
Indeed, a mixture of Bernoulli distributions is itself a Bernoulli distribution because the outcomes remain 0 and 1.  On the other hand, if the components are \emph{rescaled} Bernoulli random variables with scale factors $\alpha_i$, so that $\mu = \sum_i w_i\alpha_i\mu_i$, then we have $\sigma_i^2 = \alpha_i^2\mu_i(1 - \mu_i)$ and therefore
\begin{equation}
\sigma^2 = \sum_i w_i\alpha_i^2\mu_i - \mu^2.
\label{varmixrescaledbernoulli}
\end{equation}
We will be interested in continuous rather than discrete weights $w_i$.

In our random sampling approach, where we sample $\vec{g}, h$ before each shot, we are sampling from a mixture (with respect to $\vec{g}, h$) of rescaled Bernoulli random variables.  Let
\begin{equation}
\mu_{\vec{g}, h}\equiv \sbra{E}\widetilde{\mathcal{G}}_\text{inv}\widetilde{\mathcal{G}}_m\cdots \widetilde{\mathcal{G}}_2\widetilde{\mathcal{G}_1\mathcal{H}}\sket{\rho}.
\end{equation}
Each component distribution is specified by a choice of $\vec{g}, h$: the mean is $(\dim\phi_{\lambda'})\chi_{\lambda'}(h)\mu_{\vec{g}, h}$, and the variance is $(\dim\phi_{\lambda'})^2 \linebreak[1] \chi_{\lambda'}(h)^2 \linebreak[1] \mu_{\vec{g}, h} \linebreak[1] (1 - \mu_{\vec{g}, h})$.  Therefore, the total mean of the mixture is
\begin{equation}
\mu = (\dim\phi_{\lambda'})\mathbb{E}_{h\in H}[\chi_{\lambda'}(h)\mathbb{E}_{\vec{g}\in G^m}[\mu_{\vec{g}, h}]],
\end{equation}
and the total variance of the mixture is
\begin{align}
\sigma^2 &= (\dim\phi_{\lambda'})^2\mathbb{E}_{h\in H}[\chi_{\lambda'}(h)^2\mathbb{E}_{\vec{g}\in G^m}[\mu_{\vec{g}, h}]] - \mu^2 \\
&\leq (\dim\phi_{\lambda'})^2\mathbb{E}_{h\in H}[\chi_{\lambda'}(h)^2\mathbb{E}_{\vec{g}\in G^m}[\mu_{\vec{g}, h}]] \\
&\leq (\dim\phi_{\lambda'})^2\mathbb{E}_{h\in H}[\chi_{\lambda'}(h)^2] \\
&= (\dim\phi_{\lambda'})^2, \label{randomsamplingbound}
\end{align}
where we used character orthonormality in the last step.  It follows that for $N$ random samples, the variance of the sample mean is bounded above by $(\dim\phi_{\lambda'})^2/N$.  Note that since $\mu_{\vec{g}, h}\in [0, 1]$, the values of the mixture distribution are bounded above and below by $\pm(\dim\phi_{\lambda'})^2$; using this information alone would only allow us to conclude that the variance is bounded from above by $(\dim\phi_{\lambda'})^4$.

The fact that the above bound increases with the dimension of the target irrep already suggests that it may be more efficient to build a rank-deficient projector than a full projector onto the irrep. (As we will see, this is also true when using finite frames.) The most extreme option is to build a projector onto a 1D subspace of the target irrep.

\subsubsection{Finite Frame} \label{app:FFRB}

Using a finite frame to construct a projector eliminates the variance from $H$-averaging in favor of the variance incurred by a predetermined set of coefficients.  In this approach (which we abbreviate as FFRB), we instead write
\begin{equation}
p_m^{\lambda'}(\vec{g}) = \sum_{i=1}^{N_j} c_i\sbra{E}\widetilde{\mathcal{G}}_\text{inv}\widetilde{\mathcal{G}}_m\cdots \widetilde{\mathcal{G}}_2\widetilde{\mathcal{G}_1\mathcal{G}_i'}\sket{\rho}
\label{finiteframelincombo}
\end{equation}
for some fixed group elements $g_i'$ and real coefficients $c_i$ satisfying $\sum_{i=1}^{N_j} c_i\mathcal{G}_i' = \Pi_{\lambda'}$.  To estimate $p_m^{\lambda'}$, we thus sample from a linear combination of $N_j$ independent mixtures of Bernoulli distributions with means
\begin{equation}
\mu_i\equiv \mathbb{E}_{\vec{g}\in G^m}[\sbra{E}\widetilde{\mathcal{G}}_\text{inv}\widetilde{\mathcal{G}}_m\cdots \widetilde{\mathcal{G}}_2\widetilde{\mathcal{G}_1\mathcal{G}_i'}\sket{\rho}]
\end{equation}
($i = 1, \ldots, N_j$).  Each mixture distribution has variance bounded above as $\sigma_i^2\leq 1/4$, corresponding to the ``worst case'' $\mu_i = 1/2$.  Therefore, the total variance is bounded above as
\begin{equation}
\sigma^2\leq \frac{1}{4}\sum_{i=1}^{N_j} c_i^2.
\end{equation}
If we were to sample uniformly from the total distribution, then the variance of the sample mean would satisfy
\begin{equation}
\overline{\sigma}^2\leq \frac{1}{4N}\sum_{i=1}^{N_j} c_i^2 = \frac{N_j}{4s}\sum_{i=1}^{N_j} c_i^2,
\label{bounduniform}
\end{equation}
where the number of shots would be $s = NN_j$ because each sample requires $N_j$ shots.  A better choice is to sample non-uniformly from this distribution: if we draw $s_i$ samples from distribution $i$, then the variance of the sample mean is bounded as
\begin{equation}
\overline{\sigma}^2\leq \frac{1}{4}\sum_{i=1}^{N_j} \frac{c_i^2}{s_i},
\end{equation}
and the number of shots is $s = \sum_{i=1}^{N_j} s_i$.  We want to minimize the bound on $\overline{\sigma}^2$ with respect to the $s_i$ while holding $s$ fixed.  Using Lagrange multipliers, this corresponds to choosing
\begin{equation}
s_i = \frac{s|c_i|}{\sum_{\ell=1}^{N_j} |c_\ell|} \implies \overline{\sigma}^2\leq \frac{1}{4s}\left(\sum_{i=1}^{N_j} |c_i|\right)^2.
\label{boundnonuniform}
\end{equation}
Note that the $s_i$ are positive, whereas the $c_i$ are not necessarily positive.  By the Cauchy-Schwarz inequality, the bound \eqref{boundnonuniform} is at least as strong as \eqref{bounduniform}:
\begin{equation}
\left(\sum_{i=1}^{N_j} |c_i|\right)^2\leq N_j\sum_{i=1}^{N_j} c_i^2.
\end{equation}
If all coefficients $c_i$ are equal ($c_i = c$), then both bounds reduce to
\begin{equation}
\overline{\sigma}^2\leq \frac{(N_j c)^2}{4s}.
\end{equation}
We see that optimizing the coefficients $c_i$ means minimizing the $L^1$ norm of the coefficient vector.  Operationally, in the finite-frame approach, we sample from $N_j$ different distributions and combine the results.  By linearity of expectation, we can interpret this as sampling from a single distribution:
\begin{equation}
p_m^{\lambda'} = \sum_{i=1}^{N_j} c_i\mu_i = \mathbb{E}_{\vec{g}\in G^m}\left[\sum_{i=1}^{N_j} c_i\sbra{E}\widetilde{\mathcal{G}}_\text{inv}\widetilde{\mathcal{G}}_m\cdots \widetilde{\mathcal{G}}_2\widetilde{\mathcal{G}_1\mathcal{G}_i'}\sket{\rho}\right].
\end{equation}
The values of the overall distribution are bounded above by the sum of all positive $c_i$ and bounded below by the sum of all negative $c_i$.  Hoeffding's inequality then gives a direct estimate for the number of samples required to achieve a given confidence interval about the mean.

In light of the above discussion, let us consider a variation of the infinite-frame random sampling approach in which $h\in H$ and $\vec{g}\in G^m$ are sampled separately rather than together.  The motivation for doing so is that we compute a \emph{weighted} average over $h$ as opposed to a uniform average over $\vec{g}$.  One can choose to perform the $\vec{g}$ estimate first and thus write
\begin{equation}
p_m^{\lambda'}(h)\equiv \mathbb{E}_{\vec{g}\in G^m}[p_m^{\lambda'}(\vec{g}, h)]
\end{equation}
and $p_m^{\lambda'} = \mathbb{E}_{h\in H}[p_m^{\lambda'}(h)]$.  If $H$ were a finite group, then we could write
\begin{equation}
p_m^{\lambda'} = \sum_{h\in H} c_h\mathbb{E}_{\vec{g}\in G^m}[\mu_{\vec{g}, h}], \qquad c_h\equiv \frac{\dim\phi_{\lambda'}}{|H|}\chi_{\lambda'}(h).
\end{equation}
Then, as in the finite-frame approach, we could sample from a linear combination of $|H|$ independent mixtures of Bernoulli distributions.  The mixture labeled by a given $h$ has mean and variance
\begin{equation}
\mu_h = \mathbb{E}_{\vec{g}\in G^m}[\mu_{\vec{g}, h}], \qquad \sigma_h^2 = \mu_h(1 - \mu_h)\leq \frac{1}{4}.
\end{equation}
The total variance is a weighted sum (by $c_h^2$) of the variances of the mixture distributions.  As with a finite frame, although $\vec{g}$ is sampled uniformly, we would like to draw different numbers of samples from each distribution labeled by $h$ due to the different weightings.  Note that this is not the same as sampling over $h$ non-uniformly: for finite $H$, we simply iterate over all $h\in H$ rather than randomly sampling $h$.  By the previous considerations, for a fixed number of shots $s$, the optimal number of samples from each mixture distribution is
\begin{equation}
s_h = \frac{s|\chi_{\lambda'}(h)|}{\sum_{h'\in H} |\chi_{\lambda'}(h')|}.
\end{equation}
Note that it is important that we consider $\sum_{h\in H} |c_h|$ rather than $\sum_{h\in H} c_h$.  Indeed, by orthonormality of irreducible characters, we have
\begin{equation}
\sum_{h\in H} c_h = \frac{\dim\phi_{\lambda'}}{|H|}\sum_{h\in H} \chi_{\lambda'}(h)\chi_0(h) = (\dim\phi_{\lambda'})\delta_{\lambda', 0} = \delta_{\lambda', 0},
\end{equation}
where ``0'' labels the trivial representation.  If $H$ is an infinite group, then we cannot sum over all $h\in H$ directly.  Rather, we can write
\begin{equation}
p_m^{\lambda'}\approx \frac{\dim\phi_{\lambda'}}{|S_H|}\sum_{h\in S_H} \chi_{\lambda'}(h)\mathbb{E}_{\vec{g}\in G^m}[\mu_{\vec{g}, h}]
\end{equation}
for some $S_H\subset H$, where the approximation becomes better as $|S_H|\to\infty$.  $S_H$ can either be a predetermined uniformly distributed subset of $H$ or chosen by randomly sampling $|S_H|$ elements $h\in H$.  Either way, one can then estimate $p_m^{\lambda'}$ by sampling from a linear combination of $|S_H|$ mixture distributions, with similar optimal sample sizes as before:
\begin{equation}
s_h = \frac{s|\chi_{\lambda'}(h)|}{\sum_{h'\in S_H} |\chi_{\lambda'}(h')|}.
\end{equation}
Again, this is not the same as sampling $h$ non-uniformly from $H$.  Rather, $h$ is sampled uniformly from $H$, but each choice of $h$ further determines a distribution from which we draw a number of samples that depends nontrivially on $h$.  When $H$ is infinite, error propagation is needed to determine the overall number of shots required by this two-step procedure.  Namely, we must combine the variance from sampling finitely many elements $h\in H$ with the variance from drawing finitely many samples from each distribution labeled by $h$.

Although the finite-frame approach is more efficient than random sampling from an infinite frame with respect to constructing irrep projectors, our discussion so far shows that it is not \emph{a priori} more efficient with respect to estimating expectation values (averaged weighted survival probabilities).  This can be seen by comparing the bound on the variance of random sampling, \eqref{randomsamplingbound}, to the ``optimized'' bound \eqref{boundnonuniform} for the finite frame.

In fact, a more appropriate comparison between pure random sampling and the finite frame would involve interpreting the linear combination of random variables indicated in \eqref{finiteframelincombo} as a mixture and randomly sampling according to the corresponding distribution. (More precisely, this means defining a mixture distribution with the same mean as the original linear combination and then minimizing its variance with respect to the mixture weights.) To interpret this linear combination as a mixture, we must (re)write it as a \emph{convex} combination; we first address the ambiguity in doing so.

Any real linear combination $X$ of real-valued random variables $X_i$ can be written as a convex combination of rescaled random variables:
\begin{equation}
X\equiv \sum_{i=1}^n c_i X_i = \sum_{i=1}^n p_i\left(\frac{c_i}{p_i}X_i\right), \qquad \sum_{i=1}^n p_i = 1, \qquad p_i\geq 0.
\end{equation}
Let $X'$ be an $n$-component finite mixture distribution with weights $p_i$ and components $c_i X_i/p_i$.  By construction, $X'$ has the same mean as $X$, so it suffices to sample from $X'$ to estimate this mean.  Regardless of the $p_i$, the mean of the mixture distribution $X'$ is $\mu = \sum_{i=1}^n c_i \mu_i$, and its variance is
\begin{equation}
\sigma_{X'}^2 = \sum_{i=1}^n \frac{c_i^2}{p_i}(\sigma_i^2 + \mu_i^2) - \mu^2.
\end{equation}
We choose the $p_i$ to minimize this variance, subject to the convexity constraints.  This is the same constrained optimization problem as before: we find that the optimal choice and the corresponding variance are
\begin{equation}
p_i = \frac{|c_i|(\sigma_i^2 + \mu_i^2)^{1/2}}{\sum_{j=1}^n |c_j|(\sigma_j^2 + \mu_j^2)^{1/2}}, \qquad \sigma_{X'}^2 = \left[\sum_{i=1}^n |c_i|(\sigma_i^2 + \mu_i^2)^{1/2}\right]^2 - \mu^2.
\end{equation}
In practice, we may not know the $\mu_i$ or $\sigma_i$, in which case we assume for simplicity that $\mu_i = \tilde{\mu}$ and $\sigma_i = \tilde{\sigma}$ for all $i$, leading to
\begin{equation}
p_i = \frac{|c_i|}{\sum_{j=1}^n |c_j|}, \qquad \sigma_{X'}^2 = (\tilde{\sigma}^2 + \tilde{\mu}^2)\left(\sum_{i=1}^n |c_i|\right)^2 - \mu^2.
\end{equation}
This would be the optimal choice if the $X_i$ were independent and identically distributed.  We may compare the variance of the mixture $X'$ to the variance of the linear combination $X$ (assuming that the $X_i$ are independent), which is
\begin{equation}
\sigma_X^2 = \sum_{i=1}^n c_i^2\sigma_i^2.
\end{equation}
However, to make a fair comparison, note that sampling once from the linear combination requires sampling once from each component, while sampling once from the mixture requires sampling once from a single component.  Therefore, sampling from the linear combination is $n$ times as expensive as sampling from the mixture.  Thus we should compare
\begin{equation}
\overline{\sigma}_X^2 = n\sum_{i=1}^n c_i^2\sigma_i^2, \qquad \overline{\sigma}_{X'}^2 = \left[\sum_{i=1}^n |c_i|(\sigma_i^2 + \mu_i^2)^{1/2}\right]^2 - \mu^2.
\end{equation}
Note also that $|\vec{c}|_2\leq |\vec{c}|_1\leq \sqrt{n}|\vec{c}|_2$, where we use $|\cdot|_p$ to denote the $p$-norm: $|\vec{c}|_2\equiv (\sum_{i=1}^{N_j} c_i^2)^{1/2}$ and $|\vec{c}|_1\equiv \sum_{i=1}^{N_j} |c_i|$.

Returning to our application, we write the linear combination of interest as the following convex combination:
\begin{equation}
p_m^{\lambda'}(\vec{g}) = \sum_{i=1}^{N_j} \frac{|c_i|}{|\vec{c}|_1}\times \operatorname{sgn}(c_i)|\vec{c}|_1\mu_{\vec{g}, g_i'}, \qquad \mu_{\vec{g}, g_i'}\equiv \sbra{E}\widetilde{\mathcal{G}}_\text{inv}\widetilde{\mathcal{G}}_m\cdots \widetilde{\mathcal{G}}_2\widetilde{\mathcal{G}_1\mathcal{G}_i'}\sket{\rho}.
\end{equation}
This expression, after averaging over $\vec{g}$, is the mean of a mixture (with respect to $\vec{g}$ and $i$) of rescaled Bernoulli random variables.  The component distribution specified by $\vec{g}$ and $i$ has mean $\operatorname{sgn}(c_i)|\vec{c}|_1\mu_{\vec{g}, g_i'}$ and variance $|\vec{c}|_1^2\mu_{\vec{g}, g_i'}(1 - \mu_{\vec{g}, g_i'})$.  The total mean of the mixture is
\begin{equation}
\mu = \sum_{i=1}^{N_j} c_i\mu_i, \qquad \mu_i\equiv \mathbb{E}_{\vec{g}\in G^m}[\mu_{\vec{g}, g_i'}],
\end{equation}
and the total variance of the mixture (cf.\ \eqref{varmixrescaledbernoulli}) satisfies
\begin{equation}
\sigma^2 = |\vec{c}|_1\sum_{i=1}^{N_j} |c_i|\mu_i - \mu^2\leq |\vec{c}|_1^2 - \mu^2\leq |\vec{c}|_1^2.
\label{boundmixture}
\end{equation}
This is comparable to the ``optimized'' bound \eqref{boundnonuniform}.

\subsection{Synthetic-SPAM RB}

We now consider the sample complexity of $\text{SU}(2)$ RB with only synthetic SPAM (which we abbreviate as SSRB), using the diagonal spherical tensor operators with components \eqref{Mdefinition} as synthetic states and measurement effects.  We use $\ell$ to denote $J_z$ eigenvalues rather than $m$, reserving the latter for gate sequence length, and we also use the shorthand notation $\sket{\ell}\equiv \sket{|\ell\rangle\langle\ell|}$.  Let
\begin{equation}
\mu_{\ell, \ell', m}(\vec{g})\equiv \llangle\ell'|\widetilde{\mathcal{G}}_\text{inv}\widetilde{\mathcal{G}}_m\cdots \widetilde{\mathcal{G}}_1\sket{\ell}, \qquad \mu_{\ell, \ell', m}\equiv \mathbb{E}_{\vec{g}\in G^m}[\mu_{\ell, \ell', m}(\vec{g})].
\end{equation}
We wish to estimate
\begin{equation}
\mu_{k, m}\equiv \sum_{\ell, \ell' = -j}^j M_{k, \ell}M_{k, \ell'}\mu_{\ell, \ell', m}
\end{equation}
for all $k = 0, \ldots, 2j$.  These are the means of linear combinations of Bernoulli random variables $X_{\ell, \ell', m}$ that are themselves mixtures (with respect to $\vec{g}\in G^m$) of Bernoulli random variables.

The random variables $X_{\ell, \ell', m}$ are not all independent.  For fixed $\ell$, their means $\mu_{\ell, \ell', m} = \langle X_{\ell, \ell', m}\rangle$ satisfy
\begin{equation}
\sum_{\ell' = -j}^j \mu_{\ell, \ell', m} = 1.
\end{equation}
Moreover, their covariances satisfy
\begin{equation}
\operatorname{Cov}(X_{\ell_\text{init}, \ell_\text{final}, m}, X_{\ell_\text{init}', \ell_\text{final}', m}) = \begin{cases} 0 & \text{if $\ell_\text{init}\neq \ell_\text{init}'$}, \\ -\mu_{\ell_\text{init}, \ell_\text{final}, m}\mu_{\ell_\text{init}, \ell_\text{final}', m} & \text{if $\ell_\text{init} = \ell_\text{init}'$ and $\ell_\text{final}\neq \ell_\text{final}'$}, \\
\mu_{\ell_\text{init}, \ell_\text{final}, m}(1 - \mu_{\ell_\text{init}, \ell_\text{final}, m}) & \text{if $\ell_\text{init} = \ell_\text{init}'$ and $\ell_\text{final} = \ell_\text{final}'$}, \end{cases}
\label{SSRBcovariances}
\end{equation}
where:
\begin{itemize}
\item The first line in \eqref{SSRBcovariances} holds because measurements with different initial states are independent.
\item The second line in \eqref{SSRBcovariances} holds because each shot yields one outcome $|\ell_\text{final}\rangle\langle\ell_\text{final}|$, so if the value of $X_{\ell_\text{init}, \ell_\text{final}, m}$ is nonzero, then the value of $X_{\ell_\text{init}, \ell_\text{final}', m}$ must be zero.  Therefore, the expectation value $\langle X_{\ell_\text{init}, \ell_\text{final}, m}X_{\ell_\text{init}, \ell_\text{final}', m}\rangle$ in a single shot is always zero.  Then recall the formula $\operatorname{Cov}(X, Y) = \langle XY\rangle - \langle X\rangle\langle Y\rangle$.
\item The third line in \eqref{SSRBcovariances} follows from the standard formula $\operatorname{Cov}(X, X) = \operatorname{Var}(X) = \mu(1 - \mu)$ for Bernoulli random variables $X$ (for which $\langle X^2\rangle = \langle X\rangle$).
\end{itemize}
To be sure, let us provide a direct derivation of the second line in \eqref{SSRBcovariances}.  Letting
\begin{equation}
\sket{\rho_m}\equiv \mathbb{E}_{\vec{g}\in G^m}[\widetilde{\mathcal{G}}_\text{inv}\widetilde{\mathcal{G}}_m\cdots \widetilde{\mathcal{G}}_1\sket{\ell_\text{init}}],
\end{equation}
we can write $\llangle\ell_\text{final}|\rho_m\rrangle = \langle\ell_\text{final}|\rho_m|\ell_\text{final}\rangle = \mu_{\ell_\text{init}, \ell_\text{final}, m}$.  The physical PVM elements are $\{|\ell\rangle\langle\ell|\}$.  Suppose we measure $|\ell\rangle\langle\ell|$ in the state $\rho_m$.  With probability $\langle\ell|\rho_m|\ell\rangle$, we obtain the outcome 1 and the post-measurement state $|\ell\rangle$, while with probability $1 - \langle\ell|\rho_m|\ell\rangle$, we obtain the outcome 0 and the post-measurement state
\begin{equation}
\rho_m'\equiv \frac{(\mathds{1} - |\ell\rangle\langle\ell|)\rho_m(\mathds{1} - |\ell\rangle\langle\ell|)}{1 - \langle\ell|\rho_m|\ell\rangle}.
\end{equation}
Next, suppose we measure $|\ell'\rangle\langle\ell'|$ (with $\ell'\neq \ell$) in the state $\rho_m'$.  With probability $\frac{\langle\ell'|\rho_m|\ell'\rangle}{1 - \langle\ell|\rho_m|\ell\rangle}$, we obtain the outcome 1, while with probability $\frac{1 - \langle\ell|\rho_m|\ell\rangle - \langle\ell'|\rho_m|\ell'\rangle}{1 - \langle\ell|\rho_m|\ell\rangle}$, we obtain the outcome 0.  Therefore, after a joint measurement of $|\ell\rangle\langle\ell|$ and $|\ell'\rangle\langle\ell'|$ on $\rho_m$:
\begin{itemize}
\item With probability $\mu_{\ell_\text{init}, \ell, m}$, we obtain $X_{\ell_\text{init}, \ell, m} = 1$ and $X_{\ell_\text{init}, \ell', m} = 0$.
\item With probability $\mu_{\ell_\text{init}, \ell', m}$, we obtain $X_{\ell_\text{init}, \ell, m} = 0$ and $X_{\ell_\text{init}, \ell', m} = 1$.
\item With probability $1 - \mu_{\ell_\text{init}, \ell, m} - \mu_{\ell_\text{init}, \ell', m}$, we obtain $X_{\ell_\text{init}, \ell, m} = 0$ and $X_{\ell_\text{init}, \ell', m} = 0$.
\end{itemize}
It follows that
\begin{align}
\operatorname{Cov}(X_{\ell_\text{init}, \ell, m}, X_{\ell_\text{init}, \ell', m}) &= \mu_{\ell_\text{init}, \ell, m}(1 - \mu_{\ell_\text{init}, \ell, m})(0 - \mu_{\ell_\text{init}, \ell', m}) \nonumber \\
&+ \mu_{\ell_\text{init}, \ell', m}(0 - \mu_{\ell_\text{init}, \ell, m})(1 - \mu_{\ell_\text{init}, \ell', m}) \nonumber \\
&+ (1 - \mu_{\ell_\text{init}, \ell, m} - \mu_{\ell_\text{init}, \ell', m})(0 - \mu_{\ell_\text{init}, \ell, m})(0 - \mu_{\ell_\text{init}, \ell', m}) \\
&= -\mu_{\ell_\text{init}, \ell, m}\mu_{\ell_\text{init}, \ell', m},
\end{align}
as expected.  Altogether, for fixed $\ell_\text{init}$, the covariance matrix of the random variables $\{X_{\ell_\text{init}, \ell, m} \,|\, \ell = -j, \ldots, j\}$ has entries
\begin{equation}
\operatorname{Cov}(X_{\ell_\text{init}, \ell, m}, X_{\ell_\text{init}, \ell', m}) = \mu_{\ell_\text{init}, \ell, m}(\delta_{\ell\ell'} - \mu_{\ell_\text{init}, \ell', m}),
\end{equation}
where we drop the restriction that $\ell'\neq \ell$.

Our interest lies in the random variables
\begin{equation}
X_{k, m}\equiv \sum_{\ell, \ell' = -j}^j M_{k, \ell}M_{k, \ell'}X_{\ell, \ell', m}.
\end{equation}
We compute that
\begin{align}
\operatorname{Var}(X_{k, m}) &= \sum_{\ell, \ell', \ell'' = -j}^j M_{k, \ell}^2 M_{k, \ell'}M_{k, \ell''}\operatorname{Cov}(X_{\ell, \ell', m}, X_{\ell, \ell'', m}) \\
&= \sum_{\ell, \ell' = -j}^j M_{k, \ell}^2 M_{k, \ell'}^2\mu_{\ell, \ell', m} - \sum_{\ell, \ell', \ell'' = -j}^j M_{k, \ell}^2 M_{k, \ell'}M_{k, \ell''}\mu_{\ell, \ell', m}\mu_{\ell, \ell'', m}.
\end{align}
Since the $M_{k, \ell}$ are real but of indefinite sign, we have the following upper bound:
\begin{equation}
\operatorname{Var}(X_{k, m})\leq \sum_{\ell, \ell' = -j}^j M_{k, \ell}^2 M_{k, \ell'}^2 + \sum_{\ell, \ell', \ell'' = -j}^j M_{k, \ell}^2|M_{k, \ell'}||M_{k, \ell''}| = |\vec{M}_k|_2^2(|\vec{M}_k|_2^2 + |\vec{M}_k|_1^2),
\end{equation}
where
\begin{equation}
|\vec{M}_k|_2\equiv \left(\sum_{\ell = -j}^j M_{k, \ell}^2\right)^{1/2}, \qquad |\vec{M}_k|_1\equiv \sum_{\ell = -j}^j |M_{k, \ell}|.
\end{equation}
Using $|\vec{M}_k|_2 = 1$ and the standard inequality
\begin{equation}
|\vec{M}_k|_2\leq |\vec{M}_k|_1\leq \sqrt{2j + 1}|\vec{M}_k|_2,
\end{equation}
we obtain the simpler bound
\begin{equation}
\operatorname{Var}(X_{k, m})\leq 2j + 2.
\label{boundSSRB}
\end{equation}
We can use this bound to estimate the variance of the sample mean, which might be improved by non-uniform sampling from the individual random variables $X_{\ell, \ell', m}$ in the linear combination $X_{k, m}$.

\subsection{Synthetic-Circuit RB}

We now derive bounds on the sample complexity of RB protocols that incorporate both synthetic gates and synthetic SPAM.  This requires considering linear combinations of random variables that are no longer Bernoulli random variables.  In all cases, the dominant contribution to the bound on the variance of the estimator comes from the covariance of the random variables associated with fixed $\ell_\text{initial}$ and $\ell_\text{final}$.

\subsubsection{Synthetic-SPAM Character RB}

Consider first character RB with synthetic SPAM, or SS$\chi$RB.  We specialize to $G = H = \text{SU}(2)$ and $\lambda' = k$.  Let
\begin{equation}
\mu_{\ell, \ell', m}^k(\vec{g}, g)\equiv (2k + 1)\chi_k(g)\llangle\ell'|\widetilde{\mathcal{G}}_\text{inv}\widetilde{\mathcal{G}}_m\cdots \widetilde{\mathcal{G}}_2\widetilde{\mathcal{G}_1\mathcal{G}}\sket{\ell}, \qquad \mu_{\ell, \ell', m}^k\equiv \mathbb{E}_{\vec{g}\in G^m, g\in G}[\mu_{\ell, \ell', m}^k(\vec{g}, g)].
\end{equation}
We wish to estimate
\begin{equation}
\mu_{k, m}\equiv \sum_{\ell, \ell' = -j}^j M_{k, \ell}M_{k, \ell'}\mu_{\ell, \ell', m}^k
\end{equation}
for all $k = 0, \ldots, 2j$ by sampling from a linear combination of random variables $X_{\ell, \ell', m}^k$ that are mixtures (with respect to $\vec{g}, g$) of rescaled Bernoulli random variables.

The $X_{\ell, \ell', m}^k$ are not all independent.  For fixed $\ell$, their means $\mu_{\ell, \ell', m}^k$ no longer sum to 1 because
\begin{equation}
\sket{\rho_m}\equiv (2k + 1)\mathbb{E}_{\vec{g}\in G^m, g\in G}[\chi_k(g)\widetilde{\mathcal{G}}_\text{inv}\widetilde{\mathcal{G}}_m\cdots \widetilde{\mathcal{G}}_2\widetilde{\mathcal{G}_1\mathcal{G}}\sket{\ell}]
\end{equation}
is not a properly normalized state (a convex combination of density matrices).  However, it is still the case that
\begin{equation}
\operatorname{Cov}(X_{\ell_\text{init}, \ell, m}^k, X_{\ell_\text{init}, \ell', m}^k) = -\mu_{\ell_\text{init}, \ell, m}^k\mu_{\ell_\text{init}, \ell', m}^k
\label{SSchiRBcovariances}
\end{equation}
for $\ell\neq \ell'$, for the same reason that \eqref{SSRBcovariances} holds.  Again, let us give a direct derivation of this fact.  It is enough to consider the following toy example.  For each $\ell$, let $X_\ell$ be a finite mixture distribution with weights $w_i$ ($i = 1, \ldots, n$) of rescaled Bernoulli random variables, where component $i$ is sampled from by measuring $|\ell\rangle\langle\ell|$ in the physical state $\rho_i$ and multiplying the result by some $a_i\in \mathbb{R}$.  We have
\begin{equation}
\langle X_\ell\rangle = \sum_{i=1}^n w_i a_i\langle\ell|\rho_i|\ell\rangle.
\end{equation}
To compute $\operatorname{Var}(X_\ell)$, note that we draw the state $\rho_i$ with probability $w_i$, after which we measure 1 with probability $\langle\ell|\rho_i|\ell\rangle$ and 0 with probability $1 - \langle\ell|\rho_i|\ell\rangle$.  Hence
\begin{equation}
\operatorname{Var}(X_\ell) = \sum_{i=1}^n w_i[\langle\ell|\rho_i|\ell\rangle(a_i - \langle X_\ell\rangle)^2 + (1 - \langle\ell|\rho_i|\ell\rangle)(0 - \langle X_\ell\rangle)^2] = \sum_{i=1}^n w_i a_i^2\langle\ell|\rho_i|\ell\rangle - \langle X_\ell\rangle^2,
\end{equation}
as expected from \eqref{varmixrescaledbernoulli}.  If $a_i = 1$, then $\operatorname{Var}(X_\ell) = \langle X_\ell\rangle(1 - \langle X_\ell\rangle)$.  To compute $\operatorname{Cov}(X_\ell, X_{\ell'})$ with $\ell\neq \ell'$, we again draw $\rho_i$ with probability $w_i$.  Let us denote the results of a joint measurement of $|\ell\rangle\langle\ell|$ and $|\ell'\rangle\langle\ell'|$ on $\rho_i$ by $X_{\ell, i}$ and $X_{\ell', i}$:
\begin{itemize}
\item With probability $\langle\ell|\rho_i|\ell\rangle$, we obtain $X_{\ell, i} = 1$ and $X_{\ell', i} = 0$.
\item With probability $\langle\ell'|\rho_i|\ell'\rangle$, we obtain $X_{\ell, i} = 0$ and $X_{\ell', i} = 1$.
\item With probability $1 - \langle\ell|\rho_i|\ell\rangle - \langle\ell'|\rho_i|\ell'\rangle$, we obtain $X_{\ell, i} = 0$ and $X_{\ell', i} = 0$.
\end{itemize}
Hence
\begin{equation}
\operatorname{Cov}(X_\ell, X_{\ell'}) = \sum_{i=1}^n w_i\left[\begin{array}{c} \langle\ell|\rho_i|\ell\rangle(a_i - \langle X_\ell\rangle)(0 - \langle X_{\ell'}\rangle) + \langle\ell'|\rho_i|\ell'\rangle(0 - \langle X_\ell\rangle)(a_i - \langle X_{\ell'}\rangle) \\ {} + (1 - \langle\ell|\rho_i|\ell\rangle - \langle\ell'|\rho_i|\ell'\rangle)(0 - \langle X_\ell\rangle)(0 - \langle X_{\ell'}\rangle) \end{array}\right] = -\langle X_\ell\rangle\langle X_{\ell'}\rangle,
\end{equation}
as claimed.

Returning to our problem, we wish to consider infinite mixture distributions.  The random variables of interest are
\begin{equation}
X_{k, m}\equiv \sum_{\ell, \ell' = -j}^j M_{k, \ell}M_{k, \ell'}X_{\ell, \ell', m}^k.
\end{equation}
Using
\begin{equation}
\operatorname{Var}(X_{\ell, \ell', m}^k) = (2k + 1)^2\mathbb{E}_{g\in G}[\chi_k(g)^2\mathbb{E}_{\vec{g}\in G^m}[\llangle\ell'|\widetilde{\mathcal{G}}_\text{inv}\widetilde{\mathcal{G}}_m\cdots \widetilde{\mathcal{G}}_2\widetilde{\mathcal{G}_1\mathcal{G}}\sket{\ell}]] - (\mu_{\ell, \ell', m}^k)^2
\end{equation}
as well as \eqref{SSchiRBcovariances}, we have
\begin{align}
\operatorname{Var}(X_{k, m}) &= \sum_{\ell, \ell', \ell'' = -j}^j M_{k, \ell}^2 M_{k, \ell'}M_{k, \ell''}\operatorname{Cov}(X_{\ell, \ell', m}^k, X_{\ell, \ell'', m}^k) \\
&= (2k + 1)^2\sum_{\ell, \ell' = -j}^j M_{k, \ell}^2 M_{k, \ell'}^2\mathbb{E}_{g\in G}[\chi_k(g)^2\mathbb{E}_{\vec{g}\in G^m}[\llangle\ell'|\widetilde{\mathcal{G}}_\text{inv}\widetilde{\mathcal{G}}_m\cdots \widetilde{\mathcal{G}}_2\widetilde{\mathcal{G}_1\mathcal{G}}\sket{\ell}]] - \sum_{\ell, \ell', \ell'' = -j}^j M_{k, \ell}^2 M_{k, \ell'}M_{k, \ell''}\mu_{\ell, \ell', m}^k\mu_{\ell, \ell'', m}^k.
\end{align}
Using
\begin{equation}
\mathbb{E}_{g\in G}[\chi_k(g)^2\mathbb{E}_{\vec{g}\in G^m}[\llangle\ell'|\widetilde{\mathcal{G}}_\text{inv}\widetilde{\mathcal{G}}_m\cdots \widetilde{\mathcal{G}}_2\widetilde{\mathcal{G}_1\mathcal{G}}\sket{\ell}]]\leq \mathbb{E}_{g\in G}[\chi_k(g)^2] = 1
\end{equation}
(which follows from character orthonormality) and
\begin{align}
|\mu_{\ell, \ell', m}^k| &= (2k + 1)|\mathbb{E}_{g\in G}[\chi_k(g)\mathbb{E}_{\vec{g}\in G^m}[\llangle\ell'|\widetilde{\mathcal{G}}_\text{inv}\widetilde{\mathcal{G}}_m\cdots \widetilde{\mathcal{G}}_2\widetilde{\mathcal{G}_1\mathcal{G}}\sket{\ell}]]| \\
&\leq (2k + 1)\mathbb{E}_{g\in G}[|\chi_k(g)|] \\
&\leq (2k + 1)^2,
\end{align}
we arrive at the bound
\begin{align}
\operatorname{Var}(X_{k, m}) &\leq (2k + 1)^2\sum_{\ell, \ell' = -j}^j M_{k, \ell}^2 M_{k, \ell'}^2 + (2k + 1)^4\sum_{\ell, \ell', \ell'' = -j}^j M_{k, \ell}^2|M_{k, \ell'}||M_{k, \ell''}| \\
&= (2k + 1)^2|\vec{M}_k|_2^2(|\vec{M}_k|_2^2 + (2k + 1)^2|\vec{M}_k|_1^2).
\end{align}
Further using $|\vec{M}_k|_2 = 1$ and $|\vec{M}_k|_1\leq \sqrt{2j + 1}$ gives
\begin{equation}
\operatorname{Var}(X_{k, m})\leq (2k + 1)^2(1 + (2j + 1)(2k + 1)^2).
\label{boundSSchiRB}
\end{equation}
For large $k$, this bound on the variance of synthetic-SPAM character RB goes like the variance of synthetic-SPAM RB times the \emph{square} of the variance of character RB: in particular, it is quartic in the dimension of the target irrep.  This blowup should be regarded as an artifact of our proof method, as character orthonormality is no longer useful for simplifying the covariance terms.

\subsubsection{Synthetic-SPAM Rank-1 RB}

The idea of rank-1 RB is to use the infinite (hence overcomplete) frame consisting of all $\text{SU}(2)$ superoperators to construct 1D projectors onto the desired synthetic SPAM operations (rather than full-rank irrep projectors, as in character RB).  In this case, the weighting factor is simply the matrix element of a given $\text{SU}(2)$ superoperator between the SPAM operations.  We argue that this procedure has significantly better sample complexity than character RB due to the reduced variance in this matrix element compared to that of the character function, as the former is bounded by a constant rather than by the dimension of the irrep.

To ensure that the RB signal is acceptable, we only consider rank-1 projectors in conjunction with synthetic SPAM.  We refer to the resulting protocol by the abbreviation SSR1RB.  In pure synthetic-SPAM RB, both the synthetic state and the synthetic measurement effect are the diagonal spherical tensor operator $T^{(k)}_0$.  Consider, then, the SPAM superoperator $\mathcal{Q}^{(k)} = |T^{(k)}_0\rrangle\llangle T^{(k)}_0|$ within the irrep $k$.  Via \eqref{genburnside}, we have
\begin{equation}
(2k + 1)\mathbb{E}_{g\in \text{SU}(2)}[\Tr(\mathcal{G}^\dag\mathcal{Q}^{(k)})\mathcal{G}] = \mathcal{Q}^{(k)}, \text{ where } \Tr(\mathcal{G}^\dag\mathcal{Q}^{(k)}) = D^k_{00}(g)^\ast.
\end{equation}
The ``rank-1'' weighting factor is real since $D^k_{00}(g) = d^k_{00}(g)$.  This is a single diagonal entry of $\mathcal{G}$ in the spherical tensor basis.  One may contrast this weighting factor with that for character RB,
\begin{equation}
\Tr(\mathcal{G}^\dag\Pi_k) = \chi_k(g)^\ast,
\end{equation}
which is used to synthesize $\Pi_k = \sum_{q=-k}^k |T^{(k)}_q\rrangle\llangle T^{(k)}_q|$ and is again real.  This is a sum of diagonal entries of $\mathcal{G}$ in the spherical tensor basis.  To estimate the variance of rank-1 RB, we use
\begin{equation}
D^k_{00}(g(\alpha, \beta, \gamma)) = d^k_{00}(\beta) = P^{(0, 0)}_k(\cos\beta) = P_k(\cos\beta),
\end{equation}
where $P^{(a, b)}_n$ is a Jacobi polynomial and $P_n$ is a Legendre polynomial.  With standard normalization, the Legendre polynomials $P_n(x)$ are bounded in absolute value by 1 for $x\in [-1, 1]$.

Let us be more precise.  Let $G = \text{SU}(2)$ and fix $k$.  Given $\vec{g}\in G^m$ and $g\in G$, let
\begin{equation}
\mu_{\ell, \ell', m}^k(\vec{g}, g)\equiv (2k + 1)d^k_{00}(g)\llangle\ell'|\widetilde{\mathcal{G}}_\text{inv}\widetilde{\mathcal{G}}_m\cdots \widetilde{\mathcal{G}}_2\widetilde{\mathcal{G}_1\mathcal{G}}\sket{\ell}.
\end{equation}
We then have
\begin{equation}
\mu_{\ell, \ell', m}^k\equiv \mathbb{E}_{\vec{g}\in G^m, g\in G}[\mu_{\ell, \ell', m}^k(\vec{g}, g)] = \llangle\mathcal{E}^\dag(|\ell'\rangle\langle\ell'|)|\mathcal{Q}^{(k)}(|\ell\rangle\langle\ell|)\rrangle f_k^m,
\end{equation}
where $\mathcal{E}$ is the noise channel and $f_k$ is the corresponding quality parameter.  This is the mean of a random variable $X_{\ell, \ell', m}^k$ that is a mixture (with respect to $\vec{g}, g$) of rescaled Bernoulli random variables.  We have
\begin{equation}
\operatorname{Var}(X_{\ell, \ell', m}^k) = (2k + 1)^2\mathbb{E}_{g\in G}[d^k_{00}(g)^2\mathbb{E}_{\vec{g}\in G^m}[\llangle\ell'|\widetilde{\mathcal{G}}_\text{inv}\widetilde{\mathcal{G}}_m\cdots \widetilde{\mathcal{G}}_2\widetilde{\mathcal{G}_1\mathcal{G}}\sket{\ell}]] - (\mu_{\ell, \ell', m}^k)^2
\end{equation}
and
\begin{equation}
\operatorname{Cov}(X_{\ell, \ell', m}^k, X_{\ell, \ell'', m}^k) = -\mu_{\ell, \ell', m}^k\mu_{\ell, \ell'', m}^k
\end{equation}
for $\ell'\neq \ell''$.  Setting $X_{k, m}\equiv \sum_{\ell, \ell' = -j}^j M_{k, \ell}M_{k, \ell'}X_{\ell, \ell', m}^k$ (as before), we wish to estimate
\begin{equation}
\operatorname{Var}(X_{k, m}) = (2k + 1)^2\sum_{\ell, \ell' = -j}^j M_{k, \ell}^2 M_{k, \ell'}^2\mathbb{E}_{g\in G}[d^k_{00}(g)^2\mathbb{E}_{\vec{g}\in G^m}[\llangle\ell'|\widetilde{\mathcal{G}}_\text{inv}\widetilde{\mathcal{G}}_m\cdots \widetilde{\mathcal{G}}_2\widetilde{\mathcal{G}_1\mathcal{G}}\sket{\ell}]] - \sum_{\ell, \ell', \ell'' = -j}^j M_{k, \ell}^2 M_{k, \ell'}M_{k, \ell''}\mu_{\ell, \ell', m}^k\mu_{\ell, \ell'', m}^k.
\end{equation}
First note that
\begin{equation}
\mathbb{E}_{g\in G}[d^k_{00}(g)^2] = \frac{1}{8\pi^2}\int_0^{2\pi} d\alpha\int_0^\pi d\beta\sin\beta\int_0^{2\pi} d\gamma\, P_k(\cos\beta)^2 = \frac{1}{2}\int_{-1}^1 dx\, P_k(x)^2 = \frac{1}{2k + 1},
\end{equation}
by the orthogonality relation for Legendre polynomials.  Using
\begin{equation}
\mathbb{E}_{g\in G}[d^k_{00}(g)^2\mathbb{E}_{\vec{g}\in G^m}[\llangle\ell'|\widetilde{\mathcal{G}}_\text{inv}\widetilde{\mathcal{G}}_m\cdots \widetilde{\mathcal{G}}_2\widetilde{\mathcal{G}_1\mathcal{G}}\sket{\ell}]]\leq \mathbb{E}_{g\in G}[d^k_{00}(g)^2] = \frac{1}{2k + 1}
\end{equation}
and
\begin{align}
|\mu_{\ell, \ell', m}^k| &= (2k + 1)|\mathbb{E}_{g\in G}[d^k_{00}(g)\mathbb{E}_{\vec{g}\in G^m}[\llangle\ell'|\widetilde{\mathcal{G}}_\text{inv}\widetilde{\mathcal{G}}_m\cdots \widetilde{\mathcal{G}}_2\widetilde{\mathcal{G}_1\mathcal{G}}\sket{\ell}]]| \\
&\leq (2k + 1)\mathbb{E}_{g\in G}[|\smash{d^k_{00}}(g)|] \\
&\leq 2k + 1,
\end{align}
we have
\begin{align}
\operatorname{Var}(X_{k, m}) &\leq (2k + 1)\sum_{\ell, \ell' = -j}^j M_{k, \ell}^2 M_{k, \ell'}^2 + (2k + 1)^2\sum_{\ell, \ell', \ell'' = -j}^j M_{k, \ell}^2|M_{k, \ell'}||M_{k, \ell''}| \\
&= (2k + 1)|\vec{M}_k|_2^2(|\vec{M}_k|_2^2 + (2k + 1)|\vec{M}_k|_1^2).
\end{align}
Further using $|\vec{M}_k|_2 = 1$ and $|\vec{M}_k|_1\leq \sqrt{2j + 1}$ gives
\begin{equation}
\operatorname{Var}(X_{k, m})\leq (2k + 1)(1 + (2j + 1)(2k + 1)).
\label{boundSSR1RB}
\end{equation}
This bound is quadratic in the dimension of the target irrep.

\subsubsection{Synthetic-SPAM Finite-Frame RB}

Again, we specialize to $G = H = \text{SU}(2)$ and $\lambda' = k$.  As with ordinary finite-frame RB, finite-frame RB with synthetic SPAM (which we abbreviate as SSFFRB) can be viewed as sampling from either a mixture distribution or a linear combination of random variables.  Let us take the mixture point of view.  We wish to estimate the mean of a linear combination $X_{k, m}\equiv \smash{\sum_{\ell, \ell' = -j}^j {}} \linebreak[1] M_{k, \ell} \linebreak[1] M_{k, \ell'} \linebreak[1] \smash{X_{\ell, \ell', m}^k}$ of random variables $\smash{X_{\ell, \ell', m}^k}$ that are mixtures (with respect to $\vec{g}$ and $i$) of rescaled Bernoulli random variables.  In this case, the mean of $\smash{X_{\ell, \ell', m}^k}$ is
\begin{equation}
p_{\ell, \ell', m}^k\equiv \sum_{i=1}^{N_j} c_i^k\mathbb{E}_{\vec{g}\in G^m}[\llangle\ell'|\widetilde{\mathcal{G}}_\text{inv}\widetilde{\mathcal{G}}_m\cdots \widetilde{\mathcal{G}}_2\widetilde{\mathcal{G}_1\mathcal{G}_i'}\sket{\ell}],
\end{equation}
where we have indicated the $k$-dependence explicitly in the coefficients $c_i$ (the $g_i'$ are fixed).  To sample from $X_{\ell, \ell', m}^k$, we:
\begin{itemize}
\item select an $i$ according to the probability distribution $i\mapsto |c_i^k|/|\vec{c}^k|_1$ and a $\vec{g}$ according to the uniform measure on $G^m$,
\item measure $|\ell'\rangle\langle\ell'|$ in the state $\widetilde{\mathcal{G}}_\text{inv}\widetilde{\mathcal{G}}_m\cdots \widetilde{\mathcal{G}}_2\widetilde{\mathcal{G}_1\mathcal{G}_i'}\sket{\ell}$,
\item and multiply the result by $\operatorname{sgn}(c_i^k)|\vec{c}^k|_1$.
\end{itemize}
Therefore, we have
\begin{equation}
\operatorname{Var}(X_{\ell, \ell', m}^k) = |\vec{c}^k|_1\sum_{i=1}^{N_j} |c_i^k|\mathbb{E}_{\vec{g}\in G^m}[\llangle\ell'|\widetilde{\mathcal{G}}_\text{inv}\widetilde{\mathcal{G}}_m\cdots \widetilde{\mathcal{G}}_2\widetilde{\mathcal{G}_1\mathcal{G}_i'}\sket{\ell}] - (p_{\ell, \ell', m}^k)^2
\end{equation}
and
\begin{equation}
\operatorname{Cov}(X_{\ell_\text{init}, \ell, m}^k, X_{\ell_\text{init}, \ell', m}^k) = -p_{\ell_\text{init}, \ell, m}^k p_{\ell_\text{init}, \ell', m}^k
\end{equation}
for $\ell\neq \ell'$.  We then compute that
\begin{equation}
\operatorname{Var}(X_{k, m}) = |\vec{c}^k|_1\sum_{\ell, \ell' = -j}^j M_{k, \ell}^2 M_{k, \ell'}^2\sum_{i=1}^{N_j} |c_i^k|\mathbb{E}_{\vec{g}\in G^m}[\llangle\ell'|\widetilde{\mathcal{G}}_\text{inv}\widetilde{\mathcal{G}}_m\cdots \widetilde{\mathcal{G}}_2\widetilde{\mathcal{G}_1\mathcal{G}_i'}\sket{\ell}] - \sum_{\ell, \ell', \ell'' = -j}^j M_{k, \ell}^2 M_{k, \ell'}M_{k, \ell''}p_{\ell, \ell', m}^k p_{\ell, \ell'', m}^k.
\end{equation}
Using
\begin{equation}
\mathbb{E}_{\vec{g}\in G^m}[\llangle\ell'|\widetilde{\mathcal{G}}_\text{inv}\widetilde{\mathcal{G}}_m\cdots \widetilde{\mathcal{G}}_2\widetilde{\mathcal{G}_1\mathcal{G}_i'}\sket{\ell}]\leq 1, \qquad |p_{\ell, \ell', m}^k|\leq |\vec{c}^k|_1,
\end{equation}
we obtain
\begin{equation}
\operatorname{Var}(X_{k, m})\leq |\vec{c}^k|_1^2\left(\sum_{\ell, \ell' = -j}^j M_{k, \ell}^2 M_{k, \ell'}^2 + \sum_{\ell, \ell', \ell'' = -j}^j M_{k, \ell}^2|M_{k, \ell'}||M_{k, \ell''}|\right) = |\vec{c}^k|_1^2|\vec{M}_k|_2^2(|\vec{M}_k|_2^2 + |\vec{M}_k|_1^2).
\end{equation}
Further using $|\vec{M}_k|_2 = 1$ and $|\vec{M}_k|_1\leq \sqrt{2j + 1}$ gives
\begin{equation}
\operatorname{Var}(X_{k, m})\leq (2j + 2)|\vec{c}^k|_1^2.
\label{boundSSFFRB}
\end{equation}
This bound is quadratic in the 1-norm of the vector of coefficients in the decomposition of the irrep projector in the chosen frame.  The bounds on the variances of finite-frame and synthetic-SPAM RB essentially ``compose'' by multiplication.

We could instead take the ``linear combination'' point of view on finite-frame RB and incorporate synthetic SPAM.  Rather than repeating the analysis for this case, we consider a toy example.  Consider an unphysical linear combination of physical states
\begin{equation}
\sum_{i=1}^n a_i\rho_i,
\end{equation}
where $a_i\in \mathbb{R}$ with no restrictions.  To sample from the random variable $X_\ell$, we measure $|\ell\rangle\langle\ell|$ in each of the physical states $\rho_i$ and add up the results according to the above linear combination.  We have
\begin{equation}
\langle X_\ell\rangle = \sum_{i=1}^n a_i\langle\ell|\rho_i|\ell\rangle.
\end{equation}
We first compute $\operatorname{Var}(X_\ell) = \operatorname{Cov}(X_\ell, X_\ell)$.  Accounting for all $2^n$ possible outcomes,
\begin{equation}
\operatorname{Var}(X_\ell) = \sum_{i_n=0}^1\cdots \sum_{i_1=0}^1 \left[\prod_{j=1}^n \langle\ell|\rho_j|\ell\rangle^{i_j}(1 - \langle\ell|\rho_j|\ell\rangle)^{1-i_j}\right]\left[\sum_{k=1}^n a_k(i_k - \langle\ell|\rho_k|\ell\rangle)\right]^2.
\end{equation}
The contributions from the cross terms in the innermost squared sum vanish.  The contributions from the diagonal terms reduce to
\begin{equation}
\sum_{k=1}^n\sum_{i_k=0}^1 \langle\ell|\rho_k|\ell\rangle^{i_k}(1 - \langle\ell|\rho_k|\ell\rangle)^{1-i_k}a_k^2(i_k - \langle\ell|\rho_k|\ell\rangle)^2 = \sum_{i=1}^n a_i^2\langle\ell|\rho_i|\ell\rangle(1 - \langle\ell|\rho_i|\ell\rangle),
\end{equation}
reproducing the expected result.  Now suppose that $\ell\neq \ell'$.  After a joint measurement of $|\ell\rangle\langle\ell|$ and $|\ell'\rangle\langle\ell'|$ on $\rho_i$, we obtain the results $X_{\ell, i}$ and $X_{\ell', i}$.  There are three possibilities:
\begin{itemize}
\item With probability $\langle\ell|\rho_i|\ell\rangle$, we obtain $X_{\ell, i} = 1$ and $X_{\ell', i} = 0$.
\item With probability $\langle\ell'|\rho_i|\ell'\rangle$, we obtain $X_{\ell, i} = 0$ and $X_{\ell', i} = 1$.
\item With probability $1 - \langle\ell|\rho_i|\ell\rangle - \langle\ell'|\rho_i|\ell'\rangle$, we obtain $X_{\ell, i} = 0$ and $X_{\ell', i} = 0$.
\end{itemize}
As for the values of $X_\ell$ and $X_{\ell'}$, there are $3^n$ possibilities.  We therefore have
\begin{align}
\operatorname{Cov}(X_\ell, X_{\ell'}) &= \sum_{i_n, i_n' = 0}^1\cdots \sum_{i_1, i_1' = 0}^1 \left[\prod_{j=1}^n 0^{i_j i_j'}\langle\ell|\rho_j|\ell\rangle^{i_j(1 - i_j')}\langle\ell'|\rho_j|\ell'\rangle^{(1 - i_j)i_j'}(1 - \langle\ell|\rho_j|\ell\rangle - \langle\ell'|\rho_j|\ell'\rangle)^{(1 - i_j)(1 - i_j')}\right] \nonumber \\
&\hspace{3 cm} \times \left[\sum_{k=1}^n a_k(i_k - \langle\ell|\rho_k|\ell\rangle)\right]\left[\sum_{k'=1}^n a_{k'}(i_{k'}' - \langle\ell'|\rho_{k'}|\ell'\rangle)\right],
\end{align}
where we use $0^0 = 1$.  Letting the outermost sum be over $k$ and $k'$, we notice that only the terms with $k = k'$ survive, giving
\begin{align}
\operatorname{Cov}(X_\ell, X_{\ell'}) &= \sum_{k=1}^n a_k^2\sum_{i_k, i_k' = 0}^1 (i_k - \langle\ell|\rho_k|\ell\rangle)(i_k' - \langle\ell'|\rho_k|\ell'\rangle)0^{i_k i_k'}\langle\ell|\rho_k|\ell\rangle^{i_k(1 - i_k')}\langle\ell'|\rho_k|\ell'\rangle^{(1 - i_k)i_k'}(1 - \langle\ell|\rho_k|\ell\rangle - \langle\ell'|\rho_k|\ell'\rangle)^{(1 - i_k)(1 - i_k')} \\
&= -\sum_{k=1}^n a_k^2\langle\ell|\rho_k|\ell\rangle\langle\ell'|\rho_k|\ell'\rangle.
\end{align}
This generalizes the result for $n = 1$.

\section{Sample Complexity: Exact Results} \label{app:SCexact}

The previously derived upper bounds are quite loose.  To obtain improved sample complexity estimates for our various protocols, we work in the zero-noise limit (where the exponential decay curves are flat), which allows us to derive the leading contributions to estimator variance \emph{exactly}.  These exact results provide a starting point for perturbative treatments of noise.

\subsection{SSRB}

In our previous discussion of SSRB, we saw that
\begin{equation}
\operatorname{Var}(X_{k, m})\leq |\vec{M}_k|_2^2(|\vec{M}_k|_2^2 + |\vec{M}_k|_1^2) = 1 + |\vec{M}_k|_1^2.
\end{equation}
In practice, $|\vec{M}_k|_1^2$ does not come close to saturating the bound $|\vec{M}_k|_1^2\leq 2j + 1$.  Let us estimate $\operatorname{Var}(X_{k, m})$ rather than bounding it.  We start with the exact expression
\begin{equation}
\operatorname{Var}(X_{k, m}) = \sum_{\ell, \ell' = -j}^j M_{k, \ell}^2 M_{k, \ell'}^2\mu_{\ell, \ell', m} - \sum_{\ell, \ell', \ell'' = -j}^j M_{k, \ell}^2 M_{k, \ell'}M_{k, \ell''}\mu_{\ell, \ell', m}\mu_{\ell, \ell'', m}.
\end{equation}
Assuming that gate errors are small, we should have $\mu_{\ell, \ell, m}\approx 1$ and $\mu_{\ell, \ell', m}\approx 0$ for $\ell\neq \ell'$, i.e., $\mu_{\ell, \ell', m}\approx \delta_{\ell\ell'}$.  To this level of approximation,
\begin{equation}
\operatorname{Var}(X_{k, m})\approx \sum_{\ell = -j}^j M_{k, \ell}^4 - \sum_{\ell = -j}^j M_{k, \ell}^4 = 0.
\label{SSRBvariance}
\end{equation}
Correspondingly,
\begin{equation}
\mu_{k, m} = \sum_{\ell, \ell' = -j}^j M_{k, \ell}M_{k, \ell'}\mu_{\ell, \ell', m}\approx \sum_{\ell=-j}^j M_{k, \ell}^2 = 1.
\end{equation}
We can do better by adding error perturbatively.  Accounting for the $m$-dependence but ignoring the possible dependence on $\ell$, we set
\begin{equation}
\mu_{\ell, \ell, m} = (1 - \epsilon)^m\approx 1 - m\epsilon, \qquad \mu_{\ell, \ell'\neq\ell, m}\approx \frac{m\epsilon}{2j},
\end{equation}
which respects the condition $\sum_{\ell'=-j}^j \mu_{\ell, \ell', m} = 1$.  This gives
\begin{align}
\operatorname{Var}(X_{k, m}) &\approx \sum_{\ell=-j}^j \left(M_{k, \ell}^4(1 - m\epsilon) + \frac{m\epsilon}{2j}M_{k, \ell}^2\sum_{\ell'\neq\ell} M_{k, \ell'}^2\right) - \sum_{\ell=-j}^j \left(M_{k, \ell}^4(1 - m\epsilon)^2 + \frac{m\epsilon}{j}(1 - m\epsilon)M_{k, \ell}^3\sum_{\ell'\neq\ell} M_{k, \ell'}\right) + O(\epsilon^2) \\
&= m\epsilon\sum_{\ell=-j}^j \left(M_{k, \ell}^4 + \frac{1}{2j}M_{k, \ell}^2(1 - M_{k, \ell}^2) - \frac{1}{j}M_{k, \ell}^3\sum_{\ell'\neq\ell} M_{k, \ell'}\right) + O(\epsilon^2),
\end{align}
where we have used $\sum_{\ell'\neq\ell} M_{k, \ell'}^2 = 1 - M_{k, \ell}^2$.  Correspondingly,
\begin{equation}
\mu_{k, m}\approx \sum_{\ell=-j}^j \left(M_{k, \ell}^2(1 - m\epsilon) + \frac{m\epsilon}{2j}M_{k, \ell}\sum_{\ell'\neq\ell} M_{k, \ell'}\right) = 1 - m\epsilon\sum_{\ell=-j}^j M_{k, \ell}\left(M_{k, \ell} - \frac{1}{2j}\sum_{\ell'\neq\ell} M_{k, \ell'}\right).
\end{equation}
The point is simply that the leading contribution to the variance of the SSRB estimator is first-order in $\epsilon$.  To obtain more useful estimates, we examine protocols that combine synthetic SPAM with synthetic gates.

\subsection{SS\texorpdfstring{$\chi$}{chi}RB}

Consider first ordinary character RB with $G = \text{SU}(2)$ and fixed $k$.  In the limit of zero noise, the protocol is insensitive to the sequence length $m$: we simply sample from a mixture with respect to $g\in G$ of rescaled Bernoulli random variables.  Let
\begin{equation}
\mu_g\equiv \sbra{E}\mathcal{G}\sket{\rho}.
\end{equation}
The component distribution specified by $g$ has mean $(2k + 1)\chi_k(g)\mu_g$ and variance $(2k + 1)^2\chi_k(g)^2\mu_g(1 - \mu_g)$.  The total mean of the mixture is
\begin{equation}
\mu = (2k + 1)\mathbb{E}_{g\in G}[\chi_k(g)\mu_g] = \sbra{E}\Pi_k\sket{\rho},
\end{equation}
while the total variance of the mixture is
\begin{equation}
\sigma^2 = (2k + 1)^2\mathbb{E}_{g\in G}[\chi_k(g)^2\mu_g] - \mu^2.
\end{equation}
The $(r, r')$ entry of block $\ell$ ($\ell = 0, \ldots, 2j$) of the superoperator $\mathcal{G}$ is $D_{rr'}^\ell(g)$.  On the other hand,
\begin{equation}
\chi_k(g)^2 = \sum_{q, q' = -k}^k D_{qq}^k(g)D_{q'q'}^k(g).
\end{equation}
Using the Haar integral of a product of three Wigner $D$-matrix elements \cite{Varshalovich},
\begin{equation}
\frac{1}{8\pi^2}\int_0^{2\pi} d\alpha\int_0^\pi d\beta\sin\beta\int_0^{2\pi} d\gamma\, D_{q_1 q_1'}^{k_1}(\alpha, \beta, \gamma)D_{q_2 q_2'}^{k_2}(\alpha, \beta, \gamma)D_{q_3 q_3'}^{k_3}(\alpha, \beta, \gamma) = \frac{(-1)^{q_3 - q_3'}}{2k_3 + 1}C^{k_3, -q_3}_{k_1, q_1; k_2, q_2}C^{k_3, -q_3'}_{k_1, q_1'; k_2, q_2'},
\end{equation}
which holds when $k_1 + k_2 + k_3$ is an integer, we find that
\begin{align}
&\text{$(r, r')$ entry of block $\ell$ of } \mathbb{E}_{g\in G}[\chi_k(g)^2\mathcal{G}] \nonumber \\
&= \sum_{q, q' = -k}^k \mathbb{E}_{g\in G}[D_{qq}^k(g)D_{q'q'}^k(g)D_{rr'}^\ell(g)] = \frac{(-1)^{r - r'}}{2\ell + 1}\sum_{q, q' = -k}^k C^{\ell, -r}_{k, q; k, q'}C^{\ell, -r'}_{k, q; k, q'} = \frac{\delta_{rr'}}{2\ell + 1}\sum_{q=-k}^k (C^{\ell, -r}_{k, q; k, -q - r})^2. \label{rrpentry}
\end{align}
Note that this entry is nonzero only when $\ell\leq 2k$.  From $\mathbb{E}_{g\in G}[\chi_k(g)^2\mathcal{G}]$, we deduce $\sigma^2$ and hence the variance of the estimator for $\chi$RB---in particular, the number of shots required for the sample variance to be much less than the sample mean.

We now specialize to the case where $\rho, E$ are angular momentum eigenstates.  To set our notation, recall the procedure.  To perform $\text{SU}(2)$ character RB with noiseless operations, we repeatedly: (1) sample $g\in \text{SU}(2)$ uniformly at random, (2) apply the superoperator $\mathcal{G}$ to a fixed initial state $|\ell\rangle\langle\ell|$, and (3) measure $J_z$.  The corresponding random variable $X_{\ell, \ell'}^k$ takes the value $(2k + 1)\chi_k(g)$ if the outcome of the $J_z$ measurement is $|\ell'\rangle\langle\ell'|$ and 0 otherwise.  A $\chi$RB estimator labeled by $(\ell, \ell', k)$ is the average of $X_{\ell, \ell'}^k$ over $N$ shots.  An SS$\chi$RB estimator is a linear combination of $(2j + 1)^2$ $\chi$RB estimators, according to
\begin{equation}
X_k\equiv \sum_{\ell, \ell' = -j}^j M_{k, \ell}M_{k, \ell'}X_{\ell, \ell'}^k.
\end{equation}
A single SS$\chi$RB shot corresponds to $2j + 1$ $\chi$RB shots, distributed over the $2j + 1$ distinct initial states.

We first compute the quantities $\langle X_{\ell, \ell'}^k\rangle$ and $\langle(X_{\ell, \ell'}^k)^2\rangle$.  Using \eqref{byorthogonality}, we compute directly that
\begin{equation}
\langle X_{\ell, \ell'}^k\rangle = \llangle\ell'|\Pi_k|\ell\rrangle = M_{k, \ell}\langle\ell'|T_0^{(k)}|\ell'\rangle = M_{k, \ell}M_{k, \ell'}.
\label{meanchiRBvariable}
\end{equation}
Using \eqref{rrpentry} and \eqref{byorthogonality}, we compute that
\begin{align}
\mathbb{E}_{g\in G}[\chi_k(g)^2\llangle\ell'|\mathcal{G}|\ell\rrangle] &= \sum_{k', k'' = 0}^{2j} M_{k', \ell}M_{k'', \ell'}\llangle T_0^{(k'')}|\mathbb{E}_{g\in G}[\chi_k(g)^2\mathcal{G}]|T_0^{(k')}\rrangle \\
&= \sum_{k'=0}^{2j} \frac{M_{k', \ell}M_{k', \ell'}}{2k' + 1}\sum_{q=-k}^k (C^{k', 0}_{k, q; k, -q})^2 \\
&= \sum_{k'=0}^{2k} \frac{M_{k', \ell}M_{k', \ell'}}{2k' + 1},
\end{align}
where we have substituted
\begin{equation}
\sum_{q=-k}^k (C^{k', 0}_{k, q; k, -q})^2 = \begin{cases} 1 & \text{if $k'\leq 2k$}, \\ 0 & \text{otherwise}. \end{cases}
\end{equation}
Therefore,
\begin{equation}
\langle(X_{\ell, \ell'}^k)^2\rangle = (2k + 1)^2\mathbb{E}_{g\in G}[\chi_k(g)^2\llangle\ell'|\mathcal{G}|\ell\rrangle] = (2k + 1)^2\sum_{k'=0}^{2k} \frac{M_{k', \ell}M_{k', \ell'}}{2k' + 1}.
\label{sqchiRBvariable}
\end{equation}
We now examine the statistics of the random variable $X_k$.  From \eqref{meanchiRBvariable}, we have
\begin{equation}
\langle X_k\rangle = \sum_{\ell, \ell' = -j}^j M_{k, \ell}M_{k, \ell'}\langle X_{\ell, \ell'}^k\rangle = \left(\sum_{\ell=-j}^j M_{k, \ell}^2\right)^2 = 1.
\end{equation}
Using
\begin{equation}
\operatorname{Var}(X_{\ell, \ell'}^k) = \langle(X_{\ell, \ell'}^k)^2\rangle - \langle X_{\ell, \ell'}^k\rangle^2, \qquad \operatorname{Cov}(X_{\ell, \ell'}^k, X_{\ell, \ell''}^k) = -\langle X_{\ell, \ell'}^k\rangle\langle X_{\ell, \ell''}^k\rangle
\label{varianceandcovariance}
\end{equation}
for $\ell'\neq \ell''$, as well as \eqref{meanchiRBvariable} and \eqref{sqchiRBvariable}, we have
\begin{align}
\operatorname{Var}(X_k) &= \sum_{\ell, \ell' = -j}^j M_{k, \ell}^2 M_{k, \ell'}^2\langle(X_{\ell, \ell'}^k)^2\rangle - \sum_{\ell, \ell', \ell'' = -j}^j M_{k, \ell}^2 M_{k, \ell'}M_{k, \ell''}\langle X_{\ell, \ell'}^k\rangle\langle X_{\ell, \ell''}^k\rangle \\
&= (2k + 1)^2\sum_{k'=0}^{2k} \frac{1}{2k' + 1}\left(\sum_{\ell=-j}^j M_{k, \ell}^2 M_{k', \ell}\right)^2 - \left(\sum_{\ell=-j}^j M_{k, \ell}^4\right)\left(\sum_{\ell=-j}^j M_{k, \ell}^2\right)^2 \\
&= (2k + 1)^2\sum_{k'=0}^{2k} \frac{1}{2k' + 1}\left(\sum_{\ell=-j}^j M_{k, \ell}^2 M_{k', \ell}\right)^2 - \sum_{\ell=-j}^j M_{k, \ell}^4. \label{SSchiRBvariance}
\end{align}
Ignoring the covariance (i.e., pretending that the $X_{\ell, \ell'}^k$ are independent), we have that
\begin{align}
\operatorname{Var}(X_k) &\sim \sum_{\ell, \ell' = -j}^j M_{k, \ell}^2 M_{k, \ell'}^2\operatorname{Var}(X_{\ell, \ell'}^k) \\
&= \sum_{\ell, \ell' = -j}^j M_{k, \ell}^2 M_{k, \ell'}^2(\langle(X_{\ell, \ell'}^k)^2\rangle - \langle X_{\ell, \ell'}^k\rangle^2) \\
&= (2k + 1)^2\sum_{k'=0}^{2k} \frac{1}{2k' + 1}\left(\sum_{\ell=-j}^j M_{k, \ell}^2 M_{k', \ell}\right)^2 - \left(\sum_{\ell=-j}^j M_{k, \ell}^4\right)^2,
\end{align}
which yields only a slight overestimate of the total variance.

\subsection{SSR1RB}

As in SS$\chi$RB, in the zero-noise limit of SSR1RB, we sample from a linear combination $X_k\equiv \sum_{\ell, \ell' = -j}^j M_{k, \ell}M_{k, \ell'}X_{\ell, \ell'}^k$ of random variables $X_{\ell, \ell'}^k$, where each $X_{\ell, \ell'}^k$ is a mixture with respect to $g\in G$ of rescaled Bernoulli random variables.  In the case of SSR1RB, the component distribution specified by $g$ has mean $(2k + 1)d^k_{00}(g)\llangle\ell'|\mathcal{G}|\ell\rrangle$ and variance $(2k + 1)^2 d^k_{00}(g)^2\llangle\ell'|\mathcal{G}|\ell\rrangle(1 - \llangle\ell'|\mathcal{G}|\ell\rrangle)$.  Using that
\begin{equation}
\text{$(r, r')$ entry of block $\ell$ of } \mathbb{E}_{g\in G}[D_{00}^k(g)\mathcal{G}] = \mathbb{E}_{g\in G}[D_{00}^k(g)D_{rr'}^\ell(g)] = \frac{\delta_{k\ell}\delta_{0r}\delta_{0r'}}{2k + 1},
\end{equation}
we compute that
\begin{equation}
\langle X_{\ell, \ell'}^k\rangle = (2k + 1)\llangle\ell'|\mathbb{E}_{g\in G}[D_{00}^k(g)\mathcal{G}]|\ell\rrangle = (2k + 1)\langle\ell'|\left(\frac{M_{k, \ell}T_0^{(k)}}{2k + 1}\right)|\ell'\rangle = M_{k, \ell}M_{k, \ell'}
\label{meanR1RBvariable}
\end{equation}
(since the $\sket{\ell}$ only have support on diagonal spherical tensor operators) and therefore $\langle X_k\rangle = 1$.  We also have
\begin{equation}
\text{$(r, r')$ entry of block $\ell$ of } \mathbb{E}_{g\in G}[D_{00}^k(g)^2\mathcal{G}] = \mathbb{E}_{g\in G}[D_{00}^k(g)^2 D_{rr'}^\ell(g)] = \frac{\delta_{0r}\delta_{0r'}}{2\ell + 1}(C^{\ell, 0}_{k, 0; k, 0})^2,
\end{equation}
which is nonzero only when $\ell\leq 2k$.  Therefore,
\begin{align}
\mathbb{E}_{g\in G}[D_{00}^k(g)^2\llangle\ell'|\mathcal{G}\sket{\ell}] &= \sum_{k', k'' = 0}^{2j} M_{k', \ell}M_{k'', \ell'}\llangle T_0^{(k'')}|\mathbb{E}_{g\in G}[D_{00}^k(g)^2\mathcal{G}]|T_0^{(k')}\rrangle \\
&= \sum_{k'=0}^{2k} \frac{M_{k', \ell}M_{k', \ell'}}{2k' + 1}(C^{k', 0}_{k, 0; k, 0})^2,
\end{align}
where we have lowered the upper limit of summation from $2j$ to $2k$, and
\begin{equation}
\langle(X_{\ell, \ell'}^k)^2\rangle = (2k + 1)^2\mathbb{E}_{g\in G}[D_{00}^k(g)^2\llangle\ell'|\mathcal{G}\sket{\ell}] = (2k + 1)^2\sum_{k'=0}^{2k} \frac{M_{k', \ell}M_{k', \ell'}}{2k' + 1}(C^{k', 0}_{k, 0; k, 0})^2.
\label{sqR1RBvariable}
\end{equation}
We then have
\begin{align}
\operatorname{Var}(X_k) &= \sum_{\ell, \ell' = -j}^j M_{k, \ell}^2 M_{k, \ell'}^2\langle(X_{\ell, \ell'}^k)^2\rangle - \sum_{\ell, \ell', \ell'' = -j}^j M_{k, \ell}^2 M_{k, \ell'}M_{k, \ell''}\langle X_{\ell, \ell'}^k\rangle\langle X_{\ell, \ell''}^k\rangle \\
&= (2k + 1)^2\sum_{k'=0}^{2k} \frac{(C^{k', 0}_{k, 0; k, 0})^2}{2k' + 1}\left(\sum_{\ell=-j}^j M_{k, \ell}^2 M_{k', \ell}\right)^2 - \sum_{\ell=-j}^j M_{k, \ell}^4. \label{SSR1RBvariance}
\end{align}
This is manifestly less than the variance \eqref{SSchiRBvariance} of SS$\chi$RB because
\begin{equation}
(C^{k', 0}_{k, 0; k, 0})^2\leq \sum_{q=-k}^k (C^{k', 0}_{k, q; k, -q})^2 = 1
\end{equation}
for $k'\leq 2k$.  Note the following exact expression for $k'\leq 2k$:
\begin{equation}
\frac{(C^{k', 0}_{k, 0; k, 0})^2}{2k' + 1} = \frac{\sqrt{\pi}}{2}\frac{2^{2k - k'}}{(2k - k')!}\frac{\Gamma(\frac{k' + 1}{2})^2\Gamma(k + \frac{k'}{2} + 1)}{\Gamma(\frac{k'}{2} + 1)^2\Gamma(k + \frac{k' + 3}{2})\Gamma(-k + \frac{k' + 1}{2})^2},
\end{equation}
which vanishes if $k'$ is odd due to the pole in $\Gamma(-k + \frac{k' + 1}{2})$.

\subsection{SSFFRB}

In the limit of zero noise, ordinary FFRB with physical SPAM (from the mixture point of view) entails sampling from a finite mixture with respect to $i = 1, \ldots, N_j$ of rescaled Bernoulli random variables.  Component $i$ is chosen with probability $|c_i|/|\vec{c}|_1$. (Note that we sometimes indicate the $k$-dependence explicitly in the coefficients $c_i^k$.) Let
\begin{equation}
\mu_i\equiv \sbra{E}\mathcal{G}_i'\sket{\rho}.
\end{equation}
The component distribution specified by $i$ has mean $\operatorname{sgn}(c_i)|\vec{c}|_1\mu_i$ and variance $|\vec{c}|_1^2\mu_i(1 - \mu_i)$.  The total mean of the mixture is
\begin{equation}
\mu = \sum_{i=1}^{N_j} c_i\mu_i = \sbra{E}\left(\sum_{i=1}^{N_j} c_i\mathcal{G}_i'\right)\sket{\rho} = \sbra{E}\Pi_k\sket{\rho},
\end{equation}
while the total variance of the mixture is $\sigma^2 = |\vec{c}|_1\sum_{i=1}^{N_j} |c_i|\mu_i - \mu^2$.  To say more, we require knowledge of $\rho$ or $E$.

In SSFFRB, we set $\rho = |\ell\rangle\langle\ell|$ and $E = |\ell'\rangle\langle\ell'|$, and we wish to estimate the mean of a linear combination $X_k\equiv \sum_{\ell, \ell' = -j}^j \linebreak[1] M_{k, \ell} \linebreak[1] M_{k, \ell'} \linebreak[1] X_{\ell, \ell'}^k$ of random variables $X_{\ell, \ell'}^k$ that are mixtures (with respect to $i$) of rescaled Bernoulli random variables.  We have
\begin{equation}
\langle X_{\ell, \ell'}^k\rangle = \llangle\ell'|\Pi_k|\ell\rrangle = M_{k, \ell}M_{k, \ell'}.
\label{meanFFRBvariable}
\end{equation}
Using
\begin{align}
\llangle\ell'|\mathcal{G}_i'\sket{\ell} &= \sum_{k', k'' = 0}^{2j} M_{k', \ell}M_{k'', \ell'}\llangle T_0^{(k'')}|\mathcal{G}_i'|T_0^{(k')}\rrangle \\
&= \sum_{k'=0}^{2j} M_{k', \ell}M_{k', \ell'}\llangle T_0^{(k')}|\mathcal{G}_i'|T_0^{(k')}\rrangle \\
&= \sum_{k'=0}^{2j} M_{k', \ell}M_{k', \ell'}D_{00}^{k'}(g_i'),
\end{align}
we compute that
\begin{equation}
\langle(X_{\ell, \ell'}^k)^2\rangle = |\vec{c}|_1\sum_{i=1}^{N_j} |c_i|\llangle\ell'|\mathcal{G}_i'\sket{\ell} = |\vec{c}|_1\sum_{k'=0}^{2j} M_{k', \ell}M_{k', \ell'}\sum_{i=1}^{N_j} |c_i|D_{00}^{k'}(g_i').
\label{sqFFRBvariable}
\end{equation}
As with the other synthetic-SPAM protocols, we have $\langle X_k\rangle = 1$.  Using \eqref{varianceandcovariance}, we compute that
\begin{align}
\operatorname{Var}(X_k) &= \sum_{\ell, \ell' = -j}^j M_{k, \ell}^2 M_{k, \ell'}^2\langle(X_{\ell, \ell'}^k)^2\rangle - \sum_{\ell, \ell', \ell'' = -j}^j M_{k, \ell}^2 M_{k, \ell'}M_{k, \ell''}\langle X_{\ell, \ell'}^k\rangle\langle X_{\ell, \ell''}^k\rangle \\
&= |\vec{c}|_1\sum_{k'=0}^{2j}\left(\sum_{\ell=-j}^j M_{k, \ell}^2 M_{k', \ell}\right)^2\left(\sum_{i=1}^{N_j} |c_i|D_{00}^{k'}(g_i')\right) - \sum_{\ell=-j}^j M_{k, \ell}^4. \label{SSFFRBvariance}
\end{align}
This is identical to the zero-noise SS$\chi$RB variance \eqref{SSchiRBvariance} up to making the replacement
\begin{equation}
\frac{(2k + 1)^2}{2k' + 1}\to |\vec{c}|_1\sum_{i=1}^{N_j} |c_i|D_{00}^{k'}(g_i')
\end{equation}
in the sum over $k'$ and extending the upper limit of summation from $2k$ to $2j$.

Finally, note that finite frames can be used to construct rank-1 projectors as well as full-rank projectors, e.g., by choosing the coefficients $c_i$ so that
\begin{equation}
\sum_{i=1}^{N_j} c_i\mathcal{G}_i' = \mathcal{Q}^{(k)}.
\end{equation}
Consider the synthetic-SPAM finite-frame rank-1 RB (SSFFR1RB) protocol in the zero-noise limit.  The coefficients $c_i$ define the random variables $X_{\ell, \ell'}^k$.  Once again, we have $\langle X_{\ell, \ell'}^k\rangle = M_{k, \ell}M_{k, \ell'}$ and therefore $\langle X_k\rangle = 1$.  The computation of $\operatorname{Var}(X_k)$ proceeds as before, with the result depending on the $c_i$ and $g_i'$.  Note that, regardless of the rank, the fact that the uppermost block of any $\text{SU}(2)$ superoperator is $D_{00}^0 = 1$ implies that $\sum_{i=1}^{N_j} c_i = 0$ unless $k = 0$, in which case $\sum_{i=1}^{N_j} c_i = 1$.

\subsection{Non-Synthetic SPAM}

\subsubsection{\texorpdfstring{$\chi$}{chi}RB}

Consider $\chi$RB with non-synthetic SPAM in the zero-noise limit.  We use the pure initial state $|\ell\rangle\langle\ell|$ and measure the POVM $\{|\ell\rangle\langle\ell|, \mathds{1} - |\ell\rangle\langle\ell|\}$.  That is, we set $\rho = E = |\ell\rangle\langle\ell|$.  From \eqref{meanchiRBvariable}, the mean is
\begin{equation}
\langle X_{\ell, \ell}^k\rangle = M_{k, \ell}^2.
\end{equation}
From \eqref{sqchiRBvariable}, the variance is
\begin{equation}
\langle(X_{\ell, \ell}^k)^2\rangle - \langle X_{\ell, \ell}^k\rangle^2 = (2k + 1)^2\sum_{k'=0}^{2k} \frac{M_{k', \ell}^2}{2k' + 1} - M_{k, \ell}^4.
\label{chiRBunnormalized}
\end{equation}
Note that these expressions are invariant under taking $\ell\to -\ell$, by the symmetries of \eqref{Mdefinition}.  Importantly, since the mean is not 1 (as would be the case with synthetic SPAM), we must rescale the variance by the mean to get the estimator variance, which is
\begin{equation}
\frac{\langle(X_{\ell, \ell}^k)^2\rangle}{\langle X_{\ell, \ell}^k\rangle^2} - 1 = \frac{(2k + 1)^2}{M_{k, \ell}^4}\sum_{k'=0}^{2k} \frac{M_{k', \ell}^2}{2k' + 1} - 1.
\label{chiRBnormalized}
\end{equation}
We refer to \eqref{chiRBunnormalized} and \eqref{chiRBnormalized} as the unnormalized and normalized $\chi$RB variances, respectively.  Note that when $j$ is an integer and $k$ is odd, we have $M_{k, 0} = 0$, which makes $\ell = 0$ SPAM useless for character RB.

\subsubsection{R1RB}

Consider rank-1 RB with non-synthetic SPAM in the zero-noise limit.  We set $\rho = E = |\ell\rangle\langle\ell|$.  Again, the mean is $\langle X_{\ell, \ell}^k\rangle = M_{k, \ell}^2$ (cf.\ \eqref{meanR1RBvariable}).  From \eqref{sqR1RBvariable}, the variance is
\begin{equation}
\langle(X_{\ell, \ell}^k)^2\rangle - \langle X_{\ell, \ell}^k\rangle^2 = (2k + 1)^2\sum_{k'=0}^{2k} \frac{M_{k', \ell}^2}{2k' + 1}(C^{k', 0}_{k, 0; k, 0})^2 - M_{k, \ell}^4.
\end{equation}
After normalization, we find that the estimator variance is
\begin{equation}
\frac{\langle(X_{\ell, \ell}^k)^2\rangle}{\langle X_{\ell, \ell}^k\rangle^2} - 1 = \frac{(2k + 1)^2}{M_{k, \ell}^4}\sum_{k'=0}^{2k} \frac{M_{k', \ell}^2}{2k' + 1}(C^{k', 0}_{k, 0; k, 0})^2 - 1.
\label{R1RBnormalized}
\end{equation}
These are simple modifications of the corresponding expressions for $\chi$RB.

\subsubsection{FFRB}

Consider finite-frame RB with non-synthetic SPAM in the zero-noise limit.  With $\rho = E = |\ell\rangle\langle\ell|$, the estimator variance is
\begin{equation}
\frac{\langle(X_{\ell, \ell}^k)^2\rangle}{\langle X_{\ell, \ell}^k\rangle^2} - 1 = \frac{|\vec{c}|_1}{M_{k, \ell}^4}\sum_{k'=0}^{2j} M_{k', \ell}^2\sum_{i=1}^{N_j} |c_i|D_{00}^{k'}(g_i') - 1,
\label{FFRBnormalized}
\end{equation}
via \eqref{meanFFRBvariable} and \eqref{sqFFRBvariable}.

\section{Sample Complexity: Summary}

\subsection{Overview}

In general, we wish to optimize an RB protocol along three axes:
\begin{enumerate}
\item Minimize statistical error (maximize precision, as quantified by the variance of the estimator of an RB quality parameter).
\item Minimize systematic error (maximize accuracy, as quantified by the closeness of the mean of the estimator to the true answer).
\item Maximize efficiency (minimize sample complexity).
\end{enumerate}
In the presence of imperfect SPAM, a protocol that is not SPAM-robust cannot isolate signals from specific irreps in the limit of infinite sample size, so the mean of the estimator will never converge to the right answer.

To analytically quantify the expected performance of our synthetic RB protocols, we make the simplifying assumptions of gate-independent noise and perfect SPAM.  In the absence of SPAM error, we account for all three criteria by simply minimizing the variance of the estimator, which determines how many samples are required to estimate the mean to a given precision.

More precisely, for a fixed irrep $\lambda$ and a fixed sequence length $m$, a synthetic RB protocol samples from a random variable with mean $A_\lambda f_\lambda^m$, where $A_\lambda$ is a SPAM-dependent coefficient and $f_\lambda$ is a quality parameter. (Under the assumption of perfect SPAM, any such protocol can perfectly isolate the signal from a single irrep in the sense that the mean of the corresponding random variable is a single $A_\lambda f_\lambda^m$ rather than a sum of these quantities over different $\lambda$.) The only remnant of systematic error is that this mean differs across protocols because the coefficient $A_\lambda$ (the strength of the signal within the irrep $\lambda$) differs across protocols.  Therefore, for a given protocol, we normalize by $A_\lambda$ to compute the variance of the estimator for $f_\lambda^m$ rather than for $A_\lambda f_\lambda^m$.  This allows for a fair comparison across protocols.  For example, if $A_\lambda$ is very small for a given protocol, meaning that the signal is very weak, then merely having a small variance does not guarantee a good estimate of $f_\lambda$.

In summary, if SPAM operations are perfect, then any synthetic RB protocol allows us to sample from a random variable with mean $f_\lambda^m$.  The performance of the protocol is determined by the variance of this random variable.  In this case, we expect to find the following ranking of protocols:
\begin{equation}
\text{SSRB} > \text{SS$x$RB} > \text{$x$RB}, \qquad x\in \{\chi, \text{R1}, \text{FF}, \ldots\}.
\end{equation}
This expectation is borne out by our calculations.  In particular, in this idealized setting, combining a synthetic-gate protocol (e.g., $\chi$RB, R1RB, FFRB) with synthetic SPAM leads to worse results than pure SSRB because if one starts with a synthetic initial state that lies completely inside a target irrep and the protocol synthesizes an imperfect projector after averaging over finitely many circuits, then that imperfect projector acts as a non-identity operator within the target irrep and thereby reduces the state's overlap with the target measurement effect.  In other words, the assumption of perfect SPAM nullifies the advantage of synthetic-gate RB over SSRB in terms of SPAM-robustness. (It does not, however, render synthetic-gate RB useless: the feature of SPAM-robustness is still useful for isolating the signal from a target irrep given a non-synthetic initial state, as no physical state has support in a single nontrivial target irrep.)

If SPAM operations are imperfect, then the resulting hierarchy of protocols depends on the degree of SPAM error.  To obtain a more realistic picture of how these protocols compare, we can supplement the model of gate error to account for noisy SPAM:
\begin{itemize}
\item For synthetic-gate protocols, which are SPAM-robust, SPAM error simply affects the estimate for the SPAM coefficient.
\item For SPAM-robust protocols that incorporate synthetic SPAM, SPAM error affects the computation of the covariance in addition to shifting the estimate for the SPAM coefficient away from 1.
\item For pure synthetic-SPAM RB, which is not SPAM-robust, SPAM error generally results in contamination of the signal from other irreps (systematic error).
\end{itemize}
In brief, synthetic-SPAM RB should be used when SPAM operations are reliable; synthetic-gate RB should be used when SPAM operations are unreliable.  The goal of any protocol that uses both synthetic gates and SPAM is to achieve greater robustness to systematic error than pure synthetic-SPAM RB, at moderate cost in the sample complexity needed to achieve a desired level of precision.

\subsection{Comparison} \label{app:comparison}

We consider synthetic-gate protocols that use either an infinite frame ($\chi$RB and R1RB) or a given finite frame (FFRB) to construct projectors onto any desired irrep.  While we focus on full-rank and rank-1 projectors, these protocols can easily be adapted to synthesize projectors of any rank.  We also consider synthetic-SPAM protocols, the simplest version of which (SSRB) is efficient but not completely SPAM-robust.  Combined synthetic-gate and synthetic-SPAM RB protocols are SPAM-robust but of greater (implementation and sample) complexity than pure SSRB.  We derive sample complexity estimates for all of these protocols, including both upper bounds on the variance of the associated random variable in the presence of noise and exact expressions for the variance in the zero-noise limit.  These estimates also yield theoretical priors on error bars.

Let $G = \text{SU}(2)$.  For a fixed protocol, irrep $k$, and sequence length $m$, our estimates pertain to the variance of the estimator for the quantity $A_k f_k^m$, where $A_k$ is a SPAM coefficient and $f_k$ is an RB quality parameter.  All of our estimates assume perfect SPAM, but for any given model of SPAM error, they can be upgraded to account perturbatively for the SPAM error.  To compare protocols, we must account for the overall normalization by $A_k$:
\begin{itemize}
\item Any synthetic-SPAM protocol has $A_k = 1$ due to $|\vec{M}_k|_2 = 1$, so an estimator for $A_k f_k^m$ is automatically an estimator for $f_k^m$.
\item Any synthetic-gate protocol that uses physical SPAM and that constructs full-rank irrep projectors has $A_k = \llangle E|\mathcal{E}\Pi_k|\rho\rrangle\approx \llangle E|\Pi_k|\rho\rrangle\leq 1$; the variance of the estimator for $f_k^m$ is that for $A_k f_k^m$ divided by $A_k^2$.
\end{itemize}
In all cases, the variance of the sample mean is the variance of the estimator for $f_k^m$ divided by $N$, the number of samples from a single distribution (i.e., the number of circuit shots), assuming that sample sizes are large enough for the central limit theorem to apply.

Let us summarize and compare the crude analytical bounds from Appendix \ref{app:SCbounds}:
\begin{itemize}
\item For the SPAM-robust protocols that utilize physical SPAM, we find that
\begin{equation}
\text{the variance of the estimator for $f_k^m$ is $\leq$ } \begin{cases} (2k + 1)^2/A_k^2 & \text{for $\chi$RB and R1RB from \eqref{randomsamplingbound}}, \\ |\vec{c}|_1^2/A_k^2 & \text{for FFRB from \eqref{boundnonuniform} and \eqref{boundmixture}}. \end{cases}
\end{equation}
For $\chi$RB and R1RB, the bound is quadratic in the dimension of the target irrep. (While we only presented the argument for $\chi$RB, an essentially identical argument goes through for R1RB, without the need to invoke character orthonormality.) The worst-case bound is quadratic in $j$ since the largest irrep in the superoperator representation has dimension $4j + 1$.  For FFRB, we have suppressed an $O(1)$ constant, which is $1/4$ for non-uniform sampling from a linear combination of random variables and 1 for mixture sampling.  The bound goes like the square of the 1-norm of the coefficient vector $\vec{c}$ that describes the decomposition of the target irrep projector into the chosen frame of $\text{SU}(2)$ superoperators.  Here, $\vec{c}$ is a real vector of length $N_j$, and the $k$-dependence is implicit in $\vec{c}$.
\item For the synthetic-SPAM protocols, we find that
\begin{equation}
\text{the variance of the estimator for $f_k^m$ is $\leq$ } \begin{cases} 2j + 2 & \text{for SSRB from \eqref{boundSSRB}}, \\ O(jk^4) & \text{for SS$\chi$RB from \eqref{boundSSchiRB}}, \\ O(jk^2) & \text{for SSR1RB from \eqref{boundSSR1RB}}, \\ (2j + 2)|\vec{c}^k|_1^2 & \text{for SSFFRB from \eqref{boundSSFFRB}}. \end{cases}
\end{equation}
All of these bounds are linear in $j$; for SSRB, the bound is independent of $k$.
\end{itemize}
We expect all of these bounds to be loose compared to the exact results for the variance in the limit of zero noise.  For example, let us compare the exact result \eqref{SSchiRBvariance} for SS$\chi$RB to the bound \eqref{boundSSchiRB}.  Orthogonality of the matrix $M$ implies the na\"ive bound
\begin{equation}
|M_{k, \ell}|\leq 1,
\label{naivebound}
\end{equation}
which, when substituted into \eqref{SSchiRBvariance}, gives
\begin{equation}
\operatorname{Var}(X_k) = (2k + 1)^2\sum_{k'=0}^{2k} \frac{1}{2k' + 1}\left(\sum_{\ell=-j}^j M_{k, \ell}^2 M_{k', \ell}\right)^2 - \sum_{\ell=-j}^j M_{k, \ell}^4\leq (2j + 1)^2(2k + 1)^2\sum_{k'=0}^{2k} \frac{1}{2k' + 1} = O(j^2 k^2\log k).
\end{equation}
In the worst case, for $k$ on the order of $j$, this is already a better bound than \eqref{boundSSchiRB}.  It is difficult to make more direct comparisons in the limit of large $j$ and $k$ without precise asymptotics for Clebsch-Gordan coefficients.\footnote{As probability amplitudes, all Clebsch-Gordan coefficients obey $\big|\smash{C^{j, \ell}_{j_1, \ell_1; j_2, \ell_2}}\big| \linebreak[1] \leq \linebreak[1] 1$, which, in light of \eqref{Mdefinition}, implies that $|M_{k, \ell}|\leq \sqrt{\frac{2k + 1}{2j + 1}}$.  This is stronger than the na\"ive bound \eqref{naivebound} for $k < j$.  For large $j$, we might also use the rough approximation $\big|C^{j, \ell}_{j_1, \ell_1; j_2, \ell_2}\big|\approx \frac{1}{\sqrt{2j + 1}}$ for fixed $j_1, \ell_1, j_2, \ell_2$, which assumes that the magnitudes of the Clebsch-Gordan coefficients are asymptotically evenly distributed between all possible states.  However, this approximation is not so reliable when we vary $\ell_1$ and $\ell$ together, as in \eqref{Mdefinition}.  More precise semiclassical formulas for $C^{j, \ell}_{j_1, \ell_1; j_2, \ell_2}$ in the limit of large quantum numbers $j, j_1, j_2$ can be found in \cite{Varshalovich}.}

We can gain more insight by comparing the exact results of Appendix \ref{app:SCexact} to each other.  In the zero-noise limit, $f_k = 1$.  For the physical-SPAM protocols $\chi$RB, R1RB, and FFRB, the SPAM-dependent variance of the estimator for $f_k^m$ in the case of $J_z$-eigenstate SPAM and zero noise is given by \eqref{chiRBnormalized}, \eqref{R1RBnormalized}, and \eqref{FFRBnormalized}, respectively.  For the synthetic-SPAM protocols SSRB, SS$\chi$RB, SSR1RB, and SSFFRB, the zero-noise estimator variance is given by \eqref{SSRBvariance}, \eqref{SSchiRBvariance}, \eqref{SSR1RBvariance}, and \eqref{SSFFRBvariance}, respectively.  We summarize our findings as follows.  SSRB is highly efficient but not SPAM-robust under the most general circumstances.  The variance of pure SSRB, to zeroth order in the gate noise, is zero.  Therefore, for sufficiently small noise, the SSRB variance will always be less than that of the SPAM-robust protocols.  The SPAM-robust protocols can be listed in order of zero-noise variance (from greatest to least) as follows:
\begin{itemize}
\item The large variance of $\chi$RB comes from the SPAM-dependent signal strength and the inefficiency of constructing full-rank projectors.
\item R1RB (with physical SPAM restricted to $J_z$ eigenstates) neither degrades nor improves the signal strength relative to $\chi$RB because $J_z$ eigenstates only have support on diagonal spherical tensor operators within irreps, but it reduces the sample complexity of constructing projectors.
\item SS$\chi$RB greatly improves the signal strength relative to $\chi$RB by ensuring that the support of the SPAM is as large as possible within irreps.
\item SSR1RB both greatly improves the signal strength and reduces the sample complexity of constructing projectors relative to $\chi$RB.  Rank-1 projectors are an ideal use case for synthetic SPAM due to the need to enhance the RB signal.
\end{itemize}
Since the optimization of the finite frame is left as an open problem, we do not attempt to make quantitative comparisons between (SS)FFRB and the other protocols.

\subsection{Improving SSRB?} \label{app:improving}

SSRB is fragile to SPAM errors.  We end by commenting on a potential route toward making SSRB SPAM-robust without using synthetic gates.  Such a procedure would have the same sample complexity as SSRB itself, with the extra work relegated to classical post-processing.

In brief, SSRB works as follows.  For a given sequence length $m$, we directly estimate a $(2j + 1)\times (2j + 1)$ matrix $P$ of survival probabilities, where each entry describes the result of preparing a $J_z$ eigenstate, applying a sequence of random $\text{SU}(2)$ gates that compiles to the identity, and then measuring a $J_z$ eigenstate.  $P$ would be the identity matrix if the gates were perfect, so its off-diagonal entries reflect the amount of gate noise.  We then construct the matrix $MPM^T$, which is diagonal in the limit of perfect $\text{SU}(2)$ twirling (i.e., infinite sample size), regardless of gate noise.  The diagonal entries of $MPM^T$ are the $2j + 1$ quantities $f_k^m$.

In practice, finite sampling means that the twirling is never perfect, so the off-diagonal entries of $MPM^T$ will exhibit finite-sample fluctuations away from 0.  These fluctuations increase with increasing gate sequence length because longer sequences require more samples to twirl well.  Moreover, the off-diagonal entries of $MPM^T$ are sensitive to SPAM noise.  Their size relative to the diagonal entries can be used as a diagnostic of the reliability of SSRB.  Namely, after subtracting finite-sample fluctuations, the off-diagonal entries of $MPM^T$ can be used to estimate the amount of SPAM noise (as quantified, e.g., by the fidelity between the target initial states and the actual initial states), and in turn, how much error the SPAM noise introduces into the calculated $\text{SU}$(2) error rates.  These error bars can be used to determine whether the calculated error rates reflect gate noise or SPAM noise.

The preceding discussion suggests a possible improvement to SSRB.  In the presence of SPAM error, $M$ will not diagonalize $P$.  This problem can be circumvented by computing the eigenvalues of $P$ directly, whether separately for each $m$ or by finding the similarity transformation that comes as close as possible to diagonalizing all of the $P(m)$ simultaneously.  This approach, SSRB supplemented with (approximate simultaneous) diagonalization, treats SPAM error as an unwanted change of basis.  We abbreviate it as SSRB-D.

Several obstacles remain to be overcome for such an approach to work.  To begin, the above description of SSRB is slightly oversimplified.  SSRB can be viewed as reduced process tomography on a $(2j + 1)^2\times (2j + 1)^2$ process matrix that is known to be diagonal in the irrep basis and to have only $2j + 1$ distinct eigenvalues.  For fixed $m$, our goal is to estimate the eigenvalues $f_k^m$ of the $(2j + 1)^2\times (2j + 1)^2$ diagonal superoperator
\begin{equation}
\mathcal{P}_\text{diag}\equiv \left(\frac{1}{|G|}\sum_{g\in G} \mathcal{G}^\dag\mathcal{E}\mathcal{G}\right)^m = \sum_k f_k^m\Pi_k, \qquad f_k\equiv \frac{\Tr(\Pi_k\mathcal{E})}{\Tr(\Pi_k)},
\end{equation}
which are $(2k + 1)$-fold degenerate ($k = 0, \ldots, 2j$).  Ideally, SSRB would do so by estimating the eigenvalues of the $(2j + 1)\times (2j + 1)$ matrix $P_\text{diag}$, which consists of matrix elements of $\mathcal{P}_\text{diag}$ between synthetic states and measurement effects:
\begin{equation}
(P_\text{diag})_{kk'} = \llangle T^{(k')}_0|\mathcal{P}_\text{diag}|T^{(k)}_0\rrangle = \delta_{kk'}f_k^m
\end{equation}
(note our row/column conventions).  In the limit of infinite sample size, what we actually estimate by preparing and measuring $J_z$ eigenstates is the matrix $P$ where
\begin{equation}
P_{\ell\ell'} = p_{\ell, \ell', m} = \llangle\mathcal{E}^\dag(\ell')|\mathcal{P}_\text{diag}|\ell\rrangle, \qquad \mathcal{E}^\dag(\ell)\equiv \mathcal{E}^\dag(|\ell\rangle\langle\ell|).
\end{equation}
This is because there is always one application of the error channel left untwirled, which means that even in SSRB with perfect SPAM, the SPAM coefficient of the target irrep is not precisely 1.  Indeed, we have
\begin{equation}
(MPM^T)_{kk'} = \sum_{\ell, \ell'} M_{k, \ell}M_{k', \ell'}p_{\ell, \ell', m} = \llangle\mathcal{E}^\dag(T^{(k')}_0)|\mathcal{P}_\text{diag}|T^{(k)}_0\rrangle = \llangle\mathcal{E}^\dag(T^{(k')}_0)|T^{(k)}_0\rrangle f_k^m,
\end{equation}
which is not quite the same as $(P_\text{diag})_{kk'}$.  Therefore, for small gate noise, SSRB estimates the diagonal entries of the \emph{nearly} diagonal matrix $MPM^T$, whereas SSRB-D would estimate its eigenvalues.  These two sets of quantities are not the same, although they are close up to some error that is uncontrolled because we lack knowledge of the noise channel (however, it may be possible to bound the error \emph{a posteriori} by using the RB signal to estimate the strength of the noise channel).

With imperfect SPAM, the SSRB signal will have contamination from other irreps (on top of having SPAM coefficient $< 1$).  On the other hand, for SSRB-D to work, SPAM noise must effect a change of basis.  The entries of the $(2j + 1)\times (2j + 1)$ matrix $MPM^T$ are the matrix elements of the superoperator $\mathcal{E}\mathcal{P}_\text{diag}$ between the incomplete set of superkets $|T^{(k)}_0\rrangle$, but with noisy SPAM, there is no guarantee that the initial or final state can be written as a linear combination of \emph{diagonal} spherical tensor operators.  Thus implementing SSRB-D in a SPAM-robust way may require considering the full $(2j + 1)^2\times (2j + 1)^2$ process matrix $\mathcal{E}\mathcal{P}_\text{diag}$.  At least two questions about the feasibility of SSRB-D remain:
\begin{itemize}
\item Since errors can affect state preparation and measurement differently, can the SPAM error in totality be viewed as conjugating $P$ (or $\mathcal{E}\mathcal{P}_\text{diag}$) by a single change-of-basis matrix, which leaves the spectrum invariant?
\item Can the untwirled error channel $\mathcal{E}$ be absorbed into SPAM error?  In other SPAM-robust protocols, the untwirled error channel only affects the SPAM coefficient (so at $N = \infty$, we can fit the RB data to extract $f_k$ perfectly).  In SSRB-D, however, it affects the estimate for the eigenvalues of $\mathcal{P}_\text{diag}$ (so at $N = \infty$, we can only approximately extract $f_k$).
\end{itemize}
We leave the resolution of these questions for future work.

\section{Examples}

We give several detailed examples of how our protocols work in the case that $j = 7/2$, motivated by the prospects for coherent control of an ${}^{123}$Sb nucleus \cite{Asaad}.  Irreps are labeled by $k\in \{0, \ldots, 7\}$, while states are labeled by $\ell_\text{init}, \ell_\text{final}\in \{7/2, \ldots, -7/2\}$.  We restrict our attention to preparing $J_z$ eigenstates and to measuring the operator $J_z$, where the latter restriction reflects experimental limitations.  Note that if we can measure $J_z$, whose spectrum is nondegenerate and whose POVM elements are $|\ell\rangle\langle\ell|$, then we can also measure all of these projectors individually.

For each example, we fix a sample size $N$ (a large positive integer) and a set of sequence lengths (positive integers) $\mathcal{M}$. (Note that geometrically spaced sequence lengths are optimal for fitting exponential decays.) As a matter of notation, let $G = \text{SU}(2)$, $\vec{g}\equiv (g_1, \ldots, g_m)\in G^m$, and $g_\text{inv} = (g_m\cdots g_1)^\dag$ for any given $m\in \mathcal{M}$.  Also let $M = [M_{k, \ell}]$ with rows (resp.\ columns) indexed by $k = 0, \ldots, 7$ (resp.\ $\ell = 7/2, \ldots, -7/2$), where
\begin{equation}
M\equiv \frac{1}{2}\left[\begin{array}{cccccccc}
\frac{1}{\sqrt{2}} & \frac{1}{\sqrt{2}} & \frac{1}{\sqrt{2}} & \frac{1}{\sqrt{2}} & \frac{1}{\sqrt{2}} & \frac{1}{\sqrt{2}} & \frac{1}{\sqrt{2}} & \frac{1}{\sqrt{2}} \\
\sqrt{\frac{7}{6}} & \frac{5}{\sqrt{42}} & \sqrt{\frac{3}{14}} & \frac{1}{\sqrt{42}} & -\frac{1}{\sqrt{42}} & -\sqrt{\frac{3}{14}} & -\frac{5}{\sqrt{42}} & -\sqrt{\frac{7}{6}} \\
\sqrt{\frac{7}{6}} & \frac{1}{\sqrt{42}} & -\sqrt{\frac{3}{14}} & -\frac{5}{\sqrt{42}} & -\frac{5}{\sqrt{42}} & -\sqrt{\frac{3}{14}} & \frac{1}{\sqrt{42}} & \sqrt{\frac{7}{6}} \\
\frac{7}{\sqrt{66}} & -\frac{5}{\sqrt{66}} & -\frac{7}{\sqrt{66}} & -\sqrt{\frac{3}{22}} & \sqrt{\frac{3}{22}} & \frac{7}{\sqrt{66}} & \frac{5}{\sqrt{66}} & -\frac{7}{\sqrt{66}} \\
\sqrt{\frac{7}{22}} & -\frac{13}{\sqrt{154}} & -\frac{3}{\sqrt{154}} & \frac{9}{\sqrt{154}} & \frac{9}{\sqrt{154}} & -\frac{3}{\sqrt{154}} & -\frac{13}{\sqrt{154}} & \sqrt{\frac{7}{22}} \\
\sqrt{\frac{7}{78}} & -\frac{23}{\sqrt{546}} & \frac{17}{\sqrt{546}} & 5 \sqrt{\frac{3}{182}} & -5 \sqrt{\frac{3}{182}} & -\frac{17}{\sqrt{546}} & \frac{23}{\sqrt{546}} & -\sqrt{\frac{7}{78}} \\
\frac{1}{\sqrt{66}} & -\frac{5}{\sqrt{66}} & 3 \sqrt{\frac{3}{22}} & -\frac{5}{\sqrt{66}} & -\frac{5}{\sqrt{66}} & 3 \sqrt{\frac{3}{22}} & -\frac{5}{\sqrt{66}} & \frac{1}{\sqrt{66}} \\
\frac{1}{\sqrt{858}} & -\frac{7}{\sqrt{858}} & 7 \sqrt{\frac{3}{286}} & -\frac{35}{\sqrt{858}} & \frac{35}{\sqrt{858}} & -7 \sqrt{\frac{3}{286}} & \frac{7}{\sqrt{858}} & -\frac{1}{\sqrt{858}}
\end{array}\right].
\label{Mmatrix}
\end{equation}
This matrix relates physical states ($J_z$ eigenstates) to synthetic states.  Finally, let
\begin{equation}
\renewcommand{\arraystretch}{1.5}
F\equiv \frac{1}{8}\left[\begin{array}{cccccccc}
1 & 1 & 1 & 1 & 1 & 1 & 1 & 1 \\
1 & \frac{59}{63} & \frac{17}{21} & \frac{13}{21} & \frac{23}{63} & \frac{1}{21} & -\frac{1}{3} & -\frac{7}{9} \\
1 & \frac{17}{21} & \frac{7}{15} & \frac{1}{21} & -\frac{1}{3} & -\frac{11}{21} & -\frac{1}{3} & \frac{7}{15} \\
1 & \frac{13}{21} & \frac{1}{21} & -\frac{31}{77} & -\frac{101}{231} & \frac{1}{77} & \frac{17}{33} & -\frac{7}{33} \\
1 & \frac{23}{63} & -\frac{1}{3} & -\frac{101}{231} & \frac{1}{9} & \frac{103}{231} & -\frac{1}{3} & \frac{7}{99} \\
1 & \frac{1}{21} & -\frac{11}{21} & \frac{1}{77} & \frac{103}{231} & -\frac{33}{91} & \frac{53}{429} & -\frac{7}{429} \\
1 & -\frac{1}{3} & -\frac{1}{3} & \frac{17}{33} & -\frac{1}{3} & \frac{53}{429} & -\frac{1}{39} & \frac{1}{429} \\
1 & -\frac{7}{9} & \frac{7}{15} & -\frac{7}{33} & \frac{7}{99} & -\frac{7}{429} & \frac{1}{429} & -\frac{1}{6435}
\end{array}\right].
\label{Fmatrix}
\end{equation}
This matrix relates $\text{SU}(2)$ error rates to RB quality parameters: $\vec{f} = F\vec{p}$. (For normalized rates, we omit the factor of $1/8$.)

\subsection{SU(2) RB with Synthetic SPAM}

The simplest protocol uses only synthetic SPAM: in particular, both synthetic states and synthetic measurement effects that lie within individual irreps ($k = 0, \ldots, 7$).

We first obtain experimental estimates $p_{\ell_\text{init}, \ell_\text{final}, m}$ for the averaged survival probabilities $\mathbb{E}_{\vec{g}\in G^m}[\sbra{\ell_\text{final}}\widetilde{\mathcal{G}}_\text{inv}\widetilde{\mathcal{G}}_m\cdots \widetilde{\mathcal{G}_1}\sket{\ell_\text{init}}]$.  For each $\ell_\text{init}\in \{7/2, \ldots, -7/2\}$ and each $m\in \mathcal{M}$:
\begin{itemize}
\item Define the 8 integer-valued variables $x_{\ell_\text{init}, 7/2, m}, \ldots, x_{\ell_\text{init}, -7/2, m}$ and initialize them all to 0.
\item For $i = 1, \ldots, N$:
\begin{itemize}
\item Prepare the pure state $|\psi\rangle = |\ell_\text{init}\rangle$.
\item Sample $\vec{g}\in G^m$ uniformly at random.  Apply the gates $g_1, \ldots, g_m, g_\text{inv}$ (in that order) to $|\psi\rangle$.
\item Measure $J_z$ and denote the measurement outcome by $\ell$.  Increment $x_{\ell_\text{init}, \ell, m}$ by 1.
\end{itemize}
\item For each $\ell_\text{final}\in \{7/2, \ldots, -7/2\}$, compute $p_{\ell_\text{init}, \ell_\text{final}, m}\equiv x_{\ell_\text{init}, \ell_\text{final}, m}/N$.
\end{itemize}
This gives a list of $64|\mathcal{M}|$ numbers $\{p_{\ell_\text{init}, \ell_\text{final}, m}\}$.  Next, in classical post-processing, we do the following:
\begin{itemize}
\item For each $k\in \{0, \ldots, 7\}$ and $m\in \mathcal{M}$, compute $p_{k, m}\equiv \sum_{\ell_\text{init}, \ell_\text{final} = -7/2}^{7/2} M_{k, \ell_\text{init}}M_{k, \ell_\text{final}}p_{\ell_\text{init}, \ell_\text{final}, m}$, i.e.,
\begin{equation}
\left[\begin{array}{c} p_{0, m} \\ \vdots \\ p_{7, m} \end{array}\right]\equiv \text{diagonal of } M\left[\begin{array}{ccc}
p_{7/2, 7/2, m} & \cdots & p_{7/2, -7/2, m} \\
\vdots & \ddots & \vdots \\
p_{-7/2, 7/2, m} & \cdots & p_{-7/2, -7/2, m}
\end{array}\right]M^T,
\end{equation}
with $M$ given in \eqref{Mmatrix}.  This results in a new list of $8|\mathcal{M}|$ numbers $\{p_{k, m}\}$.
\item For each $k\in \{0, \ldots, 7\}$, fit the $|\mathcal{M}|$ numbers $p_{k, m}$ to a single decaying exponential function of the form $m\mapsto A_k f_k^m$, where $A_k$ and $f_k$ are constants. (In the limit of vanishing noise, we should have $A_k\approx 1$ for all $k$.  With arbitrary normalizations for the rows of $M$, we should have $A_k\approx$ norm of row $k$ of $M$.) Record $f_k$.
\item Compute
\begin{equation}
\left[\begin{array}{c} p_0 \\ \vdots \\ p_7 \end{array}\right]\equiv F^{-1}\left[\begin{array}{c} f_0 \\ \vdots \\ f_7 \end{array}\right],
\end{equation}
with $F$ given in \eqref{Fmatrix}.  The numbers $p_k$ ($k = 0, \ldots, 7$) are the rates of random weight-$k$ $\text{SU}(2)$ errors in the implemented $\text{SU}(2)$ rotations.
\end{itemize}

\subsection{SU(2) Character RB with Synthetic SPAM}

The following protocol uses synthetic SPAM as well as synthetic gates to ensure SPAM-robustness.

We first obtain experimental estimates $p_{\ell_\text{init}, \ell_\text{final}, m, k}$ for the averaged weighted survival probabilities $(2k + 1) \linebreak[1] \mathbb{E}_{(\vec{g}, g)\in G^{m+1}} \linebreak[1] [\chi_k(g) \linebreak[1] \sbra{\ell_\text{final}} \linebreak[1] \widetilde{\mathcal{G}}_\text{inv}\widetilde{\mathcal{G}}_m\cdots \widetilde{\mathcal{G}}_2\widetilde{\mathcal{G}_1\mathcal{G}} \linebreak[1] \sket{\ell_\text{init}}]$.  For each $\ell_\text{init}\in \{7/2, \ldots, -7/2\}$ and each $m\in \mathcal{M}$:
\begin{itemize}
\item For each $k\in \{0, \ldots, 7\}$, define the 8 real-valued variables $x_{\ell_\text{init}, 7/2, m, k}, \ldots, x_{\ell_\text{init}, -7/2, m, k}$ and initialize them all to 0.
\item For $i = 1, \ldots, N$:
\begin{itemize}
\item Prepare the pure state $|\psi\rangle = |\ell_\text{init}\rangle$.
\item Sample $(\vec{g}, g)\in G^{m+1}$ uniformly at random.  Apply the gates $g_1 g, g_2, \ldots, g_m, g_\text{inv}$ (in that order) to $|\psi\rangle$.  Note that $g$ and $g_1$ are compiled into a single gate.
\item Measure $J_z$ and denote the measurement outcome by $\ell$.  For each $k\in \{0, \ldots, 7\}$, increment $x_{\ell_\text{init}, \ell, m, k}$ by $\chi_k(g)$, which can be written as a sum of diagonal Wigner $D$-matrix elements:
\begin{equation}
\chi_k(g)\equiv \sum_{q=-k}^k D_{qq}^k(g).
\end{equation}
\end{itemize}
\item For each $\ell_\text{final}\in \{7/2, \ldots, -7/2\}$ and $k\in \{0, \ldots, 7\}$, compute $p_{\ell_\text{init}, \ell_\text{final}, m, k}\equiv (2k + 1)x_{\ell_\text{init}, \ell_\text{final}, m, k}/N$.
\end{itemize}
This gives a list of $512|\mathcal{M}|$ numbers $\{p_{\ell_\text{init}, \ell_\text{final}, m, k}\}$.  Next, in classical post-processing, we do the following:
\begin{itemize}
\item For each $k\in \{0, \ldots, 7\}$ and $m\in \mathcal{M}$, compute $p_{k, m}\equiv \sum_{\ell_\text{init}, \ell_\text{final} = -7/2}^{7/2} M_{k, \ell_\text{init}}M_{k, \ell_\text{final}}p_{\ell_\text{init}, \ell_\text{final}, m, k}$, i.e.,
\begin{equation}
p_{k, m}\equiv \text{$(k, k)$ component of } M\left[\begin{array}{ccc}
p_{7/2, 7/2, m, k} & \cdots & p_{7/2, -7/2, m, k} \\
\vdots & \ddots & \vdots \\
p_{-7/2, 7/2, m, k} & \cdots & p_{-7/2, -7/2, m, k}
\end{array}\right]M^T,
\end{equation}
with $M$ given in \eqref{Mmatrix}.  This results in a new list of $8|\mathcal{M}|$ numbers $\{p_{k, m}\}$.
\item The remaining steps are the same as in the standard SSRB protocol.
\end{itemize}

\subsection{SU(2) Rank-1 RB with Synthetic SPAM}

This synthetic protocol improves upon the sample complexity of synthetic-SPAM character RB while aiming not to sacrifice too much SPAM-robustness.

We first obtain experimental estimates for the averaged weighted survival probabilities $(2k + 1) \linebreak[1] \mathbb{E}_{(\vec{g}, g)\in G^{m+1}} \linebreak[1] [D_{00}^k(g) \linebreak[1] \sbra{\ell_\text{final}} \linebreak[1] \widetilde{\mathcal{G}}_\text{inv} \linebreak[1] \widetilde{\mathcal{G}}_m \linebreak[1] \cdots \linebreak[1] \widetilde{\mathcal{G}}_2 \linebreak[1] \widetilde{\mathcal{G}_1\mathcal{G}} \linebreak[1] \sket{\ell_\text{init}}]$.  The procedure is identical to that for SS$\chi$RB, except that instead of using the character function (a sum of diagonal Wigner $D$-matrix elements) as a weighting factor, one uses the single diagonal Wigner $D$-matrix element $D_{00}^k(g)$.  In other words, instead of using the trace of the $k^\text{th}$ block in the $\text{SU}(2)$ superoperator, one simply uses the middle diagonal entry of the $k^\text{th}$ block (assuming that the basis of spherical tensor operators $T^{(k)}_q$ is ordered with $q$ ranging from $k$ to $-k$ in either descending or ascending order).  The classical post-processing is exactly the same as in SS$\chi$RB.

\subsection{Sample Complexity Comparison}

Note that all of the synthetic-SPAM protocols, as presented above, involve running independent sets of circuits for each of the 8 distinct initial states.  An alternative option, when settings are expensive, is to use the same collection of random circuits for each initial state.  We quantify only the sample complexity of shots, not settings.

From \eqref{chiRBnormalized}, the zero-noise $\chi$RB variance is (row $=$ SPAM, column $=$ irrep):
\begin{center}
\begin{tabular}{|c|c|c|c|c|c|c|c|c|} \hline
& $k = 0$ & $k = 1$ & $k = 2$ & $k = 3$ & $k = 4$ & $k = 5$ & $k = 6$ & $k = 7$ \\ \hline
$\ell = \pm 7/2$ & 7 & 28.6816 & 91.8386 & 451.654 & 4073.14 & 76502.2 & $3.74866\times 10^6$ & $8.43448\times 10^8$ \\ \hline
$\ell = \pm 5/2$ & 7 & 70.447 & 155094 & 1360.17 & 268.103 & 514.734 & 4711.6 & 276013 \\ \hline
$\ell = \pm 3/2$ & 7 & 480.6 & 1581.9 & 308.139 & 85720.9 & 1560.08 & 404.56 & 3077.44 \\ \hline
$\ell = \pm 1/2$ & 7 & 39815 & 197.953 & 8131.02 & 1014.03 & 2469.18 & 4082.36 & 381.656 \\ \hline
\end{tabular}
\end{center}
Therefore, among $J_z$ eigenstates, the best $\chi$RB SPAM for $k = 1, \ldots, 7$ is
\begin{equation}
\ell = \pm 7/2, \pm 7/2, \pm 3/2, \pm 5/2, \pm 5/2, \pm 3/2, \pm 1/2,
\end{equation}
respectively, which is consistent with our discussion around \eqref{Mdefinition}.  For example, the components of a $j = 7/2$ stretched state (or spin coherent state) inside the irreps $k = 0, \ldots, 7$ have the squared norms
\begin{equation}
(M_{0, \pm 7/2}^2, \ldots, M_{7, \pm 7/2}^2) = \left(\frac{1}{8}, \frac{7}{24}, \frac{7}{24}, \frac{49}{264}, \frac{7}{88}, \frac{7}{312}, \frac{1}{264}, \frac{1}{3432}\right),
\end{equation}
where the small components in the high-spin irreps are reflected in the especially large variances in the $\ell = \pm 7/2$ row.

From \eqref{R1RBnormalized}, the zero-noise R1RB variance is (row $=$ SPAM, column $=$ irrep):
\begin{center}
\begin{tabular}{|c|c|c|c|c|c|c|c|c|} \hline
& $k = 0$ & $k = 1$ & $k = 2$ & $k = 3$ & $k = 4$ & $k = 5$ & $k = 6$ & $k = 7$ \\ \hline
$\ell = \pm 7/2$ & 7 & 7.52245 & 12.5807 & 43.3217 & 303.615 & 4642.29 & 191700 & $3.72854\times 10^7$ \\ \hline
$\ell = \pm 5/2$ & 7 & 16.257 & 28940.8 & 153.982 & 21.0241 & 32.779 & 257.413 & 13036.4 \\ \hline
$\ell = \pm 3/2$ & 7 & 152.067 & 250.717 & 42.3744 & 7764.81 & 110.824 & 23.2173 & 157.019 \\ \hline
$\ell = \pm 1/2$ & 7 & 15623 & 45.3069 & 1267.85 & 102.154 & 200.906 & 279.119 & 21.6442 \\ \hline
\end{tabular}
\end{center}
The best SPAM follows the same pattern as for $\chi$RB.

With synthetic SPAM, we compute using \eqref{SSchiRBvariance} and \eqref{SSR1RBvariance} that the zero-noise variance is (column $=$ irrep):
\begin{center}
\begin{tabular}{|c|c|c|c|c|c|c|c|c|} \hline
& $k = 0$ & $k = 1$ & $k = 2$ & $k = 3$ & $k = 4$ & $k = 5$ & $k = 6$ & $k = 7$ \\ \hline
SS$\chi$RB & 0 & 1.07619 & 3.23842 & 6.15572 & 10.4498 & 15.668 & 23.0531 & 34.0697 \\ \hline
SSR1RB & 0 & 0.269048 & 0.540816 & 0.773292 & 1.02387 & 1.28994 & 1.62223 & 2.11888 \\ \hline
\end{tabular}
\end{center}
Unlike with physical SPAM, the variance increases monotonically with irrep dimension.  If we sacrifice SPAM-robustness, then we obtain from \eqref{SSRBvariance} (in the zero-noise limit):
\begin{center}
\begin{tabular}{|c|c|c|c|c|c|c|c|c|} \hline
& $k = 0$ & $k = 1$ & $k = 2$ & $k = 3$ & $k = 4$ & $k = 5$ & $k = 6$ & $k = 7$ \\ \hline
SSRB & 0 & 0 & 0 & 0 & 0 & 0 & 0 & 0 \\ \hline
\end{tabular}
\end{center}
These numbers can be compared directly to simulation results.  We summarize some of our analytical results in Figure \ref{su2rbvariance}.

\begin{figure}[!htb]
\centering
\includegraphics[width=0.7\textwidth]{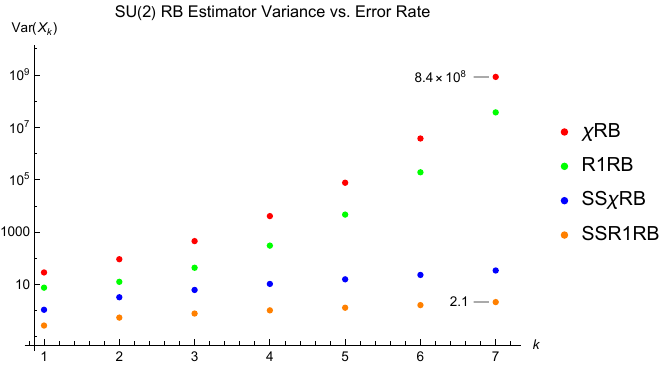}
\caption{High-dimensional spins provide an elegant platform for quantum error correction, as they can encode a logical qubit with natural ``transversal'' gates (finite subgroups of $\text{SU}(2)$).  Until now, the group $\text{SU}(2)$ has been infeasible to benchmark because it is infinite and nonabelian, so its superoperator representation is highly reducible and its irreps have large dimension.  We solve this problem using synthetic RB.  This plot compares the sample complexities of various RB protocols for the example of a spin-$7/2$ system, which is the smallest spin supporting a $2I$ (icosahedral) spin code \cite{Gross}.  Error rates are labeled by nontrivial irreps ($k = 1, \ldots, 7$).  The red, green, blue, and orange trends show analytical results for protocols that incorporate successively more synthetic RB techniques.  Red: character RB, where the initial state is a spin coherent state.  Green: synthetic RB with only synthetic gates (enhanced relative to character RB).  Blue: synthetic RB with only synthetic SPAM, where the effective initial state is a diagonal spherical tensor operator.  Orange: full synthetic RB with both synthetic gates and SPAM.  For the most extreme error rate, we obtain an improvement of eight orders of magnitude in efficiency between red and orange.}
\label{su2rbvariance}
\end{figure}

\section{Numerical Results} \label{app:numerics}

We corroborate our analytical estimates for the performance of synthetic RB protocols with simulation data.  Due to their improved performance over physical-SPAM protocols, we focus on the synthetic-SPAM protocols SSRB, SS$\chi$RB, and SSR1RB.  For a fixed number of randomly sampled circuits, we extract uncertainties for the quality parameters $f_k$ from curve fits as well as uncertainties for the final error rates $p_k$.  To compute uncertainties $\sigma_{p_k}$ for the $p_k$ from uncertainties $\sigma_{f_k}$ for the $f_k$, we use
\begin{equation}
(\sigma_{p_0}, \ldots, \sigma_{p_{2j}}) = \sqrt{\operatorname{diag}^{-1}(F^{-1}\operatorname{diag}(\sigma_{f_0}^2, \ldots, \sigma_{f_{2j}}^2)F^{-T})},
\end{equation}
where the operation ``$\operatorname{diag}$'' converts a list of numbers to a diagonal matrix, the operation ``$\operatorname{diag}^{-1}$'' converts the diagonal of a (not necessarily diagonal) matrix to a list of numbers, and $F^{-T}\equiv (F^{-1})^T = (F^T)^{-1}$.  In the limit of infinite sample size, the $\text{SU}(2)$ error rates for any gate-independent noise channel $\mathcal{E}$ can be computed from the quality parameters
\begin{equation}
f_k = \frac{1}{2k + 1}\sum_{q=-k}^k \llangle T_q^{(k)}|\mathcal{E}(T_q^{(k)})\rrangle.
\label{exactfk}
\end{equation}
Any SPAM-robust protocol should produce error rates that converge to those computed via \eqref{exactfk}.

We consider the following gate error channels:
\begin{enumerate}
\item Coherent errors $\rho\mapsto U\rho U^\dag$ with $U = e^{-iH}$ and Hamiltonian $H = \gamma J_z^2$.  Such errors are supported only in even irreps $k$.  Perturbatively in $\gamma$, coherent $J_z^2$ errors produce predominantly quadrupole ($k = 2$) $\text{SU}(2)$ errors.
\item An energy-dependent dephasing channel produced by a fluctuating $J_z$ Hamiltonian, which suppresses off-diagonal matrix elements of $\rho$ in the $J_z$ eigenbasis according to\footnote{This kind of error arises from a Lindblad equation describing energy diffusion, in which one of the jump operators is the Hamiltonian itself.  The corresponding dissipative term is quadratic in the Hamiltonian and gives rise to a factor of $e^{-\gamma t(E_i - E_j)^2}$ in the time evolution of the density matrix, when written in the energy eigenbasis.}
\begin{equation}
\langle\ell|\rho|\ell'\rangle\mapsto e^{-\gamma(\ell - \ell')^2}\langle\ell|\rho|\ell'\rangle.
\end{equation}
This channel produces error rates $p_k$ that decrease monotonically with $k$, where the $k$-dependence becomes more uniform as the strength $\gamma$ grows.  Perturbatively in $\gamma$, it produces predominantly dipole ($k = 1$) $\text{SU}(2)$ errors.
\end{enumerate}
Note that coherent gate errors generally do not produce $\text{SU}(2)$ error rates that are monotonic in the irrep $k$ beyond the perturbative regime, i.e., as the strength or duration of the noise increases.  We regard the ``perturbative regime'' as that where the rate of weight-0 errors (i.e., no errors) is large.

In Figures \ref{fig_randpovm_coherent}, \ref{fig_randpovm_dephasing}, \ref{fig_rotateeffects_coherent}, \ref{fig_rotateeffects_dephasing}, we plot for $j = 7/2$ the simulated results of running synthetic RB protocols with varying SPAM noise, $10^4$ randomly chosen circuits, and infinitely many shots per circuit.  We include the exponential decay curves used to compute $f_k$ and $p_k$, where colors indicate irreps.  The uncertainties for $f_k$ are shown explicitly; negative error rates are spurious and should be regarded as implicitly cut off at 0.  We consider two kinds of gate noise:
\begin{itemize}
\item The coherent noise channel is $\rho\mapsto U\rho U^\dag$ for $U = \exp(-i\gamma J_z^2)$ and $\gamma = 0.04$.
\item The dephasing noise channel is an elementwise rescaling $\rho_{\ell\ell'}\mapsto \exp(-i\gamma|\ell - \ell'|^2)\rho_{\ell\ell'}$ for $\gamma = 0.01$.
\end{itemize}
For these two channels, the error rates computed from \eqref{exactfk} are as follows:
\begin{center}
\begin{tabular}{|c|c|c|c|c|c|c|c|c|} \hline
& $p_0$ & $p_1$ & $p_2$ & $p_3$ & $p_4$ & $p_5$ & $p_6$ & $p_7$ \\ \hline
coherent & 0.9668 & 0 & 0.03301 & 0 & $1.434\times 10^{-4}$ & 0 & $1.110\times 10^{-7}$ & 0 \\ \hline
dephasing & 0.9068 & 0.08787 & 0.005118 & $1.991\times 10^{-4}$ & $5.315\times 10^{-6}$ & $9.504\times 10^{-8}$ & $1.039\times 10^{-9}$ & $5.297\times 10^{-12}$ \\ \hline
\end{tabular}
\end{center}
In all cases, state preparation error is modeled by replacing the nominal $|\ell\rangle\langle\ell|$ with $V_{\ell}|\ell\rangle\langle\ell|V_{\ell}^\dag$, where $V_\ell = \exp(-i\phi\vec{n}_{\ell}\cdot \vec{J})$ for varying angles $\phi$ and a random unit vector $\vec{n}_{\ell}$.  Measurement error is modeled by either:
\begin{itemize}
\item a coherent rotation of all measurement effects about a random axis by the same angle $\phi$
\item or a random (but fixed) permutation of the measurement effects. 
\end{itemize}
Our models of SPAM error are chosen to test the performance of our protocols rather than for physical realism.  It is easier to specify such ``adversarial'' SPAM error via lists of initial states and measurement effects than as quantum channels.  Measurement errors must preserve the condition that the POVM effects sum to the identity (one simple option is to map the set of effects $\{|\ell\rangle\langle\ell|\}$ to a permutation thereof).  Simple options for state preparation error include coherent (unitary) errors such as random $\text{SU}(2)$ rotations (this is the error model for spin codes) and incoherent (stochastic) errors that map the initial state to a stochastic mixture of $J_z$ eigenstates.  Note that the effect of random $\text{SU}(2)$ errors averages out after many shots, but the effect of biased noise does not.  Correspondingly, we use per-prep $\text{SU}(2)$ errors where the axis of the $\text{SU}(2)$ rotation is fixed for each prepared state rather than chosen randomly in each shot.

In the absence of SPAM error, we find the performance hierarchy SSRB $>$ SSR1RB $>$ SS$\chi$RB, as expected from analytical results for the variance (which assume \emph{both} perfect SPAM and noiseless gates).  We also observe that SPAM error contaminates the estimated gate error in the non-SPAM-robust SSRB protocol.  The SPAM-robustness and sample complexity advantages of SSR1RB allow it to outperform both SSRB and SS$\chi$RB when faced with sufficiently large SPAM error.

We still isolate the desired single exponential decays if either the state preparations or the measurement operations are supported entirely on the correct irrep.  For example, $\text{SU}(2)$ rotations that act identically on all initial states or on all measurement effects will only affect the SPAM coefficients, not the observed exponential decay rates.  The same is true of arbitrary errors on either the states or the measurements alone.  This is because these SPAM errors reduce the overlap between synthetic states and effects within the same irrep (thus decreasing the signal strength) without mixing different irreps.  For this reason, SSRB is fairly robust to SPAM error: it can fail only when \emph{both} state preparation and measurement are noisy.


\begin{figure}[!htb]
\centering
\includegraphics[width=\textwidth, trim={0 0 0 3.4cm}, clip]{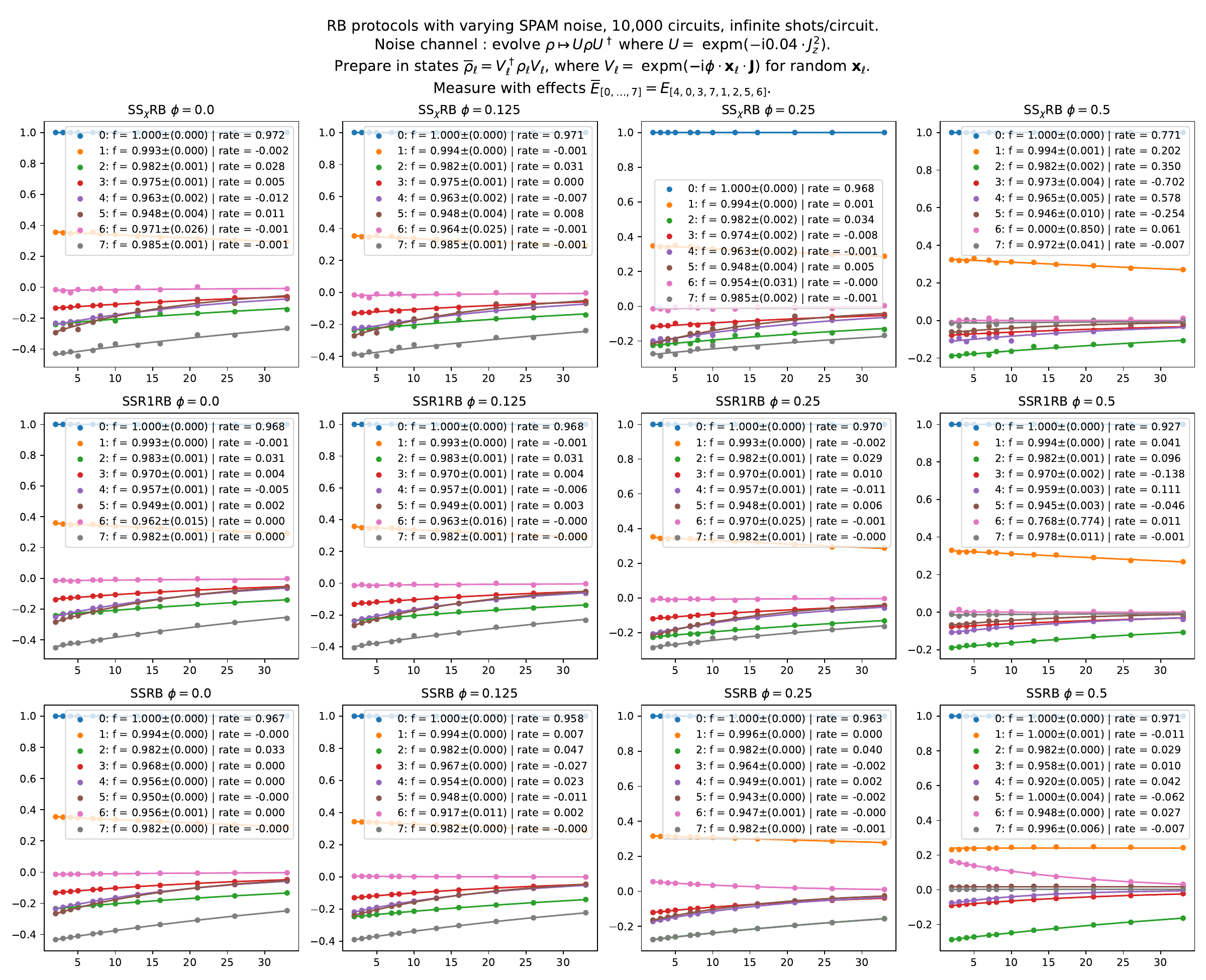}
\caption{Simulated data for synthetic RB protocols with \textbf{coherent} gate noise and measurement error corresponding to a \textbf{random permutation} of POVM effects.  In this case, $\phi$ quantifies the amount of state preparation error.  SSRB produces both accurate and precise results in the absence of state preparation error, even in the presence of measurement error (bottom left).}
\label{fig_randpovm_coherent}
\end{figure}

\begin{figure}[!htb]
\centering
\includegraphics[width=\textwidth, trim={0 0 0 3.4cm}, clip]{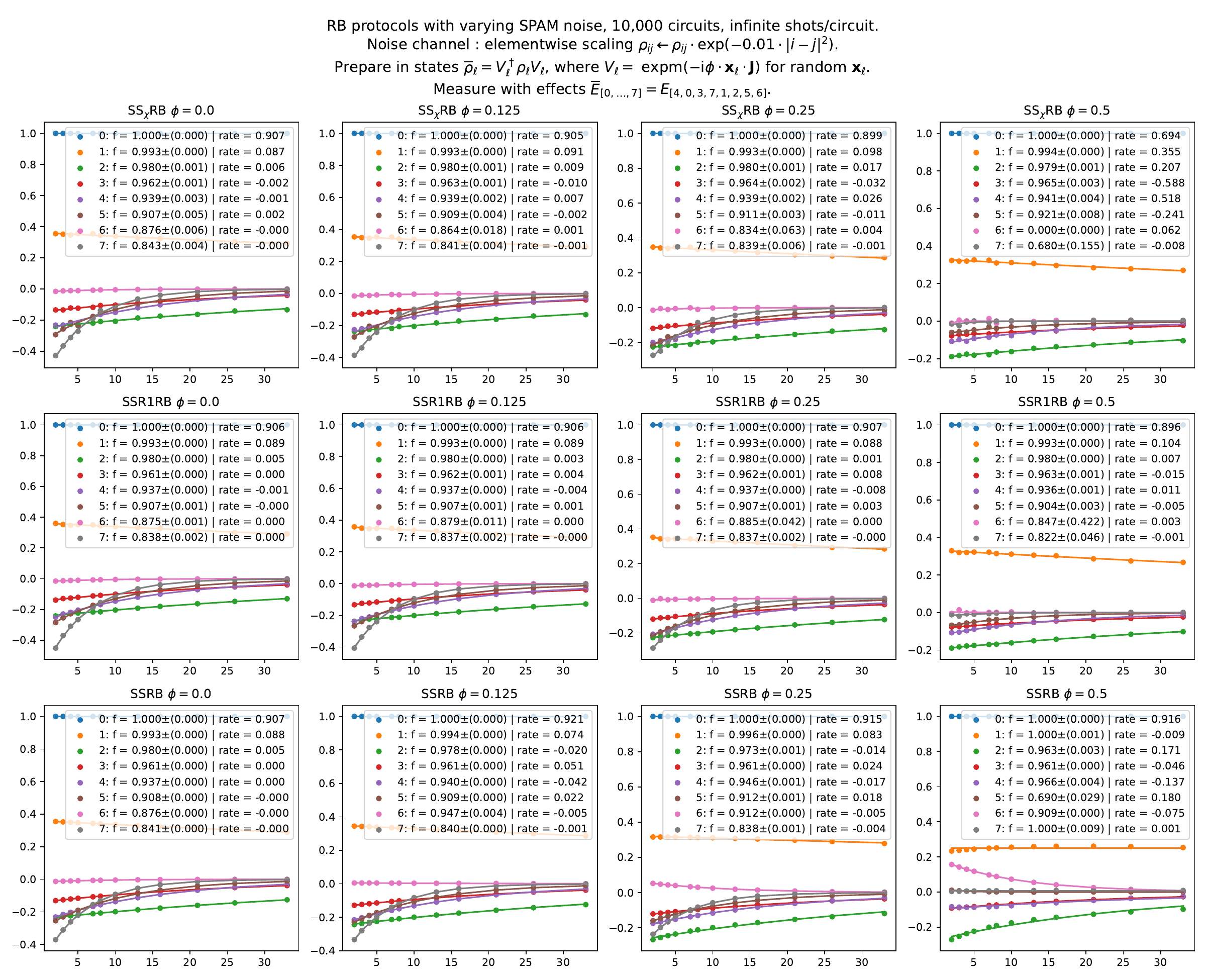}
\caption{Simulated data for synthetic RB protocols with \textbf{dephasing} gate noise and measurement error corresponding to a \textbf{random permutation} of POVM effects.  In this case, $\phi$ quantifies the amount of state preparation error.  As $\phi$ increases, the results of SSRB become noticeably less accurate compared to those at the bottom left.  This loss in accuracy is mitigated by SS$\chi$RB and to an even greater extent by SSR1RB, at some cost in precision (rightmost column).}
\label{fig_randpovm_dephasing}
\end{figure}

\begin{figure}[!htb]
\centering
\includegraphics[width=\textwidth, trim={0 0 0 3.4cm}, clip]{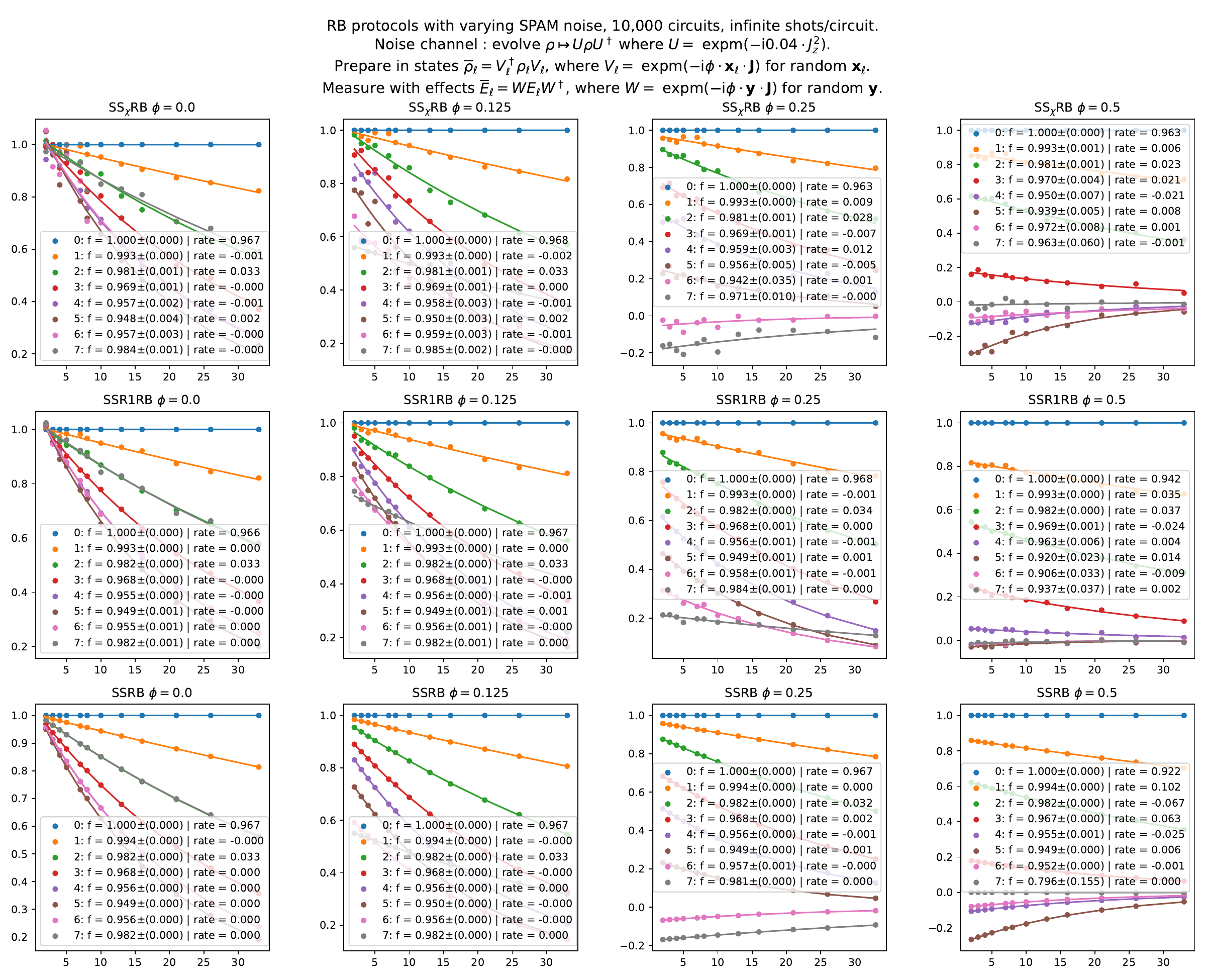}
\caption{Simulated data for synthetic RB protocols with \textbf{coherent} gate noise and measurement error corresponding to a \textbf{coherent rotation} of POVM effects.  In this case, $\phi$ quantifies the amount of error in state preparation \emph{and} measurement.  Pure SSRB is robust to this kind of SPAM error for a range of $\phi$ (bottom row), but the results of the SPAM-robust protocols remain more reliable for large $\phi$ (rightmost column).}
\label{fig_rotateeffects_coherent}
\end{figure}

\begin{figure}[!htb]
\centering
\includegraphics[width=\textwidth, trim={0 0 0 3.4cm}, clip]{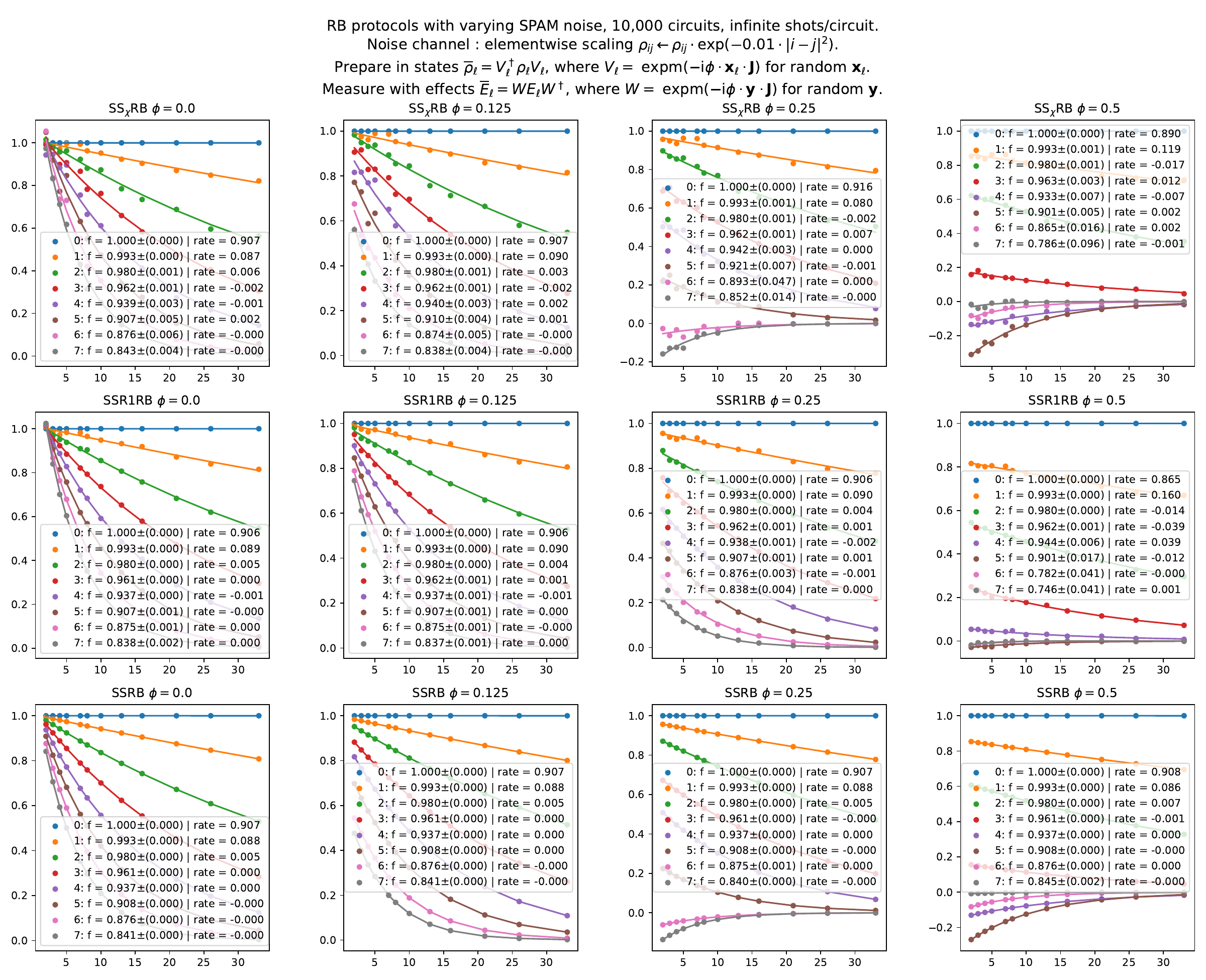}
\caption{Simulated data for synthetic RB protocols with \textbf{dephasing} gate noise and measurement error corresponding to a \textbf{coherent rotation} of POVM effects.  In this case, $\phi$ quantifies the amount of error in state preparation \emph{and} measurement.  For this gate error channel, the results of SSRB are quite robust to the specified SPAM error (bottom row).}
\label{fig_rotateeffects_dephasing}
\end{figure}

\section{Connections to Other Protocols}

\subsection{Comparison to Pauli and Clifford Benchmarking} \label{app:comparisontoPC}

Here, we provide an illustrative comparison of $\text{SU}(2)$ RB to Pauli and Clifford (character) RB via analyses of their sample complexity.  For simplicity, we focus on the case of a single qubit.

The superoperator (Pauli transfer matrix) representation of the Pauli group is insensitive to global phases and is therefore not a faithful representation, as the 16 elements of the single-qubit Pauli group map to only 4 distinct superoperators.  When we twirl over the ``Pauli group,'' we really mean that we twirl over its superoperator representation.  The superoperator representation of the Pauli group is abelian and multiplicity-free.  Its 1D irreps are spanned by the Pauli operators $I, X, Y, Z$ (or, more generally, the $4^n$ $n$-qubit Pauli matrices $\{I, X, Y, Z\}^{\otimes n}$).  An arbitrary quantum channel twirled over the Pauli group becomes a Pauli channel whose eigenvectors are Pauli operators.  A Pauli channel can be described by a list of Pauli eigenvalues or by a list of Pauli error rates.  For a single qubit, the Pauli Choi unit
\begin{equation}
\rho\mapsto P\rho Q
\end{equation}
(where $P, Q$ are Pauli operators), when twirled over the Pauli group, becomes
\begin{equation}
\rho\mapsto \frac{1}{4}\sum_{P'} (P'PP')\rho(P'QP') = \delta_{P, Q}P\rho P.
\end{equation}
An arbitrary Pauli-twirled channel takes the form
\begin{equation}
T = \operatorname{diag}(a, b, c, d)
\end{equation}
as a superoperator, which acts within the irreps spanned by $\sket{I}, \sket{X}, \sket{Y}, \sket{Z}$.  The Pauli eigenvalues $a, b, c, d$ are the quality parameters in regular RB, as
\begin{equation}
\sbra{E}T^n\sket{\rho} = \alpha a^n + \beta b^n + \gamma c^n + \delta d^n,
\end{equation}
and inserting projectors isolates the individual terms.  On the other hand, we have
\begin{equation}
\mathcal{G}_I = \left[\begin{array}{cccc}
1 & 0 & 0 & 0 \\
0 & 1 & 0 & 0 \\
0 & 0 & 1 & 0 \\
0 & 0 & 0 & 1
\end{array}\right], \qquad
\mathcal{G}_X = \left[\begin{array}{cccc}
1 & 0 & 0 & 0 \\
0 & 1 & 0 & 0 \\
0 & 0 & -1 & 0 \\
0 & 0 & 0 & -1
\end{array}\right], \qquad
\mathcal{G}_Y = \left[\begin{array}{cccc}
1 & 0 & 0 & 0 \\
0 & -1 & 0 & 0 \\
0 & 0 & 1 & 0 \\
0 & 0 & 0 & -1
\end{array}\right], \qquad
\mathcal{G}_Z = \left[\begin{array}{cccc}
1 & 0 & 0 & 0 \\
0 & -1 & 0 & 0 \\
0 & 0 & -1 & 0 \\
0 & 0 & 0 & 1
\end{array}\right],
\end{equation}
where $\mathcal{G}_O : \rho\mapsto O\rho O$.  Linear combinations of these superoperators, weighted by characters ($\pm 1$), give projectors onto irreps.  As a Pauli channel, we can write
\begin{equation}
T = p_I\mathcal{G}_I + p_X\mathcal{G}_X + p_Y\mathcal{G}_Y + p_Z\mathcal{G}_Z,
\end{equation}
where the Pauli error rates $p_i$ satisfy $\sum_i p_i = 1$.  The two sets of coefficients (Pauli eigenvalues and error rates) are related by
\begin{equation}
\left[\begin{array}{c} a \\ b \\ c \\ d \end{array}\right] = \left[\begin{array}{cccc} 1 & 1 & 1 & 1 \\ 1 & 1 & -1 & -1 \\ 1 & -1 & 1 & -1 \\ 1 & -1 & -1 & 1 \end{array}\right]\left[\begin{array}{c} p_I \\ p_X \\ p_Y \\ p_Z \end{array}\right].
\end{equation}
The matrix on the right is an unnormalized Walsh-Hadamard transform.  It is, in addition, the character table of the Pauli group.  The generalization to multiple qubits is straightforward.

The Clifford group is often defined as the normalizer of the Pauli group in the unitary group.  By this definition, the single-qubit Clifford group is generated by $\{H, S\}$, up to phases.  Another common definition of the Clifford group is as the quotient of the aforementioned group by its center.  By this definition, the Clifford group is isomorphic to the 24-element octahedral group (which acts by proper rotations of the octahedron formed by the signed Pauli matrices on the Bloch sphere) \cite{Barends}, and its superoperator representation is faithful.  Note that an alternative definition of the Clifford group as the binary octahedral group $2O$ \cite{Gross} lies in between these two definitions, as $2O$ has nontrivial center $\mathbb{Z}_2$.  Our interest lies in the superoperator representation of the Clifford group, which is the same for any of these definitions.  In standard RB, the superoperator representation of the Clifford group on a single-qubit state space splits as $\frac{1}{2}\otimes \frac{1}{2} = 0\oplus 1$.  A Clifford-twirled channel can therefore be described by a single nontrivial ``traceless operators'' decay rate or by a single nontrivial ``traceless Kraus operators'' error rate.  The Pauli transfer matrix representations of the 24 equivalence classes of single-qubit Clifford operations (up to global phase) are precisely the signed permutation matrices that fix $|I\rrangle$ and have unit determinant.  We denote them by $\Gamma_{PQR}^{\epsilon\epsilon'}$, where $\epsilon, \epsilon'\in \{+1, -1\}$ and $(P, Q, R)$ is some permutation of $(X, Y, Z)$:
\begin{alignat}{2}
\Gamma_{XYZ}^{\epsilon\epsilon'} &\equiv
\left[\begin{array}{cccc}
1 & 0 & 0 & 0 \\
0 & \epsilon & 0 & 0 \\
0 & 0 & \epsilon\epsilon' & 0 \\
0 & 0 & 0 & \epsilon'
\end{array}\right], \qquad &
\Gamma_{XZY}^{\epsilon\epsilon'} &\equiv
\left[\begin{array}{cccc}
1 & 0 & 0 & 0 \\
0 & \epsilon & 0 & 0 \\
0 & 0 & 0 & \epsilon' \\
0 & 0 & -\epsilon\epsilon' & 0
\end{array}\right], \\
\Gamma_{YZX}^{\epsilon\epsilon'} &\equiv
\left[\begin{array}{cccc}
1 & 0 & 0 & 0 \\
0 & 0 & 0 & \epsilon' \\
0 & \epsilon & 0 & 0 \\
0 & 0 & \epsilon\epsilon' & 0
\end{array}\right], \qquad &
\Gamma_{YXZ}^{\epsilon\epsilon'} &\equiv
\left[\begin{array}{cccc}
1 & 0 & 0 & 0 \\
0 & 0 & -\epsilon\epsilon' & 0 \\
0 & \epsilon & 0 & 0 \\
0 & 0 & 0 & \epsilon'
\end{array}\right], \\
\Gamma_{ZXY}^{\epsilon\epsilon'} &\equiv
\left[\begin{array}{cccc}
1 & 0 & 0 & 0 \\
0 & 0 & \epsilon\epsilon' & 0 \\
0 & 0 & 0 & \epsilon' \\
0 & \epsilon & 0 & 0
\end{array}\right], \qquad &
\Gamma_{ZYX}^{\epsilon\epsilon'} &\equiv
\left[\begin{array}{cccc}
1 & 0 & 0 & 0 \\
0 & 0 & 0 & \epsilon' \\
0 & 0 & -\epsilon\epsilon' & 0 \\
0 & \epsilon & 0 & 0
\end{array}\right].
\end{alignat}
Note that $\mathcal{G}_I = \Gamma_{XYZ}^{++}$, $\mathcal{G}_X = \Gamma_{XYZ}^{+-}$, $\mathcal{G}_Y = \Gamma_{XYZ}^{--}$, $\mathcal{G}_Z = \Gamma_{XYZ}^{-+}$.

To build intuition for the relative advantages of character RB and other methods of constructing irrep projectors, we consider single-qubit character RB with the Clifford group as the benchmarking group (``Clifford character RB''), and we assess how different choices of character group affect the efficiency of the protocol.  Similar considerations apply to Pauli character RB, as any subgroup of the Pauli group is also a subgroup of the Clifford group.  The Clifford group is already a 2-design, so character RB is not strictly necessary in this case.  This discussion is simply meant to build intuition for the case where the benchmarking group is not a 2-design.

\subsubsection{Building Projectors}

To separate the two irreps in the superoperator representation of the Clifford group using character RB, it is most efficient to use the smallest possible character group ($\mathbb{Z}_2$).  The Pauli group also works well as a character group because its superoperator representation has 1D irreps, so the characters have unit modulus.  Using the Clifford group as its own character group is inefficient because the variance of the estimator around the mean is much larger: the characters of the Clifford group are large in magnitude because (in the case of $n$ qubits) the nontrivial irrep has dimension $4^n - 1$, leading to significant cancellations in the character sum.  The characters also vanish on many elements, leading to circuits that contribute trivially to the character sum.  Here, we use the size of the coefficients as a rough proxy for the sample complexity, which we make more precise later by considering the 1-norm of the coefficient vector.

\paragraph{Clifford character group.}

With this choice of character group, the projector $\Pi_\text{traceless} = \operatorname{diag}(0, 1, 1, 1)$ onto the nontrivial irrep is constructed as the following sum of 24 terms:
\begin{equation}
\Pi_\text{traceless} = \frac{3}{24}\Bigg[{\underbrace{3\Gamma_{XYZ}^{++} - \Gamma_{XYZ}^{+-} - \Gamma_{XYZ}^{-+} - \Gamma_{XYZ}^{--}}_{\operatorname{diag}(0, 4, 4, 4)}} + {\underbrace{\sum_\epsilon \epsilon\sum_{\epsilon'} \Gamma_{XZY}^{\epsilon\epsilon'}}_{\operatorname{diag}(0, 4, 0, 0)}} + {\underbrace{\sum_{\epsilon'} \epsilon'\sum_\epsilon \Gamma_{YXZ}^{\epsilon\epsilon'}}_{\operatorname{diag}(0, 0, 0, 4)}} + {\underbrace{\sum_{\epsilon, \epsilon'} (-\epsilon\epsilon')\Gamma_{ZYX}^{\epsilon\epsilon'}}_{\operatorname{diag}(0, 0, 4, 0)}} + 0\sum_{\epsilon, \epsilon'} (\Gamma_{YZX}^{\epsilon\epsilon'} + \Gamma_{ZXY}^{\epsilon\epsilon'})\Bigg],
\label{pitraceless}
\end{equation}
where the characters of the nontrivial irrep (traces of the lower $3\times 3$ block) take the values $3, \pm 1, 0$.  On the other hand, the projector onto the trivial irrep is given by averaging over all 24 transfer matrices with uniform weight (character 1).  The inefficiency of this approach is evident from the character formula \eqref{pitraceless} for the projector: even the terms that contribute with nonzero coefficient are highly redundant and lead to a result that could have been obtained with far fewer terms.

\paragraph{Pauli character group.}

With this choice of character group, we construct the projectors onto the irreps of the Pauli group as follows:
\begin{align}
\Pi_I &= \operatorname{diag}(1, 0, 0, 0) = \frac{1}{4}(\Gamma_{XYZ}^{++} + \Gamma_{XYZ}^{+-} + \Gamma_{XYZ}^{-+} + \Gamma_{XYZ}^{--}), \\
\Pi_X &= \operatorname{diag}(0, 1, 0, 0) = \frac{1}{4}(\Gamma_{XYZ}^{++} + \Gamma_{XYZ}^{+-} - \Gamma_{XYZ}^{-+} - \Gamma_{XYZ}^{--}), \\
\Pi_Y &= \operatorname{diag}(0, 0, 1, 0) = \frac{1}{4}(\Gamma_{XYZ}^{++} - \Gamma_{XYZ}^{+-} - \Gamma_{XYZ}^{-+} + \Gamma_{XYZ}^{--}), \\
\Pi_Z &= \operatorname{diag}(0, 0, 0, 1) = \frac{1}{4}(\Gamma_{XYZ}^{++} - \Gamma_{XYZ}^{+-} + \Gamma_{XYZ}^{-+} - \Gamma_{XYZ}^{--}).
\end{align}
The characters are $\pm 1$, which are smaller in magnitude than those in the Clifford case.  Using the Pauli group as the character group is more efficient in two ways: the projector sum involves fewer terms, and those terms contribute with coefficients of smaller magnitude.  However, this approach is only capable of constructing projectors onto 1D subspaces of the nontrivial Clifford superoperator irrep rather than the full projector, so the resulting RB signal will be more SPAM-dependent.

\paragraph{$\mathbb{Z}_2$ character group.}

If settings are expensive, an alternative option that requires sampling from even fewer distributions than with Pauli character group (two rather than four) is to use the $\mathbb{Z}_2$ character group generated by a given Pauli (super)operator.  In this case, we construct projectors onto 2D subspaces of the nontrivial Clifford irrep:
\begin{align}
\operatorname{diag}(0, 0, 1, 1) = \frac{1}{2}(\Gamma_{XYZ}^{++} - \Gamma_{XYZ}^{+-}), \\
\operatorname{diag}(0, 1, 0, 1) = \frac{1}{2}(\Gamma_{XYZ}^{++} - \Gamma_{XYZ}^{--}), \\
\operatorname{diag}(0, 1, 1, 0) = \frac{1}{2}(\Gamma_{XYZ}^{++} - \Gamma_{XYZ}^{-+}),
\end{align}
which are equivalently projectors onto the isotypic component of the multiplicity-two nontrivial irrep of the chosen $\mathbb{Z}_2$ subgroup.

\paragraph{Finite frame: Pauli superoperators.}

One could instead take the following non-character-based approach.  The Clifford superoperators are overcomplete for the space of superoperators with $1 + 3$ block structure (in the same way that $\text{SU}(2)$ su\-per\-op\-er\-a\-tors are overcomplete for the space of superoperators with $1 + 3 + 5 + \cdots$ block structure).  Indeed, the Pauli superoperators, which are the purely diagonal Clifford superoperators, already form a basis for the space of diagonal superoperators.  Therefore, we can use them to directly construct the full projector onto the nontrivial Clifford irrep:
\begin{equation}
\Pi_\text{traceless} = \operatorname{diag}(0, 1, 1, 1) = \frac{1}{4}(3\Gamma_{XYZ}^{++} - \Gamma_{XYZ}^{+-} - \Gamma_{XYZ}^{-+} - \Gamma_{XYZ}^{--}).
\label{pitracelesspauli}
\end{equation}
This formula requires only four terms, in contrast to \eqref{pitraceless}.

\paragraph{Finite frame: non-Pauli Clifford superoperators.}

Examining \eqref{pitraceless} suggests an alternative non-character-based approach.  By using exclusively non-Pauli Clifford superoperators, we can write
\begin{equation}
\Pi_\text{traceless} = \operatorname{diag}(0, 1, 1, 1) = \frac{1}{4}\left[\sum_\epsilon \epsilon\sum_{\epsilon'} \Gamma_{XZY}^{\epsilon\epsilon'} + \sum_{\epsilon'} \epsilon'\sum_\epsilon \Gamma_{YXZ}^{\epsilon\epsilon'} + \sum_{\epsilon, \epsilon'} (-\epsilon\epsilon')\Gamma_{ZYX}^{\epsilon\epsilon'}\right].
\end{equation}
This scheme for constructing a full projector requires 12 terms rather than the four of \eqref{pitracelesspauli}, albeit with coefficients of uniformly small magnitude.

\paragraph{More qubits.}

The observation behind \eqref{pitracelesspauli} clearly generalizes to $n$ qubits: linear combinations of Pauli superoperators can be used to construct the projector onto the nontrivial Clifford irrep while requiring far fewer terms than Clifford character RB with Clifford character group does.  In fact, such linear combinations can be used to construct projectors of any dimension.  However, since large coefficients magnify errors (and hence increase the sample complexity of estimating RB quality parameters to a given precision), we would like to compare the magnitudes of the coefficients in the linear combinations required to construct different projectors.

Roughly, there is a tradeoff between the size of the desired projector (the dimension of its image) and the variation in magnitude of the coefficients required to construct it from Pauli superoperators.  The variation in the coefficients decreases as the size of the projector decreases, at the cost of also decreasing SPAM-robustness (as can be seen from the structure of the Walsh-Hadamard matrix).  For example, to construct the full projector onto the nontrivial irrep requires a linear combination where the identity superoperator appears with coefficient of order 1 while the $4^n - 1$ nontrivial Pauli superoperators appear with coefficients that are exponentially small in the number of qubits:
\begin{equation}
\operatorname{diag}(0, 1, \ldots, 1) = \left(1 - \frac{1}{4^n}\right)\mathcal{G}_I - \frac{1}{4^n}\sum_{P\neq I} \mathcal{G}_P.
\end{equation}
This is because the projector onto the nontrivial irrep equals the identity superoperator minus the projector onto the trivial irrep, where the latter is a uniform average over all $4^n$ Pauli superoperators. [Consequently, the errors or variations in the circuits that implement the contribution of the identity superoperator will be exponentially overweighted relative to those of the other circuits, and the signal from this non-character-based approach will be dominated by the contributions of the identity circuits.  This is consistent with standard Clifford RB, in which we only run circuits that compile to the identity, whose signal consists of a constant that is exponentially small in $n$ (corresponding to the trivial irrep) plus an exponential decay (corresponding to the nontrivial irrep).  The constant term is what we subtract by using character RB to isolate the nontrivial irrep.] At the other extreme, to construct any 1D projector, we must add up $4^n$ terms with coefficients of magnitude $4^{-n}$ (but of indefinite sign).

\subsubsection{Sample Complexity}

To quantify the sample complexity of these different protocols, we borrow from our analysis of finite-frame RB in Appendix \ref{app:FFRB}.  Any one of these protocols constructs a projector as a finite linear combination of $N_\text{group}$ group element superoperators with coefficients $\vec{c}$.  Let $|\vec{c}|_p$ denote the $p$-norm of the length-$N_\text{group}$ coefficient vector $\vec{c}$.  As in our analysis of finite-frame RB, uniform sampling across all circuits that compile to a given group element yields the following bound on the variance of the sample mean:
\begin{equation}
\overline{\sigma}^2\leq \frac{N_\text{group}|\vec{c}|_2^2}{4s},
\label{groupbounduniform}
\end{equation}
where $s$ is the number of shots (so that $s/N_\text{group}$ samples are drawn from the distribution corresponding to each group element).  With optimal (non-uniform) sampling, in which $s_i = s|c_i|/|\vec{c}|_1$ samples are drawn from distribution $i$, the variance of the sample mean is bounded as
\begin{equation}
\overline{\sigma}^2\leq \frac{|\vec{c}|_1^2}{4s}.
\label{groupboundoptimal}
\end{equation}
In writing \eqref{groupbounduniform} and \eqref{groupboundoptimal}, we have used that each protocol samples from $N_\text{group}$ separate Bernoulli distributions, each of which has variance bounded above as $\sigma_i^2\leq 1/4$.  By leaving $\sigma_i$ explicit, we obtain the fine-grained versions
\begin{equation}
\overline{\sigma}^2\leq \frac{N_\text{group}}{s}\sum_{i=1}^{N_\text{group}} (c_i\sigma_i)^2, \qquad \overline{\sigma}^2\leq \frac{1}{s}\left(\sum_{i=1}^{N_\text{group}} |c_i\sigma_i|\right)^2
\end{equation}
of \eqref{groupbounduniform} and \eqref{groupboundoptimal}, respectively, which we will not need.  To interpret the variances in \eqref{groupbounduniform} and \eqref{groupboundoptimal} as pertaining to the estimation of an RB quality parameter, we must normalize by the appropriate SPAM coefficient (see Appendix \ref{app:comparison}); we leave this step implicit.

Below, we write the components $\{c_i\}$ of $\vec{c}$ as a multiset, where superscripts denote multiplicities.  In comparing protocols, we use
\begin{equation}
\text{variance of $\left\{\begin{array}{c} \text{uniform} \\ \text{optimal} \end{array}\right\}$ sampling}\lesssim \left\{\begin{array}{c} N_\text{group}|\vec{c}|_2^2 \\[2 pt] |\vec{c}|_1^2 \end{array}\right\},
\end{equation}
where we omit the dependence on sample size as well as the overall factor of $1/4$ in \eqref{groupbounduniform} and \eqref{groupboundoptimal}.

\paragraph{Clifford character group.}

For single-qubit Clifford character RB with Clifford character group, we have
\begin{equation}
N_\text{group} = 24, \qquad \{c_i\} = \left\{\frac{3}{8}, -\frac{1}{8}^9, \frac{1}{8}^6, 0^8\right\}, \qquad N_\text{group}|\vec{c}|_2^2 = 9, \qquad |\vec{c}|_1^2 = \frac{81}{16} = 5.0625.
\label{cliffordcharbounds}
\end{equation}
This protocol synthesizes the full (rank-3) projector onto the nontrivial Clifford irrep.

\paragraph{Pauli character group.}

For single-qubit Clifford character RB with Pauli character group, we have
\begin{equation}
N_\text{group} = 4, \qquad \{c_i\} = \left\{\frac{1}{4}^2, -\frac{1}{4}^2\right\}, \qquad N_\text{group}|\vec{c}|_2^2 = |\vec{c}|_1^2 = 1.
\label{paulicharbounds}
\end{equation}
This option is significantly more efficient than \eqref{cliffordcharbounds} and has the same performance as standard RB, for which $N_\text{group} = 1$ and $\{c_i\} = \{1\}$.  The downside is that it only constructs rank-1 projectors inside the nontrivial Clifford irrep.

\paragraph{$\mathbb{Z}_2$ character group.}

For single-qubit Clifford character RB with $\mathbb{Z}_2$ character group, we have
\begin{equation}
N_\text{group} = 2, \qquad \{c_i\} = \left\{\frac{1}{2}, -\frac{1}{2}\right\}, \qquad N_\text{group}|\vec{c}|_2^2 = |\vec{c}|_1^2 = 1.
\end{equation}
This scheme has the same performance as \eqref{paulicharbounds} but greater robustness to SPAM error due to constructing rank-2 projectors inside the nontrivial Clifford irrep.

\paragraph{Finite frame: Pauli superoperators.}

Going beyond character RB by using only Paulis to construct the full nontrivial irrep projector, we have
\begin{equation}
N_\text{group} = 4, \qquad \{c_i\} = \left\{\frac{3}{4}, -\frac{1}{4}^3\right\}, \qquad N_\text{group}|\vec{c}|_2^2 = 3, \qquad |\vec{c}|_1^2 = \frac{9}{4} = 2.25.
\label{ffpaulibounds}
\end{equation}
These bounds are the square roots of those from \eqref{cliffordcharbounds}.

\paragraph{Finite frame: non-Pauli Clifford superoperators.}

Using only non-Pauli Cliffords (excepting those with no diagonal entries in the lower $3\times 3$ block and that hence contribute with coefficient 0) to construct the full nontrivial irrep projector, we have
\begin{equation}
N_\text{group} = 12, \qquad \{c_i\} = \left\{\frac{1}{4}^6, -\frac{1}{4}^6\right\}, \qquad N_\text{group}|\vec{c}|_2^2 = |\vec{c}|_1^2 = 9.
\end{equation}
This is a worse option than both \eqref{cliffordcharbounds} and \eqref{ffpaulibounds}.

\paragraph{More qubits.}

Finally, we generalize \eqref{ffpaulibounds} to $n$ qubits.  Using Paulis to construct the full nontrivial irrep projector, we have
\begin{equation}
N_\text{group} = 4^n, \qquad \{c_i\} = \left\{1 - \frac{1}{4^n}, -\frac{1}{4^n}^{4^n - 1}\right\}, \qquad N_\text{group}|\vec{c}|_2^2 = 4^n - 1, \qquad |\vec{c}|_1^2 = 4\left(1 - \frac{1}{4^n}\right)^2.
\end{equation}
On the other hand, using Paulis to construct a rank-1 projector inside the nontrivial irrep (which amounts to $n$-qubit Clifford character RB with Pauli character group), we have
\begin{equation}
N_\text{group} = 4^n, \qquad \{c_i\} = \left\{\frac{1}{4^n}^{2^{2n - 1}}, -\frac{1}{4^n}^{2^{2n - 1}}\right\}, \qquad N_\text{group}|\vec{c}|_2^2 = |\vec{c}|_1^2 = 1.
\end{equation}
Hence optimal sampling for the full nontrivial irrep projector is only worse by a constant factor than optimal sampling for a rank-1 projector rather than exponentially worse (as would be the case for uniform sampling).

\subsection{From \texorpdfstring{$\text{SU}(2)$}{SU(2)} to the Heisenberg-Weyl Group} \label{app:contraction}

To describe errors meaningfully, we should choose an error basis adapted to the symmetries of the system under consideration: not all qudits are created equal.  Our scheme differs from RB based on the generalized Pauli and Clifford groups for qudits \cite{Sanders, Jafarzadeh}, which assumes that the appropriate phase space is a discrete torus rather than a sphere.

Nonetheless, one can interpolate between $\text{SU}(2)$ RB for spins and Pauli RB for qubits by going to continuous variables.  A particle on a line (or a bosonic mode) has Hilbert space $L^2(\mathbb{R})$.  The continuous analogue of the Pauli group is the Heisenberg-Weyl group of phase space translations, which has a natural representation (the Schr\"odinger representation) on $L^2(\mathbb{R})$.  The 1D irreps of the superoperator representation of the Heisenberg-Weyl group are plane waves in phase space indexed by $\vec{k}\in \mathbb{R}^2$.

The Heisenberg-Weyl group, when supplemented with rotations in phase space, is a $j\to\infty$ analogue of $\text{SU}(2)$.  To obtain this ``augmented'' single-mode Heisenberg-Weyl group (with two generators for translations and one for rotations) from $\text{SU}(2)$, we perform the group contraction at the level of the algebra $[J_i, J_j] = i\hbar\epsilon_{ijk}J_k$ or, equivalently,
\begin{equation}
[J_3, J_\pm] = \pm\hbar J_\pm, \qquad [J_+, J_-] = 2\hbar J_3,
\label{su2algebra}
\end{equation}
in a representation-dependent way \cite{Arecchi}.  Here, we make $\hbar$ explicit.

Setting $(x, p, M) = (\epsilon J_1, \epsilon J_2, J_3)$ and taking $\epsilon\to 0$ (where $\epsilon$ has dimensions of $1/\sqrt{\hbar}$) contracts the algebra \eqref{su2algebra} to that of the Euclidean group $\text{ISO}(2)$:
\begin{equation}
[x, p] = 0, \qquad [M, x] = i\hbar p, \qquad [M, p] = -i\hbar x.
\label{iso2algebra}
\end{equation}
The subalgebra of translations is $[x, p] = 0$.

On the other hand, suppose we fix $j\in \frac{1}{2}\mathbb{Z}_{\geq 0}$ as well as a ``north'' or ``south'' convention and then set
\begin{equation}
t_\pm = \frac{J_\pm}{\sqrt{2j}}, \qquad t_3 = \begin{cases} J_3 - \hbar j & \text{in the north convention}, \\ J_3 + \hbar j & \text{in the south convention}, \end{cases}
\end{equation}
where $\hbar j$ means $\hbar j\mathds{1}$.  The operator $t_3$ is chosen so that in the north convention, it annihilates the highest-weight state $|j, j\rangle$ of the spin-$j$ irrep of $\text{SU}(2)$, while in the south convention, it annihilates the lowest-weight state $|j, -j\rangle$.  In the limit $j\to\infty$, we obtain from \eqref{su2algebra} that
\begin{equation}
[t_3, t_\pm] = \pm\hbar t_\pm, \qquad [t_-, t_+] = \begin{cases} -\hbar^2 & \text{in the north convention}, \\ \hbar^2 & \text{in the south convention}. \end{cases}
\end{equation}
We may therefore make the identifications
\begin{equation}
(t_-, t_+) = \hbar\times \begin{cases} (a^\dag, a) & \text{in the north convention}, \\ (a, a^\dag) & \text{in the south convention}, \end{cases} \qquad t_3 = \hbar N,
\end{equation}
where $a$ and $a^\dag$ are the usual bosonic ladder operators satisfying
\begin{equation}
[a, a^\dag] = 1, \qquad [N, a] = -a, \qquad [N, a^\dag] = a^\dag,
\end{equation}
which are related to the quadratures as follows:
\begin{equation}
a = \frac{x + ip}{\sqrt{2\hbar}}, \qquad a^\dag = \frac{x - ip}{\sqrt{2\hbar}}, \qquad N = a^\dag a.
\end{equation}
Equivalently, setting
\begin{equation}
(t_1, t_2) = \sqrt{\frac{\hbar}{2}}\times \begin{cases} (x, p) & \text{(north)}, \\ (x, -p) & \text{(south)}, \end{cases} \qquad t_3 = \begin{cases} M & \text{(north)}, \\ -M & \text{(south)}, \end{cases}
\end{equation}
we obtain the algebra
\begin{equation}
[x, p] = i\hbar, \qquad [M, x] = i\hbar p, \qquad [M, p] = -i\hbar x,
\label{augmentedalgebra}
\end{equation}
which has a noncommutative subalgebra of translations.  As $j\to\infty$, the unitary irreps of $\text{SU}(2)$ go over to unitary irreps of the ``augmented'' Heisenberg-Weyl group with generators $x, p, M$ and algebra \eqref{augmentedalgebra}.  This is a large-spin limit of the Holstein-Primakoff transformation.

Note that the result of the group contraction depends on ``where'' the contraction is performed, in the sense of which distinguished state is chosen in the Hilbert space of the $\text{SU}(2)$ irrep.  For large $j$, in a neighborhood of the state $|j, j\rangle$ or $|j, -j\rangle$ (i.e., the north or south pole of the Bloch sphere), $\text{SU}(2)$ looks like the Heisenberg-Weyl group (as in \eqref{augmentedalgebra}).  However, when performing the group contraction around the equator of the Bloch sphere, $\text{SU}(2)$ contracts to $\text{ISO}(2)$, as in \eqref{iso2algebra} \cite{Akhtar}.

It would be fascinating to make contact with bosonic RB \cite{BosonicRB-2019, BosonicRB-2020, BosonicRB-2024-1, BosonicRB-2024-2, BosonicRB-2024-3} by taking the large-spin limit of a spin qudit, obtaining a noncompact group of phase space symmetries via contraction.

\section{Native Operations} \label{app:native}

A primary motivation for realizing nuclear spin qudits is the implementation of novel quantum error-correcting codes \cite{Gross}.  Here, we examine the native experimental operations on such systems in light of the error model for these codes.

\subsection{Qudit Control}

For notational convenience, we set $|m\rangle\equiv |j, m\rangle$ where $m = -j, \ldots, j$.  In the spin-$j$ representation of $\text{SU}(2)$, we have
\begin{align}
J_x &= \textstyle \frac{1}{2}\sum_{m=-j}^j (c_{m+1}|m + 1\rangle\langle m| + c_m|m - 1\rangle\langle m|), \\
J_y &= \textstyle \frac{1}{2}\sum_{m=-j}^j (-ic_{m+1}|m + 1\rangle\langle m| + ic_m|m - 1\rangle\langle m|), \\
J_z &= \textstyle \sum_{m=-j}^j m|m\rangle\langle m|,
\end{align}
where $c_m\equiv \sqrt{(j + m)(j - m + 1)}$.  Global $\text{SU}(2)$ rotations are parametrized by a unit vector $\vec{n}$ as $e^{-i\theta\vec{n}\cdot\vec{J}}$.

The experimentally realizable native operations are linear combinations of Pauli operators that couple neighboring levels:
\begin{align}
X_k &\equiv |j - k + 1\rangle\langle j - k| + |j - k\rangle\langle j - k + 1|, \label{nativeX} \\
Y_k &\equiv -i|j - k + 1\rangle\langle j - k| + i|j - k\rangle\langle j - k + 1|, \label{nativeY} \\
Z_k &\equiv |j - k + 1\rangle\langle j - k + 1| - |j - k\rangle\langle j - k|, \label{nativeZ}
\end{align}
where $k = 1, \ldots, 2j$.  A general control Hamiltonian takes the form
\begin{equation}
H = \sum_{k=1}^{2j} [(B_k\cos\phi_k)X_k + (B_k\sin\phi_k)Y_k + \delta_k Z_k].
\end{equation}
$H$ implements a global $\text{SU}(2)$ rotation when it can be written as a (real) linear combination of $J_x, J_y, J_z$.  We can parametrize such a linear combination as $\theta\vec{n}\cdot\vec{J}$ where $\vec{n} = (\cos\alpha\sin\beta, \sin\alpha\sin\beta, \cos\beta)$.  So $H$ implements a global rotation if and only if
\begin{equation}
B_k e^{i\phi_k} = \frac{1}{2}c_{j-k+1}\theta\sin\beta e^{i\alpha}, \qquad \delta_k = \left[kj - \frac{k(k - 1)}{2}\right]\theta\cos\beta
\label{target}
\end{equation}
for $k = 1, \ldots, 2j$ and some real parameters $\theta, \alpha, \beta$.  The first equation in \eqref{target} requires that either $B_k = \frac{1}{2}c_{j - k + 1}\theta\sin\beta$ and $\phi_k = \alpha$ or $B_k = -\frac{1}{2}c_{j - k + 1}\theta\sin\beta$ and $\phi_k = \alpha + \pi$.  This constrains the $3\cdot 2j$ real parameters down to 3.  The rotation angle $\theta$ reflects the length of time over which the Hamiltonian is applied.

The operators \eqref{nativeX}--\eqref{nativeZ} can be contrasted with those in the Gell-Mann basis for the fundamental representation of $\mathfrak{su}(d)$ ($d = 2j + 1$).  The $X$-type and $Y$-type Gell-Mann matrices are hermiticized matrix units:
\begin{equation}
X_{jk} = |j\rangle\langle k| + |k\rangle\langle j|, \qquad Y_{jk} = -i|j\rangle\langle k| + i|k\rangle\langle j|.
\end{equation}
The diagonal $Z$-type Gell-Mann matrices take the form
\begin{equation}
Z_\ell = \sum_{j=1}^\ell |j\rangle\langle j| - \ell|\ell + 1\rangle\langle\ell + 1|
\end{equation}
for $\ell = 1, \ldots, d - 1$.

\subsection{Spin Codes}

Given a logical qubit encoded in a physical system, one would like to implement logical gates from the group $\text{SU}(2)$ of single-qubit unitaries.  In looking for systems on which $\text{SU}(2)$ transformations are easy to realize, spins are natural candidates because their Hilbert spaces carry irreps of $\text{SU}(2)$.

Starting from the assumption that $\text{SU}(2)$ rotations of a single physical spin correspond to easily implementable unitaries as well as physically natural errors, Gross \cite{Gross} proposed the following procedure for constructing quantum error-correcting codes:
\begin{enumerate}
\item Choose an irrep of $\text{SU}(2)$ (i.e., a spin $j$).
\item Choose a finite subgroup $H\subset \text{SU}(2)$ (a set of ``transversal'' logical gates).
\item Decompose the spin-$j$ Hilbert space into irreps of $H$.
\end{enumerate}
If the spin-$j$ Hilbert space contains a faithful two-dimensional irrep of $H$, and if this candidate ``code space'' further satisfies the Knill-Laflamme conditions for $\text{SU}(2)$ displacement errors to first order in $J_x, J_y, J_z$, then it comprises a logical qubit on which $H$ acts as a group of logical operations.  These logical operations are ``transversal'' in the sense that they are implemented by global $\text{SU}(2)$ rotations of the entire spin, and thus completely analogous to transversal gates for qubit codes.\footnote{The analogy between spin codes and multiqubit codes becomes a precise mathematical statement in the case of permutation-invariant multiqubit codes, for which the mapping between spin states and Dicke states identifies the spin ``weight'' or ``distance'' (in the sense of spherical tensor rank) with the corresponding notions for qubits \cite{Kubischta-Teixeira}.}

The ``sporadic'' finite subgroups of $\text{SU}(2)$ correspond to (improper) symmetry groups of Platonic solids.  These are the binary tetrahedral, octahedral, and icosahedral groups ($2T, 2O, 2I\subset \text{SU}(2)$) of order 24, 48, and 120, respectively.  $2O$ is isomorphic to the single-qubit Clifford group.  $2I$ contains half of the Clifford group $2O$ because it has $2T$ as a subgroup.

The smallest spin that supports a $2O$ (transversal Clifford) code satisfying the error correction conditions is $j = 13/2$.  On the other hand, the smallest spin that supports a $2I$ (icosahedral) code satisfying the error correction conditions is $j = 7/2$.  The largest available atomic nuclei have $j = 9/2$, making the $j = 7/2$ code an ideal target.

The $j = 7/2$ code is ``perfect'' in the sense that the two-dimensional code space and its images under the errors $J_x, J_y, J_z$ are mutually orthogonal and span the entire eight-dimensional physical Hilbert space.  In the absence of a fault-tolerant syndrome extraction protocol (a four-outcome measurement that distinguishes the four error spaces), the current experimental prospect is to use the $j = 7/2$ code as an error-detecting code in combination with postselection.  Note that the authors of \cite{Ardavan} decode the error spaces for this code into non-overlapping angular momentum eigenstates.  However, devising a scheme for measuring the syndrome fault-tolerantly remains an open problem.

\subsection{Icosahedral Spin Code}

A possible concern is that the native experimental operations realize gates from $\text{SU}(2j + 1)$ that are not global $\text{SU}(2)$ rotations and therefore violate the assumptions of Gross's error model. (They are, however, universal for $\text{SU}(2j + 1)$, as an arbitrary unitary can be compiled from a sequence of Givens rotations between adjacent levels \cite{Cybenko, Brennen}.) To determine whether spin codes (which are designed to protect against small random $\text{SU}(2)$ rotations) are still effective in this setting, we specialize to the $j = 7/2$ icosahedral spin code and examine how the recovery channel for this code fares against control errors.

The normalized codewords are
\begin{equation}
|\overline{0}\rangle = \sqrt{\frac{3}{10}}\left|\frac{7}{2}\right\rangle + \sqrt{\frac{7}{10}}\left|-\frac{3}{2}\right\rangle, \qquad |\overline{1}\rangle = \sqrt{\frac{7}{10}}\left|\frac{3}{2}\right\rangle - \sqrt{\frac{3}{10}}\left|-\frac{7}{2}\right\rangle.
\end{equation}
The error set is $\{\mathds{1}, J_x, J_y, J_z\}$ with $\vec{J}^2 = j(j + 1) = 63/4$.  The eight states $\{|\overline{i}\rangle\}\cup \{J_w|\overline{i}\rangle\}$ are nonzero and mutually orthogonal \cite{Ardavan}, which implies that the error correction conditions are satisfied:
\begin{equation}
\langle\overline{i}|\overline{j}\rangle = \delta_{ij}, \qquad \langle\overline{i}|J_w|\overline{j}\rangle = 0, \qquad \langle\overline{i}|J_w J_{w'}|\overline{j}\rangle = \frac{21}{4}\delta_{ij}\delta_{ww'},
\end{equation}
where $i, j\in \{0, 1\}$ and $w, w'\in \{x, y, z\}$.  The recovery map $\mathcal{R}$ is a CPTP map on the eight-dimensional physical Hilbert space that reverses any correctable error $\mathcal{N}$ applied to a logical state $\overline{\rho} = \sum_{i, j} \overline{\rho}_{ij}|\overline{i}\rangle\langle\overline{j}|$ (with $\overline{\rho}_{00} + \overline{\rho}_{11} = 1$, $\overline{\rho}_{ij} = \overline{\rho}_{ji}^\ast$, and $\overline{\rho}_{00}\overline{\rho}_{11} - |\overline{\rho}_{01}|^2\geq 0$):
\begin{equation}
\mathcal{R}\circ \mathcal{N}(\overline{\rho}) = \overline{\rho}.
\end{equation}
It admits the Kraus decomposition
\begin{equation}
\mathcal{R}(\rho) = \sum_w R_w\rho R_w^\dag + R_\perp\rho R_\perp, \qquad R_w\equiv \frac{2}{\sqrt{21}}\sum_i |\overline{i}\rangle\langle\overline{i}|J_w, \qquad R_\perp\equiv \mathds{1} - \sum_w R_w^\dag R_w,
\end{equation}
where $\sum_w R_w^\dag R_w = \frac{4}{21}\sum_{w, i} J_w|\overline{i}\rangle\langle\overline{i}|J_w$ is the projector onto the space of states that can be reached by errors acting on codewords and $R_\perp$ is the projector onto the orthogonal complement ($R_\perp^\dag = R_\perp^2 = R_\perp$).

The recovery channel can invert any noise channel whose Kraus operators are linear combinations of correctable errors:
\begin{equation}
\mathcal{N}(\rho) = \sum_i N_i\rho N_i^\dag, \qquad N_i = c_i\mathds{1} + c_{x, i}J_x + c_{y, i}J_y + c_{z, i}J_z.
\label{generalnoise}
\end{equation}
Indeed, ignoring the normalization condition $\sum_i N_i^\dag N_i = \mathds{1}$ (which imposes some constraints on the coefficients $c_i$ and $c_{w, i}$), we compute that
\begin{equation}
\mathcal{R}\circ \mathcal{N}(\overline{\rho}) = \left(|\vec{c}|^2 + \frac{j(j + 1)}{3}\sum_w |\vec{c}_w|^2\right)\overline{\rho}\propto \overline{\rho}.
\end{equation}
For example, a unitary $\text{SU}(2)$ rotation error
\begin{equation}
\mathcal{U}_\epsilon(\rho)\equiv e^{-i\epsilon\vec{n}\cdot\vec{J}}\rho e^{i\epsilon\vec{n}\cdot\vec{J}} = \rho + i\epsilon[\rho, \vec{n}\cdot\vec{J}] + O(\epsilon^2)
\end{equation}
can be written in the form \eqref{generalnoise} up to $O(\epsilon^2)$ terms, which justifies why the recovery channel can correct such errors to first order in $\epsilon$: $\mathcal{R}\circ \mathcal{U}_\epsilon(\overline{\rho}) = \overline{\rho} + O(\epsilon^2)$.

On the other hand, we can ask how the recovery channel performs with respect to the most general noise channel for \emph{control} errors.  Setting
\begin{equation}
N' = c\mathds{1} + \sum_{k=1}^7 (\epsilon_k^X X_k + \epsilon_k^Y Y_k + \epsilon_k^Z Z_k)
\label{generalcontrol}
\end{equation}
as a representative Kraus operator, we compute that
\begin{equation}
\mathcal{R}(N'\overline{\rho}N'^\dag) = |c|^2\overline{\rho} + \sum_{i, j} \epsilon_{ij}\overline{\rho}_{ij}|\overline{i}\rangle\langle\overline{j}| + O(\epsilon^2)
\label{partiallyrecovered1}
\end{equation}
where
\begin{equation}
\left[\begin{array}{cc} \epsilon_{00} & \epsilon_{01} \\ \epsilon_{10} & \epsilon_{11} \end{array}\right]\equiv \left[\begin{array}{cc} \epsilon_0 + \epsilon_0^\ast & \epsilon_0 + \epsilon_1^\ast \\ \epsilon_1 + \epsilon_0^\ast & \epsilon_1 + \epsilon_1^\ast \end{array}\right], \qquad \left[\begin{array}{c} \epsilon_0 \\ \epsilon_1 \end{array}\right]\equiv c^\ast\left[\begin{array}{c} 3\epsilon_{1}^Z - 7\epsilon_{5}^Z + 7\epsilon_{6}^Z \\ -7\epsilon_{2}^Z + 7\epsilon_{3}^Z - 3\epsilon_{7}^Z \end{array}\right].
\label{errorcombos}
\end{equation}
Interestingly, the terms in \eqref{partiallyrecovered1} that are linear in $\epsilon$ depend only on $Z$ control errors.  As an example, consider the unitary error channel
\begin{equation}
\mathcal{U}_H(\rho)\equiv e^{-iH}\rho e^{iH}, \qquad H = \sum_{k=1}^7 (\epsilon_k^X X_k + \epsilon_k^Y Y_k + \epsilon_k^Z Z_k)
\end{equation}
acting on a logical state.  In this case, the recovery channel acts as follows:
\begin{equation}
\mathcal{R}\circ \mathcal{U}_H(\overline{\rho}) = \overline{\rho} - \frac{i}{10}(3\epsilon_1^Z + 7\epsilon_2^Z - 7\epsilon_3^Z - 7\epsilon_5^Z + 7\epsilon_6^Z + 3\epsilon_7^Z)(\overline{\rho}_{01}|\overline{0}\rangle\langle\overline{1}| - \overline{\rho}_{10}|\overline{1}\rangle\langle\overline{0}|) + O(\epsilon^2).
\label{partiallyrecovered2}
\end{equation}
(Here, the first-order terms depend only on the off-diagonal components of the logical state $\overline{\rho}$, and therefore vanish if $\overline{\rho}$ is a classical mixture of the pure states $|\overline{0}\rangle\langle\overline{0}|$ and $|\overline{1}\rangle\langle\overline{1}|$.) In both the general case \eqref{partiallyrecovered1}--\eqref{errorcombos} and the special case \eqref{partiallyrecovered2}, the first-order error in the post-recovery state depends only on specific combinations of the $Z$ control errors $\epsilon_{k\neq 4}^Z$ that vanish if the total $Z$ error is a global $Z$-rotation:
\begin{equation}
\sum_{k=1}^7 \epsilon_k^Z Z_k\propto J_z.
\end{equation}
In particular, the code can correct all $X$ and $Y$ control errors $\epsilon_k^X$ and $\epsilon_k^Y$ to first order.

Therefore, we find that the spin-$7/2$ code performs better than expected.  It was designed to correct $\text{SU}(2)$ rotation errors to first order in $J_x, J_y, J_z$---or, by linearity, all errors expressible as linear combinations of $\mathds{1}, J_x, J_y, J_z$. (Note that being able to correct $\text{SU}(2)$ rotations to \emph{arbitrary} order is tantamount to being able to correct arbitrary errors, as any $(2j + 1)\times (2j + 1)$ complex matrix can be written as a linear combination of images of global $\text{SU}(2)$ rotations under the spin-$j$ irrep.) However, the code also exhibits robustness against small $X$ and $Y$ control errors that cannot be written in this form.

A partial explanation may be found by asking whether certain control errors happen to satisfy the Knill-Laflamme conditions for this code.  We compute that
\begin{equation}
\langle\overline{i}|X_k|\overline{j}\rangle = \langle\overline{i}|Y_k|\overline{j}\rangle = 0
\end{equation}
for all $i, j$ as well as $\langle\overline{0}|Z_k|\overline{1}\rangle = \langle\overline{1}|Z_k|\overline{0}\rangle = 0$, whereas
\begin{align}
\langle\overline{0}|(Z_1, \ldots, Z_7)|\overline{0}\rangle &= \left(\frac{3}{10}, 0, 0, 0, -\frac{7}{10}, \frac{7}{10}, 0\right), \\
\langle\overline{1}|(Z_1, \ldots, Z_7)|\overline{1}\rangle &= \left(0, -\frac{7}{10}, \frac{7}{10}, 0, 0, 0, -\frac{3}{10}\right).
\end{align}
The fact that the $X_k$ and $Y_k$ satisfy the first-order Knill-Laflamme conditions, while not all of the $Z_k$ do, is consistent with the fact that the recovery channel can correct errors of the form \eqref{generalcontrol} to first order in $\epsilon_k^X$ and $\epsilon_k^Y$.  On the other hand, the matrix elements $\langle\overline{i}|P_k P_\ell'|\overline{j}\rangle$ are generally nonvanishing for $P, P'\in \{X, Y, Z\}$.  Therefore, control errors violate the second-order Knill-Laflamme conditions.

\end{document}